\documentclass[useAMS,usenatbib]{mn2e}
\pdfoutput=1
\usepackage[T1]{fontenc}
\usepackage{aecompl}
\usepackage[pdftex]{graphicx}
\usepackage{amsmath}
\usepackage{mathtools}
\usepackage[utf8]{inputenc}
\usepackage{pdfpages}
\usepackage{aas_macros}
\usepackage[toc,page]{appendix}
\usepackage{graphicx}
\usepackage{epstopdf}
\usepackage{subfig}

\newcommand{\ignore}[1]{}
\usepackage{color}
\long\def\symbolfootnote[#1]#2{\begingroup%
\def\thefootnote{\fnsymbol{footnote}}\footnote[#1]{#2}\endgroup}

\title[The 3D structure of the Galactic Bulge]
{A parametric description of the 3D structure of the Galactic Bar/Bulge
using the VVV survey}

\author[I.T. Simion et al.]
  {I. T.~Simion$^{1,2}$\thanks{email: isimion@ast.cam.ac.uk},
  V.~Belokurov$^1$, M.~Irwin$^1$, S. E.~Koposov$^{1,3}$,
  \newauthor
   C.~Gonzalez-Fernandez$^1$, A. C.~Robin$^4$, J.~Shen$^{2, 5}$,  Z.-Y. ~Li$^{2, 5}$ \\ 
$^{1}$Institute of Astronomy, Madingley Road, Cambridge, CB3 0HA, UK\\
$^{2}$Key Laboratory for Research in Galaxies and Cosmology, Shanghai Astronomical Observatory, 80 Nandan Road, Shanghai 200030, \\
China\\
$^{3}$Carnegie Mellon University, 5000 Forbes Avenue, Pittsburgh \\
 $^{4}$Institut UTINAM, CNRS UMR6213, Université de Bourgogne-Franche-Comté, OSU THETA, Observatoire de Besançon, \\
Besançon, France\\
$^{5}$School of  Astronomy and Space Science, University of Chinese Academy of Sciences, 19A Yuquan Road, Beijing 100049, China
}

\date{Accepted 2017 July 18; in original form 2017 January 27}

\pagerange{\pageref{firstpage}--\pageref{lastpage}} \pubyear{2017}

\def\LaTeX{L\kern-.36em\raise.3ex\hbox{a}\kern-.15em
    T\kern-.1667em\lower.7ex\hbox{E}\kern-.125emX}

\begin{document}

\label{firstpage}

\maketitle

\begin{abstract} We study the structure of the inner Milky Way (MW) using
the latest data release of the Vista Variables in Via Lactea (VVV) survey. The
VVV is a deep near-infrared, multi-colour photometric survey with a
coverage of 300 square degrees towards the Bulge/Bar. We use Red Clump
(RC) stars to produce a high-resolution dust map of the VVV's field of
view. From de-reddened colour-magnitude diagrams we select Red Giant
Branch stars to investigate their 3D density distribution within the central 4 kpc. We demonstrate that our best-fit
parametric model of the Bulge density provides a good description of the VVV data, with a median percentage residual of 5 \% over the fitted region. The strongest of the otherwise low-level residuals
are overdensities associated with a low-latitude structure as well
as the so-called X-shape previously identified using the split RC. These
additional components contribute only \textit{$\sim5\%$} and \textit{$\sim7\%$} respectively to the Bulge 
mass budget. The best-fit Bulge is ``boxy'' with an axis ratio of
[1:0.44:0.31] and is rotated with respect to the Sun-Galactic Centre
line by at least $\sim20^{\circ}$. We provide an estimate of the total, full sky, mass of the Bulge of $M_\mathrm{Bulge}^{\mathrm{Chabrier}} = 2.36 \times 10^{10} M_{\odot}$ for a Chabrier initial mass function. 
We show there exists a strong degeneracy between the viewing angle and the dispersion of the Red Clump (RC) absolute magnitude distribution. 
The value of the latter is strongly dependent on the assumptions made about the intrinsic luminosity function of the Bulge.

\end{abstract}

\section{Introduction}

Integrated light, star counts, gas kinematics and microlensing studies
over the past 20 years have provided plenty of evidence that the heart
of our Galaxy harbors a triaxial entity that is distinct from the disc
\citep[see e.g.][]{Blitz1991,Binney1991,Dwek95,Bi97,St97,Fr98,Alcock2000,
Bi02,Evans2002,Babu2005,Mc10,Na10}.
Dominating
the dynamics in the inner regions of the Milky Way, this centrally
concentrated and elongated structure resembles both a bar and a bulge,
although not in the classical sense for the latter.
\citep[see][for details]{Ko04}. However, as the
intention of this paper is not to re-enact the decades old battle between the
Bar and the Bulge, we will refer to the central 4 kpc
region of the Galaxy as the ``Bulge'', as have many authors before
us \citep[see e.g.][]{Zoca2016}.

Near Infrared (NIR) bands, where the extinction is only around 10\% of
that in the optical, are ideal for studying the dust-obscured Galactic
center. Therefore, infrared maps from surveys such as COBE/DIRBE have
long been used to construct 3D models of the Bulge: \citet{Dwek95} and
\citet{Fr98} have fitted parametric models while \citet{Bi97} and
\citet{Bi02} have obtained non-parametric models using the same
data. The 2MASS \citep[][]{Skru2006} catalogue added depth and resolution 
to the picture of the inner Milky Way and helped to provide tight 
constraints on models of the Galactic centre \citep{Ro12}. More
recently the VISTA Variables in the Via Lactea (VVV) project
\citep{Mi10} has staked a claime as the deepest wide-coverage NIR
survey aimed at delivering an accurate 3D map of the Galactic
Bulge. The VVV survey has already been used to study the split Red Clump
(RC) in the Bulge \citep[][]{Saito2011}, create a high resolution
extinction map of the region \citep[][]{Gonz2012}, discover new
Galactic star clusters \citep[][]{Bori2011} and, most relevant to this
study, probe the structure of the Galactic Bar \citep[][]{We13}.

The work of \citet{We13} is one of the latest in the long line of
Red Giant Branch density modelling exercises that have helped to
reveal the structural properties of the Bulge region
\citep[e.g.][]{Stanek1994,St97,Nikolaev1997,Unavane1998,Lopez2000,Ben2005,Nishi2005,Ratten2007}.
Many of the works above relied in particular on the Red Clump population to
disentangle the Bulge from the disc, taking advantage of the well-defined
location of the RC in Color-Magnitude Diagrams and its superb
performance as a standard candle \citep[][]{Pa1998}. Importantly, the
RC star counts analysis has so far produced the most consistent range
of values for the viewing angle of the ellipsoid in the Bulge: between
$20^{\circ}$ and $30^{\circ}$, while the same quantity estimated with
other tracers appears to have a much larger uncertainty, with values
as low as $10^{\circ}$ and as high as $50^{\circ}$ quoted 
\cite[see e.g.][for a summary]{Va09}. Of course it is quite likely that
this apparent dramatic variation in the basic property of the
Bar/Bulge is not solely the sign of inconsistency between the
methods but also an indication that the angle does change
depending on the tracer used and the range of Galactic $l$ and $b$
explored. For example, there exists ample evidence for a long bar,
clearly seen outside the Bulge at larger $|l|$, with a viewing angle in the range $28^{\circ}$ - $45^{\circ}$ \citep[for a recent study see][and references therein]{We15}. 
  Similarly, the structure traced with the RR Lyrae
might have very little triaxiality, if any at all, thus most likely
representing the actual (classical) old Bulge of the Milky Way
\citep[see][]{Dekany2013, Pi15, Ku16}. 

In this study we use the most recent data release of the VVV and
strive to find the most appropriate analytic function which best
describes the full 3D Bulge density distribution. Therefore, our adopted
approach is different to that employed by e.g. \citet{We13} who
applied instead a symmetrised non-parametric modelling procedure to an
earlier version of the same VVV data. There are clearly advantages and 
disadvantages
associated with both parametric and non-parametric methods. On the one
hand, describing the stellar distribution in the Bulge with an
analytic expression clearly gives a more portable solution which
can be straightforwardly used in subsequent dynamical modelling of
the influences of the Bar/Bulge. On the other hand, the actual density
field in the central Galaxy might not be of a simple triaxial
shape. This of course does indeed take place in the Milky Way and is
examplified by the recently discovered cross-like component of the
Bulge as manifested by the split RC \citep[see e.g.][]{Mc10,Na10}.

The paper is organised as follows: the VVV data used is described in 
Section~\ref{sec:data}; the procedure used to build a high resolution 
extinction map is detailed in Section~\ref{sec:extinction}; 
Section~\ref{sec:model} contains a description of the parametric approach 
used to investigate the Bulge stellar density law using a mock catalogue; 
in Section~\ref{sec:fitting} we outline the fitting procedure; in 
Section~\ref{sec:datafit} we explain the models used and provide 
the best fit results. Finally, in Sections~\ref{sec:discussion} 
and~\ref{sec:conclusions} we discuss our results and draw the conclusions.
\section{The VVV survey}
\label{sec:data}

In 2016 the VISTA Variables in the Via Lactea (VVV) survey finished
the 6 year long observing campaign and although many publications 
have already stemmed from it, it remains a largely unexplored resource.
The VVV survey contains a stellar sample large enough to examine the
structural properties of the Bulge in unprecedented detail, in the $J,
H, K_{s}$ NIR bands and reaches up to four magnitudes deeper than
2MASS. The depth of the survey allows us to select the Red Giant branch 
population visible across the whole survey footprint, apart from some 
highly reddened regions close to the Galactic plane, and thus build a 3D 
parametric model to describe their density distribution towards the Bulge. 

The VVV survey covers $\sim$315 square degrees of the inner Milky Way,
i.e. a region with Galactic longitude $-10^{\circ} < l < +10^{\circ}$
and latitude $-10^{\circ} < b < 5^{\circ}$, as well as 220 square
degrees of the Galactic Plane, spanning between $295^{\circ} < l <
350^{\circ}$ and $-2^{\circ} < b < 2^{\circ}$. In this work, we
exclusively use the Galactic Bulge part of the survey. VISTA data are
processed through the VISTA Data Flow System (VDFS) by the Cambridge
Astronomical Survey Unit (CASU\footnote{http://casu.ast.cam.ac.uk}).

The photometric and astrometric calibration of the VVV survey is performed 
relative to the 2MASS point source catalogue (PSC) ensuring that the
the calibration stars are measured at the same time as the target sources (see e.g. Gonzalez-Fernandez et al. 2017, in preparation). 
For the $J$, $H$ and $K_{s}$ passbands employed here typical overall calibration errors are around 1--2\%. However, the accuracy of these calibrations are field dependent and can be more uncertain in very crowded fields with high levels of spatially varying extinction, where the derived zero points can vary from field to field by up to 0.1 magnitudes. We therefore performed a full global quality check using overlap regions for every tile to faciliate correcting tiles with offset zero points and/or unusual colour-magnitude diagrams. 
\section{Extinction map}
\label{sec:extinction}

\begin{figure}
\centering
\includegraphics[scale = 0.195]{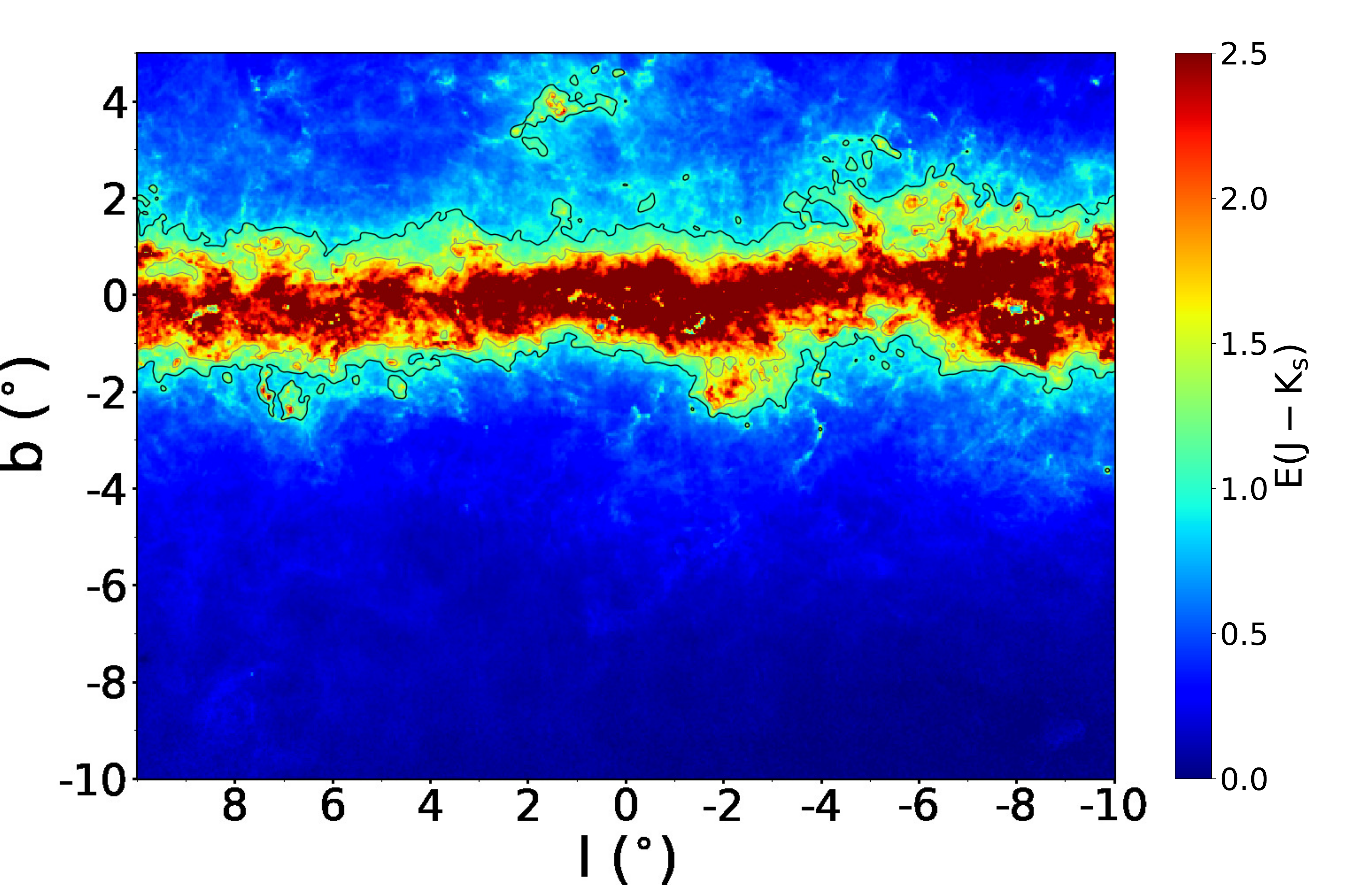}
\caption{Reddening map in the Bulge area covered by the VVV survey, sampled at
1$'$ x 1$'$  with an effective resolution of 2$'$. Overlaid on the map are contours 
of constant reddening $E(J'-K'_{s}) = 1.5$ (thin gray line) and 
$E(J'-K'_{s}) = 1.0$ (thick gray line). The region inside the 
$E(J'-K'_{s}) = 1.0$ contour, where the high reddening can cause incompleteness
issues, is omitted in the VVV data 
modelling procedure explained in Section~\ref{sec:fitting}.}
 \label{ext}
\end{figure}

The biggest challenge to building an accurate map of the Bulge stellar
density arises from the rapid extinction variations across different
lines of sight. Following a similar approach to the one described in
\citet{Go11}, we use the RC giants in the VVV to build a high resolution
(1$'$ x 1$'$ sampling) reddening map (see Figure~\ref{ext}) by comparing the
mean colour of the RC stars observed in each field with that of the RC
population in Baade's window, $(J'-K'_{s})_{\mathrm{BW}} = 0.89$ 
(throughout the paper we denote with $J'$ and $K'_{s}$ the observed VVV
magnitudes and with $J$ and $K_{s}$ the extinction-corrected
magnitudes), under the assumption that the intrinsic colour of the RC
remains, on average, the same across the Bulge i.e. $J-K_{s} = 0.62$.
We adopt the \citet{Ni09} interstellar extinction law derived for the 
central Bulge region since it was shown to reproduce most accurately the 
extinction of RC stars in the VVV \citep{Go12}.

To build the extinction map we combine the 2MASS-VVV colour transformations 
$J' = J'_{\mathrm{2M}} - 0.065(J'-K'_{s})_{\mathrm{2M}}$ and $K'_{s} = K'_{s\mathrm{2M}} + 0.010(J'-K'_{s})_{\mathrm{2M}}$
and obtain the \citet{Ni09} law for VVV photometry: $A_{J} = 1.351E(J'-K'_{s})$
and $A_{Ks} = 0.482E(J'-K'_{s})$. We then calculate the reddening 
$ E(J' - K'_{s})_{\mathrm{RC}}$ by comparing the observed RC colour in each field 
with its known extinction-free value in Baade's window, $J-K_{s} = 0.62$.
We are able to use this technique in all Bulge fields covered by the survey 
as VVV photometry reaches the RC even in fields close to the Galactic 
Plane. Correction for extinction then proceeds on a star by star basis
interpolating between the nearest four extinction map values.

The problem with using two dimensional maps such as ours, is the
assumption that the extinction arises mainly from foreground material
while in reality the sources are intermixed with dust at a range of distances,
rendering the use of 3D reddening maps necessary especially at low latitudes
($|b| < 3^{\circ}$). 
To compare the accuracy of 2D versus 3D dust maps, \citet{Sc14b} used a large 
spectroscopic sample of stars from the Apache Point Observatory Galactic 
Evolution Experiment (APOGEE) towards the MW Bulge, to conclude that for 
sources at distances larger than $\sim$ 4 kpc from the Sun, 2D maps can be 
applied without significant systematic offset. Moreover, it was noticed that 
\textit{none} of the existing 3D maps agrees with independent extinction 
calculations based the APOGEE spectra towards the MW Bulge. Their study 
suggests that there is still room for improvement in the construction of
3D dust maps and that they do not outperform 2D maps for distances
larger than 4 kpc from the Sun.

\section{Model ingredients}
\label{sec:model}
 
In this Section we describe the main ingredients of our model which consists of
synthetic thin and thick discs and a purely analytic Bulge. The model takes 
into account the disc contribution to the observed luminosity function as a 
function of field of view and magnitude and therefore has an advantage over 
recent VVV studies which either fitted a polynomial background to the discs 
and RGB (from both Bulge and discs) populations \citep{We13} or assumed a 
uniform disc contamination throughout the Bulge area \citep{Va16}.

The disc populations are built with \textit{Galaxia}\footnote{http://galaxia.sourceforge.net} \citep{Sh11}, a C++
code able to generate synthetic surveys of the Milky Way, implementing
the \textit{Besancon}\footnote{http://model.obs-besancon.fr/} Galaxy model \citep{Ro03}.
\subsection{Isochrones}

We add to \textit{Galaxia} a recent set of
PARSEC isochrones in the VISTA photometric system
\citep{Br12}, built using the default parameters of the online
\textit{Padova} CMD interface
(we used PARSEC v1.2S + COLIBRI PR16 \footnote{http://stev.oapd.inaf.it/cgi-bin/cmd}, described in \citealt{Marigo2017}). For the table of isochrones we choose 177 age bins and 39 metallicity bins,
within the parameters range used in Galaxia for a
previous PARSEC release \citep{Ma08}. 

VVV zero point measurements (section 2.2 in \citealt{Rubele2012}) found a 
constant offset between the VISTA system calibration and the Vegamag system 
of the stellar models of 0.021 mag and 0.001 mag for the $J$ and $K_{s}$ bands respectively. This offset was implemented in the PARSEC version we use.
\subsection{The discs}
\begin{figure}
\centering
\includegraphics[scale = 0.195]{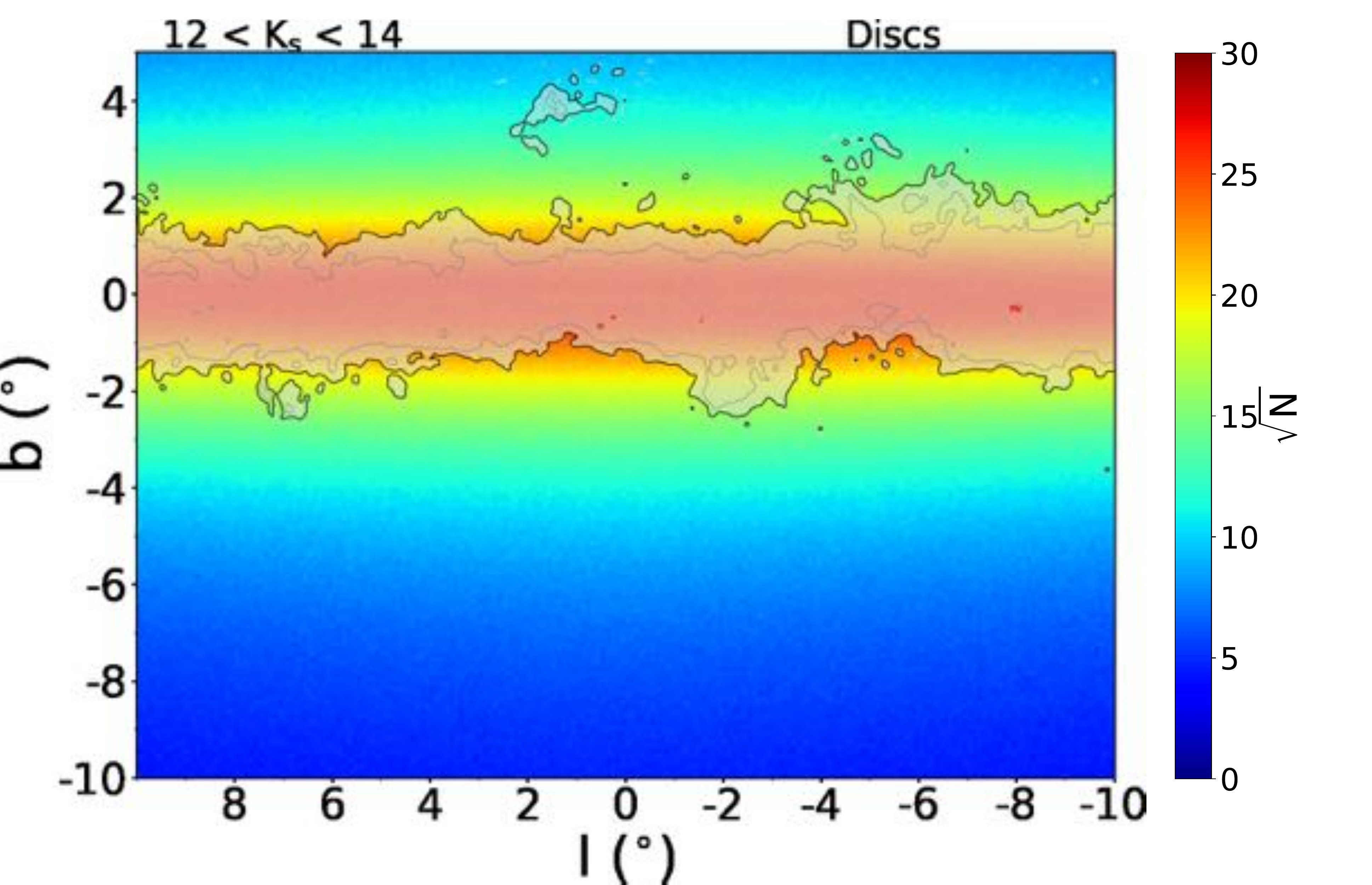}
\caption{Number density distribution of mock thin and
  thick disc stars generated with \textit{Galaxia}, in
  Galactic coordinates. The contours delineate the region with high reddening (see Figure~\ref{ext}) that will be excluded from the fitting procedure
  described in Sections~\ref{sec:fitting} and~\ref{sec:datafit}. The integrated number of stars outside the masked area is equal to the number of disc stars in the VVV found in this work ($=N_{1}^{-}+N_{2}^{-}$ for model $E$ in Table \ref{bestresults}). This figure is obtained averaging several mock samples to minimise shot noise.}
 \label{discs}
 \end{figure} 

The thin and thick disc populations are an important component of the
central regions; they are implemented by averaging several repeated
mock catalogue realisations with \textit{Galaxia} to minimise sampling
noise. The disc models are shown in Figure~\ref{discs}, in Galactic
coordinates - these are mock stars within the same range of extinction-corrected
magnitudes and colours, $12 < K_{s} < 14$, $0.4 < J - K_{s} < 1.0$, 
used for the selected VVV sample (see Section~\ref{sec:fitting}). Overlaid we 
mark two contours of constant reddening $E(J'-K'_{s}) = 1.5$ (light gray) and 
$E(J'-K'_{s}) = 1.0$ (thick gray), as in Figure~\ref{ext}. The initial mass 
function (IMF), star formation rate (SFR), ages, metallicity and density laws 
of the discs implemented in $Galaxia$ are described in tables 1, 2 and 3 in 
\citet{Ro03}. The thin and thick discs are assumed to have a warp and a flare 
that is modelled following the prescription of \citet{Ro03}.\\
\subsection{The Bulge luminosity function (LF)}
\subsubsection{Analytic description of the Bulge LF}
We assume a 10 Gyrs old \citep{Br97} Bulge population with solar metallicity 
[Fe/H]$ = 0$ dex \citep[e.g.][]{Zo03} and a metallicity dispersion of 
$\sigma_{[Fe/H]} = 0.4$ \citep{Zo08} with no metallicity gradients (default 
values of the Besancon Galaxy Model). 

\begin{figure*}
\hspace{-0.8cm}
\includegraphics[width=110mm]{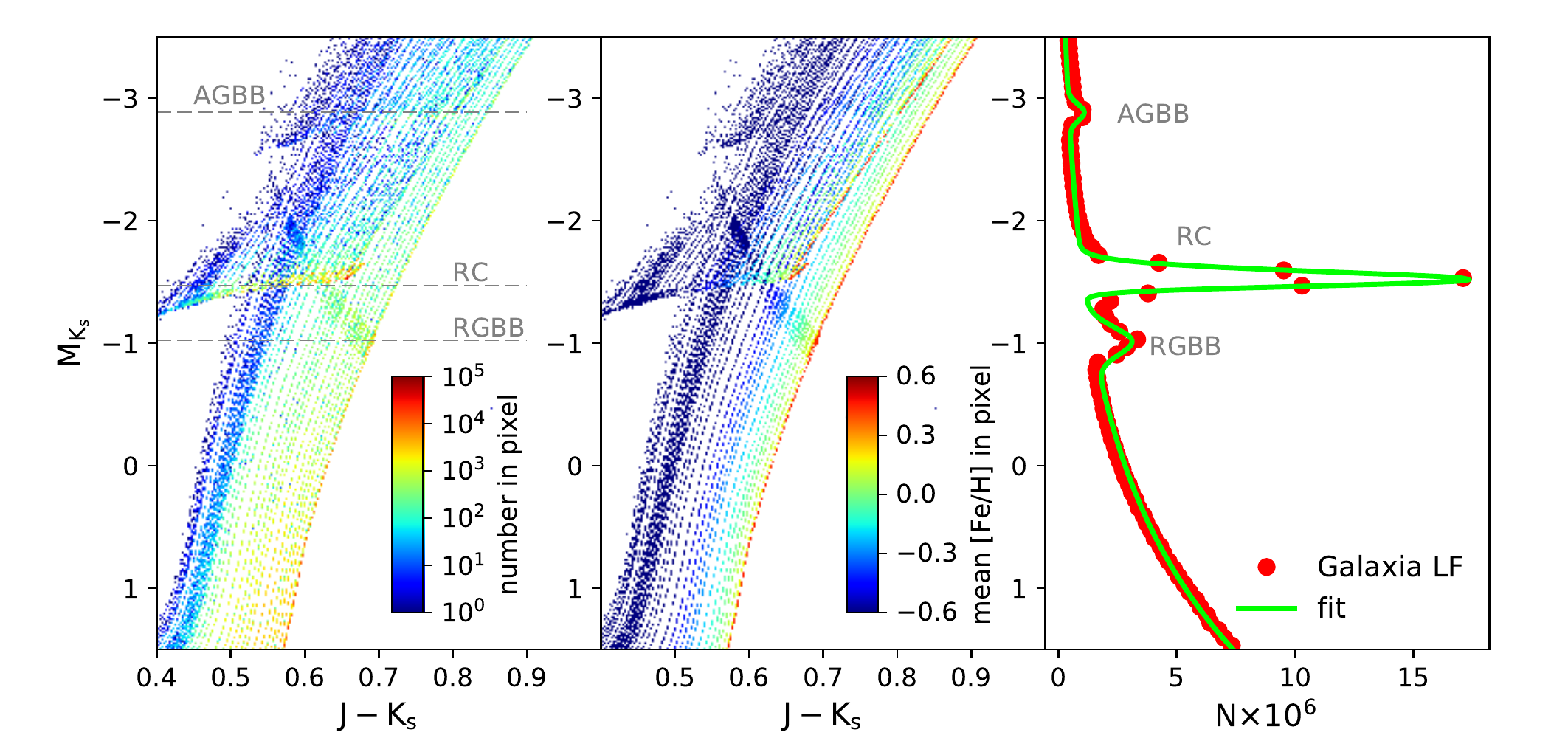}
\includegraphics[width=60mm]{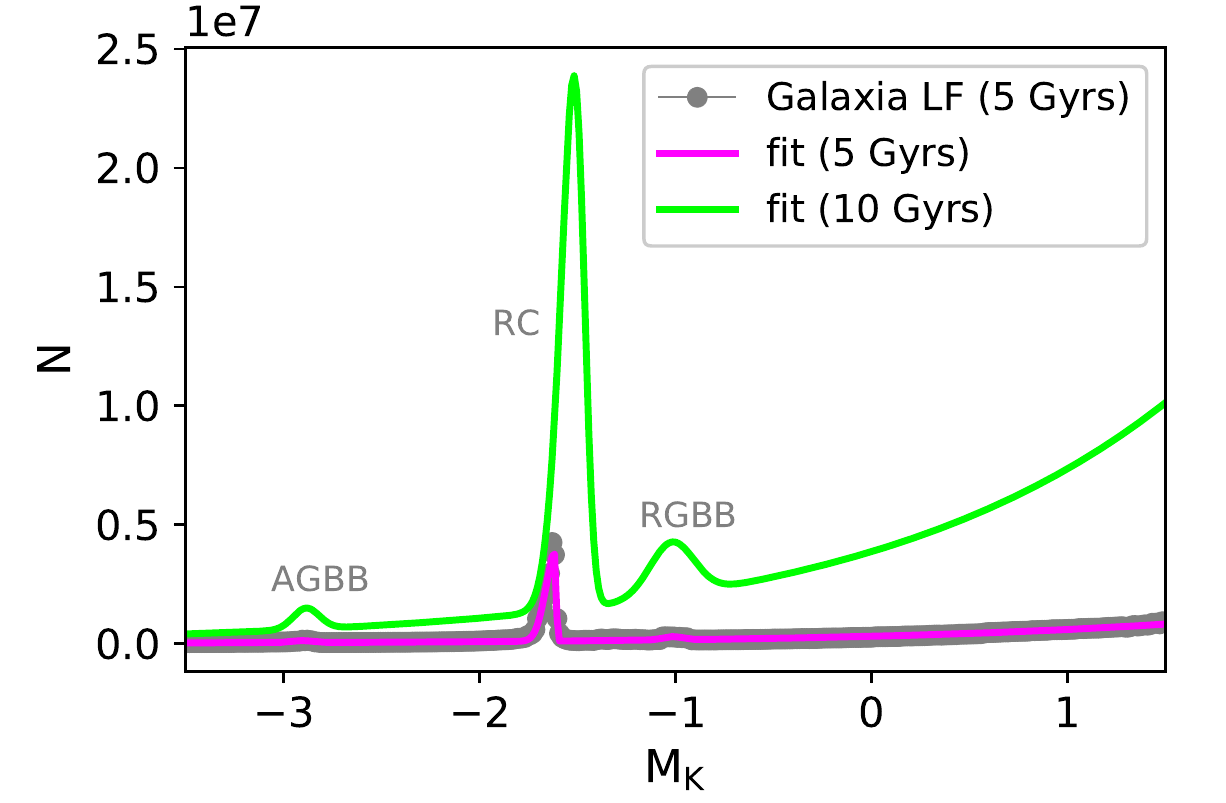}
\caption[Synthetic Luminosity Function for the Bulge
  population]{$First$ $panel$: Number density distribution of \textit{Galaxia} generated stars using PARSEC isochrones in a colour magnitude diagram with [Fe/H] = 0 dex, $\sigma_{[Fe/H]} = 0.4$ and age $=10$ Gyrs. $Second$ $panel:$ mean metallicity in each of the pixels in the first panel.  $Third$ $panel$: \textit{Galaxia} synthetic LF (red) and model (green) with 3 Gaussians and an exponential, following the \citet{Na15} description (the parameters are given in the top row of Table \ref{LFparams}). $Fourth$ $panel$: Model LF for a 5 Gyrs old (purple) and a 10 Gyrs old (green) population. The RC of 5 Gyrs old population is 0.1 mag brighter than the RC of the 10 Gyrs old population. The integrated area under each curve gives the total number of stars we found for each component in the results section (see the $N_{3}$ values in Table \ref{bestresults} for components $S$ and $E$, in model $S+E$)}.
 \label{LF}
\end{figure*}

With these specifications, we simulate 
the LF with \textit{Galaxia} using a Chabrier IMF, within the 
colour range $0.4 < J-K_{s} < 1.0$, adopted to reduce contamination from 
non-giants in the VVV data. The Chabrier IMF is chosen over the $Galaxia$ default Salpeter IMF as it was shown to describe more accurately
observations and model predictions in the Galactic Bulge (see \citealt{Ch03} and \citealt{Po15} respectively). The resulting synthetic LF, $\phi_{B}(M_{K};$ 10 Gyrs), is shown in the third panel of Figure $\ref{LF}$ in red. 
We model it following the description in section 3.1 of \citet{Na15},
with three Gaussians to represent the Asymptotic Giant Branch Bump (AGBB), the RC and the Red Giant Branch Bump (RGBB) 
and an exponential function for the Red Giant Branch (RGB) stars. The RC was modelled with either a Gaussian or a skew-Gaussian
distribution, with the latter providing a slightly better fit; the addition of the skewness parameter affects only the $\mu^{\mathrm{RC}}$ and $\sigma^{\mathrm{RC}}$ values,
 as expected (see the results in Table \ref{LFparams}). The only constraint was on the mean RGBB magnitude, set to $\mu^{\mathrm{RGBB}} = -1.02$, where the peak of the LF  lies in the RGBB region. The fitted LF, with a skew-Gaussian model for the RC, is shown in green in the third panel of Figure $\ref{LF}$,
overlapping the histogrammed LF simulated with \textit{Galaxia} (in red). We find $\mu^{\mathrm{RC}} = -1.53$, consistent with previous determinations of the mean RC magnitude in the K band (see table 1 in the RC review paper by \citealt{Gi16}).
 
The first panel of the Figure shows the 
number density distribution of the Galaxia generated stars in a colour magnitude diagram with the specifications mentioned above, and the second panel the mean metallicity in each of those pixels. Notice the theoretical models predict the AGBB, RC and RGBB mean magnitudes depend on metallicity; this variation is responsible for the magnitude dispersions in the mock LF. The dependence of the RC magnitude on the age and metallicity of the population has also been recorded empirically, e.g. the globular cluster 47 Tuc (age $=11$ Gyrs, [Fe/H] $= $-0.7 dex) has $\mu_{K}^{\mathrm{RC}}=-1.28$ mag while the much younger open cluster NGC 2204 (age = $1.7$ Gyrs, [Fe/H] $=$ -0.38 dex) has $\mu_{K}^{\mathrm{RC}}=-1.67$ (see table 1 in \citealt{Pe03} and top panel of figure 6 in\citealt{Gi16} which shows the RC in open clusters as a function of age and metallicity). It is therefore important, for building a reliable LF, to have accurate assumptions on the metallicity distribution function and the age of the Bulge population as they influences both the mean magnitude of each Gaussian (AGBB/RC/RGBB) and the magnitude dispersion.

The photometric metallicity maps derived by  \citet{Go13}  measure a metallicity gradient in the Bulge of $\sim$0.04 dex/deg. The metal rich stars ([Fe/H] $\sim$ 0) dominate the inner regions with little to no metallicity gradient within $|b|<5^{\circ}$, while more metal poor stars prevail at distances further from the plane. The absolute majority of stars in the VVV survey are found at $|b|<5^{\circ}$, so our assumption of solar metallicity (adopted by \citealt{Ro03} and after, by \citealt{We13}) with 0.4 dex dispersion should be a good approximation. Recent measurements have also shown that the metallicity distribution function of RC stars is bimodal \citep[e.g.][]{Zo17} with two peaks, one slightly more metal rich and one less metal rich than the Sun; however the two mean metallicities and their ratio vary across the 26 sparse fields observed by the GIRAFFE survey (see figure 4 in \citealt{Zo17}), making it difficult to implement these variations in a spatially varying LF over 315 square degrees.

An age variation in the Bulge stellar population is another factor that could shift the mean magnitude and increase the magnitude dispersion $\sigma$ of the RC, AGBB and RGBB populations: for example, according to theoretical models, the RC of a 5 Gyrs population is 0.1 mag brighter than that of a 10 Gyrs population (see the violet LF in the right panel of Figure \ref{LF} and the last row of Table \ref{LFparams} which gives the model parameters with a skew-Gaussian RC distribution). In literature there is overall good consensus on the age (10 Gyrs) of the Bulge (see review by \citealt{Zoca2016} and references therein), with some studies postulating the existence of a younger Bulge component \citep[e.g.][]{Ne14, Ca16, Ha16}.

In the next subsection we take into account 
\textit{other} factors, such as the extinction residuals and the VVV photometric errors, that influence the magnitude dispersion of the RC, RGBB and AGBB (more often we just mention $\sigma^{\mathrm{RC}}$, as the RC is the main population in our VVV sample), which are characteristic to the survey and our methods -rather
 than to the intrinsic Bulge population. The LF is further discussed in 
Section~\ref{sec:discussion}.

\setcounter{table}{0}
\begin{table*}
 \centering
 \begin{minipage}{177mm}
  \caption{Parameters describing the  \textit{Galaxia} generated synthetic LF modelled with three Gaussians to represent the RC, AGBB and RGBB and an exponential function for the RGB (see equation in section 3.1 of \citealt{Na15} and the caption of Figure \ref{LF}). For a 10 Gyrs old population (age in the first column), we provide the parameters for both a skew-Gaussian and a Gaussian RC distribution, in the first and second row respectively. The parameters for a 5 Gyrs population, where the RC is modelled with skew-Gaussian distribution, are given in the bottom row.}
   \label{LFparams}
  \footnotesize
  \centering
  \begin{tabular}{@{}lccclccccr@{}}
 \hline
& age & $a$ & $b$ & $(f^{\mathrm{AGBB}}, \mu^{\mathrm{AGBB}}, \sigma^{\mathrm{AGBB}})$ &  $(f^{\mathrm{RC}}, \mu^{\mathrm{RC}}, \sigma^{\mathrm{RC}}, \gamma^{\mathrm{RC}} ) $& $ (f^{\mathrm{RGBB}}, \mu^{\mathrm{RGBB}}, \sigma^{\mathrm{RGBB}})$ \\
   \hline
   \hline
& 10 & 0.199  & 0.642& 0.0077 ,-2.886, 0.067 &  0.174, -1.472, 0.091, -2.1&   0.032, -1.02, 0.112 \\
& $\pm$ & 0.003   & 0.012 & 0.0030,   0.029, 0.029 &    0.003, 0.002, 0.003, 0.2& 0.004, fixed, 0.015  \\
& 10 & 0.198  & 0.642& 0.0077 , -2.886, 0.068 &   0.174, -1.528, 0.062, ---&   0.032, -1.02, 0.108 \\
& $\pm$ & 0.004   & 0.017 & 0.0031,   0.031, 0.031 &    0.003, 0.002, 0.001,---& 0.004, fixed, 0.016  \\
   \hline
& 5          & 0.204   & 0.653 & 0.0076  ,-2.908, 0.055 &   0.166, -1.610, 0.052, -9.8&   0.011, -1.018, 0.050 \\
& $\pm$ & 0.003   & 0.013 & 0.0023,   0.019,  0.019 &    0.002, 0.003, 0.001, -0.9&   0.002, 0.011, 0.011  \\
 \hline
    \end{tabular}
 \end{minipage}
\end{table*}

\begin{figure}
\centering
\includegraphics[width=80mm]{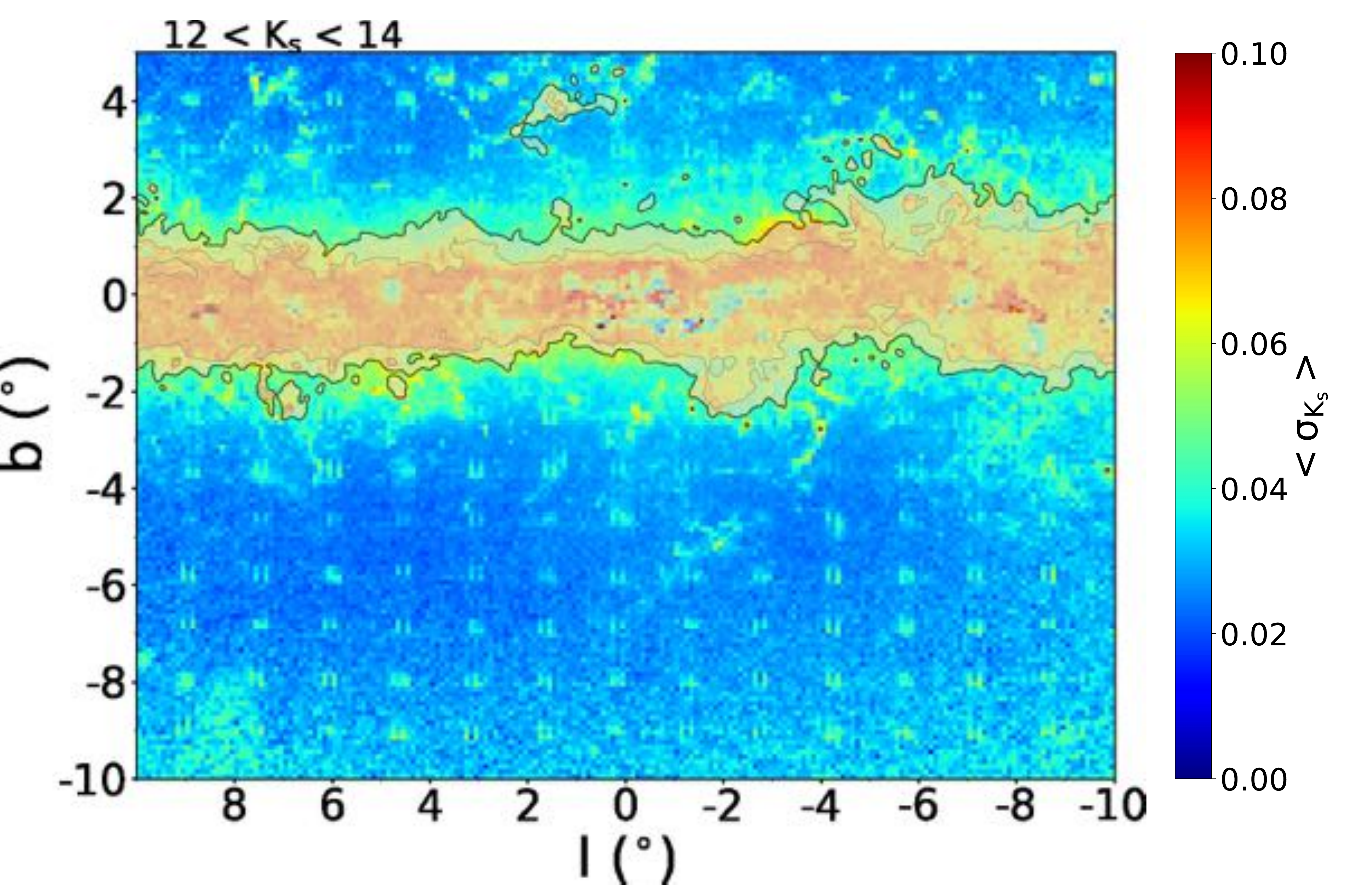}
\caption{Magnitude dispersion of the LF in the $K_{s}$ band, $\sigma_{K_{s}}(l,b)$, due to VISTA photometric errors and residual 
  extinction. The dispersion is dominated by residuals in the extinction and 
  its distribution follows closely the reddening map. The photometric
  errors increase up to $\sigma_{K_{s}}^{\mathrm{phot}} \sim 0.03$ in
  the very central region. The vertical stripes in correspondence with
  the edges of the tiles are caused by known issues with
  one of the VISTA detectors (\#16).}
 \label{errK}
\end{figure} 
\subsubsection{VISTA photometric errors and residual extinction}

The observed RC in the VVV colour-magnitude diagram
(CMD), shown in Figure \ref{cmd}, will be more spread compared to the 
theoretical LF because of variations in distance, extinction, age and metallicity in the Bulge population, coupled with survey 
specific photometric errors. We calculate the total magnitude dispersion
$\sigma_{K_{s}}(l,b)$ caused by photometric errors and extinction corrections 
(see equation 6 in \citealt{We13}) and show its spatial variation in Galactic 
coordinates in Figure~\ref{errK}. To account for the effects of photometric 
errors and residual extinction correction we convolve the theoretical LF (green curve 
in Figure~\ref{LF}) with a Gaussian kernel with 
$\sigma_{K_{s}}(l,b)$, for models $E$ and $S$ described in the next session.
However, the effects of the convolution are negligible on the number of counts predicted by our model: convolving the intrinsic LF with a Gaussian with the maximum dispersion, max($\sigma_{K_{s}}(l,b)) = 0.065$, found close to the 
exclusion boundary $E(J'-K'_{s})$ = 0.1, causes a
variation in the peak of the observed magnitude distribution 
$N(K_{s})$ smaller than 1.8\%. As the convolution slows down our fitting procedure and has negligible effects, it will not be performed for models with two components and the tests in the next sections.
%
%
\begin{figure}
\centering
\includegraphics[width=55mm]{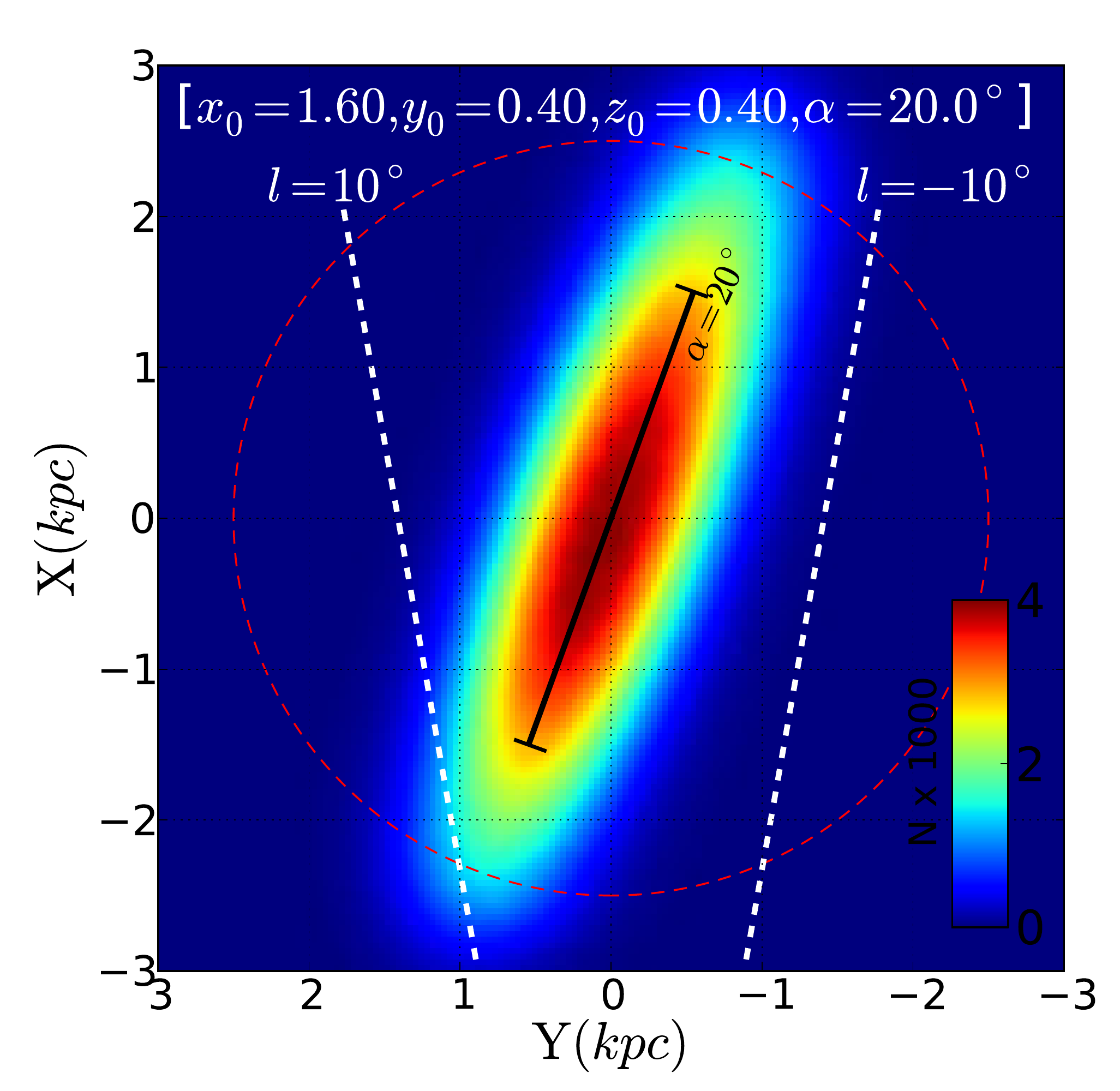}
\caption{X-Y star number density map for the
  Bulge mock catalogue generated by \textit{Galaxia} with $-5^{\circ}
  < b < 5^{\circ}$. This realisation has a $G$ model
  (Equation~\ref{G}) density law, with [$x_{0}$,$y_{0}$, $z_{0}$,
    $|\alpha|$] = [1.6, 0.4, 0.4, 20$^{\circ}$] and cut-off radius,
  $R_{c} = $ 2.5 kpc (red circle). Notice that for a very long bar
  (large values of $x_{0}$), the near end would be cut off by the
  limiting longitude of the VVV survey $l = 10^{\circ}$, preventing us
  to estimate the correct extent of the semi-major axis. The angle between the major axis direction and the Sun-Galactic centre direction gives rise to a longitudinal asymmetry in the
  $(l,b)$ density distributions.}
 \label{xy_bulge}
\end{figure} 
\subsection{The Bulge density distribution}
Several earlier Bulge studies have shown that the distribution of
sources in the Bulge is triaxial, while \citet{Dwek95} found that
Gaussian-like functions provide a better fit to the COBE/DIRBE
observations of the Bulge projected surface brightness at 2.2 $\mu$m
rather than other classes of functions. In particular, they showed that a
`boxy' Gaussian, 
\begin{flalign}
\rho_{B} = \rho_{0} \mathrm{exp} ^{-0.5r_{s}^2} \qquad (\mathrm{model} \quad
G)
\label{G}
\end{flalign}
where,
\begin{flalign*}
r_{s}^{2} = \sqrt{\bigg[ \bigg(\frac{X}{x_{0}} \bigg)^{2} +
\bigg(\frac{Y}{y_{0}}\bigg)^{2} \bigg]^{2} +
\bigg(\frac{Z}{z_{0}}\bigg)^{4}} \quad ,
\end{flalign*}
best described their data and this is the model
adopted by \citet{Ro03} for the Besancon Galaxy model which is incorporated in the
\textit{Galaxia} code.

In the next section, we fit the data with a more general form of
Equation~\ref{G}, namely an exponential-type model (model \textit{E}),
with 7 free parameters, including the viewing angle, $\alpha$:
\begin{flalign}
\rho_{B} = \rho_{0} \mathrm{exp} ^{-0.5r_{s}^n} \qquad (\mathrm{model} \quad
E)
\label{E}
\end{flalign}
where,
\begin{flalign*}
r_{s}^{c_{\| }} = \bigg[ \bigg(\frac{X}{x_{0}} \bigg)^{c_{\perp}} +
\bigg(\frac{Y}{y_{0}}\bigg)^{c_{\perp}} \bigg]^{ \frac{c_{\|}}{c_{\perp}} }
+ \bigg(\frac{Z}{z_{0}}\bigg)^{c_{\|}} \quad .
\end{flalign*}
\citet{Fr98} preferred a hyperbolic secant density distribution
\begin{flalign}
\rho_{B} = \mathrm{sech}^{2}(r_{s}) \qquad (\mathrm{model} \quad S)
\label{S}
\end{flalign}
to the exponential-type model, which has the advantage of having one
parameter less, the power $n$. This function will also be used to fit the Bulge density distribution.

The free parameters of the Bulge model, are:
\begin{itemize}
\item \textbf{$\alpha$}, the angle between the Bulge major axis and
  the line connecting the Sun to the Galactic Centre; we fix
  equal to 0 the angles $\beta$, the tilt angle between the Bulge
  plane and the Galactic plane and $\gamma$, the roll angle around the
  Bulge major axis ;
\item \textbf{$x_{0}$}, the scale-length on the major axis, and
  \textbf{$y_{0}$} and \textbf{$z_{0}$} the scale lengths on the
  in-plane and off-plane minor axes;
\item \textbf{$c_{\perp}$} and \textbf{$c_{||}$}, the face-on and
  edge-on shape parameters: the Bulge has an in-plane elliptical shape
  when $c_{\perp} = $2 (see Figure~\ref{xy_bulge}), diamond shape when
  $c_{\perp} < 2$ and boxy shape when $c_{\perp} > 2$;
\item the power \textbf{$n$};
\end{itemize}
$\rho_{0}$ is the density normalisation and $R_{c}$ is a cutoff radius, kept
fixed during the fitting procedure. The cutoff radius, $R_{c}$, is implemented by multiplying the density distribution $\rho_{B}$ by the tapering function
\begin{flalign*}
f(R) = 1 \qquad  \quad \quad  \mathrm{if}  \quad R < R_{c} \\
 = \mathrm{exp}^{ -2(R -R_{c})^{2}  } \quad \mathrm{if} \quad R > R_{c}
\end{flalign*}
to produce a smooth transition to zero for the source number density
beyond $R_{c}$ (also used for the Bulge density law in \citealt{Sh11}, see their table 1).

We do not attempt to build a parametric model of the Bulge X-shape, which we discuss in
detail in Section~\ref{sec:discussion}.
\subsection{Other stellar populations in the inner MW}

We assume the halo population contributes a negligible fraction of RC stars 
to the whole VVV RC sample (see e.g. column 4 of table 3 in \citealt{Zo17}). 
Our colour magnitude selection box does not include stars with 
$J-K_{s} < 0.4$ mag, where we would expect to see the halo old metal poor 
stars. We also assume that the ring, the spiral arms and
the thin long bar \citep[e.g.][]{Gonz14, We15} do not significantly
contribute to the VVV star counts in the fields selected for the
analysis (regions of low reddening, with $E(J'-K'_{s}) < $1.0) as they
are confined to the Galactic Plane. 

We therefore assume that in the regions probed by our analysis, the thin 
disc, thick disc and the Bulge make up the bulk of the stellar density and 
the contribution of other populations is negligible.

\section{The fitting procedure. Test on mock data}
\label{sec:fitting}

In this section we outline the fitting procedure and demonstrate its
robustness by testing it on a mock Bulge dataset generated with 
\textit{Galaxia} pre-defined parameters. These tests help assess the quality of 
our results when fitting the VVV dataset in Section~\ref{sec:datafit}.
\subsection{The mock data}

The mock data ($D^{\mathrm{mock}}$) is a catalogue of stars generated
with \textit{Galaxia} \citep{Sh11} using a specified set of 
parameters.  It consists of a synthetic disc ($N_{d}^{\mathrm{mock}}$)
multiplied by a constant ($Scale$) and a mock sample of Bulge/Bar
stars ($N_{B}^{mock}$) so that the number of stars in a given
sightline at a given magnitude $K_{s}$ is $D^{\mathrm{mock}} =
N_{d}^{\mathrm{mock}}\times Scale + N_{B}^{\mathrm{mock}}$, where
$Scale = 0.4$ (a value close to the observed value reported in Table
\ref{bestresults}). The parameter $Scale$, introduced to match the
total number of disc stars, is a free parameter in our model and is
needed to account for variations in the number density introduced by
changing the isochrones set. We generate with \textit{Galaxia} a mock
Bulge catalogue of $\sim$13 million stars, to roughly match the number
of VVV stars, following model $G $ (Equation~\ref{G}) density
distribution with parameters [$x_{0}$, $y_{0}$, $z_{0}$, $|\alpha|$] =
[1.6, 0.4, 0.4, 20$^{\circ}$]. This distribution is shown in
Figure~\ref{xy_bulge}, projected onto the XY-plane, where the Galactic
Centre is at [$X,Y$] = [0,0].  Only the latitude range $-5^{\circ} < b
< 5^{\circ}$ is plotted but we fit $-10^{\circ} < b < 5^{\circ}$, as
in the data.
%

%
\begin{figure*}
\centering
\hspace*{-0.75cm}
\includegraphics[scale =0.105]{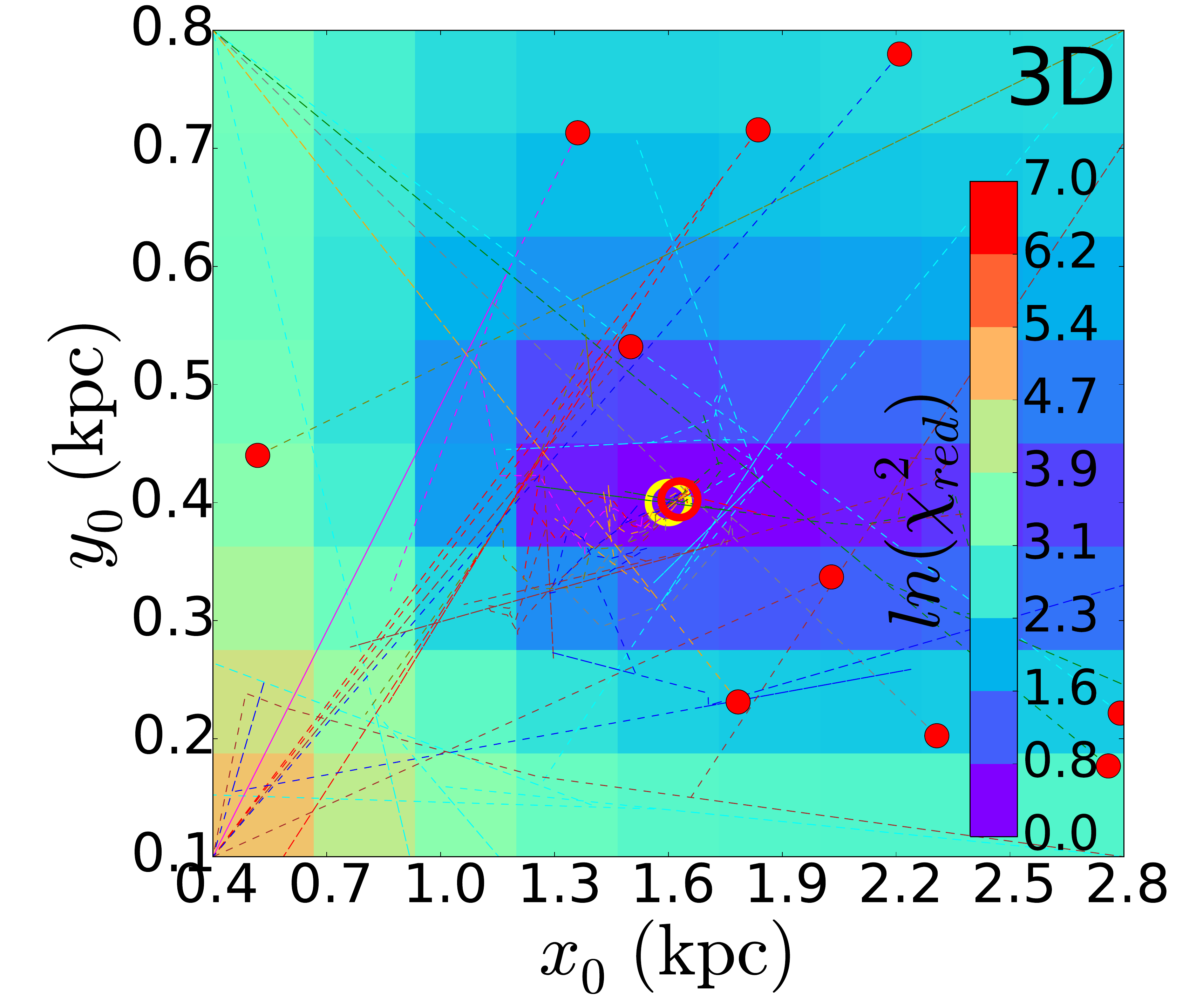}
\includegraphics[scale =0.105]{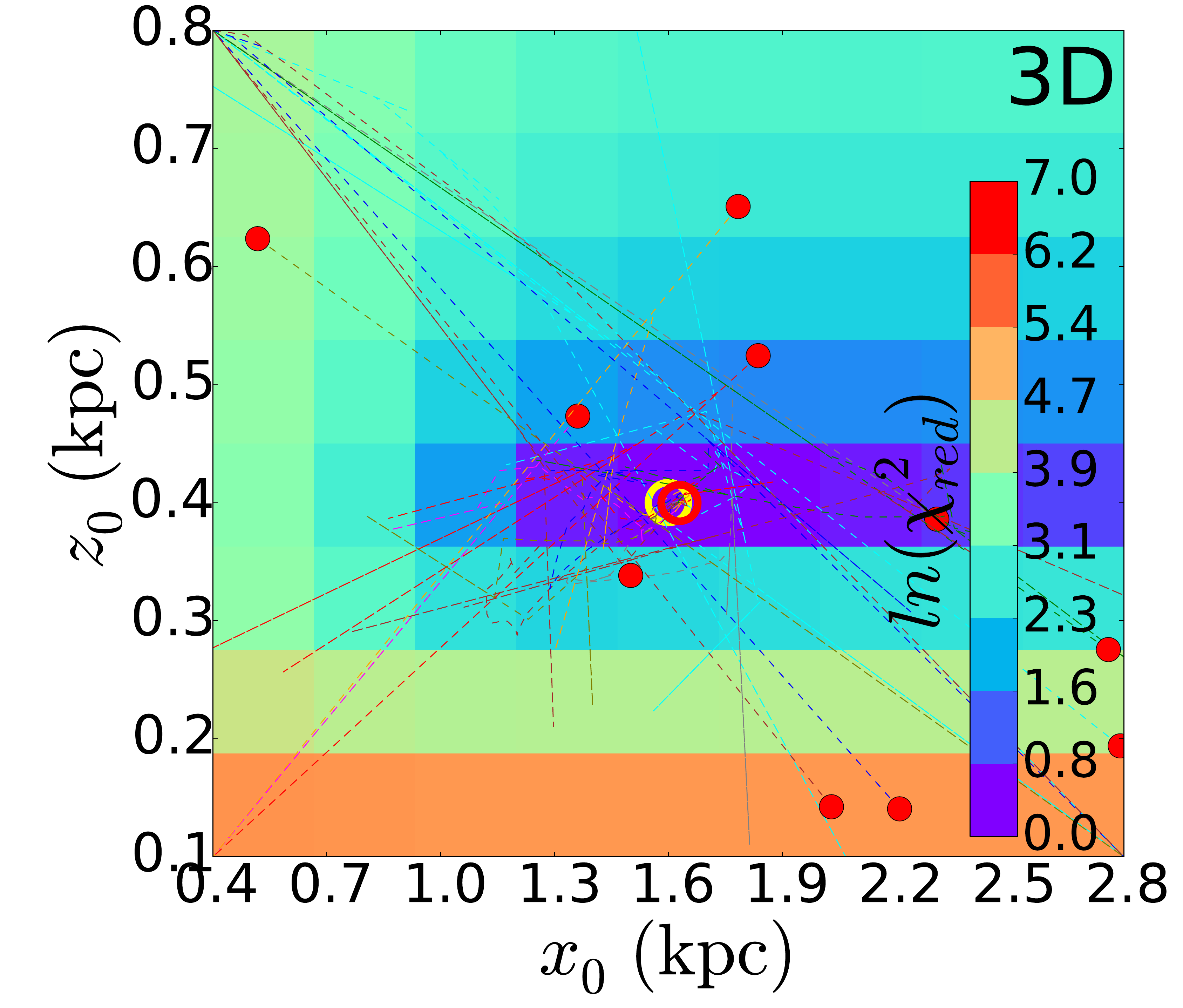}
\includegraphics[scale =0.105]{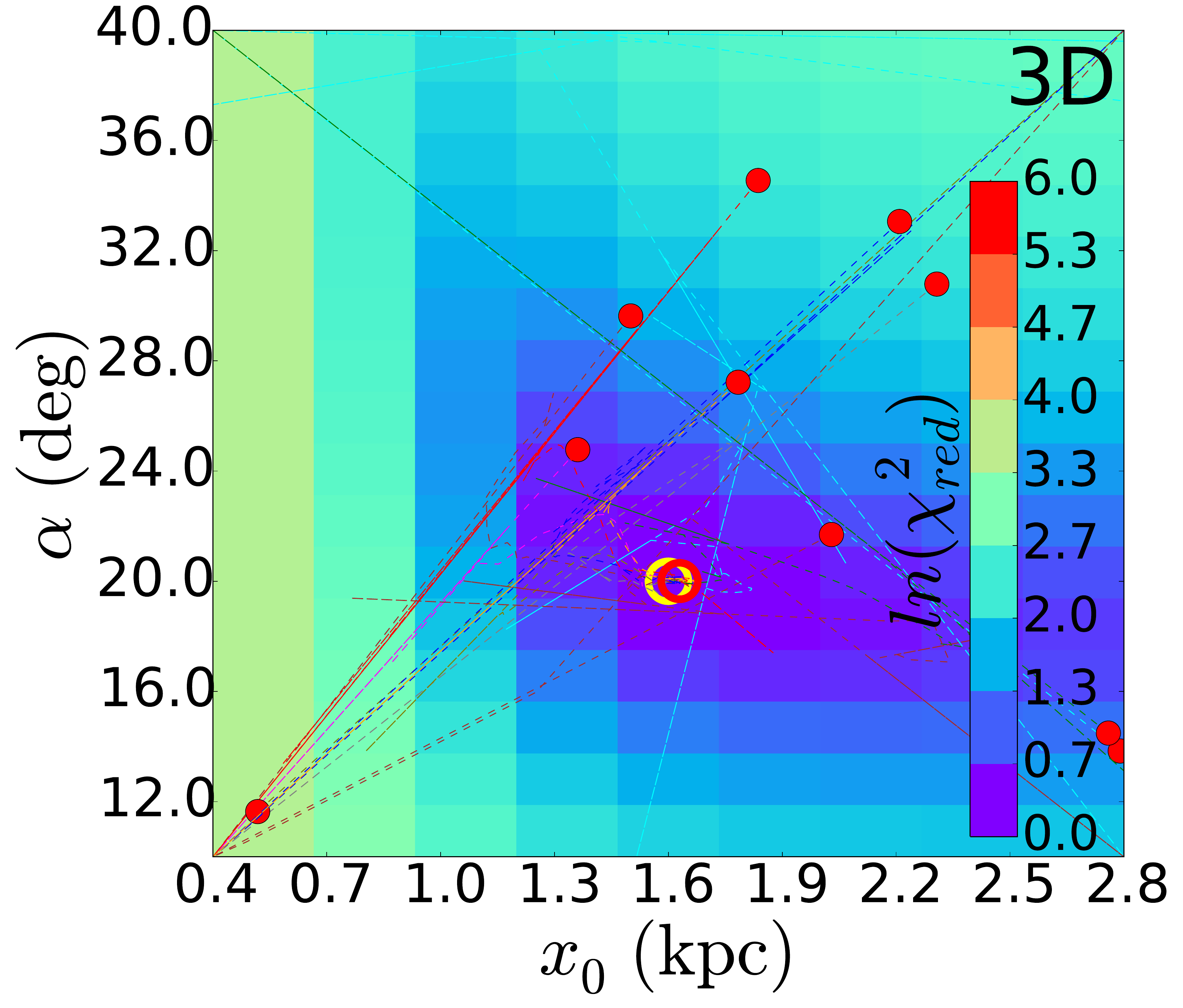} 
\includegraphics[scale =0.105]{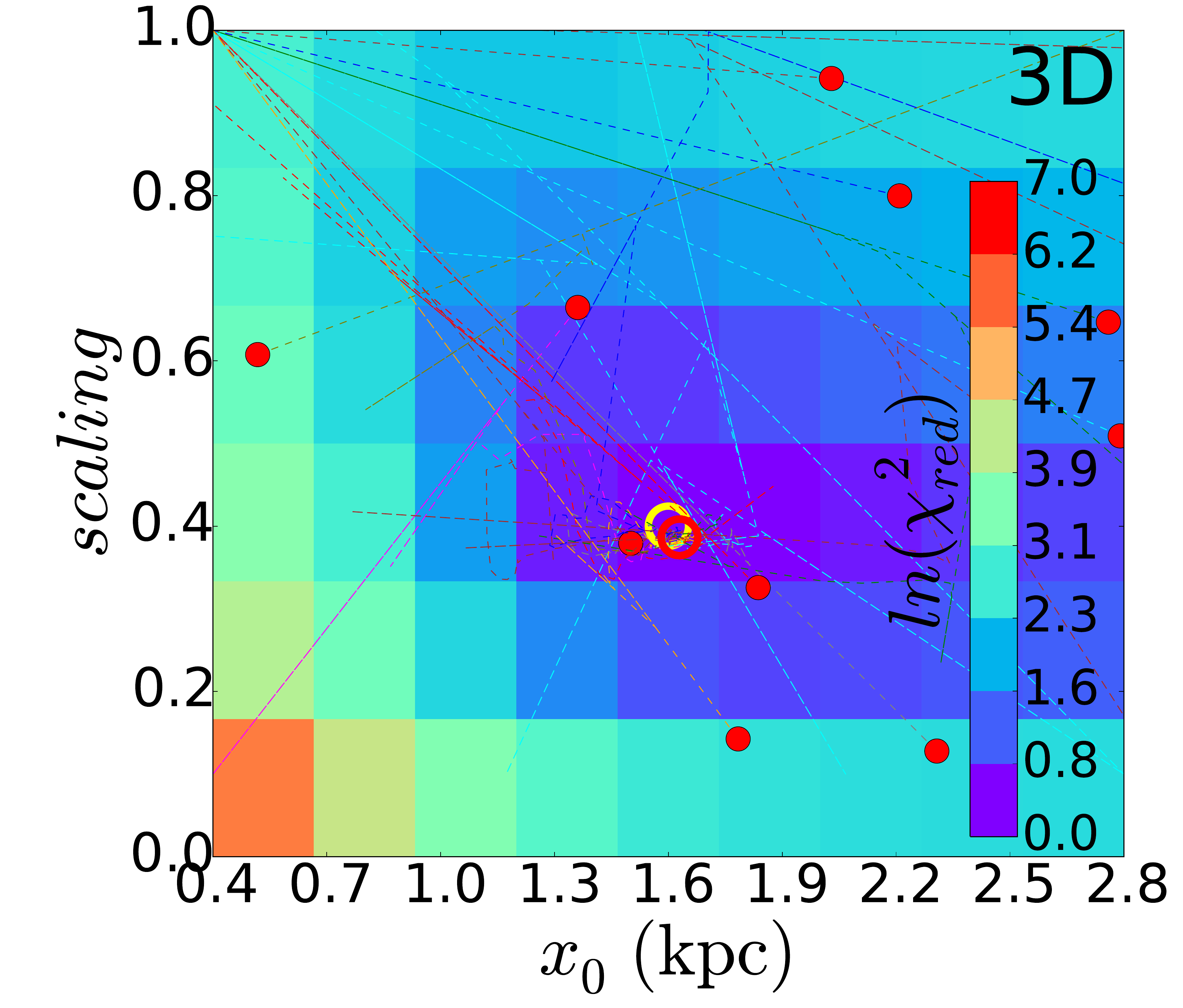}
\includegraphics[scale =0.105]{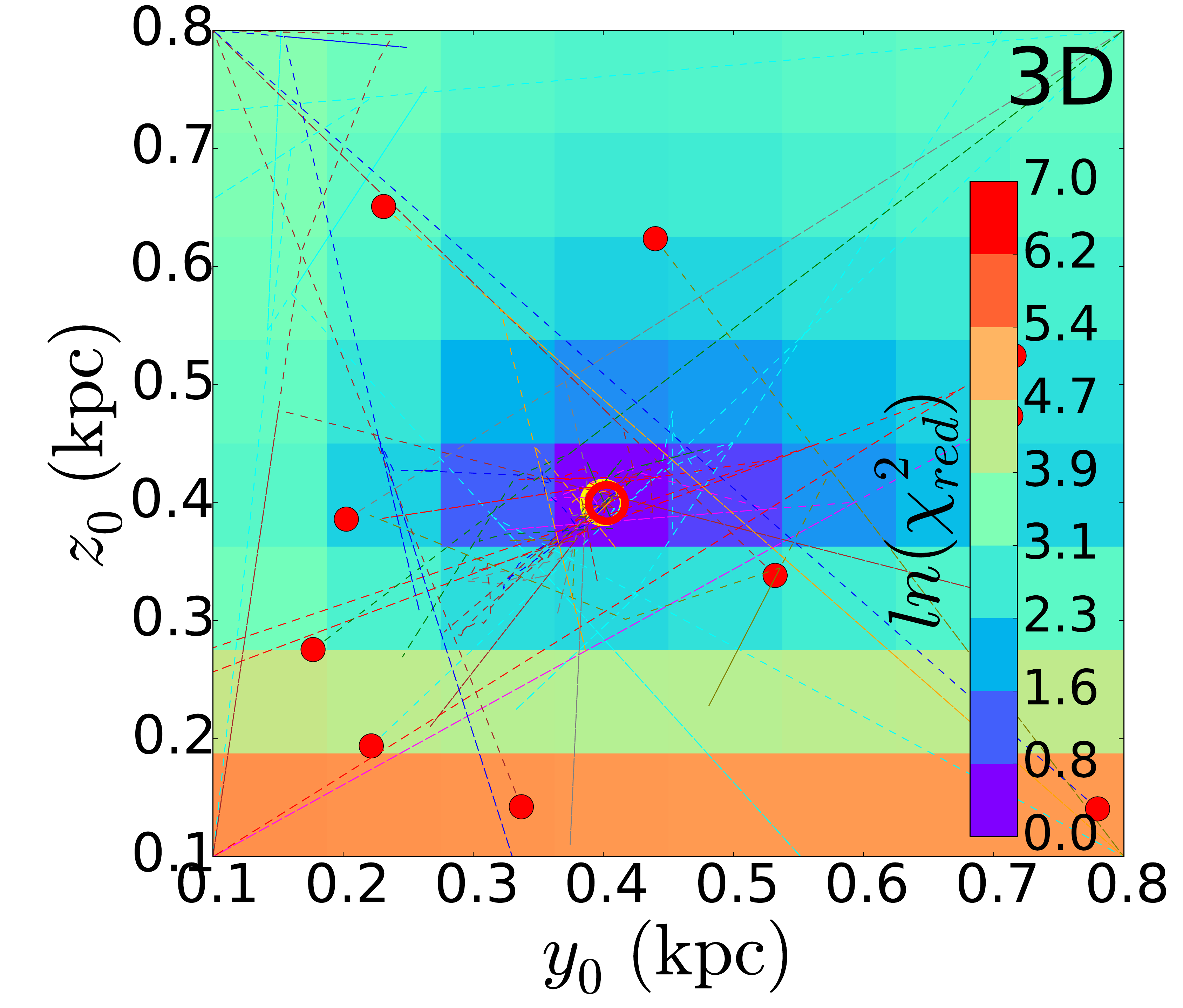}\\
\hspace*{-0.75cm} \includegraphics[scale
  =0.105]{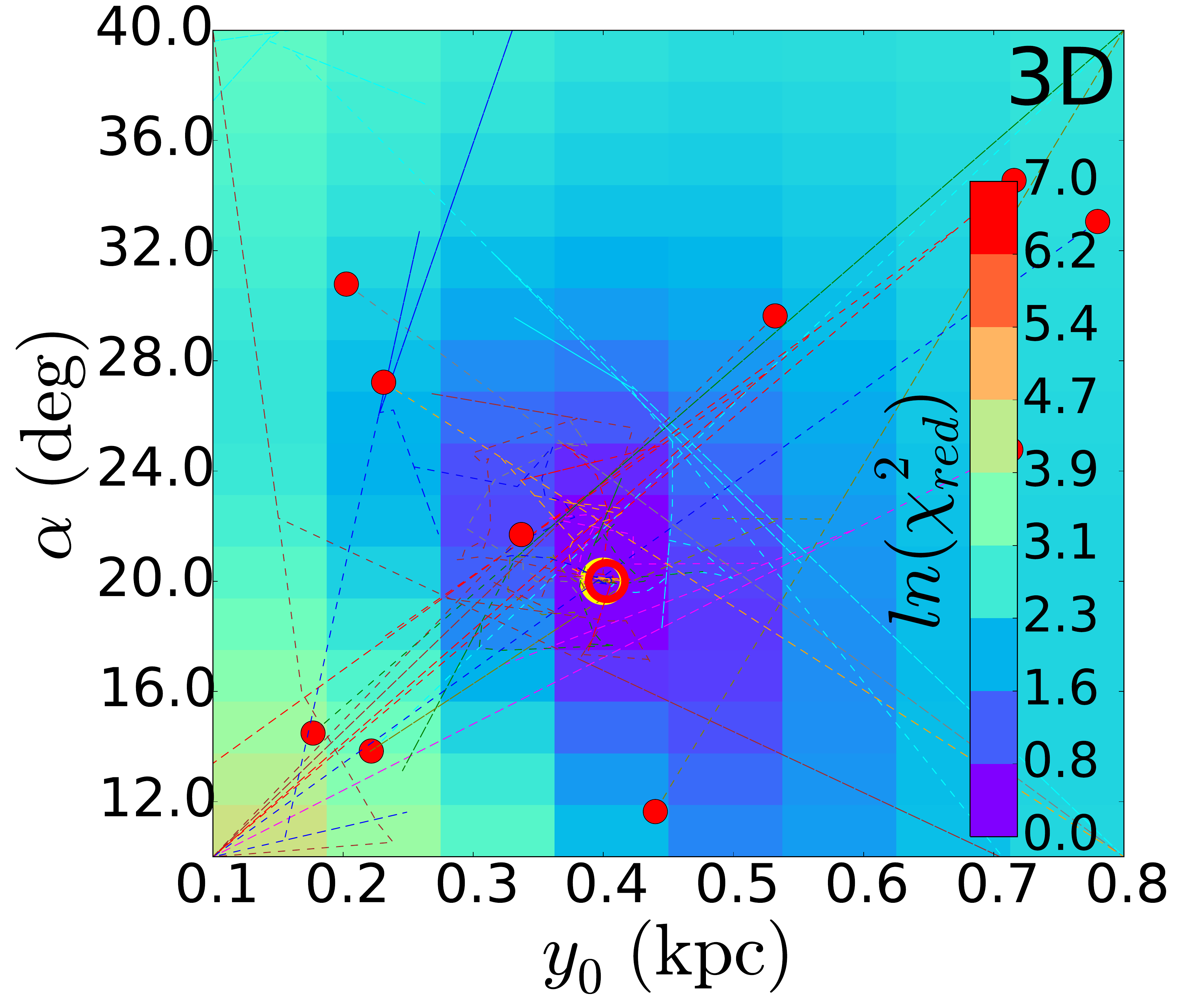} \includegraphics[scale
  =0.105]{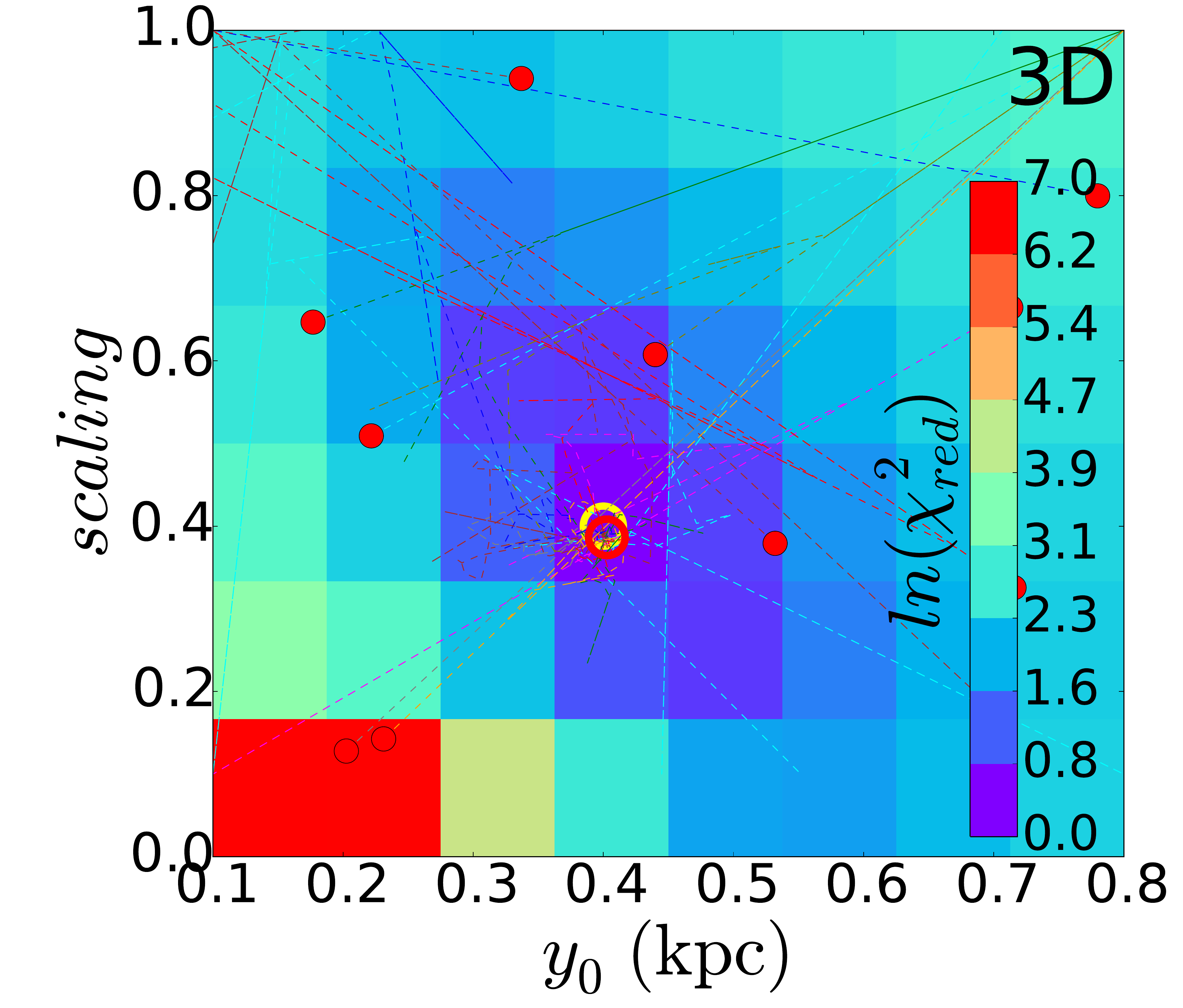} \includegraphics[scale
  =0.105]{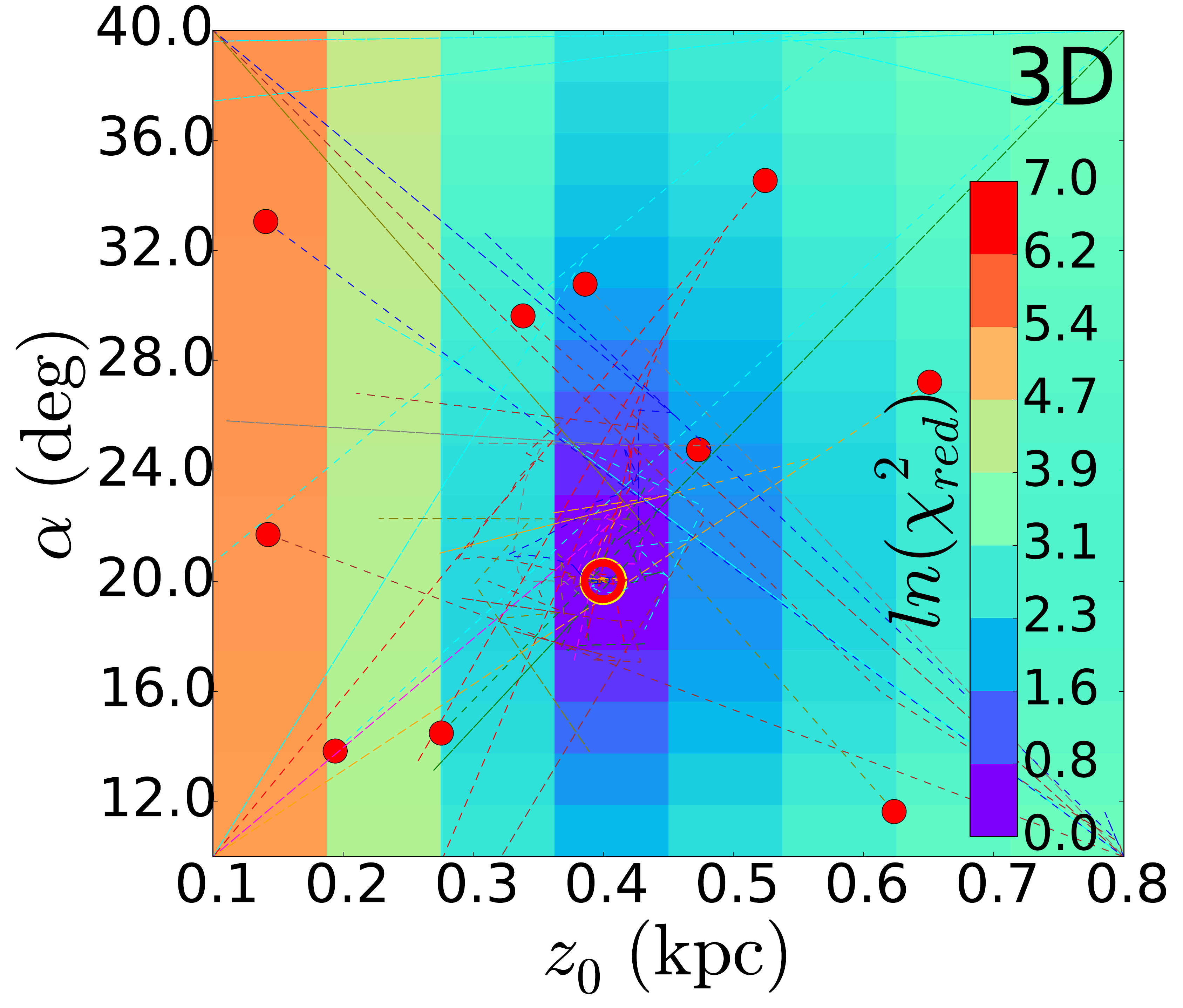} \includegraphics[scale
  =0.105]{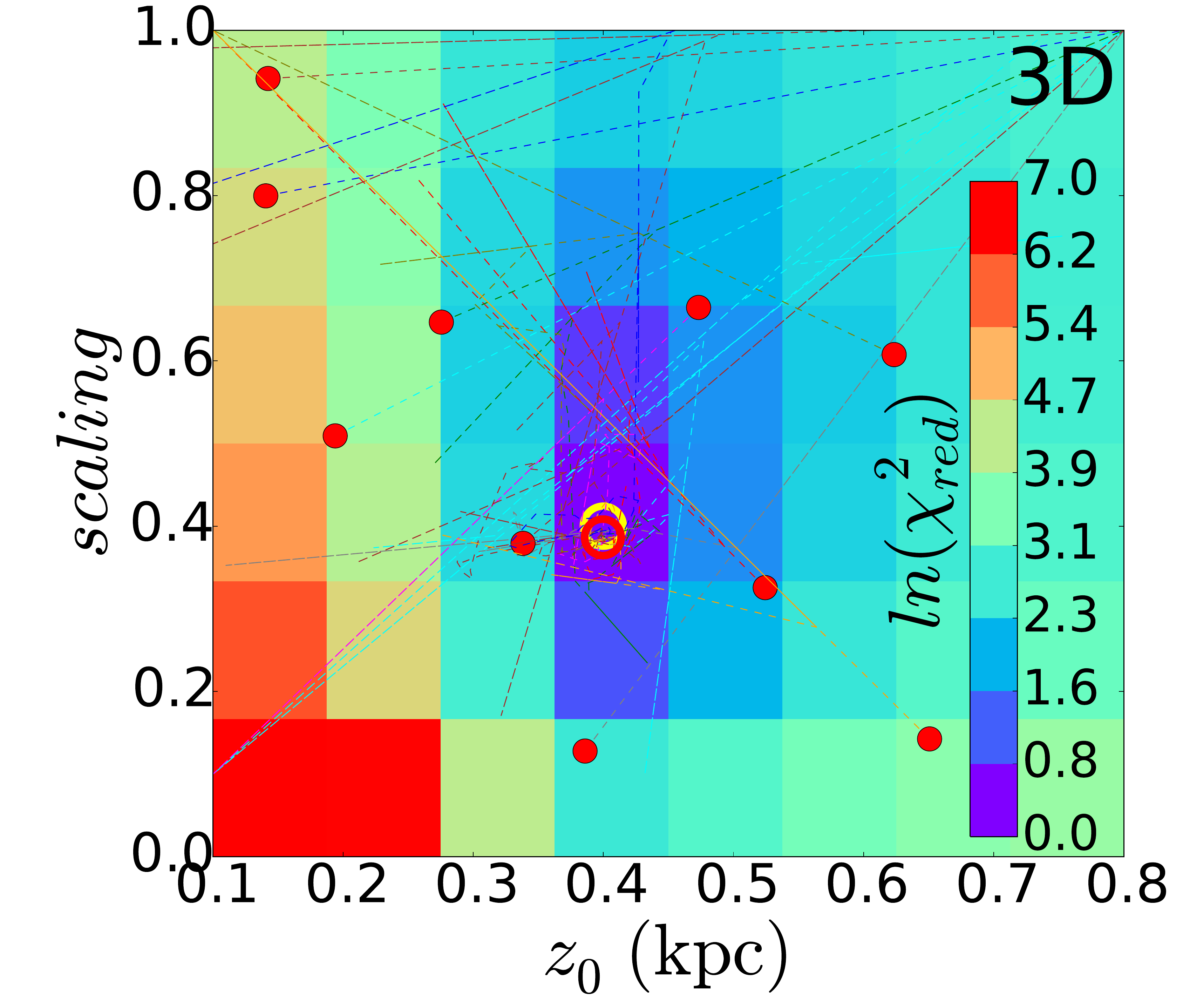} \includegraphics[scale
  =0.105]{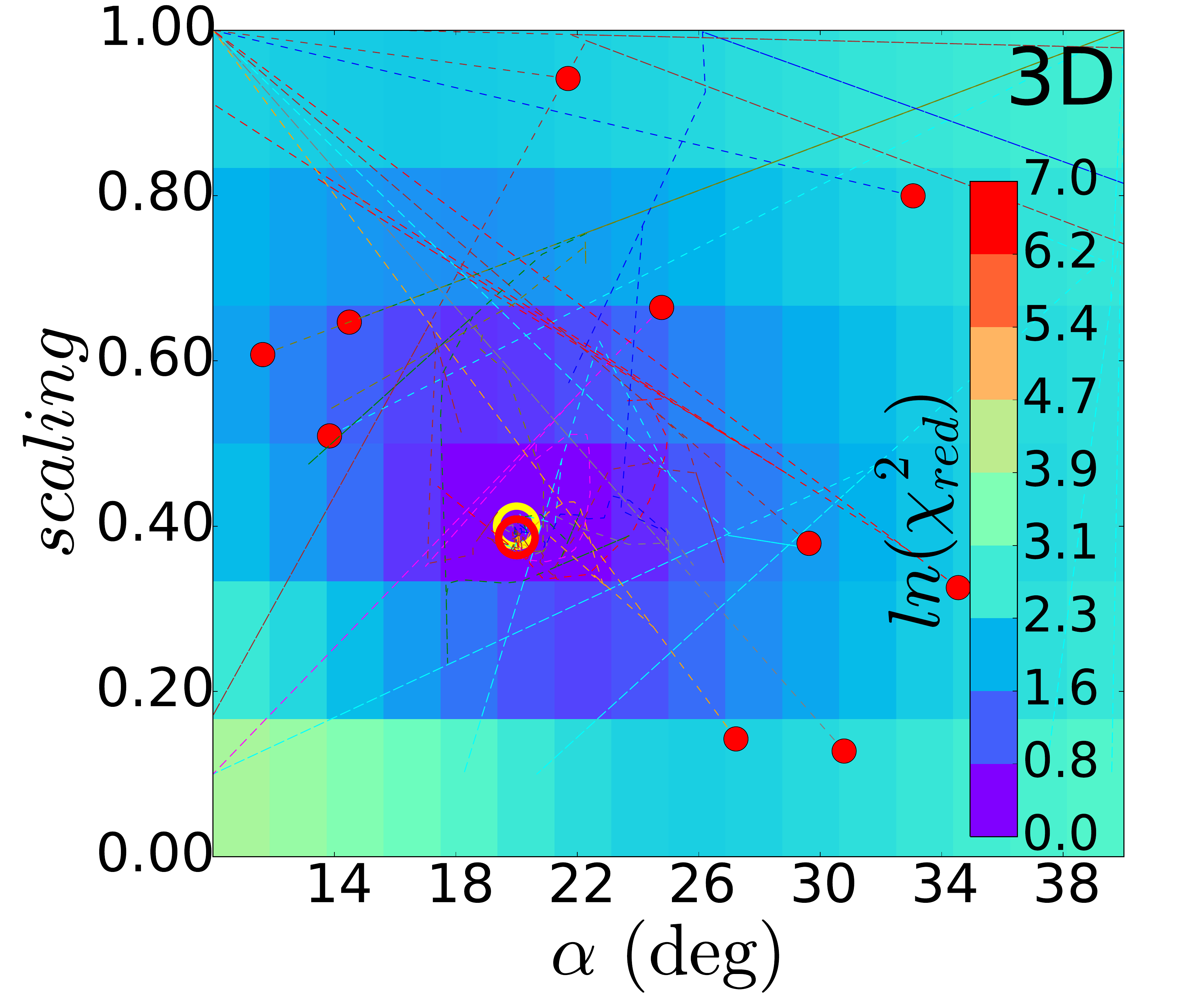}\caption[Slices of
  $\ln$($\chi_{red}^{2}$) maps as a function of two varying parameters
]{Slices of $\ln$($\chi_{red}^{2}$) maps as a function of two varying
  parameters while the other three are fixed to their true value. The absolute
  minimum found by the grid search (darkest violet bin) and the 10 BFGS runs (red open
  circle) coincides with the true value (yellow circle). The BFGS runs
  were initialised in random points (full red circles) covering the
  full parameter space. }
\label{chimap3D_test20}
\end{figure*} 
\subsection{The model}

The model ($M$) predicting the number of stars in a given line of
sight at a given magnitude, has the general form:
\begin{equation}
M = < N_{d} > \times Scale + N_{\mathrm{B}}
\label{model_total}
\end{equation}
where $< N_{d} > $ is an average over several $N_{d}$ disc realisations
(see Figure \ref{discs}). We calculate $N_{\mathrm{B}}$, the Bulge apparent
magnitude distribution, analytically. For each field with Galactic
coordinates $(l,b)_{i}$, d$N_{\mathrm{B}}(K_{s}) = N_{\mathrm{B}}(K_{s})$d$K_{s}$ is the
differential star counts with apparent magnitudes in the range
($K_{s}$, $K_{s}+dK_{s}$)
\begin{equation*}
\begin{split}
\mathrm{d}N_{\mathrm{B}}(K_{s})= \rho_{\mathrm{B}}(D) \phi_{\mathrm{B}}(M_{K_{s}}) D^{2} \cos(b) \Delta l \Delta
b \mathrm{d}D,
\label{N}
\end{split}
\end{equation*}
where the absolute magnitude is a function of the apparent magnitude
and distance modulus $M_{K_{s}} = K_{s}-5\log_{10}D
+5$. $N_{B}(K_{s})$, the apparent magnitude distribution, in a given
direction $(l,b)_{i}$ (and a solid angle $\Omega$) is the integral of
the LF $\phi_{\mathrm{B}}$ times the density law $\rho_{\mathrm{B}}(D)$
\begin{equation}
N_{\mathrm{B}}(K_{s})= \int_{4}^{12} \rho_{\mathrm{B}}(D) \phi_{B}(K_{s}-5\log_{10}D +5 )
D^{2} \Omega \mathrm{d}D.
\label{Bulge_model}
\end{equation}

In the fitting process, we calculate the predicted number of counts
$N_{\mathrm{B}}(K_{s})$ for 20 different values of $K_{s}$ between $12 <
K_{s}/$mag$ < 14$ so that the
total number of counts is $N_{\mathrm{B}}(12 < K_{s} < 14) = N_{\mathrm{B}}
=\sum\limits_{j = 0}^{20} (N_{\mathrm{B}})_{j}(K_{s})$. The limits of the integral in Equation~\ref{Bulge_model} are set to 4 and 12 kpc as $\rho_{\mathrm{B}}$ approaches zero outside this range for $12< K_{s}< 14$. 
\subsection{The fitting procedure}
\subsubsection{Method}

We follow a Bayesian approach which requires us to
define the likelihood function of our data and the prior distributions
of the parameters $\vartheta$. The likelihood of the entire data set,
$L$, is the product of the probability of each number count $N_{i}$ in
the $n_{\mathrm{fov}}$ fields of view $(l,b)_{i}$, given our model
prediction $M$ (Equation~\ref{model_total}):
\begin{flalign*}
L  =  \prod\limits_{i=1}^{n_{\mathrm{fov}}} \mathrm{p}(N_{i}|M(\vartheta)) \quad .
\end{flalign*}
Assuming the measurements $N_{i}$ follow a Poisson
distribution we search for the parameters
$\vartheta$ that maximise the logarithm of the likelihood for the whole data set:
\begin{flalign}
 \mathrm{ln} L = \sum \limits_{i=1}^{n_{\mathrm{fov}}} [N_{i} \mathrm{ln} M_{i}(\vartheta) - M_{i}(\vartheta)] + \mathrm{constant}
\label{prob}
\end{flalign}

\setcounter{table}{1}
\begin{table*}
 \centering
 \begin{minipage}{177mm}
  \caption[Best fit results for the mock catalogue generated with
    \textit{Galaxia}]{Best fit results for the mock catalogue
    generated with \textit{Galaxia}. The first column specifies the
    model, the second column the shape of the density distribution, the third column the reduced chi squared $\tilde{\chi}^{2}$ as an indication of the 'goodness' of the fit,
    the fourth column the log-likelihood, the fifth column the best fit
    parameters of the density distribution and finally, the sixth column, gives the discs
    scaling, $Scale$. In the last two rows we list the results obtained
    from fitting an $S$ model and an exponential form of the density
    distribution, where $r$ is the Galactocentric radius. From the
    comparison of the results, we can conclude that the errors on the
    parameters will be dominated by systematics.}
   \label{testmock}
  \footnotesize
  \centering
  \begin{tabular}{@{}lccclccccr@{}}
 &  & &      & Test \textit{Galaxia} &     \\
 \hline
& $\rho_{b}$ & $\tilde{\chi}^{2}$ & -ln($L_{G}$) & [$x_{0}$,
$y_{0}$,$z_{0}$, $\alpha$, $c_{//}$, $c_{\perp}$,$n$] & $Scale$ \\
   \hline
   \hline
$Galaxia$ model & $\propto \mathrm{exp} ^{-0.5r_{s}^2} $&- & - & [1.600,
0.400, 0.400, -20.00, 4.000, 2.000, 2.000] & 0.400 \\
   \hline
fitted model &$ \propto \mathrm{exp} ^{-0.5r_{s}^n} $ &1.00 & 384,408 &
[1.629, 0.403, 0.400, -20.01, 3.877, 1.923, 1.993] & 0.386 \\
fitted model &$\propto$ sech$^{2}(r_{s})$ &1.15 & 416,195 & [1.834, 0.456,
0.429, -19.79, 3.279, 1.717, -] & 0.271 \\
fitted model &$\propto \mathrm{exp} ^{-0.5 r^{2}} $ &1.10 & 406,338 &
[1.961, 0.468, 0.454, -20.30, -, -, -] & 0.317 \\
 \hline
    \end{tabular}
 \end{minipage}
\end{table*}
We use the $Python$ implemented limited-memory BFGS method
\citep[L-BFGS,][]{Br95} that calculates an estimate of the inverse
Hessian matrix to facilitate search through the model parameter space. This
method is particularly well suited for optimisation problems with a
large number of parameters, e.g. including the ones governing the
shape of the Bulge density, $n$, $c_{\perp}$, $c_{||}$. Moreover, it
allows the user to constrain the parameters within reasonable physical
values, equivalent to adopting a uniform Bayesian prior. We draw the initial 
(guess) values of the free parameters from uniform distributions, with the 
same limits as our grid search, and 
add three extra parameters, $n$, $c_{\perp}$, $c_{||}$, also drawn
from random uniform distributions, with limits as in
Figure~\ref{chimap3D_test20}.

\subsubsection{Optimal integration step estimation}

Before fitting, we need to find the most appropriate integration steps
($\Delta l, \Delta b$ and $\Delta D$) for the Bulge model ($N_{B}$),
the only population we generate analytically and whose parameters we
want to determine. Using the same density function and true values in
both the synthetic Bulge and the mock catalogue, we made several tests
varying ($\Delta l, \Delta b$ and $\Delta D$) in our model to find the
resolution that produces the best match to the mock catalogue. We
found that small-size fields ($\Delta$$l$, $\Delta$$b$) =
(0.1$^{\circ}$,0.1$^{\circ}$) produced a reduced chi-squared
$\tilde{\chi}^{2}$ $\approx$ 1 while the distance step can be
relatively large (e.g. $400$ pc) given that we calculate
$N_{B}(K_{s})$ for magnitude step sizes of $dK_{s} = 0.1$ mag. With
this choice, the model gave a perfect match to the mock catalogue,
with a reduced chi-squared of $\tilde{\chi}^{2} = 1.02$. The number of
sources varies rapidly from pencil beam to pencil beam in the inner
Galaxy, so it is not surprising that in order to have an accurate
model prediction, we need to calculate parameters over a very fine
grid of Galactic coordinates. 

To fit the VVV data we increase the magnitude resolution to $\mathrm{d} K_{s} = 0.05$ mag,  for which we need to employ a finer distance integration step of $\Delta D = 50$ pc.

\subsection{Test results}

In the test run we compute chi-squared maps varying five
parameters $\vartheta_{0} =$ [$x_{0}$,$y_{0}$, $z_{0}$, $|\alpha|$,S], 
over a wide range of values on a coarse grid [$\Delta
  x_{0}$,$\Delta y_{0}$, $\Delta z_{0}$, $\Delta |\alpha|$, $\Delta
  S$] = [0.3 kpc, 0.1 kpc, 0.1 kpc, 2$^{\circ}$, 0.2 ]. The high
reddening region where $E(J'-K'_{s}) >1.0$ (gray mask in
Figure~\ref{errK}) is excluded from the test, to reproduce the data
fitting procedure. We compute $N_{B} (12< K_{s}<14 )$ for
fields of 0.1$^{\circ} \times $ 0.1$^{\circ}$ and in each pencil beam
we calculate $N_{B}$ for 20 values of magnitude $K_{s}$
(Equation~\ref{Bulge_model}). We group the model $M$ (Equation
\ref{model_total}) in 5 magnitude bins to be fitted on 5 magnitudes
bins in the mock catalogue (VVV data in Section~\ref{sec:datafit}), in order 
to extract information about the depth distribution of sources in the Bulge.

Figure~\ref{chimap3D_test20} shows the resulting slices of
$\ln(\chi^{2})$ maps as a function of the two varying parameters,
while the other three are fixed to their true value. The absolute
minima (violet bin in the figure) coincide with the true values used
as an input of the model (yellow open circles in the figure),
proving that the method works well. These maps also reveal the correlations between parameters, see
e.g. $\alpha$ versus $x_{0}$.

The mock catalogue can also be used to test the accuracy of the L-BFGS method
employed in Section~\ref{sec:datafit} to find the parameters of the real Bulge 
density distribution. The 10 filled red circles in Figure~\ref{chimap3D_test20}
are the initial parameters of 10 L-BFGS independent runs with initial values 
drawn from uniform distributions.
The identical final solutions of each run (red open circles) where the 
algorithm converges, coincide with the fiducial value (yellow circles) 
$\vartheta=$[1.629, 0.4026, 0.3996, 20.01, 3.88,1.925,1.993,0.3865] $\pm$
[0.004, 0.0009, 0.0009, 0.02, 0.015, 0.007, 0.004, 0.0009] and are very
close to the true value $\vartheta_{0}=[1.6,0.4,0.4,20,4,2,2,0.4]$,
included in Table \ref{testmock}. The statistical errors in the parameters, 
obtained from the covariance matrix, are extremely small and we consider 
them to be negligible in comparison to the systematic errors 
(see Section~\ref{sec:discussion}). 
However, the test on the mock data also provides an indication of how well the 
fitting method recovers the true parameters of the Bulge density distribution 
(see the first two lines in Table \ref{testmock}, which shows the true values
and the best fit values found using the L-BFGS minimisation method). The
agreement here leads us to conclude that the random error component is well-captured by the method.

An estimate of the systematic errors can be obtained by adopting 
different density distributions in the models as compared to the mock data.
In Table \ref{testmock} we list the results obtained from fitting a hyperbolic 
secant model and an exponential form of the density
distribution, with $r$ the Galactocentric radius (last two rows of
the table) and from comparing the results of the three fits to the
true values $\vartheta_{0}$, we notice that the best fit parameters
$\vartheta$ of the correct density law are close to their true value
$\vartheta_{0}$ ( $\delta \vartheta = [x, y, z] - [x_{0}, y_{0},
  z_{0}] = [0.029, 0.003, 0.000]$) while when fitting the `wrong'
density laws, the results diverge significantly from the true value
($\delta \vartheta = [0.234, 0.056, 0.029]$ and $\delta \vartheta =
[0.361, 0.068, 0.054]$ for the two `wrong' models). In conclusion, the choice of the density distribution
alone introduces systematic errors much larger than the standard
errors.
\section{Fitting the VVV data}
\label{sec:datafit}
%
%
\begin{figure}
\centering
\includegraphics[width=36mm]{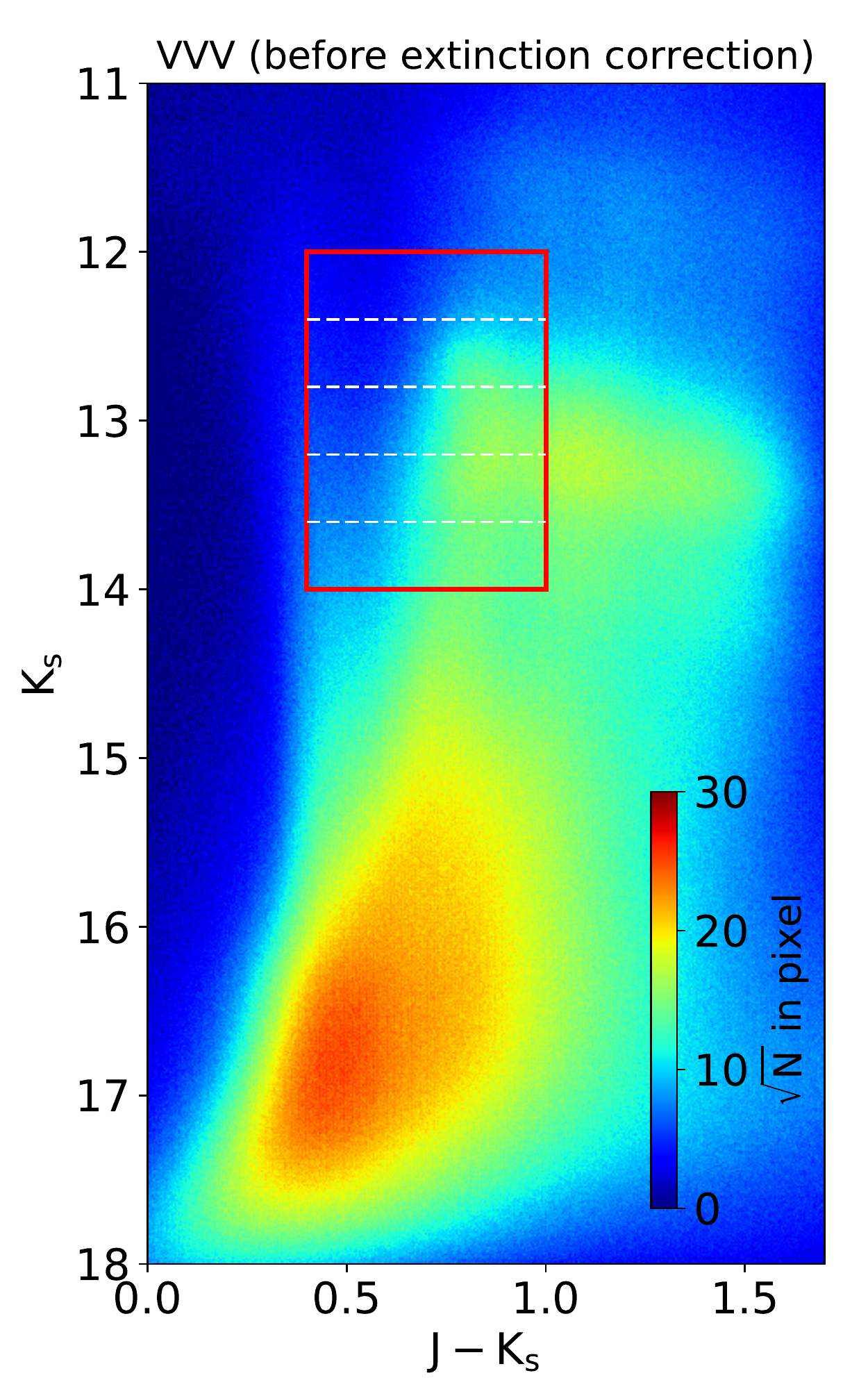}
\includegraphics[width=36mm]{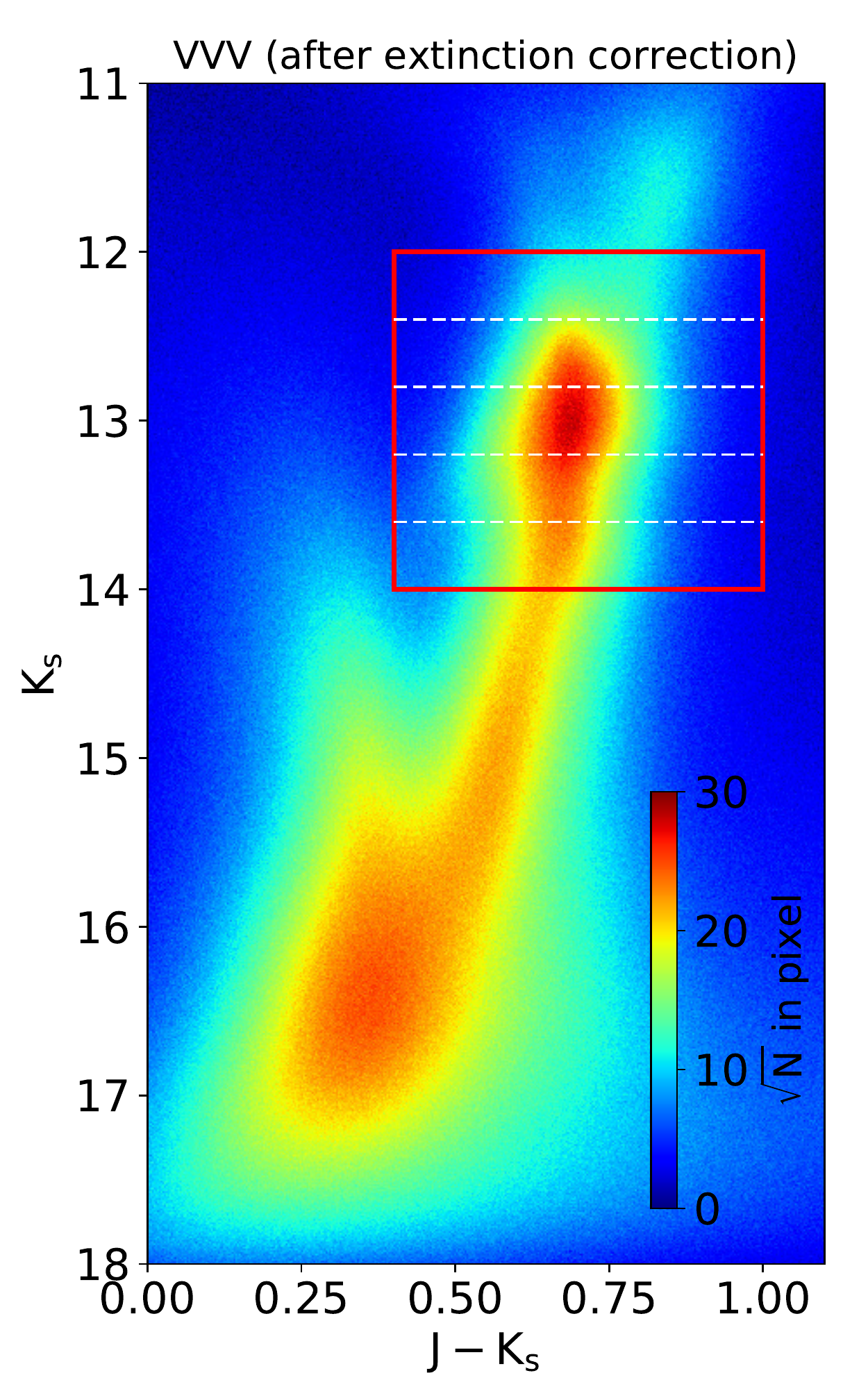}
\caption{CMD of all the Bulge VVV targets with class K $< 0$ and $E(J'-K'_{s}) < 1.0$, totalling $\sim$ 96$\times10^{6}$ sources, both before ($left$ $panel$) and after ($right$ $panel$) correcting for reddening (shown in Figure \ref{ext}). The red square marks our working sample colour - magnitude selection (their spatial distribution shown in Figure~\ref{data}). The dotted white lines show the five magnitude slices used for the plots in Figures \ref{models} and \ref{diff}. The spatial distribution of the stars in each slice is shown in the left panels of Figure~\ref{models}.
}
 \label{cmd}
\end{figure} 
\begin{figure}
\centering
\includegraphics[scale=0.2]{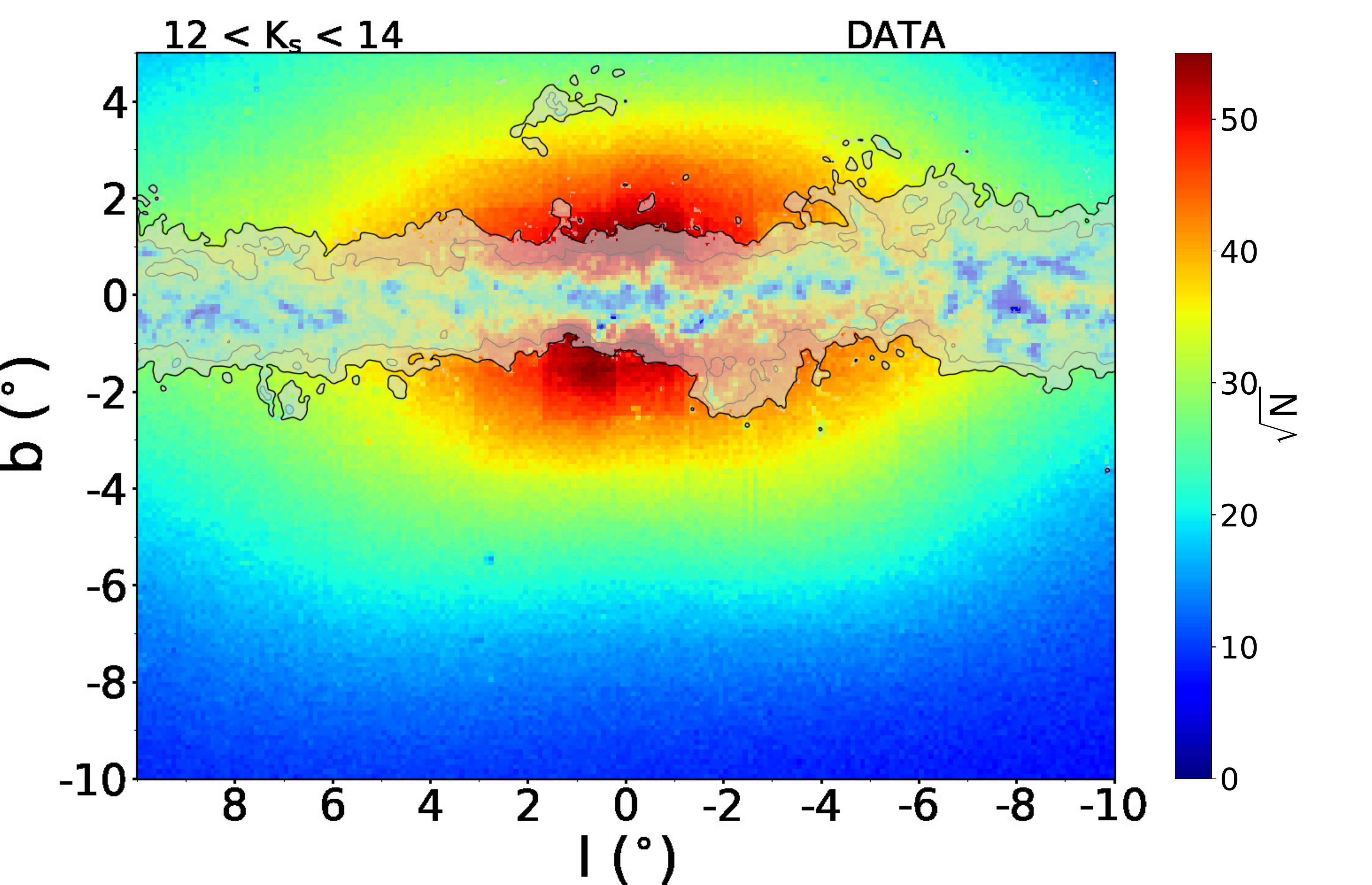}
\caption{The number density distribution of VVV stars with $12 < K_{s}$/mag$
  < 14$ and $0.4 < J-K_{s} < 1.0$ (red selection box in Figure
  \ref{cmd}), in Galactic coordinates. The contours delineate the
  highly reddened regions $E(J'-K'_{s}) = 1.5$ (light gray) and
  $E(J'-K'_{s}) = 1.0$ (thick gray). Close to the Galactic Plane
  (e.g. inside the $E(J'-K'_{s}) = 1.0$ reddening contour) the dust causes incompleteness issues. Notice also the variation in the number of counts in adjacent
  tiles in the inner regions.}
 \label{data}
\end{figure} 
In this section we fit a triaxial density model to the observed 
VVV magnitude distribution in the RC region of the colour-magnitude diagram. 
We focus on describing the inner Milky Way large scale structure by finding a model capable of reproducing the distributions seen in the majority 
of the fields at the expense of discrepancy in a small number of fields, which include fields close to the Galactic Plane/Centre and those at 
$b < -5^{\circ}$, where a double RC is observed \citep[e.g.][]{Mc10, Sa12}.
\subsection{Data sample}

The photometry of every source in the single band tile catalogues
produced by CASU is labelled with a morphological classification
flag. In the following sections, we use detections classified as
`stellar', `borderline stellar' or `saturated stars' in the $K_{s}$
magnitude band, which has the best seeing. We avoid applying a strict
cleaning procedure in order to work with a sample of Bulge stars as
complete as possible. 

We choose sources with photometric errors 
$\sigma_{J}$, $\sigma_{K_{s}} < 0.2$ mag and colours $(J-K_{s})>0.4$,
to minimise contamination from disc Main Sequence stars and $(J-K_{s})
< 1.0 $ to separate Red Giant stars from spurious objects and highly
reddened K and M dwarfs. In our analysis, we apply a faint end magnitude cut of $K_{s} = 14$ mag, which we consider adequate to 
minimise the effects of incompleteness given the substantial extinction 
over the field-of-view, and a bright end limit of $K_{s}=12$ mag to avoid 
saturation and non-linearity \citep{Go13} which sets in around 
$K'_{s} < 11.5$. These cuts form our colour-magnitude selection for the VVV working sample and are marked with a red box in Figure \ref{cmd}. The two panels show the distribution of all the VVV sources  with $E(J'-K'_{s}) < 1$, totalling 96 milion objects, in a CMD before (left panel) and after (right panel) extinction 
correction. The observed $K_{s}^{'}$-band magnitudes for each star are corrected for
the effects of extinction, $K_{s} = K_{s}^{'} - A_{K_{s}} (l, b)$,
using the reddening maps $E(J'-K'_{s})$ constructed from the VVV data (see
Section~\ref{sec:extinction}). In the left panel, most of the RC stars are outside our selection box due to extinction and reddening effects which cause the stars to appear fainter and cooler. The  number density distribution of all the stars within the red selection box is then shown
in Galactic coordinates in Figure~\ref{data}. Again, the contours delineate the
highly reddened regions $E(J'-K'_{s}) = 1.5$ (light gray) and
$E(J'-K'_{s}) = 1.0$ (thick gray). 
 
We illustrate the completeness of the survey by plotting the variation of the observed magnitude distributions with latitude in Figure \ref{completeness} of the Appendix, both before (top row) and after (bottom row) extinction correction. The stars were selected in $1.0^{\circ}$ x $0.4^{\circ}$ fields at constant longitude $l=0^{\circ}$ and decreasing latitude $b$. The fields within the $E(J'-K'_{s})=1$ mag contour are incomplete with the high extinction causing a depletion of stars in the 13 to 15 $K'_{s}$ magnitude range and a rapid drop in the star
counts for $K'_{s} > 15$ mag (see orange dotted line in the figure).  However, further from the plane, the incompleteness limit increases and at $b \sim -3^{\circ}$ the distribution is complete up to $K'_{s} \sim 15.2$ or $K_{s} \sim 15$, making our working magnitude range (red vertical lines in the figure) a fairly conservative cut.

\subsection{An example data fit}
%
%
\begin{figure}
\centering
\includegraphics[width=95mm]{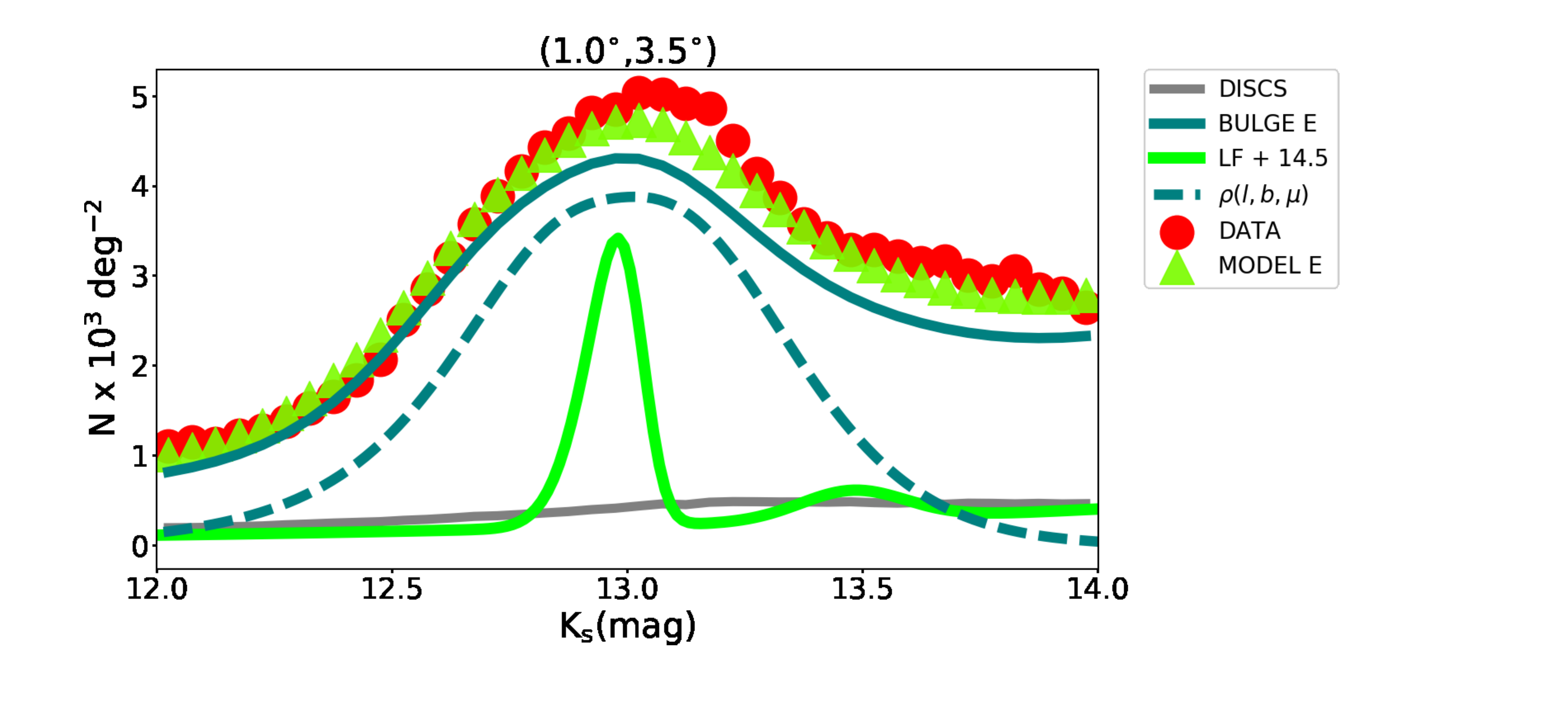}
\caption[Apparent magnitude distributions for the data and a global
  best fit models.]{Apparent magnitude distributions for the data
  (red) and a global best fit model, $M =
  < N_{d}>\times Scale + N_{B}$  (green triangles).  The apparent magnitude
  distribution $N_{\mathrm{B}}$ of the Bulge (dark green,
  Equation~\ref{Bulge_model}) is a convolution between the absolute
  magnitude LF (green, labeled `LF+14.5', where 14.5
  is the distance modulus added to the absolute magnitudes
  $M_{K_{s}}$) and the density law $\rho_{B}(l,b,D)$ (dotted dark green,
  Equation~\ref{E}). The discs $< N_{d}>\times Scale$ (labeled `DISCS') are shown in gray.}
 \label{LF_shift}
\end{figure} 
%

%
\begin{figure*}
\centering
\hspace*{-0.0in}
\includegraphics[scale = 0.14]{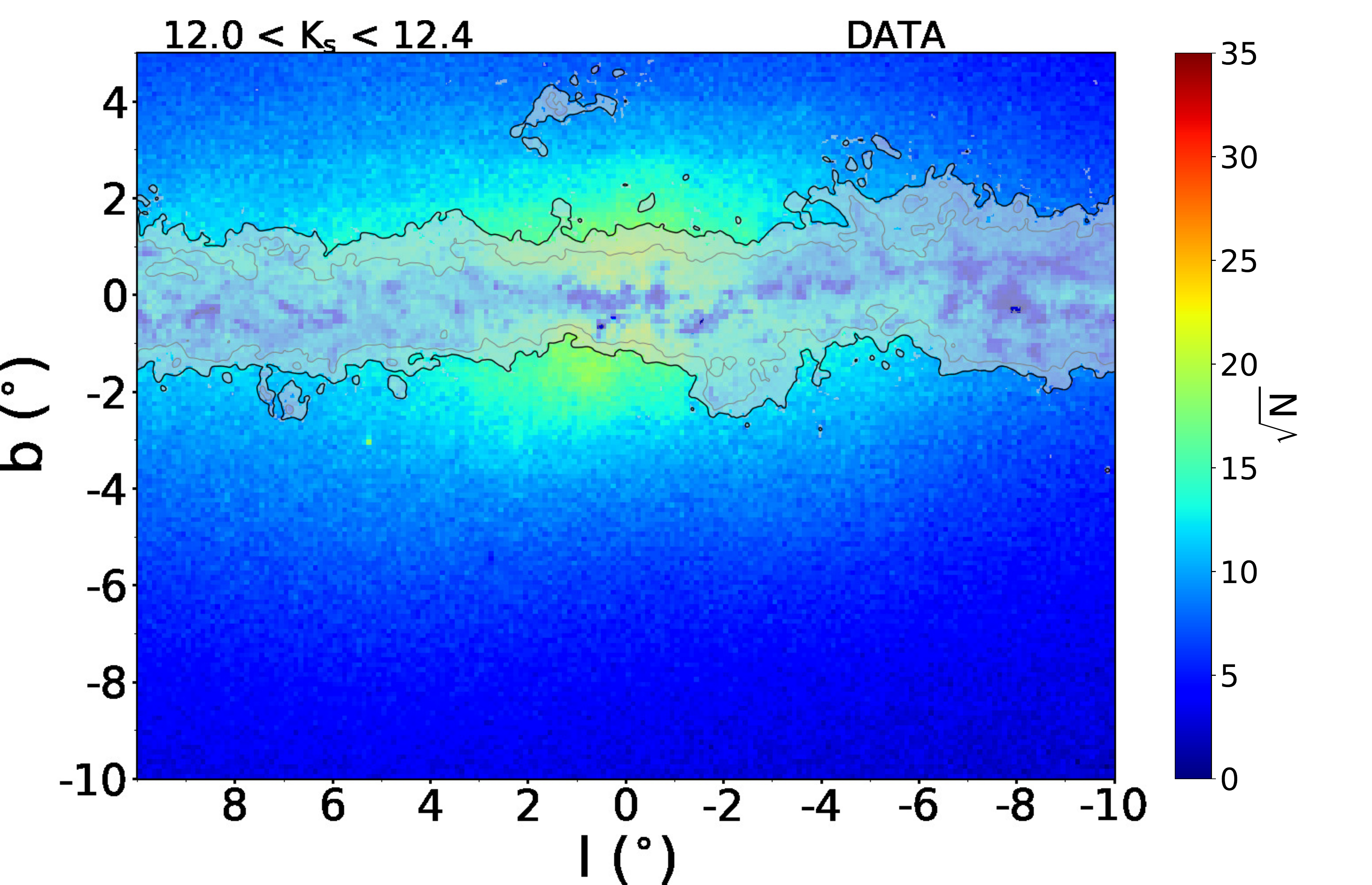}
\hspace*{-0.4in}
\includegraphics[scale=0.14]{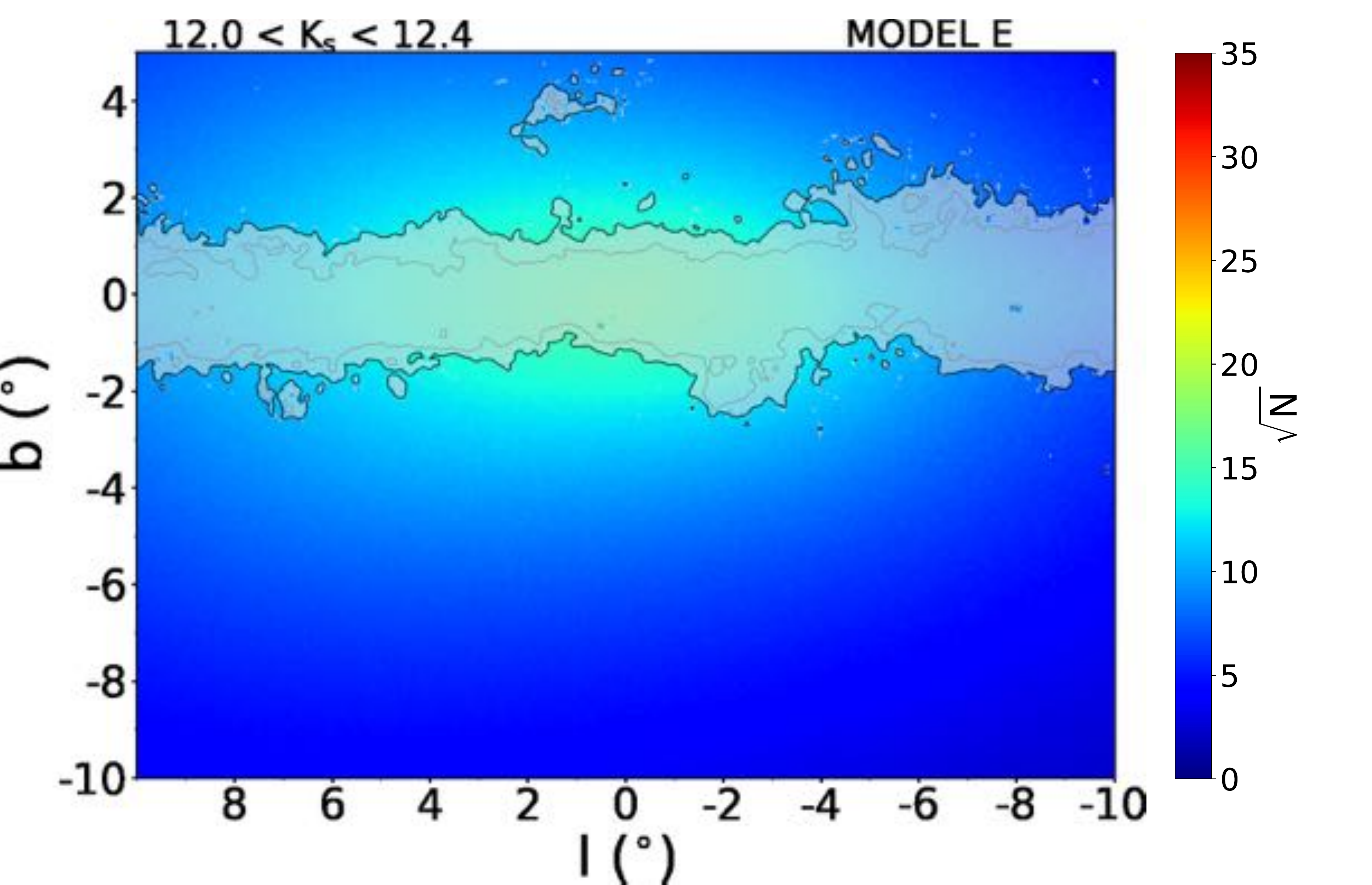}
\hspace*{-0.4in}
\includegraphics[scale=0.14]{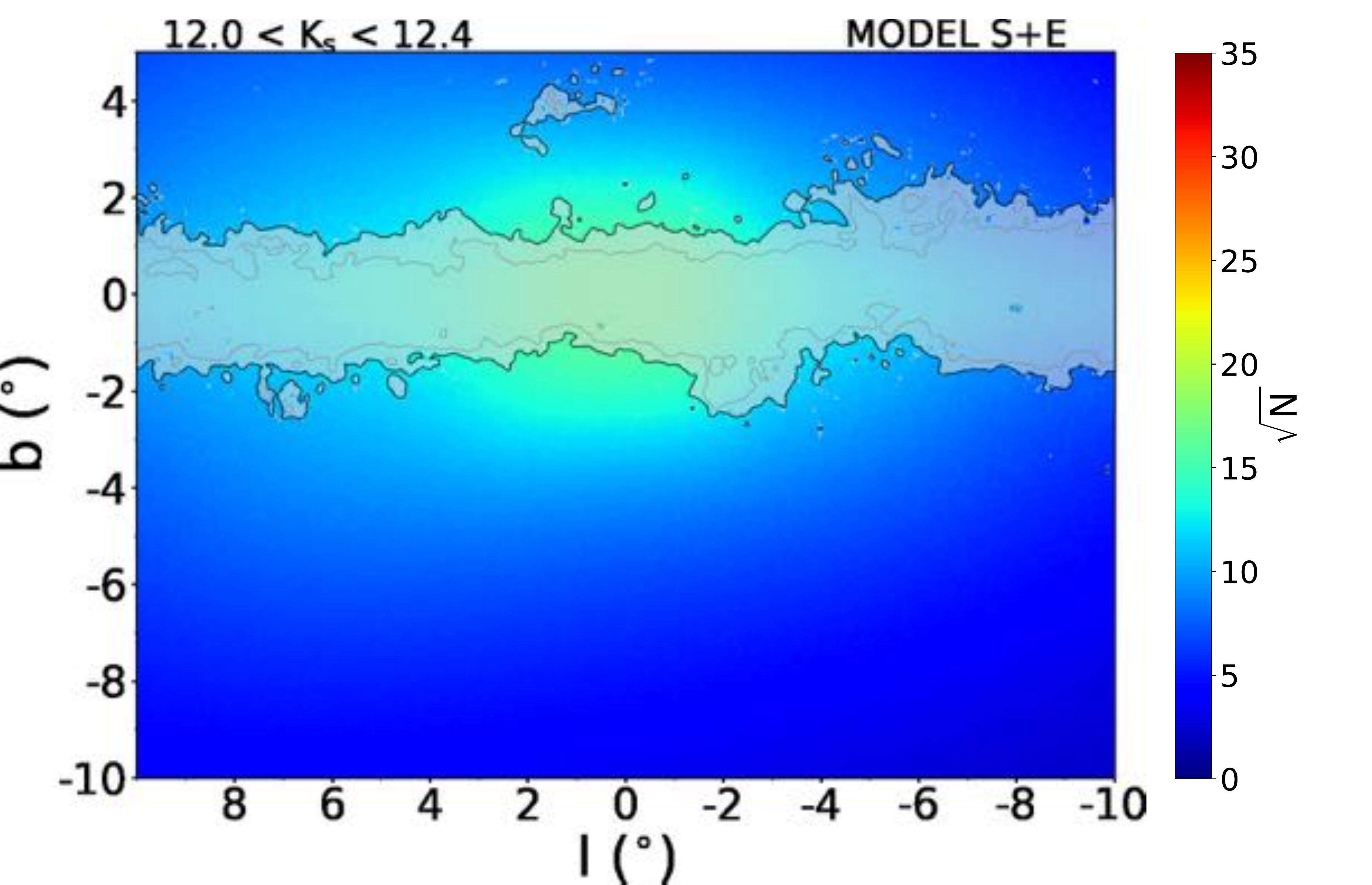} \\
\hspace*{-0.0in}
\includegraphics[scale = 0.14]{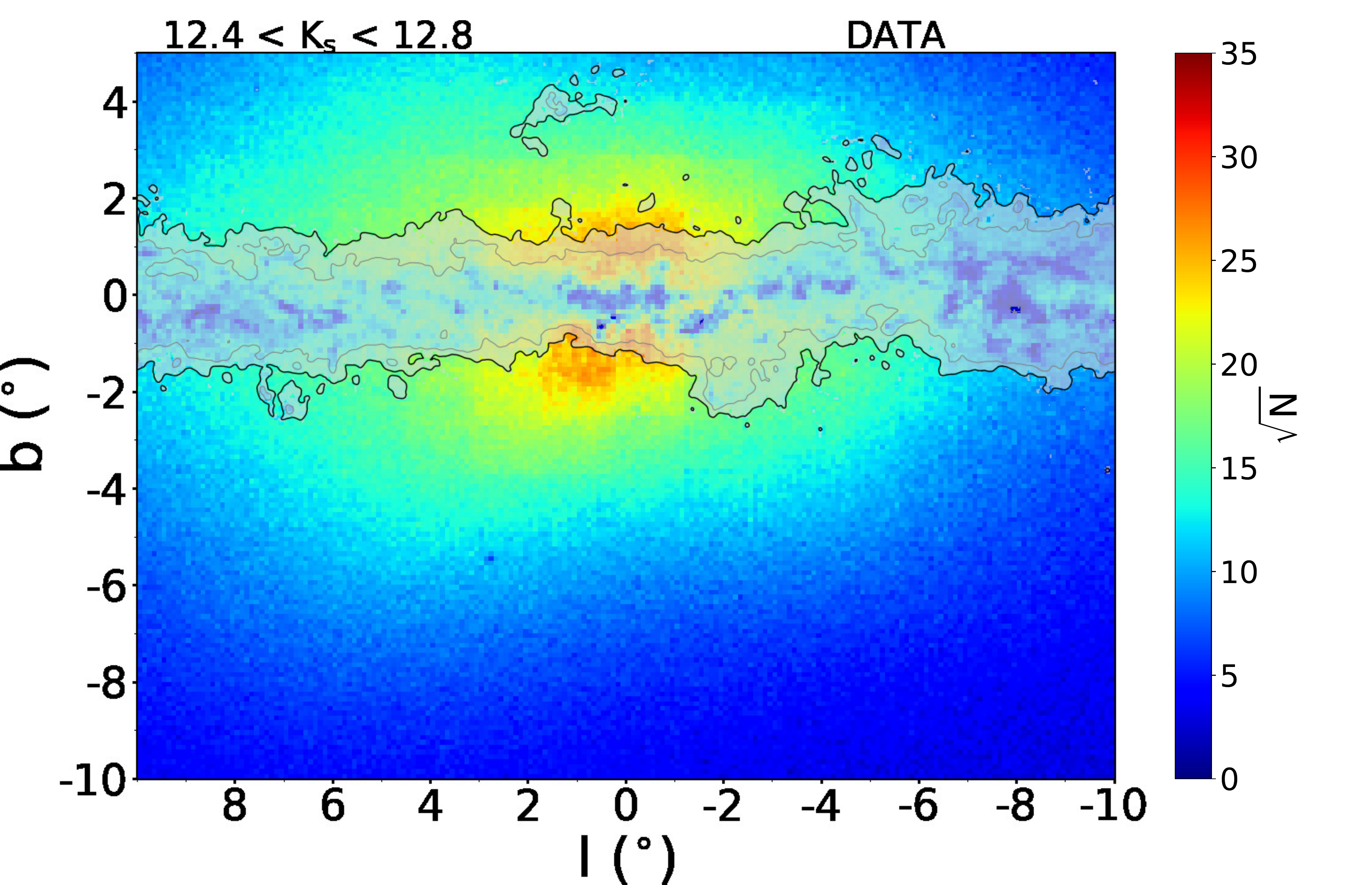}
\hspace*{-0.4in}
\includegraphics[scale=0.14]{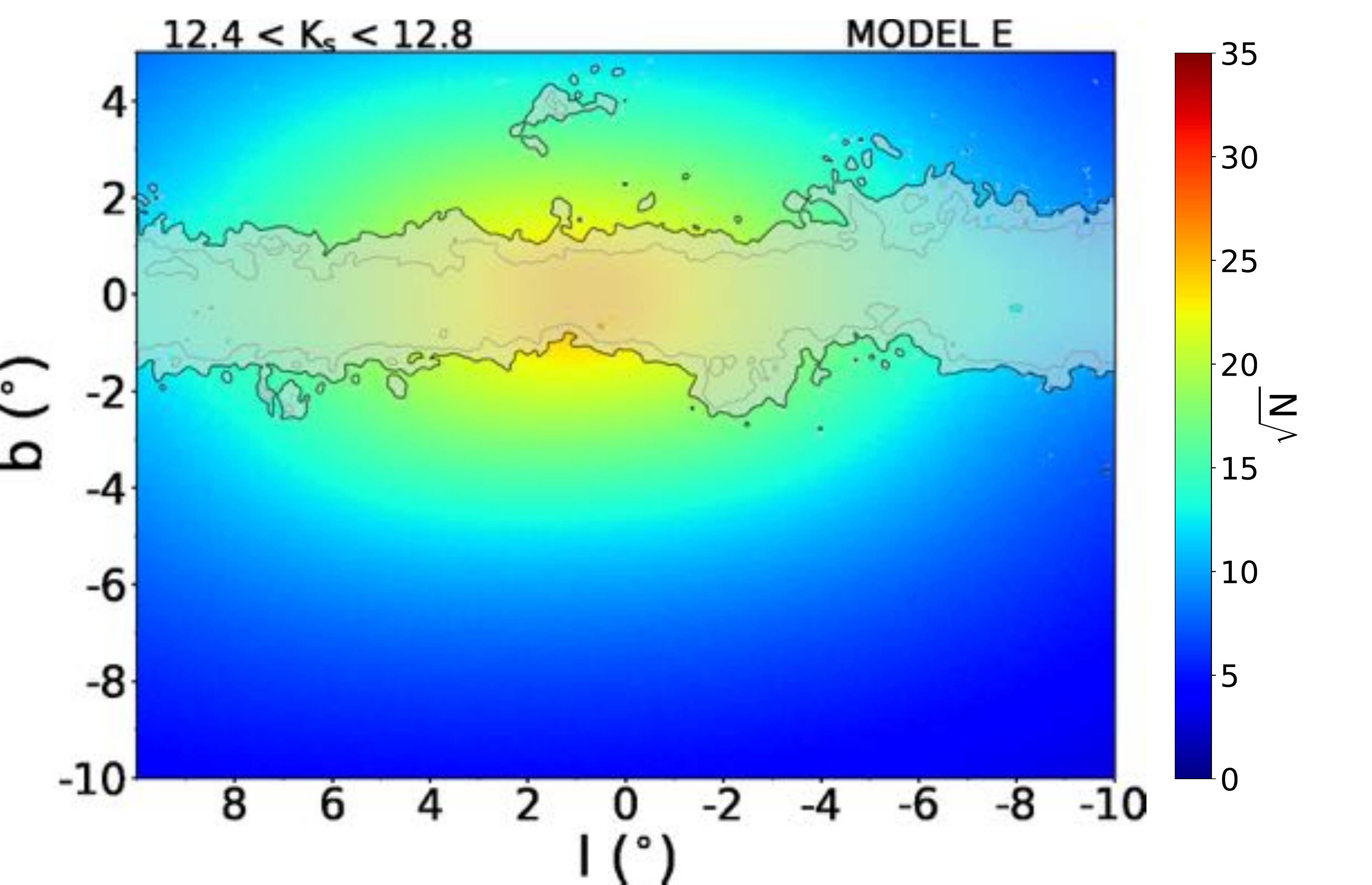}
\hspace*{-0.4in}
\includegraphics[scale=0.14]{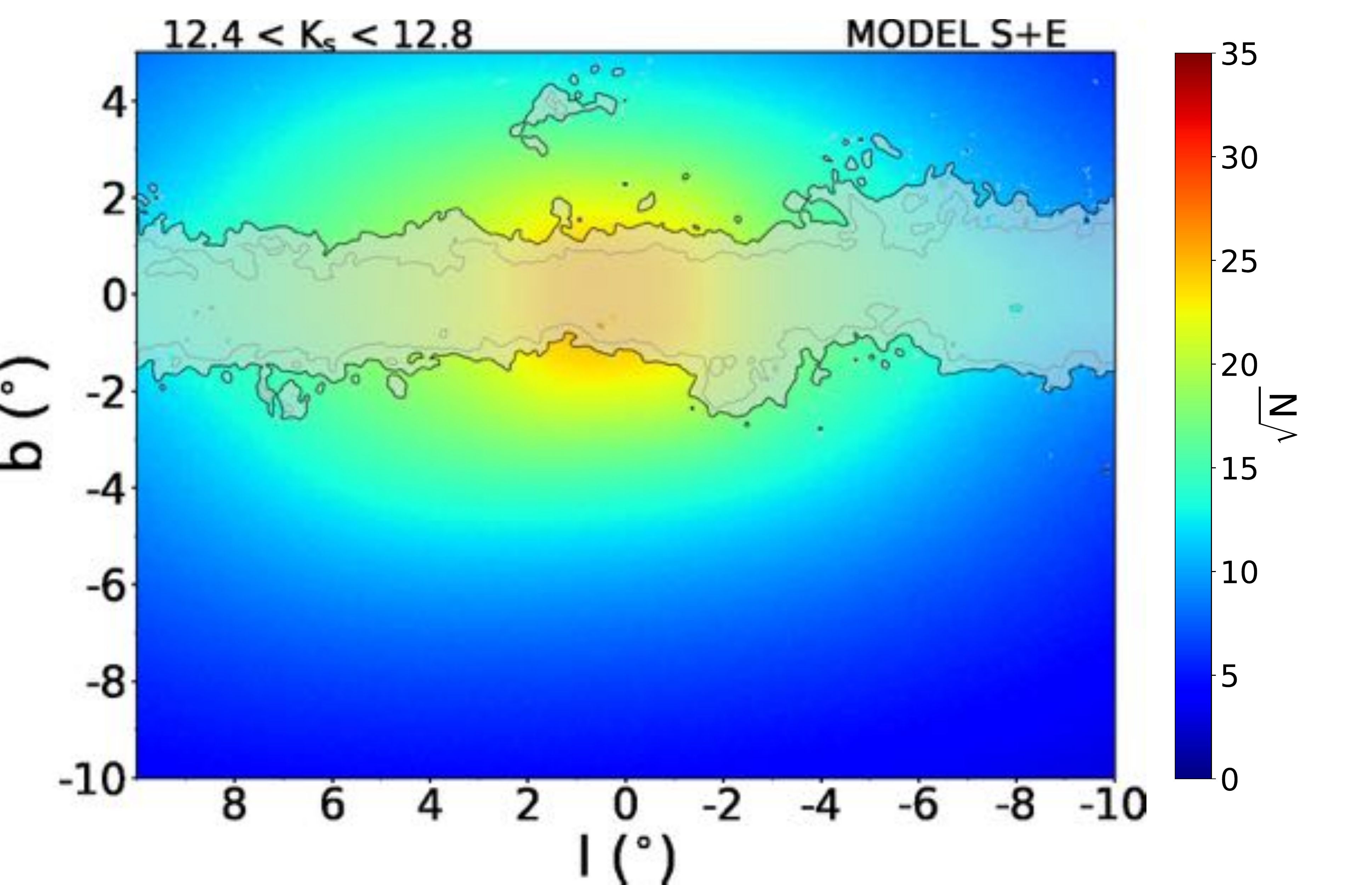} \\
\hspace*{-0.0in}
\includegraphics[scale = 0.14]{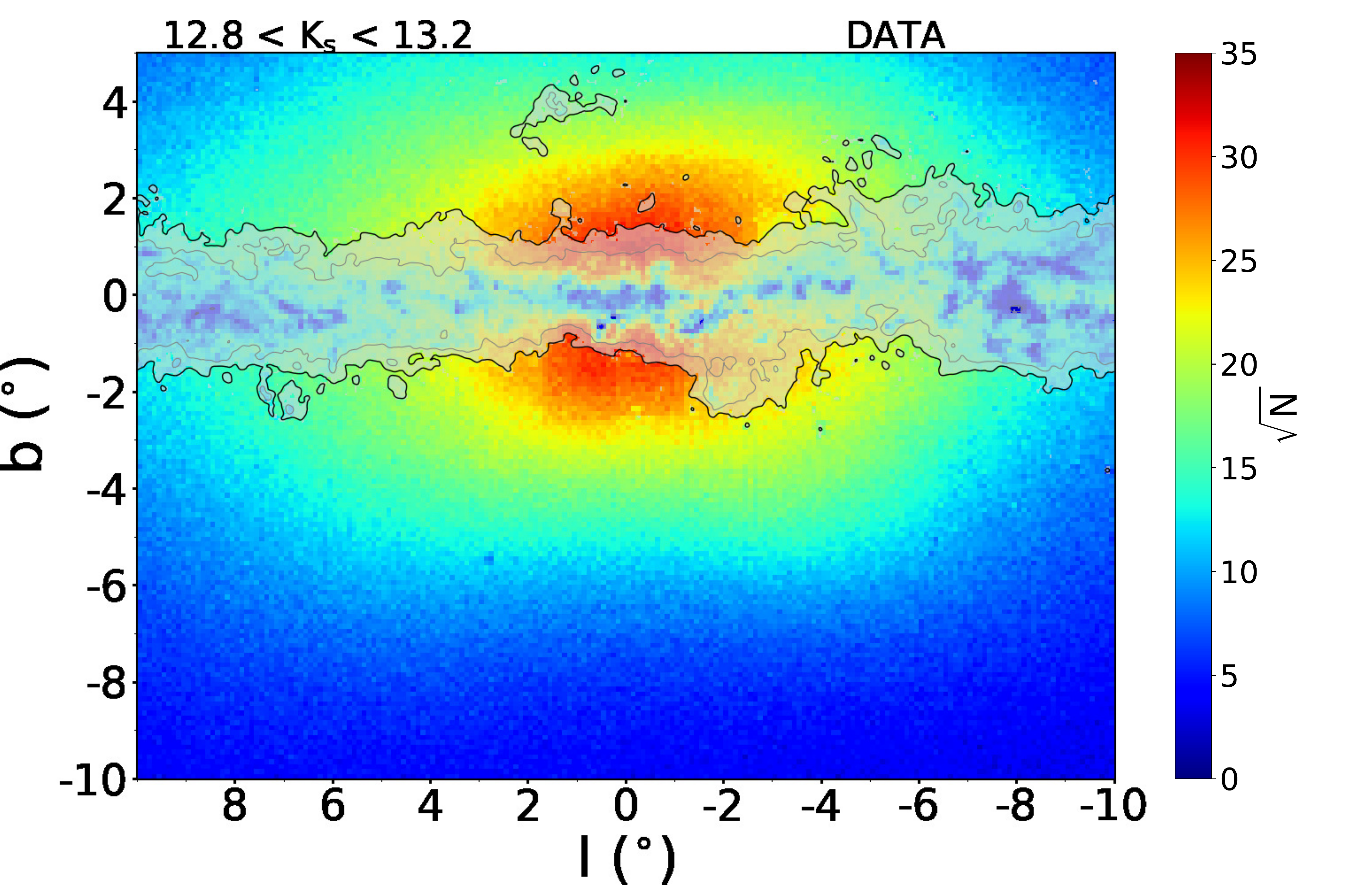}
\hspace*{-0.4in}
\includegraphics[scale=0.14]{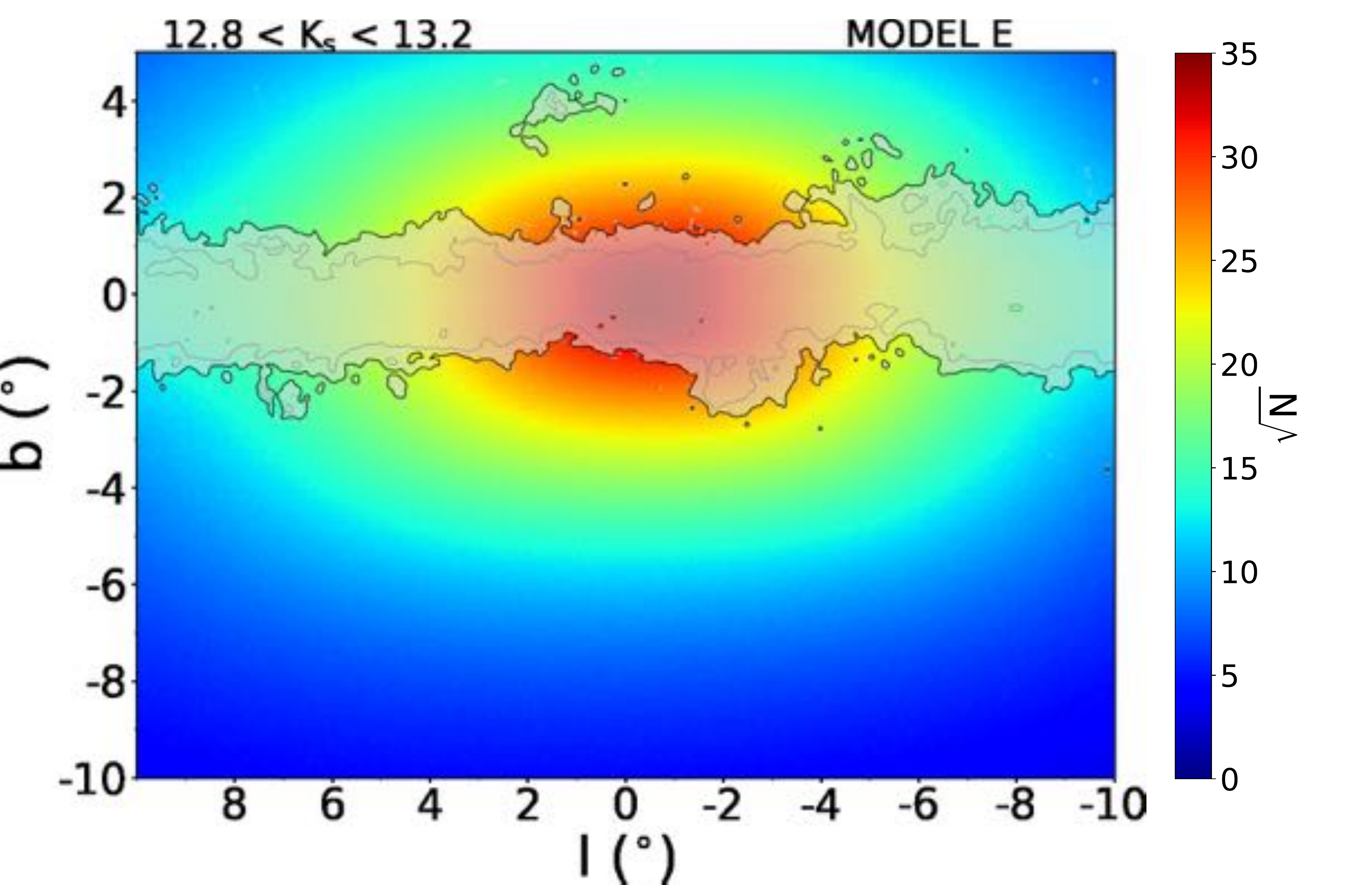}
\hspace*{-0.4in}
\includegraphics[scale=0.14]{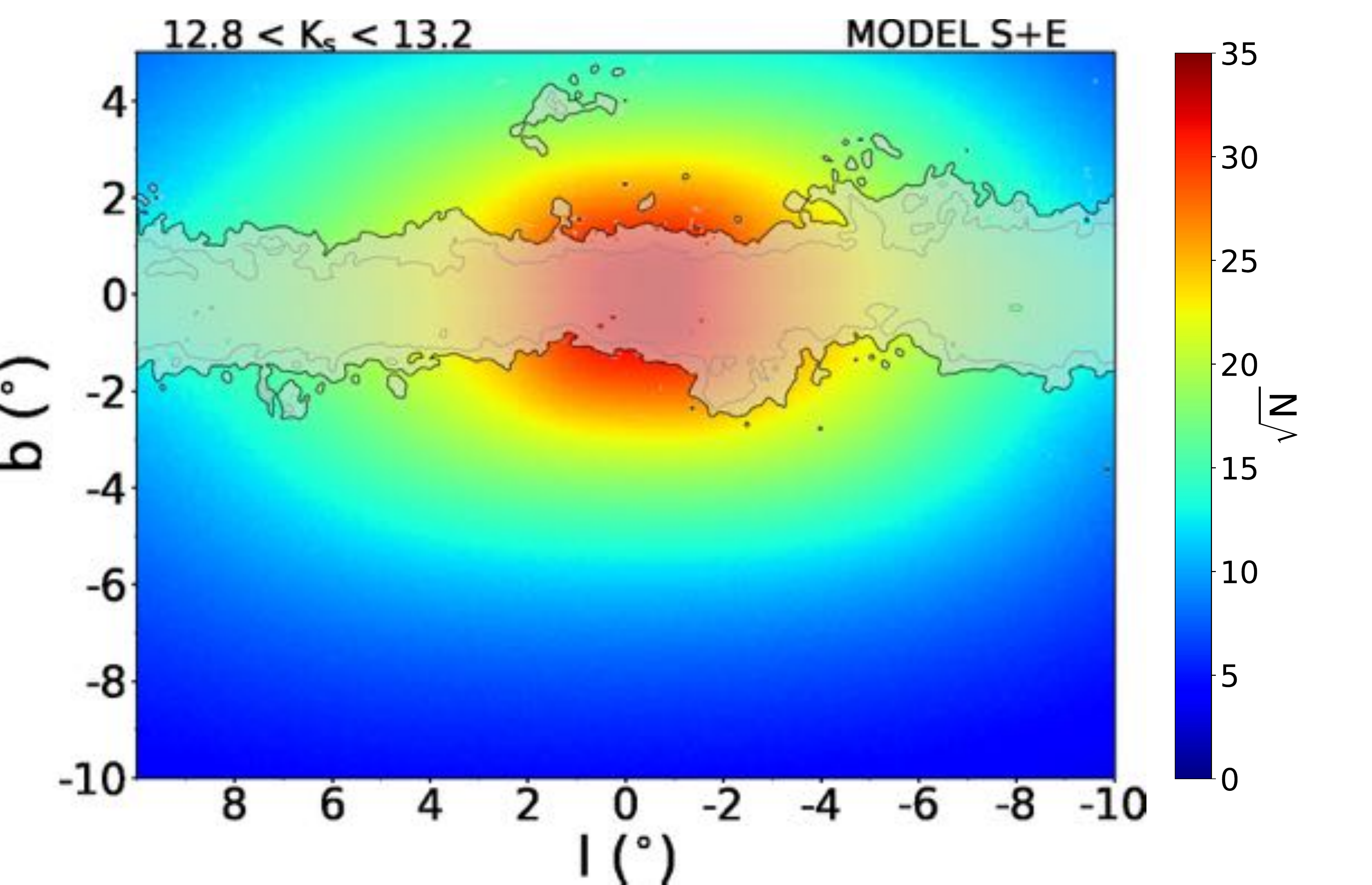} \\
\hspace*{-0.0in}
\includegraphics[scale = 0.14]{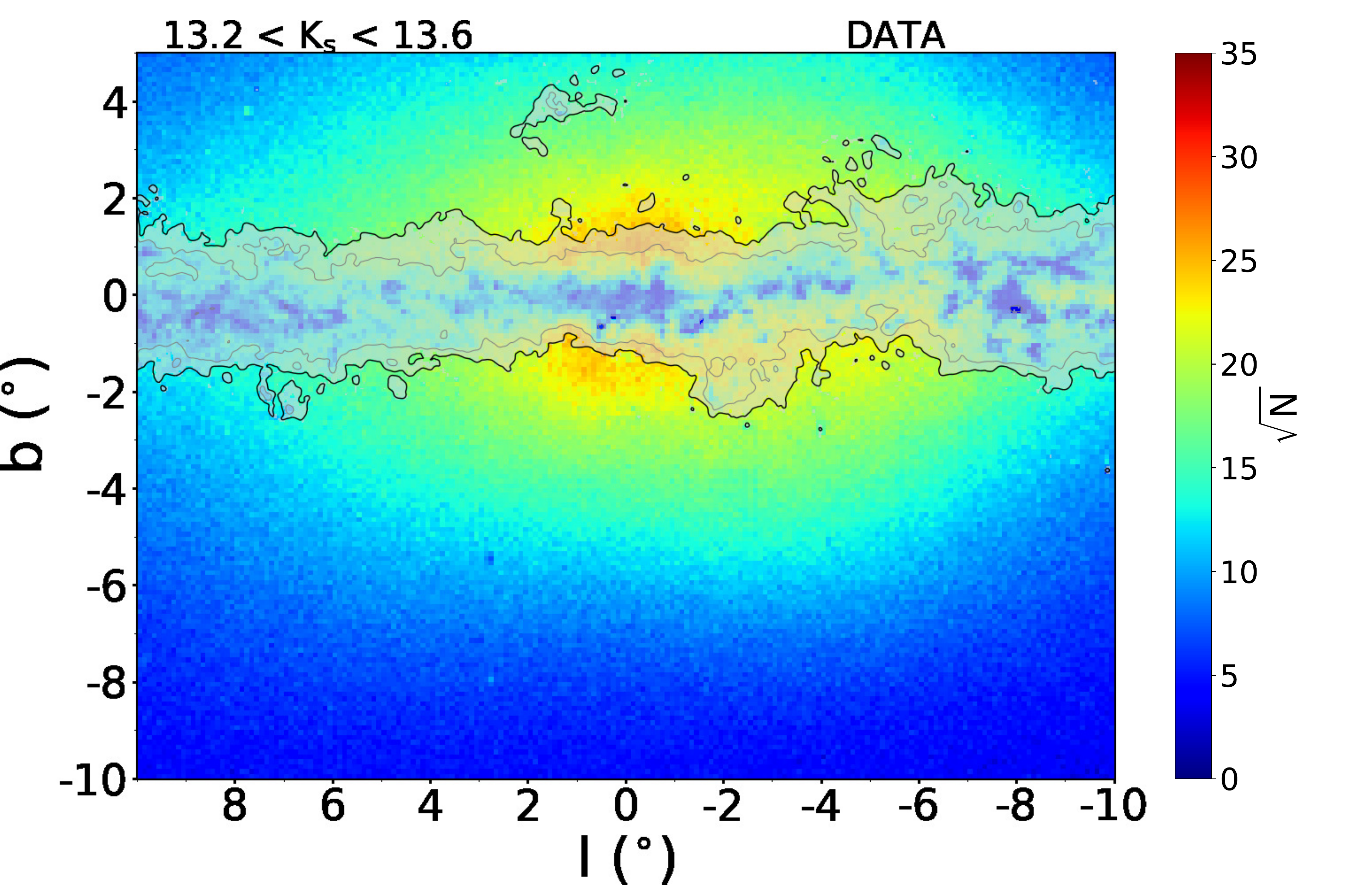}
\hspace*{-0.4in}
\includegraphics[scale=0.14]{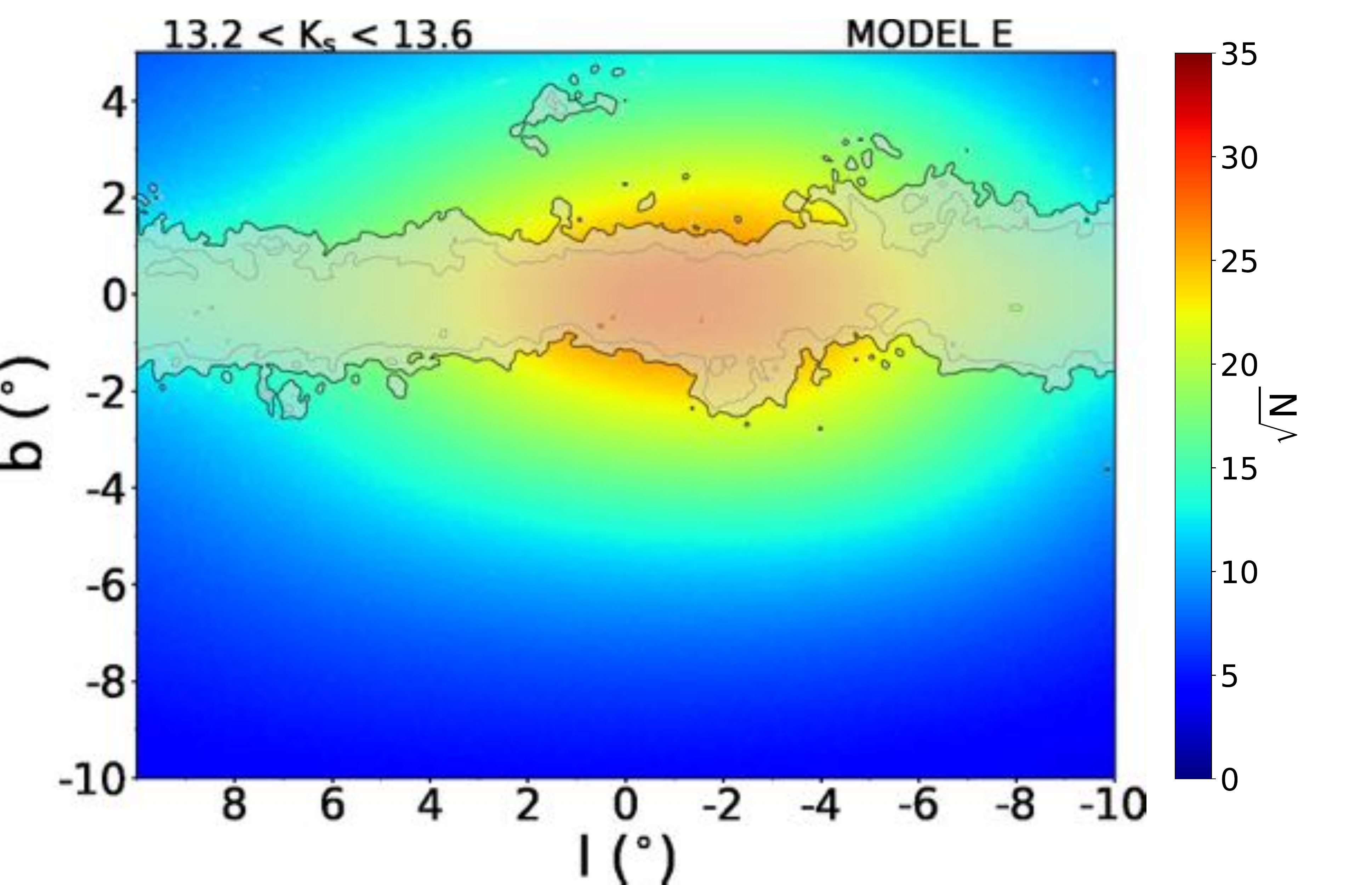}
\hspace*{-0.4in}
\includegraphics[scale=0.14]{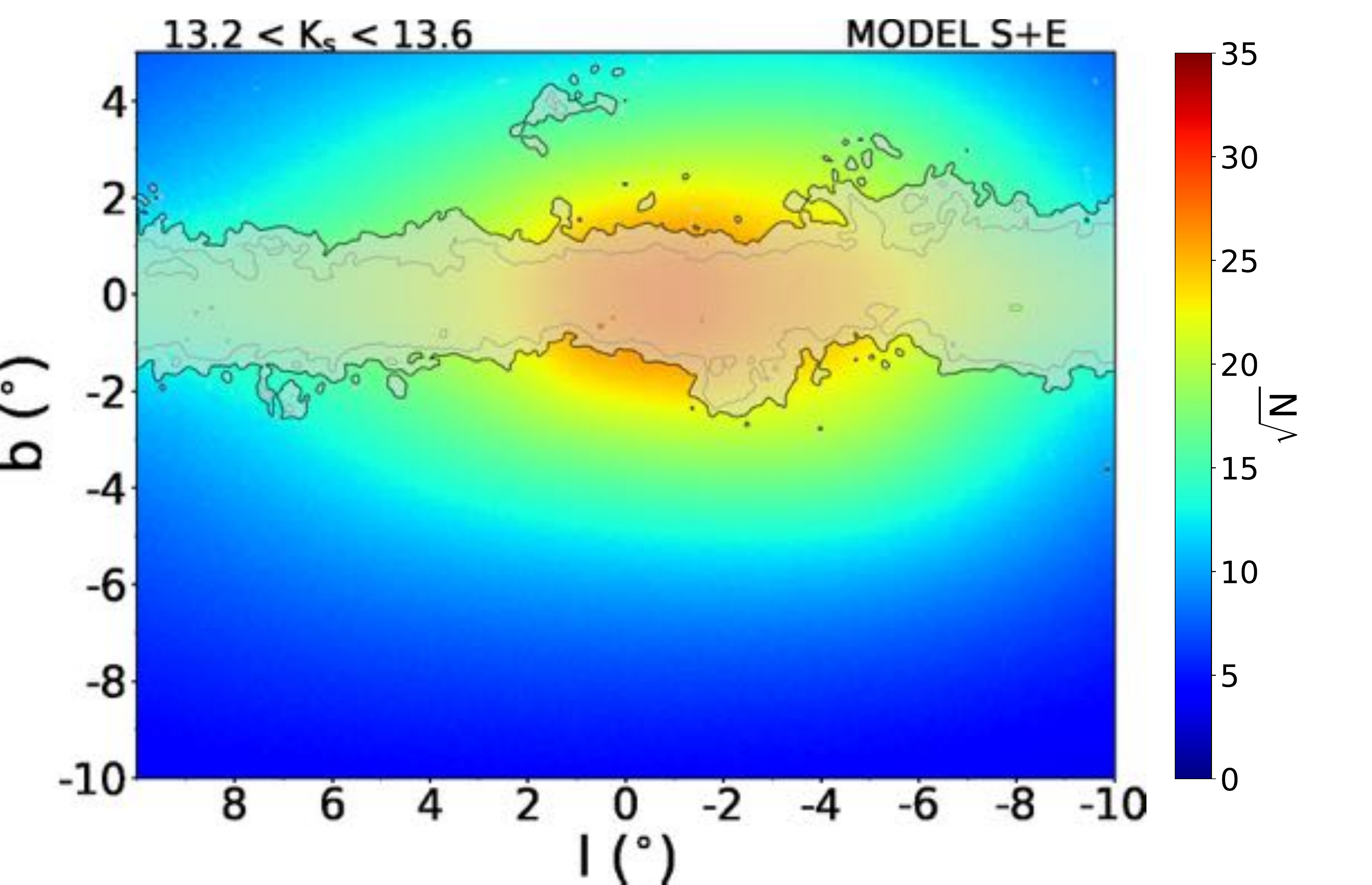} \\
\hspace*{-0.0in}
\includegraphics[scale = 0.14]{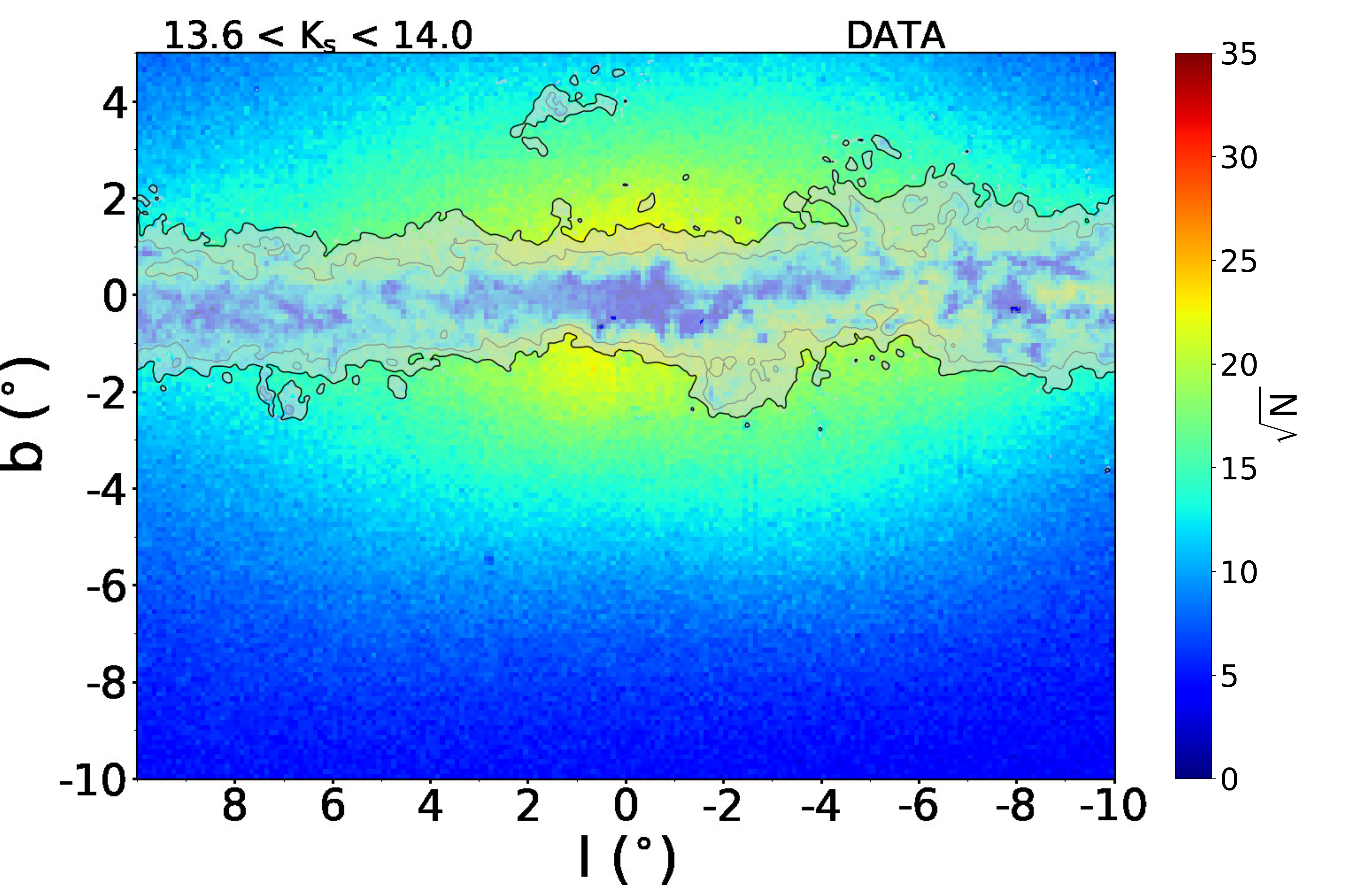}
\hspace*{-0.4in}
\includegraphics[scale=0.14]{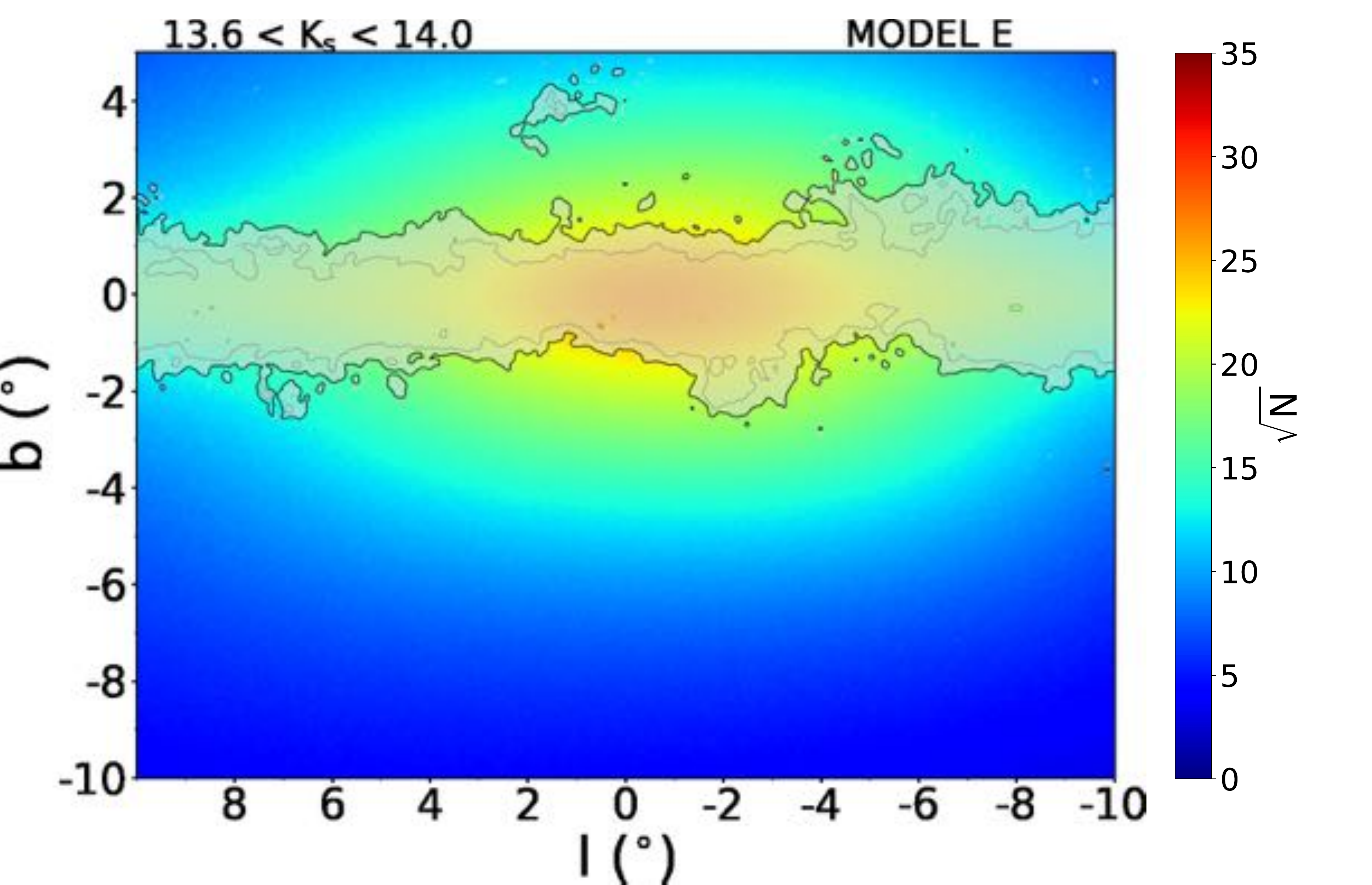}
\hspace*{-0.4in}
\includegraphics[scale=0.14]{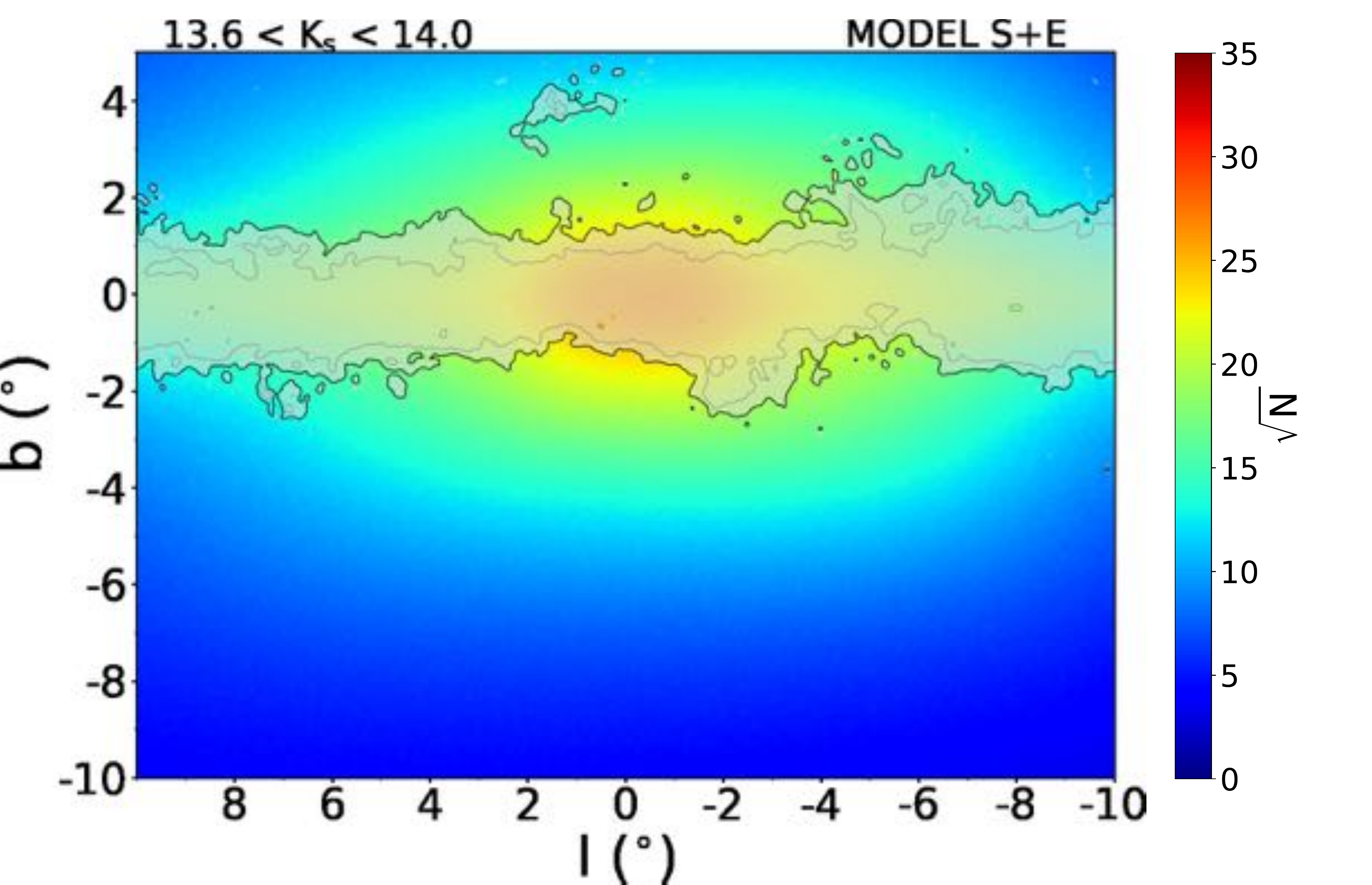} \\
\caption[Data and best fits for models $E$ and $(S+E)$ in Galactic
  coordinates ]{Data number density distribution ($left$ $panels$) and best fit models $E$ ($middle$) and $(S+E)$ ($right$) in Galactic
  coordinates. The 5 rows are slices of $K_{s}$ magnitudes between
  $12< K_{s}< 14$ with the brightest magnitude slice on the top row.
  High extinction causes incompleteness close to the Galactic
  plane as it can be seen in the data, therefore
  the region within the $E(J'-K'_{s})=1.0$ contour (thick gray) was excluded from
  all analysis. The fitted model appears good in nearly all regions (see discussion in Section 6.5);
  the residuals between the data and each model are shown Figure~\ref{diff}.}
\label{models}
\end{figure*}

\begin{figure*}
\centering
\hspace*{-0.0in}
\includegraphics[scale=0.14]{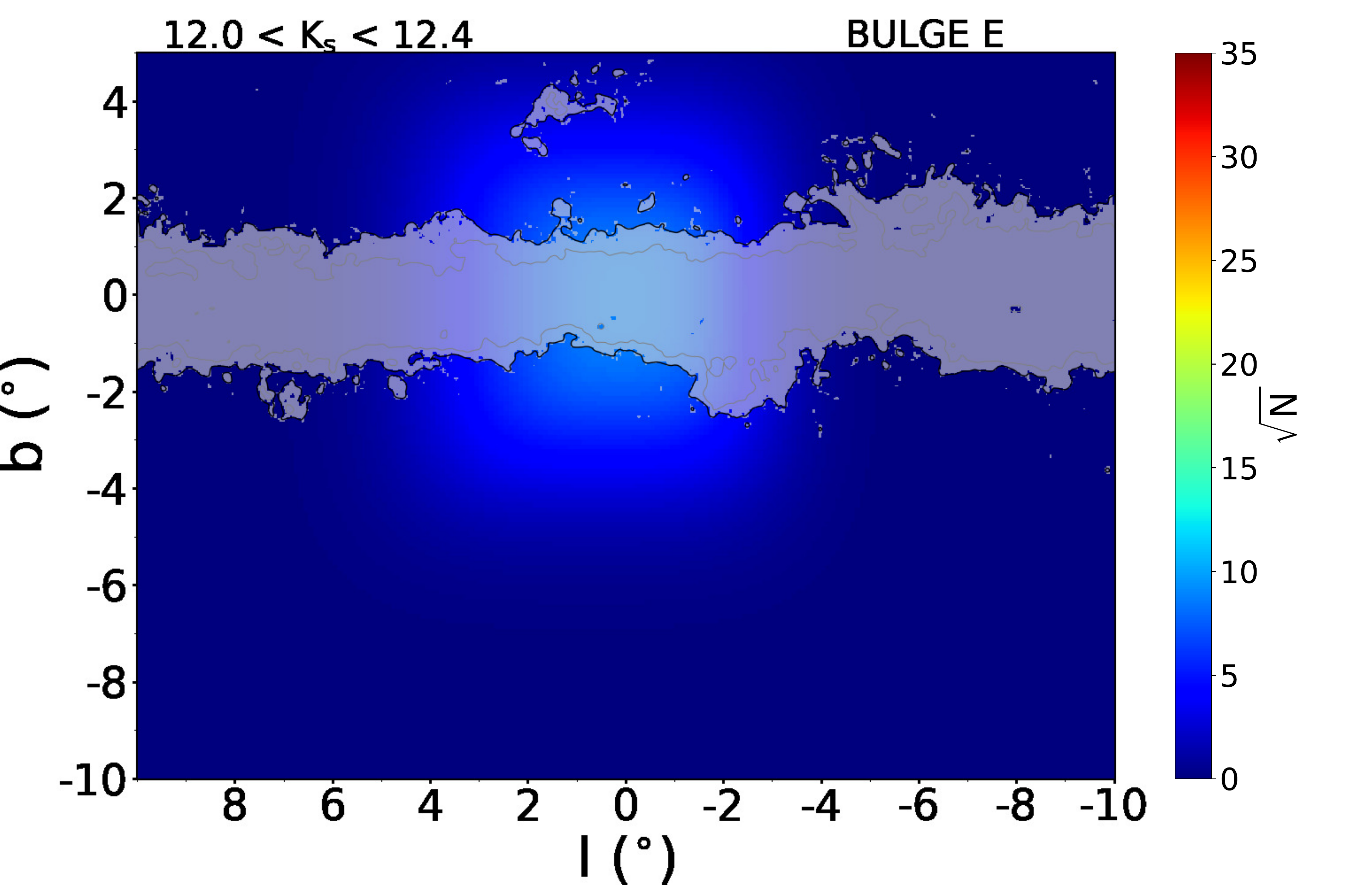}
\hspace*{-0.2in}
\includegraphics[scale=0.14]{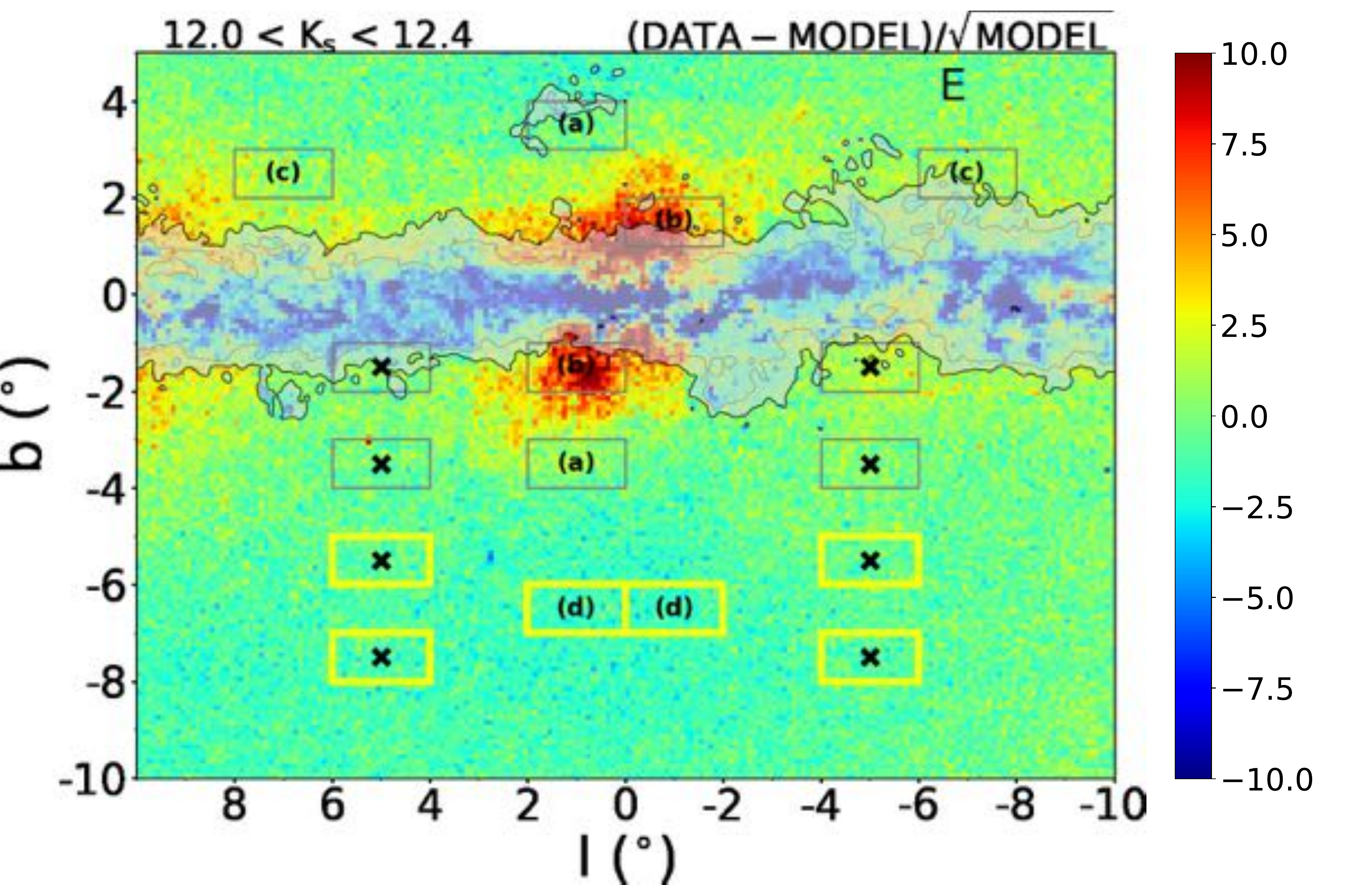}
\hspace*{-0.4in}
\includegraphics[scale=0.14]{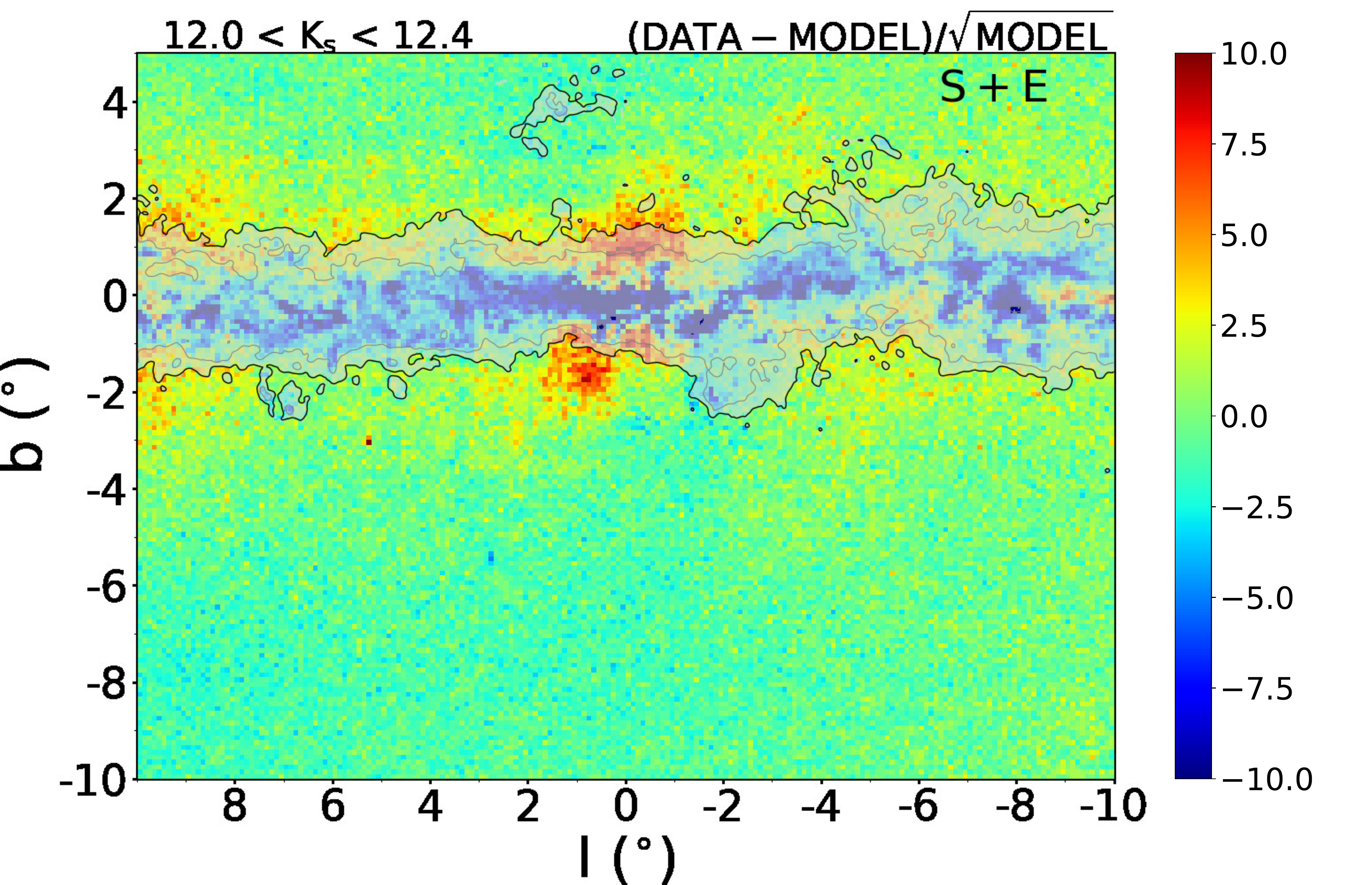} \\
\hspace*{-0.0in}
\includegraphics[scale=0.14]{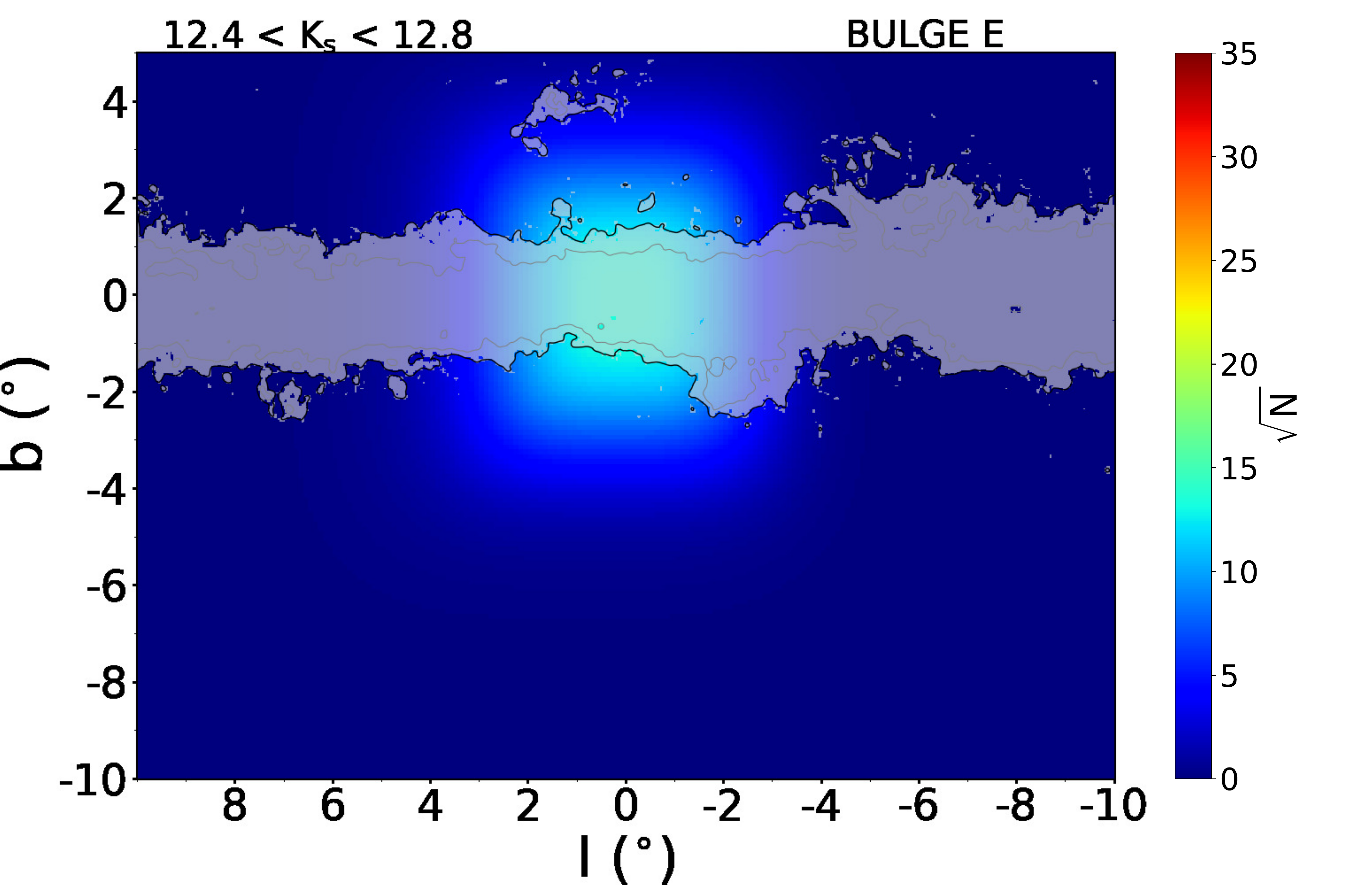}
\hspace*{-0.2in}
\includegraphics[scale=0.14]{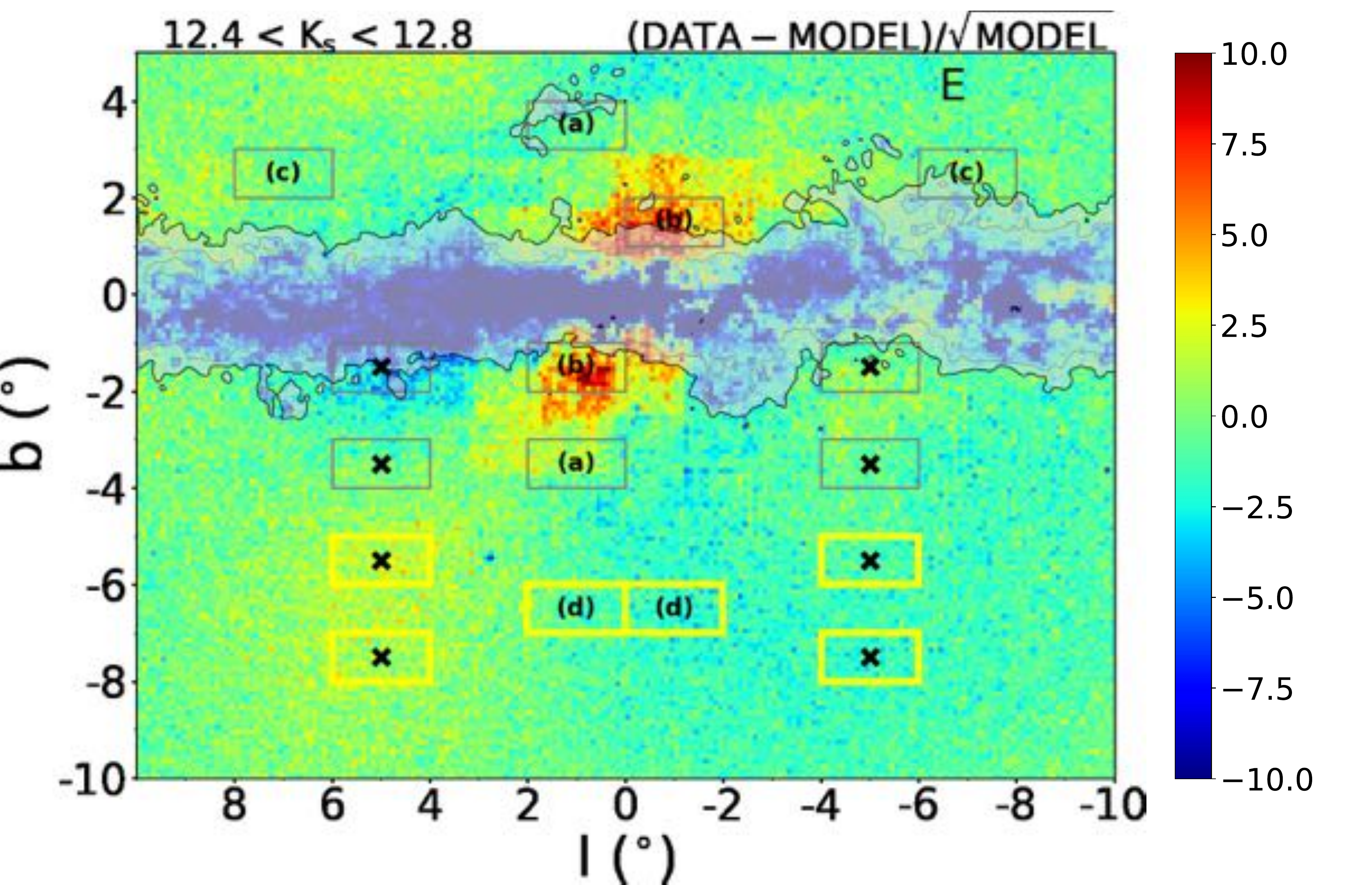}
\hspace*{-0.4in}
\includegraphics[scale=0.14]{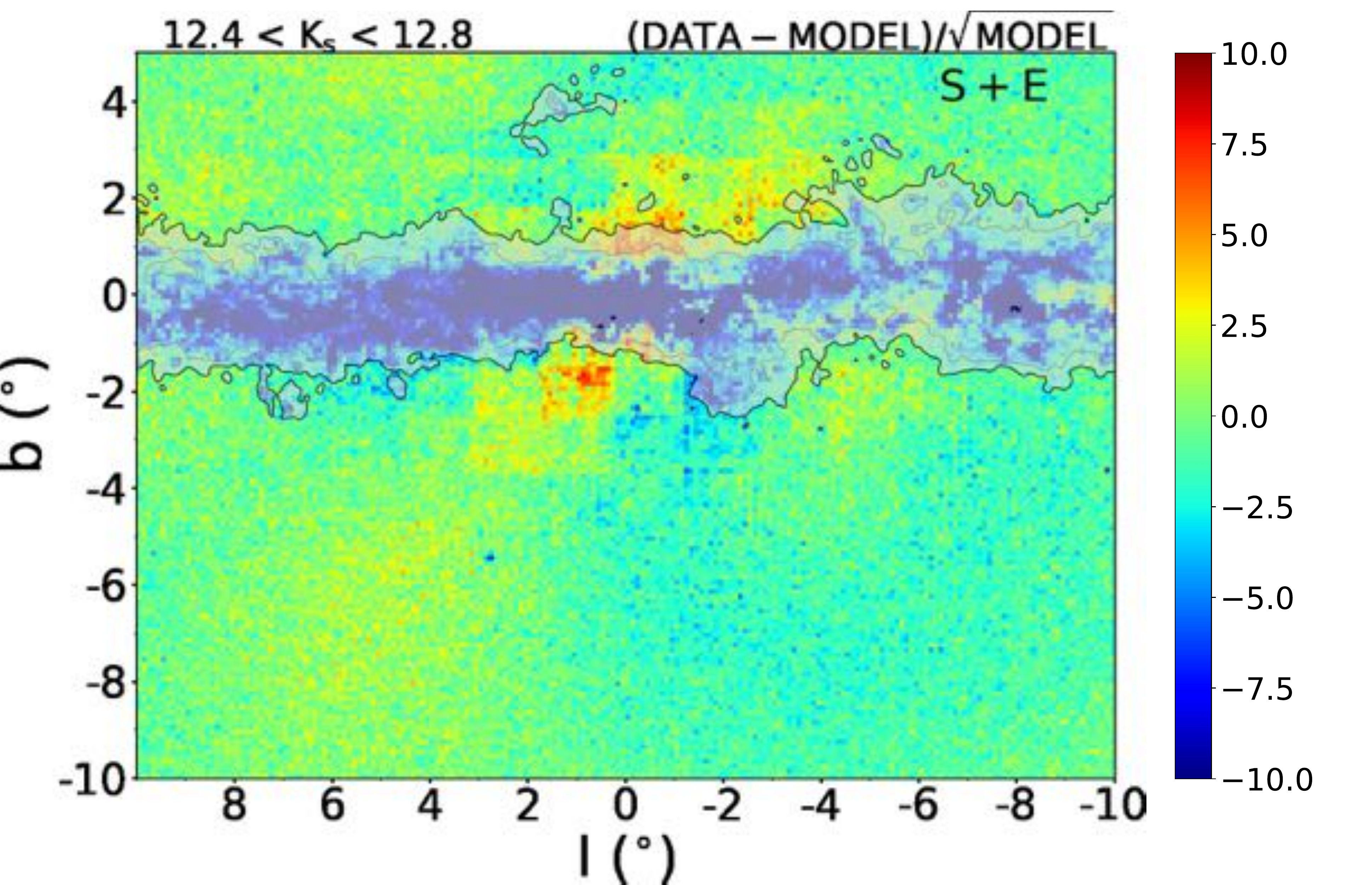} \\
\hspace*{-0.0in}
\includegraphics[scale=0.14]{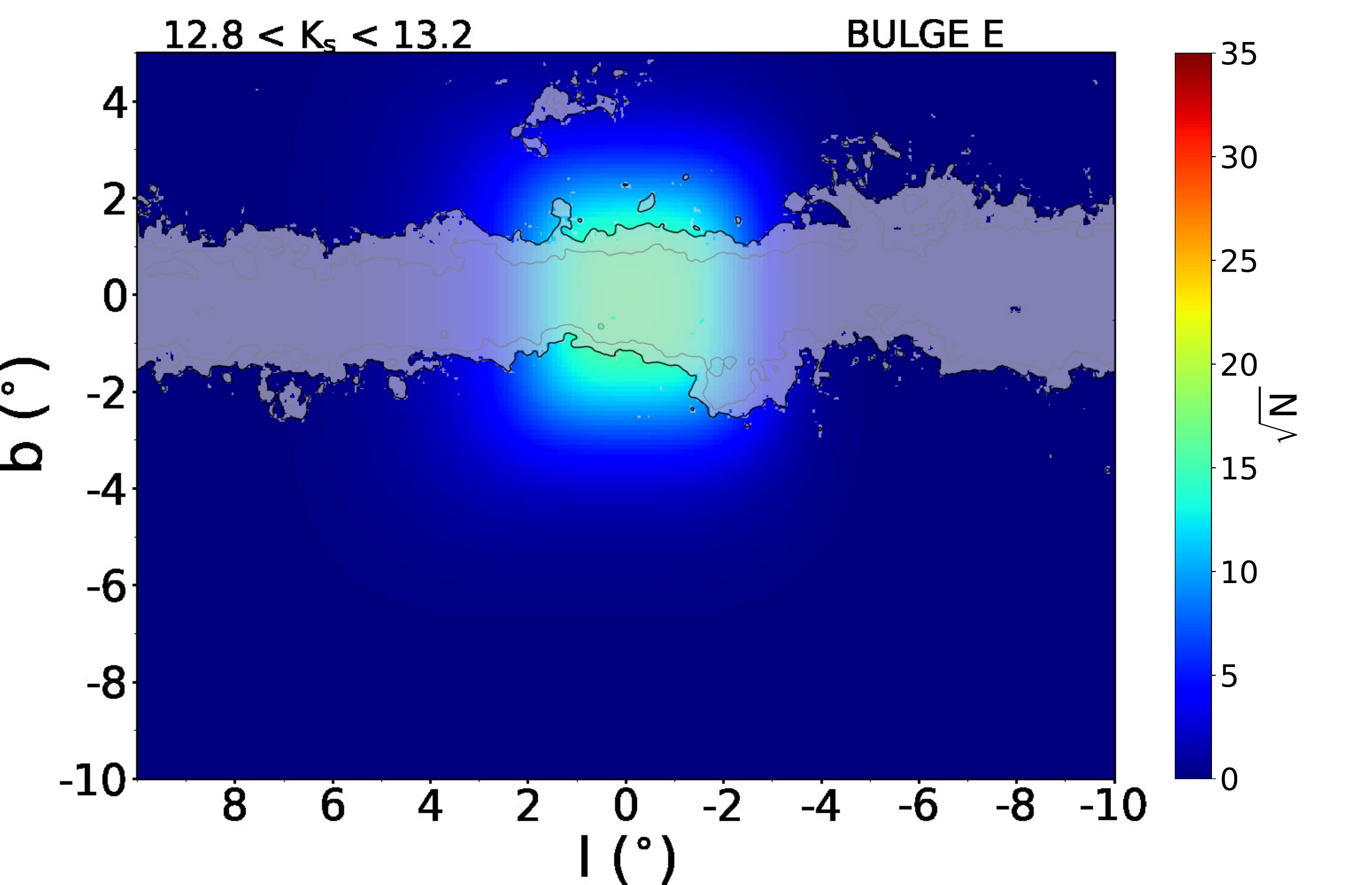}
\hspace*{-0.2in}
\includegraphics[scale=0.14]{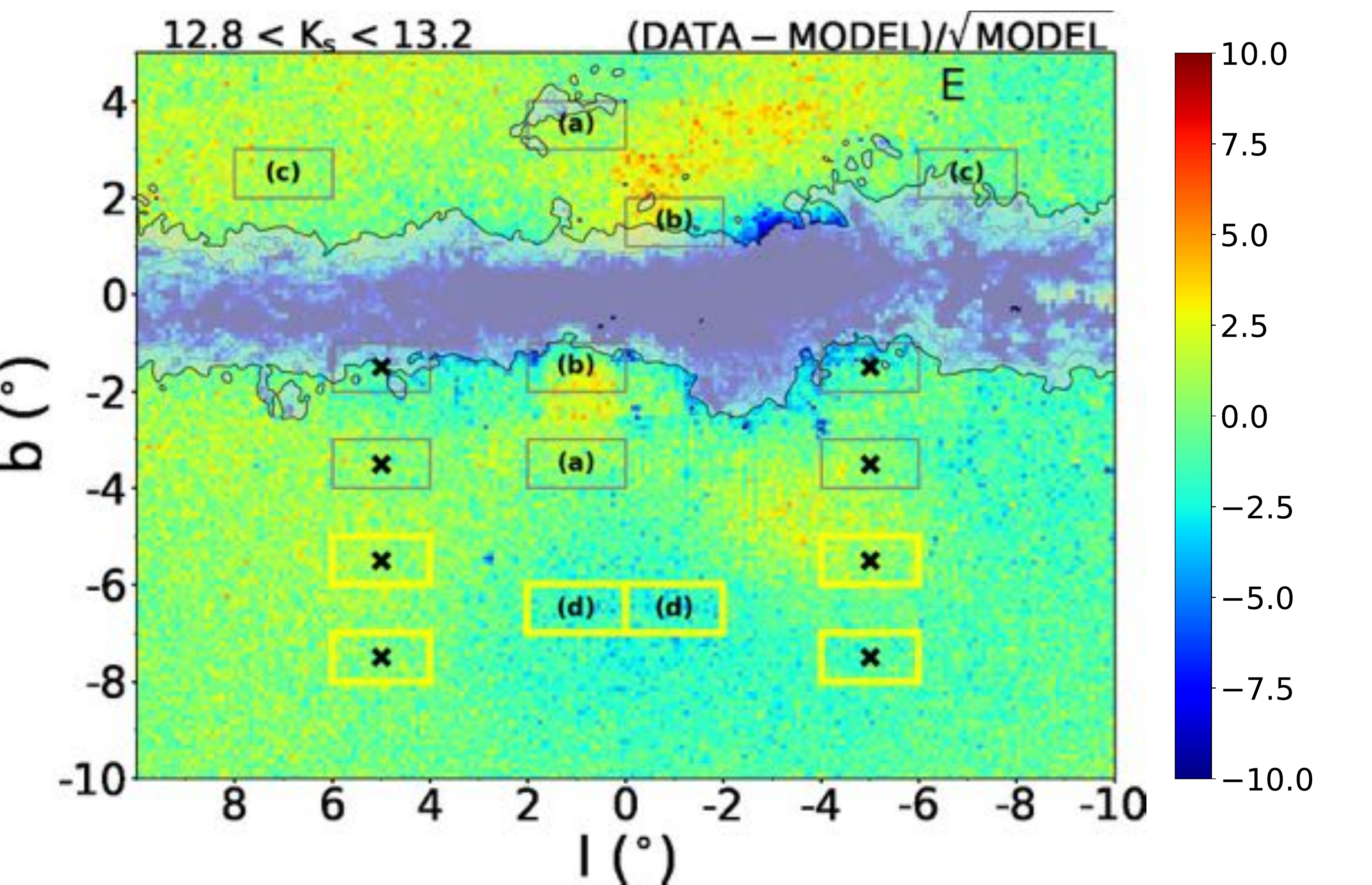}
\hspace*{-0.4in}
\includegraphics[scale=0.14]{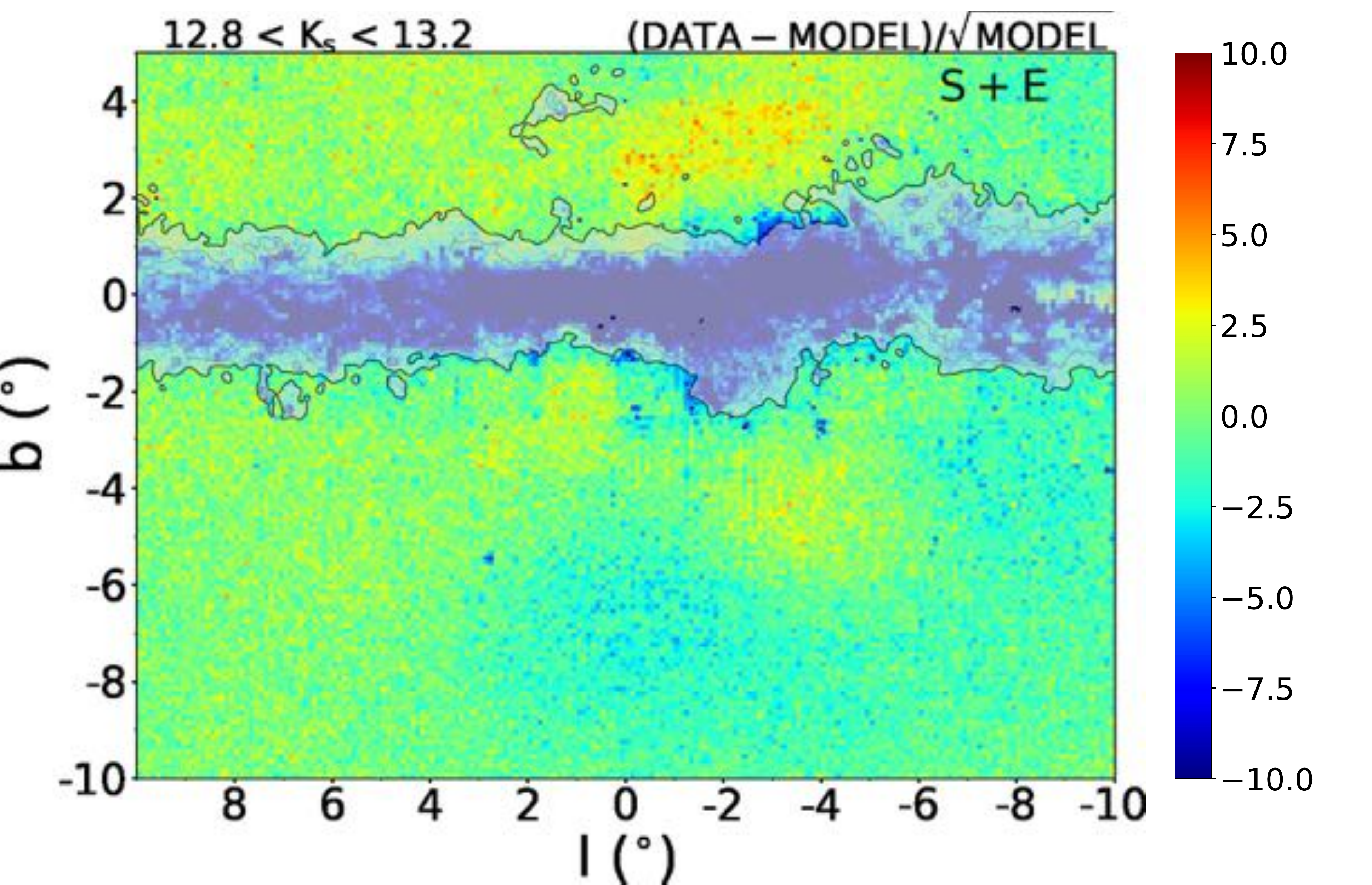} \\
\hspace*{-0.0in}
\includegraphics[scale=0.14]{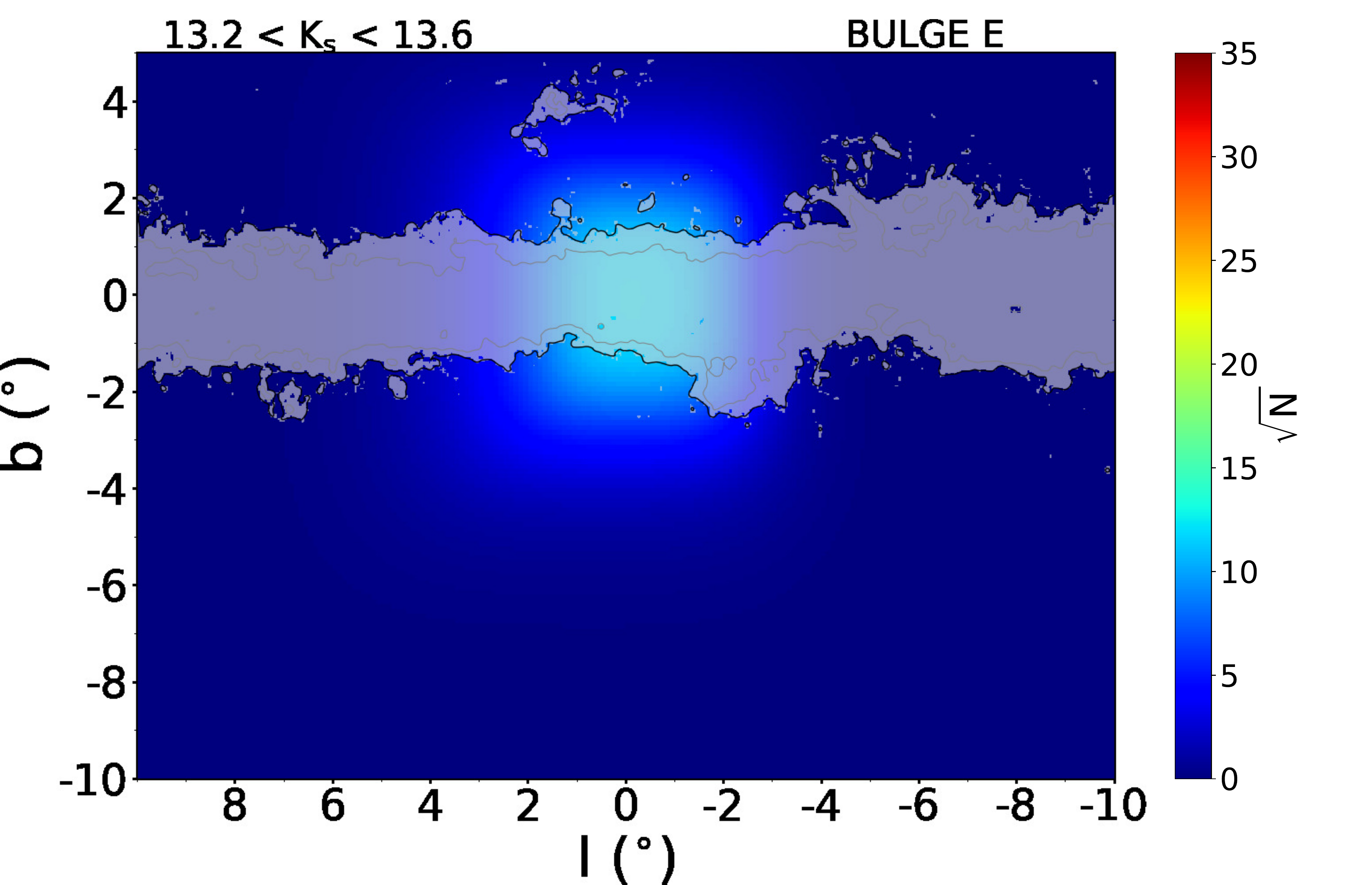}
\hspace*{-0.2in}
\includegraphics[scale=0.14]{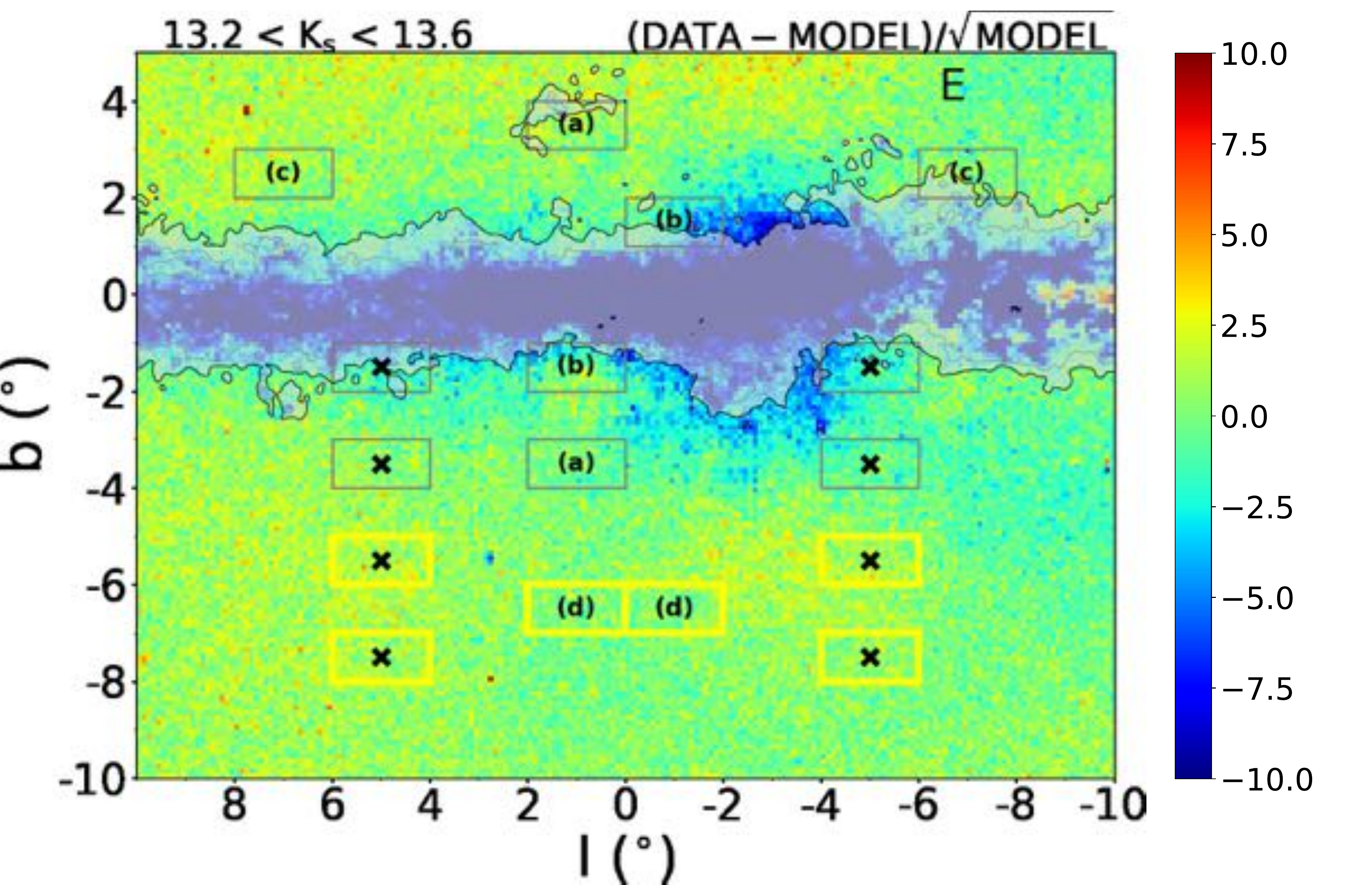}
\hspace*{-0.4in}
\includegraphics[scale=0.14]{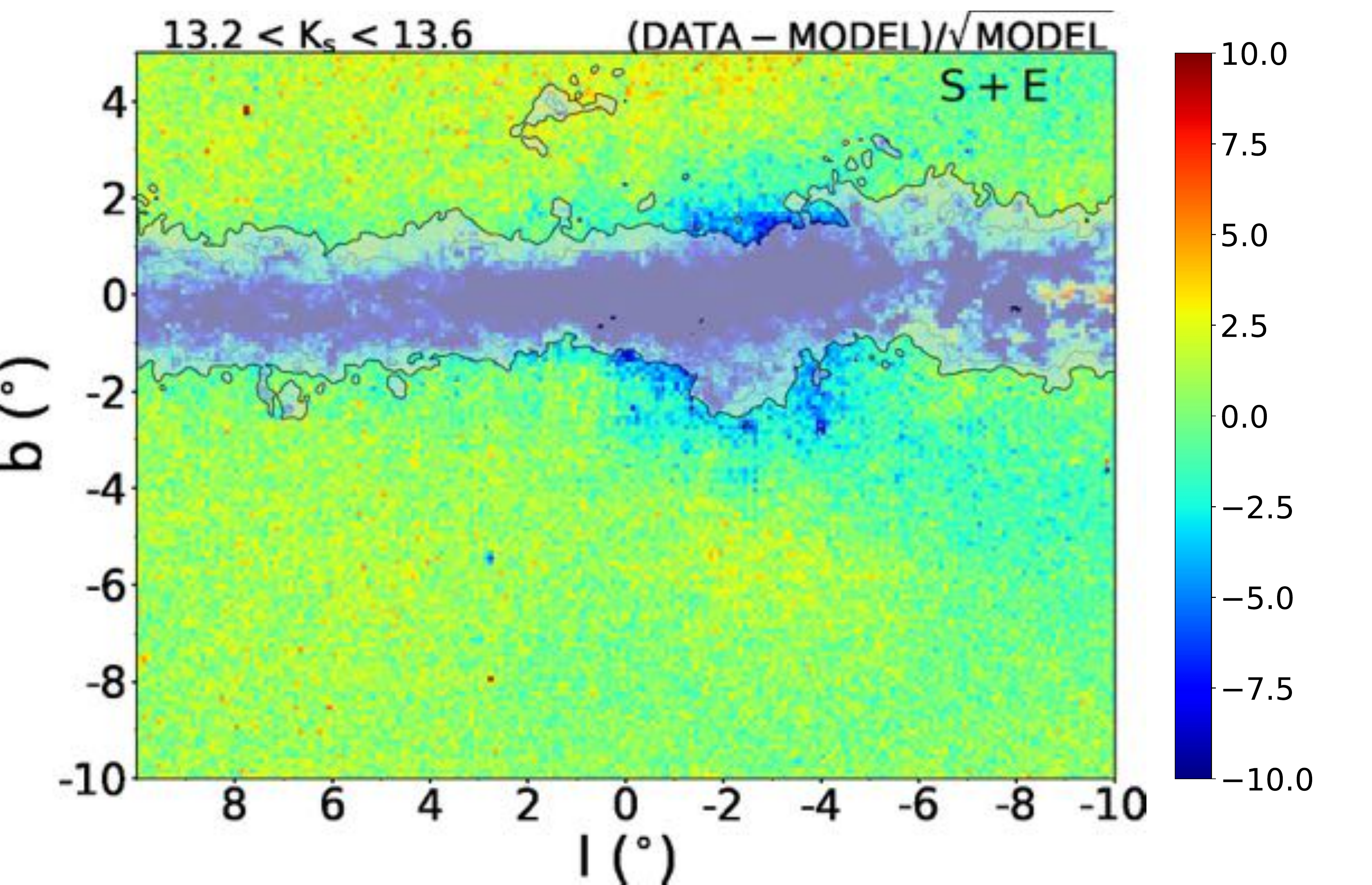} \\
\hspace*{-0.0in}
\includegraphics[scale=0.14]{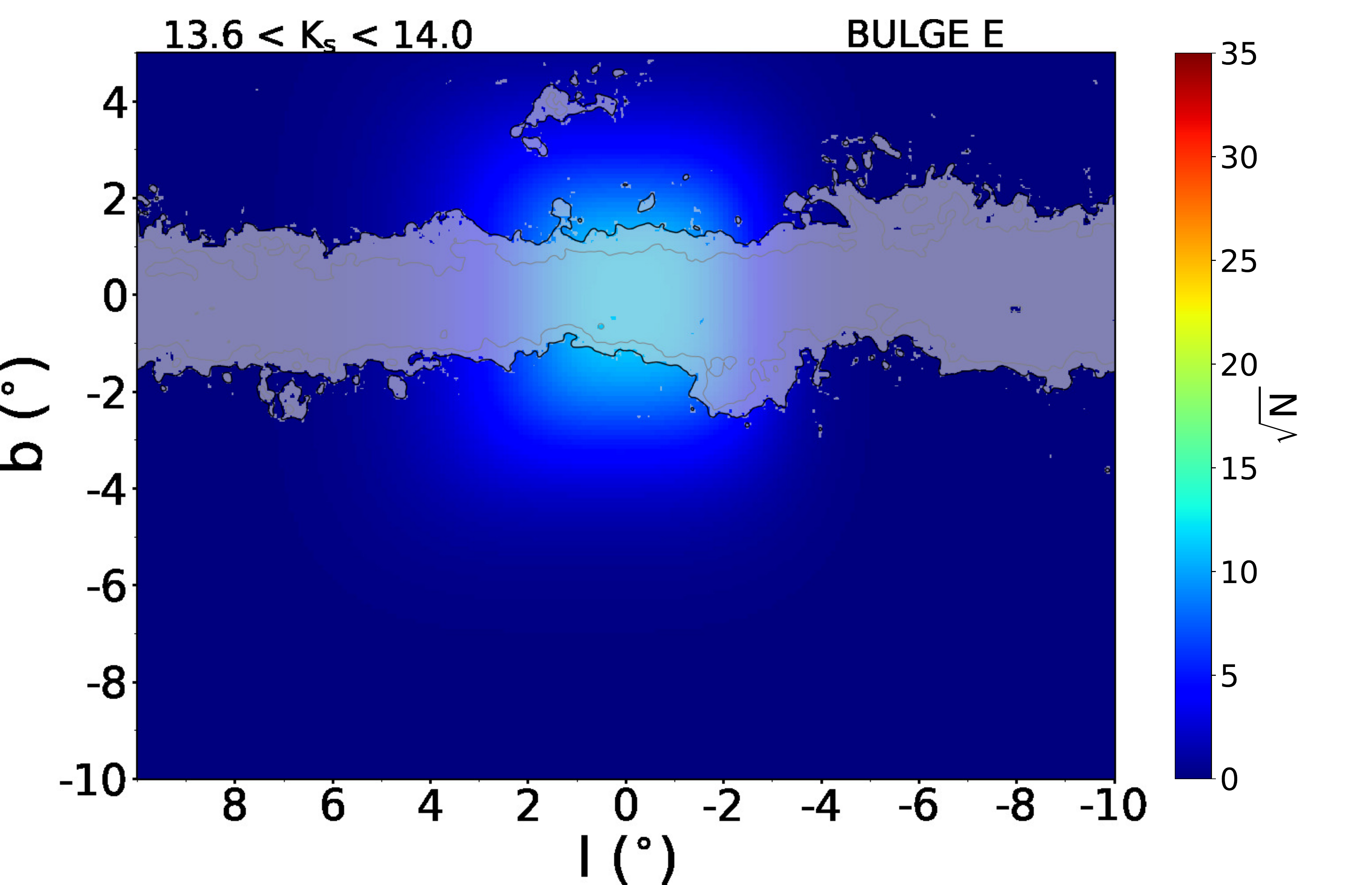}
\hspace*{-0.2in}
\includegraphics[scale=0.14]{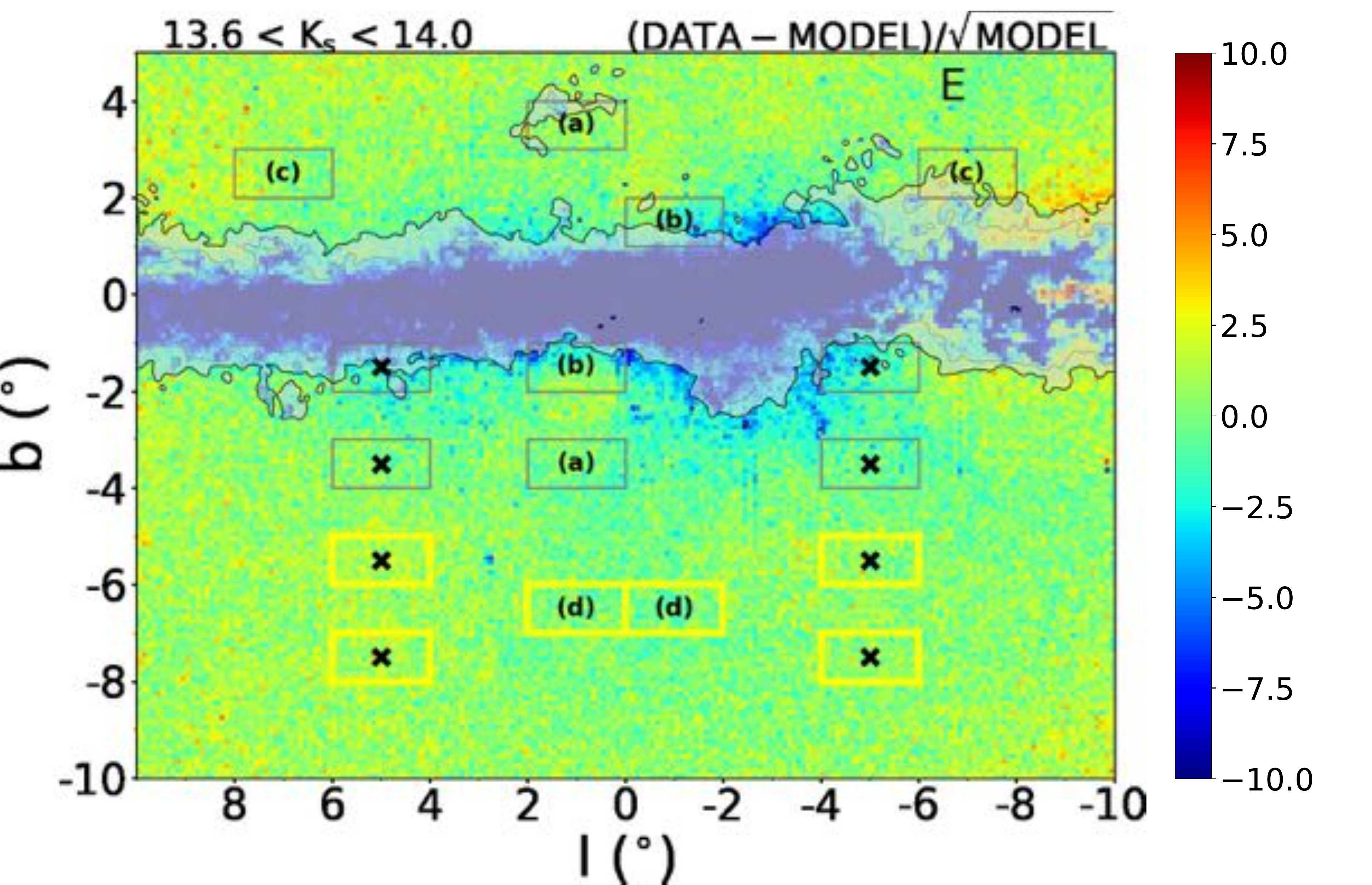}
\hspace*{-0.4in}
\includegraphics[scale=0.14]{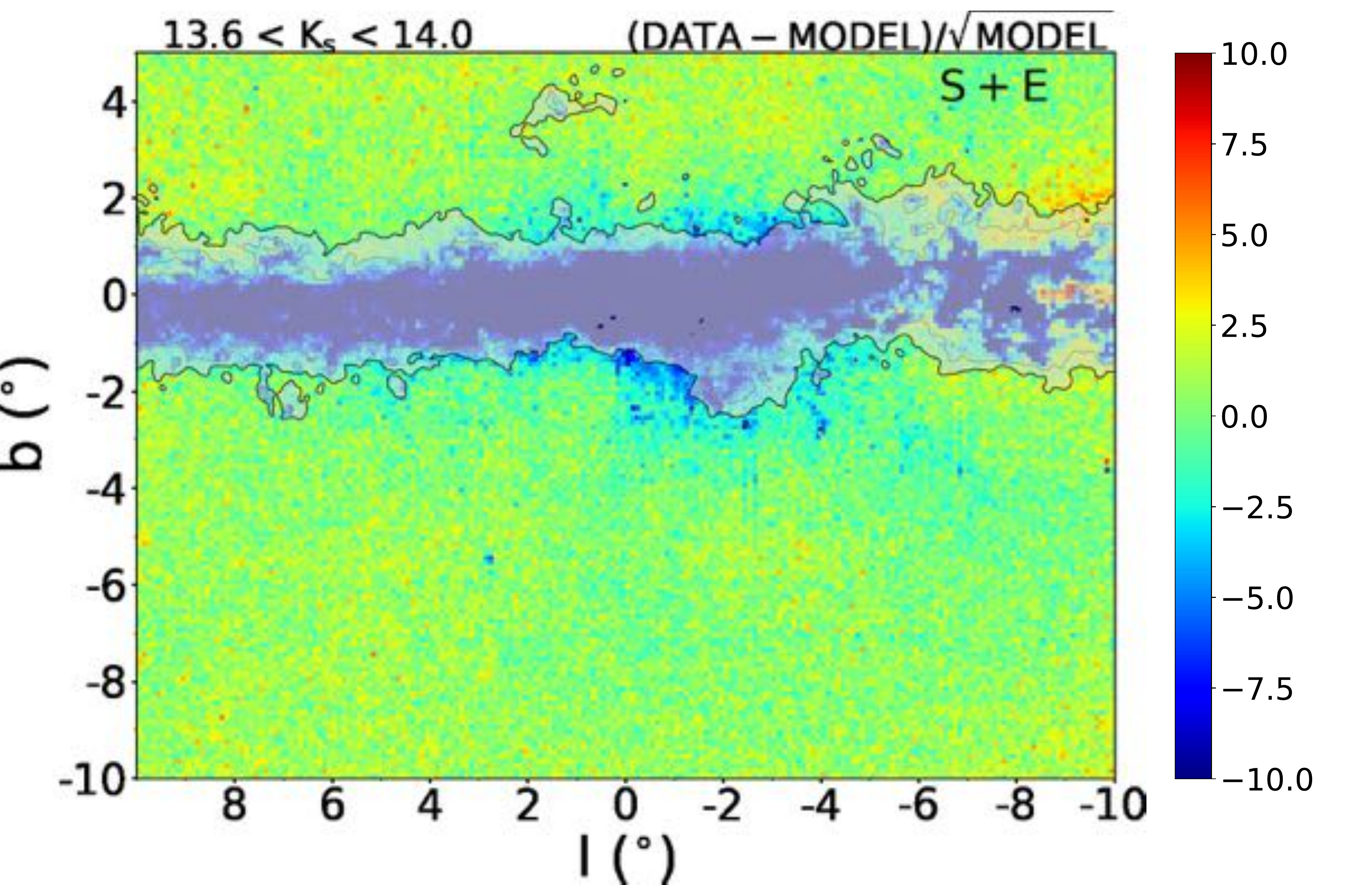} \\
\caption[Component $E$ of the $(S+E)$ model and difference between the
  data and two best fit models divided by the Poisson noise, in
  Galactic Coordinates]{\textit{Left panels:} component $E$ of the
  $(S+E)$ model with axis ratio [1:0.14:0.17] and viewing angle
  $|\alpha| = 1^{\circ}$.  \textit{Middle panels}: differences between
  data and model $E$, divided by the Poisson noise in the model. The model does not perform well in the central regions, $|l|<2^{\circ}$ and $|b|< 3^{\circ}$, where an excess is visible in the data, especially in the brightest magnitude bins (see the discussion in Section 7.4). \textit{Right
    panels:} same as in the middle panels but for model
  $(S+E)$.  The
  addition of the extra component (shown in the left panels) improves the quality of the fit in the inner regions. The magnitude distributions in the $2^{\circ} \times
  1^{\circ}$ fields marked with either letters or crosses, are provided
  in Figures \ref{LF_all} and \ref{LF_all2} respectively. We mark in yellow the fields with $b<-5^{\circ}$ for which we show the \citet{Lo2017} X-shaped model in the same figures.}
\label{diff}
\end{figure*}

The functions involved in the data fitting procedure are illustrated in 
Figure~\ref{LF_shift}. The apparent magnitude
distribution $N_{\mathrm{B}}$ of the Bulge (thick blue, Equation~\ref{Bulge_model}) is 
the integral of the absolute magnitude LF (green, labeled 
`LF+14.5') with a distance modulus of 14.5 added to the absolute magnitudes, 
$M_{K_{s}}$, times the density law $\rho_{\mathrm{B}}(l,b,D)$ (in the figure we show
$\rho_{\mathrm{B}}(l,b,\mu)$ in magenta, Equation~\ref{E}). The total apparent
magnitude distribution $M = < N_{\mathrm{d}}>\times Scale + N_{\mathrm{B}}$ (thin dotted
line, labeled `MODEL E') is given by the sum of the Bulge $N_{\mathrm{B}}$ (labeled
Bulge E) and the discs $< N_{\mathrm{d}}>$ (labeled `DISCS'). To obtain $N_{B}$
(thick blue line), Equation~\ref{Bulge_model} was calculated on a grid
of 20 x 10 $(l,b)_{i}$ fields of view between $0^{\circ}< l <
2^{\circ}$ and $3^{\circ}< b < 4^{\circ}$ for 50 $K_{s}$ apparent magnitudes bins between
$12< K_{s}< 14$ mag.

We do not fit directly for the Sun-Galactic centre distance, which is
a fixed parameter, $R_{\sun}=8$ kpc (see table 1 in \citealt{Val2017}), but we allow the mean
magnitude of the RC to vary by $\Delta \mu^{\mathrm{RC}}$. Here $\Delta \mu^{\mathrm{RC}}$ can 
translate to a new value for $R_{\sun}$, as it is effectively a distance modulus
shift, or a new value for the mean magnitude of the RC. For all models we found 
$\Delta \mu^{\mathrm{RC}} = -0.1$, or $\mu_{10}^{\mathrm{RC}} = -1.63$ at $R_{\sun}=8$ kpc, 
similar to \citealt{La12} and \citealt{Ha17} who found
$M_{K}^{\mathrm{RC}} = -1.61$ using nearby Hipparcos and Gaia stars respectively, and \citealt{We15}
who found $M_{K}^{\mathrm{RC}} = -1.67$ at $R_{\odot} =$8 kpc (note their LF were 
constructed interpolating the BASTI isochrones in the 2MASS photometric system, \citealt{Pi04}). 

The model plotted in Figure~\ref{LF_shift} is a result of the 
full global fit of the whole VVV dataset, discussed in the next subsection.  We 
performed the fits for two cut-off radii, $R_{\mathrm{c}} = $2.5 and 4.5 kpc and found 
that $R_{\mathrm{c}} = 4.5$ kpc provides a better fit, therefore in the remainder of
this work it is kept constant.
\subsection{The full Bulge sample fit}

We compute the model on a continuous grid of 6$' \times 6'$ fields
covering the whole area of the VVV survey amounting to 200 $\times$ 150
$(l,b)_{i}$ lines of sight, for 40 magnitude bins $K_{s}$ (green triangles in Figure
\ref{LF_shift}) to be fitted to the data (red circles in the same
figure). We first consider a single Bulge population and next, we add
a second component to improve the agreement in the central
regions at low latitudes where the residuals highlight an overdensity. 

The three models we consider are:
\begin{itemize}
\item model $E$: the density model is described by an exponential-type
  model (Equation~\ref{E}) and has a 10 Gyr-old input LF (see Figure~\ref{LF}). The model $M$ has 10 free
  parameters, 7 describing the density law, a scaling for the thin
  discs and one for the thick disc and the RC peak magnitude shift
  $\Delta \mu^{\mathrm{RC}}$;
\item model $S$: this density law is described by a hyperbolic secant
  density distribution (Equation~\ref{S}) and has a 10 Gyr-old input
  luminosity function, the same as model $E$. The model has 9 free
  parameters, 6 describing the density law, a scaling factor for the
  thin discs and one for the thick disc and $\Delta \mu^{\mathrm{RC}}$;
\item model $S+E$: \citet{Ro12} using 2MASS data found that a
  combination of an exponential-tye ($E$) and a sech$^{2}$ ($S$) density
  law provide the best description of the Bulge (see their table 2)
  and we use the same description for the VVV data. In total, the
  model has 16 free parameters, 7 for model $E$, 6 for model $S$, two
  scaling factor for the discs and one for the ratio between the two
  Bulge components. The
  magnitude shift is not a free parameter, we set to
  $\Delta_{\mu}^{\mathrm{RC}} = -0.1$ in accordance with the results found in the single component fit.
\end{itemize}

We perform several runs of the L-BFGS minimisation starting from
random initial points to test whether the final result is dependent on
the initialisation position and to avoid being trapped in a local
minimum. We consider our best solution to be the one that maximises the
log-likelihood, (Equation~\ref{prob}) across all trials;
note that several runs may actually converge to the same best-fit
solution. We do not have tight constrains on the parameters, limiting
them to an initial set of random values that would have a physical
meaning as described in the previous section. The model $(S+E)$ has 16
free parameters so to avoid getting not meaningful results we choose
the starting point for the main component S to be the best fit
solution of the one population fit (model $S$), while the starting
point for the E component is drawn from a uniform random distribution.

\begin{table*}
 \centering
 \hspace*{0cm}
 \begin{minipage}{176mm}
  \caption[Summary of the best fit parameters for three models]{Best
    fit results for the three density models we have used to reproduce
    the VVV data. The $first$ $column$ assigns a label for each result, for easier reference in the text; the $second$ $column$ specifies the model $S$, $E$ or $S+E$; $third$ $column$ the logarithm of the likelihood, which is an indication of the `goodness' of the fit. The $fourth$ $column$ lists three values: the $\Delta_{\mu}^{\mathrm{RC}}$ that is a free parameter except in one case where we underline it (label `$R_{\odot}$ free'), $R_{\odot}$ that is set to 8 kpc except in the row labelled `$R_{\odot}$ free' and $\sigma^{\mathrm{RC}}$. $\sigma^{\mathrm{RC}}$ by default takes the value in Table \ref{LFparams}; in the first two rows, labelled `main', the LF is convolved with a Gaussian kernel with $\sigma_{K_{s}} (l,b)$ shown in Figure \ref{errK};  in the rows labelled with `$\sigma^{\mathrm{RC}}$ free', it is a free parameter. The $fifth$ $column$ gives the best fit  parameters of the density distribution (Equations~\ref{E} and~\ref{S}) and finally, the $sixth$ $column$, provides the number of stars in the thin disc, thick disc and Bulge including/excluding the region of high extinction e.g. $N_{1}/N_{1}^{-} =$ 4.63/1.78 means that the number of stars predicted by model E in the thin disc is 4.63 $\times 10^{6}$ and, excluding the high extinction region (which is the number we fit), is 1.78 $\times 10^{6}$. The table provides the results using the Besancon discs (top rows) and the updated discs discussed in Section 7.2 (bottom rows). } 
   \label{bestresults}
  \footnotesize
  \centering
  \begin{tabular}{@{}lccclccccr@{}}

label & Model &  ln($L_{P}$) & [$\Delta_{\mu}^{\mathrm{RC}}$, $R_{\odot}$, $\sigma^{\mathrm{RC}}$] & [$x_{0}$ , $y_{0}$, $z_{0}$, $\alpha$, $c_{//}$, $c_{\perp}$, $n$] & $[N_{1}/N_{1}^{-}, N_{2}/N_{2}^{-}, N_{3}/N_{3}^{-}] \times 10^{6}$ \\
    \hline
    \hline
        & &  & &Results using the Besancon discs:&  \\
         \hline
main & $E$  &  36,737,730 &  [0.1, -, *$\sigma_{K_{s}}(l,b)$] & [0.55, 0.24, 0.17, -19.49, 2.83, 1.76, 1.10] & [4.63/1.78, 1.84/1.27, 21.43/13.44] \\
main &\textit{S}  & 36,732,251 & [0.1, -,*$\sigma_{K_{s}}(l,b)$] &[1.47, 0.63, 0.47, -19.57,  3.04, 1.88] & [5.29/2.03, 2.31/1.59, 20.05/12.87] \\
\hline
main &$S$  + &  36,754,040 & [0.1, -, -] &[1.65, 0.71, 0.50, -21.10, 2.89, 1.49] & [4.67/1.75, 2.04/1.40, 19.51/12.66] \\
 & $E$ & & & [1.52, 0.24, 0.27, -2.11, 3.64, 3.54, 2.87] & \qquad \qquad \qquad \qquad \qquad [1.52/0.67] \\
\hline
$\sigma^{\mathrm{RC}}$ free&\textit{S}  & 36,743,487 & [0.08, -, 0.26] &[1.10, 0.45, 0.45, -34.4,  3.58, 2.09] & [5.72/2.20, 2.18/1.49, 15.88/12.80] \\
$\sigma^{\mathrm{RC}}$ free&\textit{E}  & 36,749,046 & [0.08, -, 0.26] &[0.42, 0.17, 0.16, -33.5,  3.50, 2.02, 1.09] & [5.17/1.98, 1.86/1.28, 20.82/13.22] \\
$R_{\odot}$ free & \textit{S}  & 36,731,560 & [\underline{0.0}, 7.65, - ] &[1.42, 0.61, 0.45, -19.45,  3.73, 2.39] & [5.30/2.03, 2.29/1.57, 20.06/12.88] \\
 \hline
 \hline
     & &  & &Results using the updated discs: & \\
    \hline
  & E  & 36,730,036 &  [0.1, -, -] & [0.57, 0.25, 0.16, -18.86,  2.41, 1.75, 1.08] & [2.53/0.90, 1.72/1.18, 23.50/14.41] \\
& S & 36,722,458 &  [0.1, -, -] & [1.61, 0.69, 0.48, -19.16,  2.50, 1.86] & [3.11/1.10, 2.21/1.52, 22.05/13.86] \\
 \hline
& S  + &  36,747,069 &  [0.1, -, -] & [1.77, 0.78, 0.51, -20.83, 2.45, 1.47] &  [2.59/0.92, 1.91/1.31, 21.25/13.54]  \\
& E & & & [1.47, 0.24, 0.26, -2.22, 3.64, 3.55, 2.80] & \qquad \qquad \qquad \qquad \qquad [1.65/0.72]\\
    \hline
    \end{tabular}
 \end{minipage}
\end{table*}

\begin{figure*}
\centering
\hspace*{-0.35in}
\includegraphics[scale=0.12]{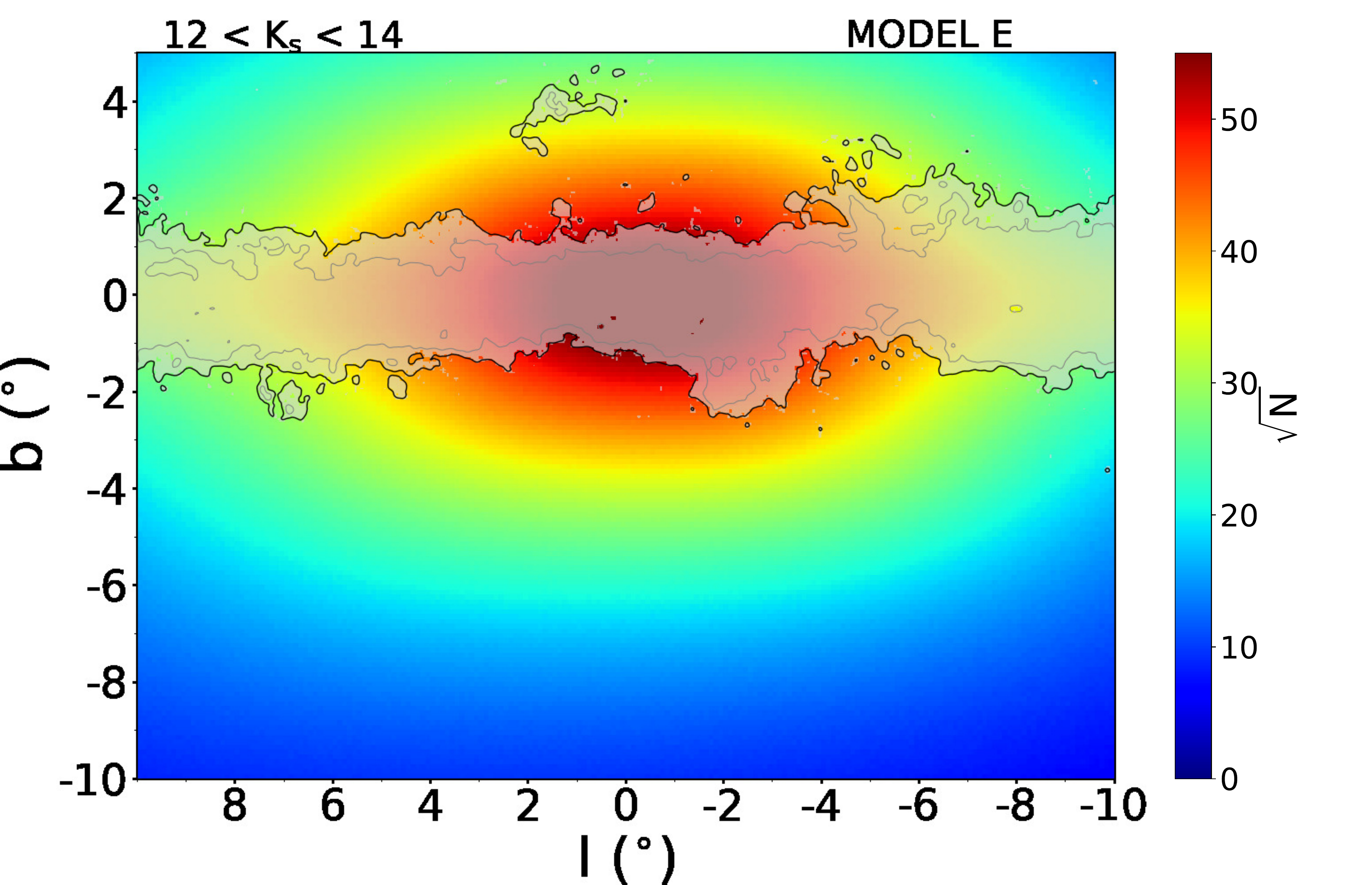}
\hspace*{-0.35in}
\includegraphics[scale=0.12]{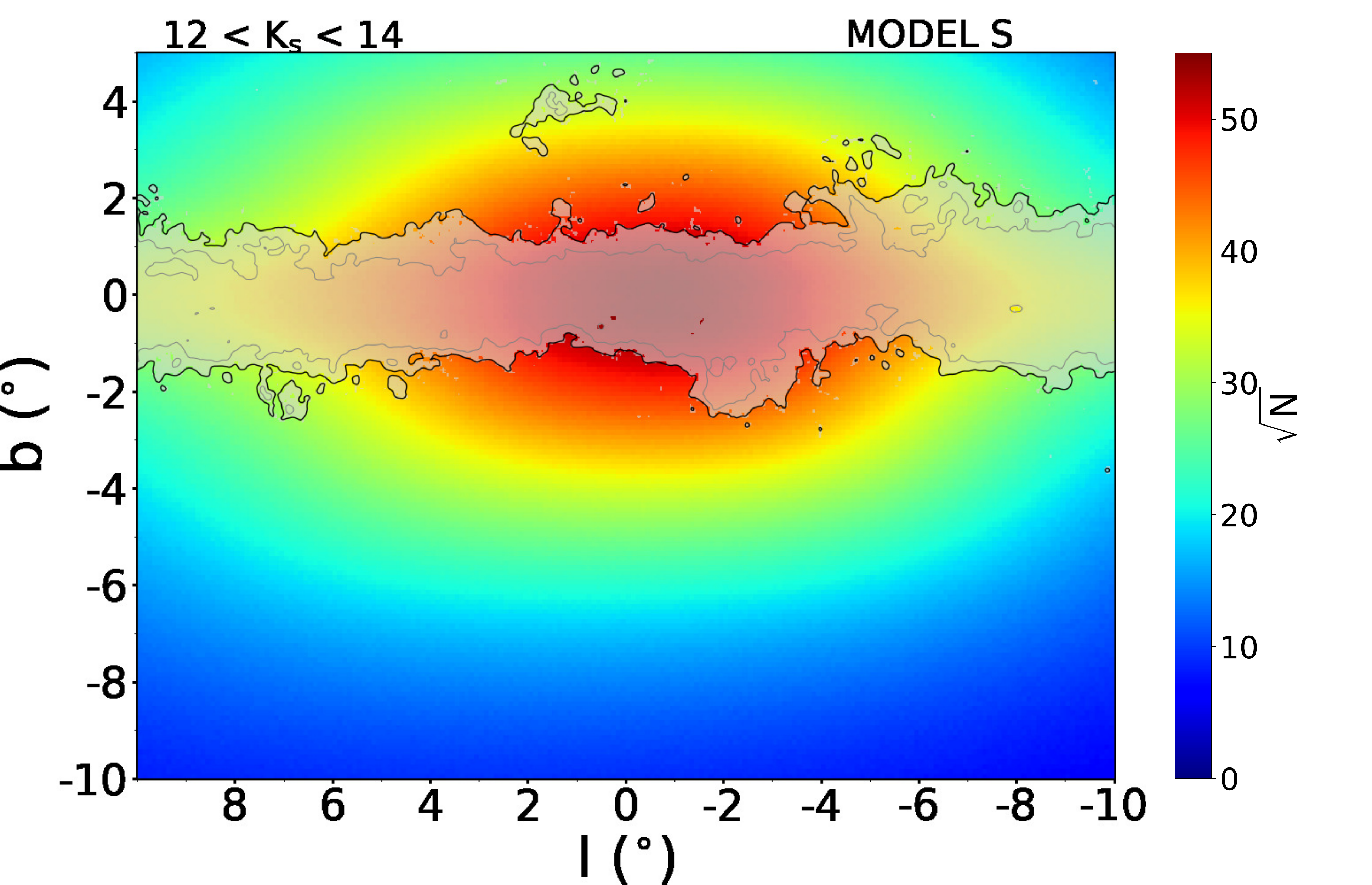}
\hspace*{-0.35in}
\includegraphics[scale=0.12]{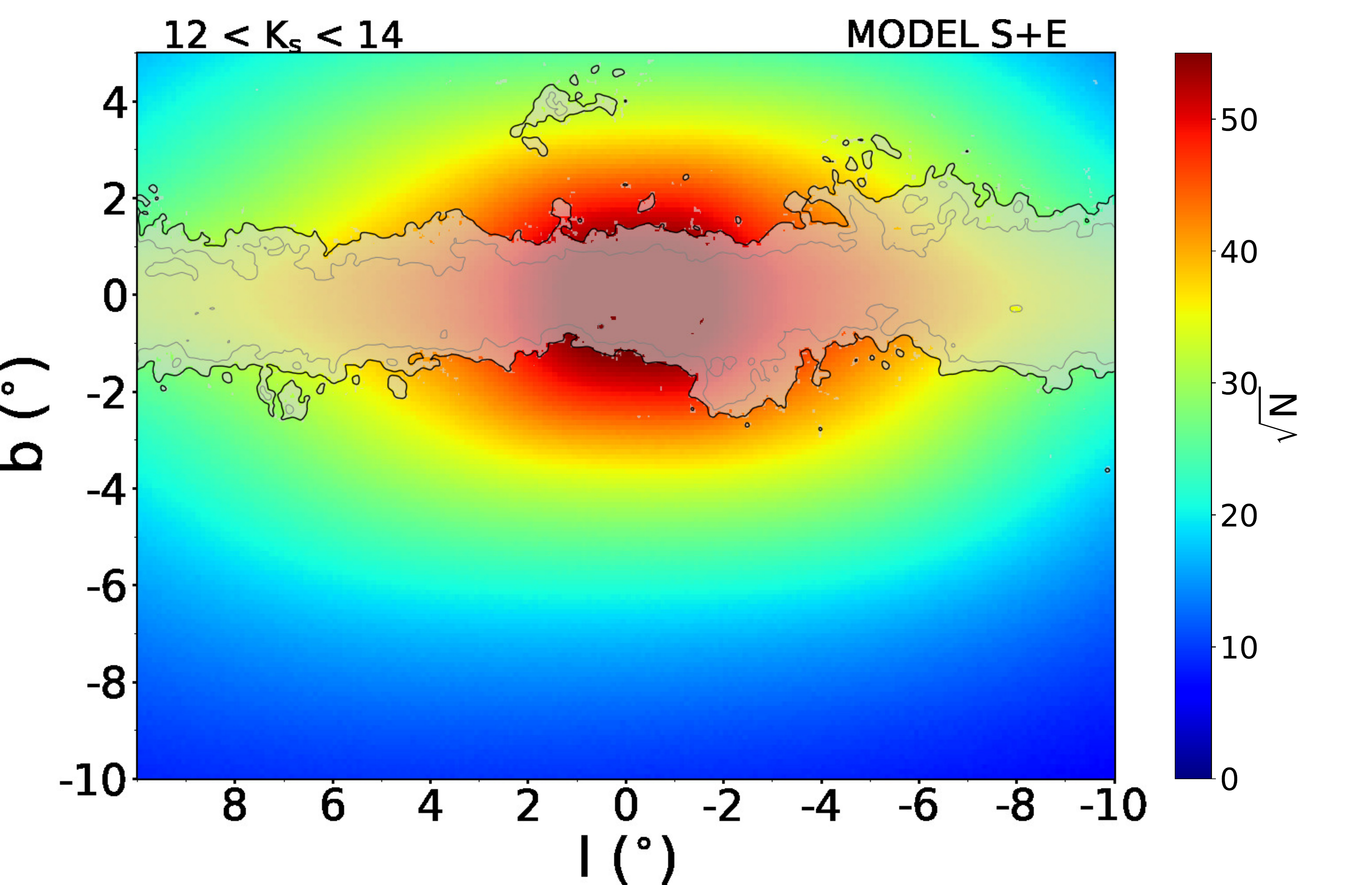}
 \\
\hspace*{-0.35in}
\includegraphics[scale=0.12]{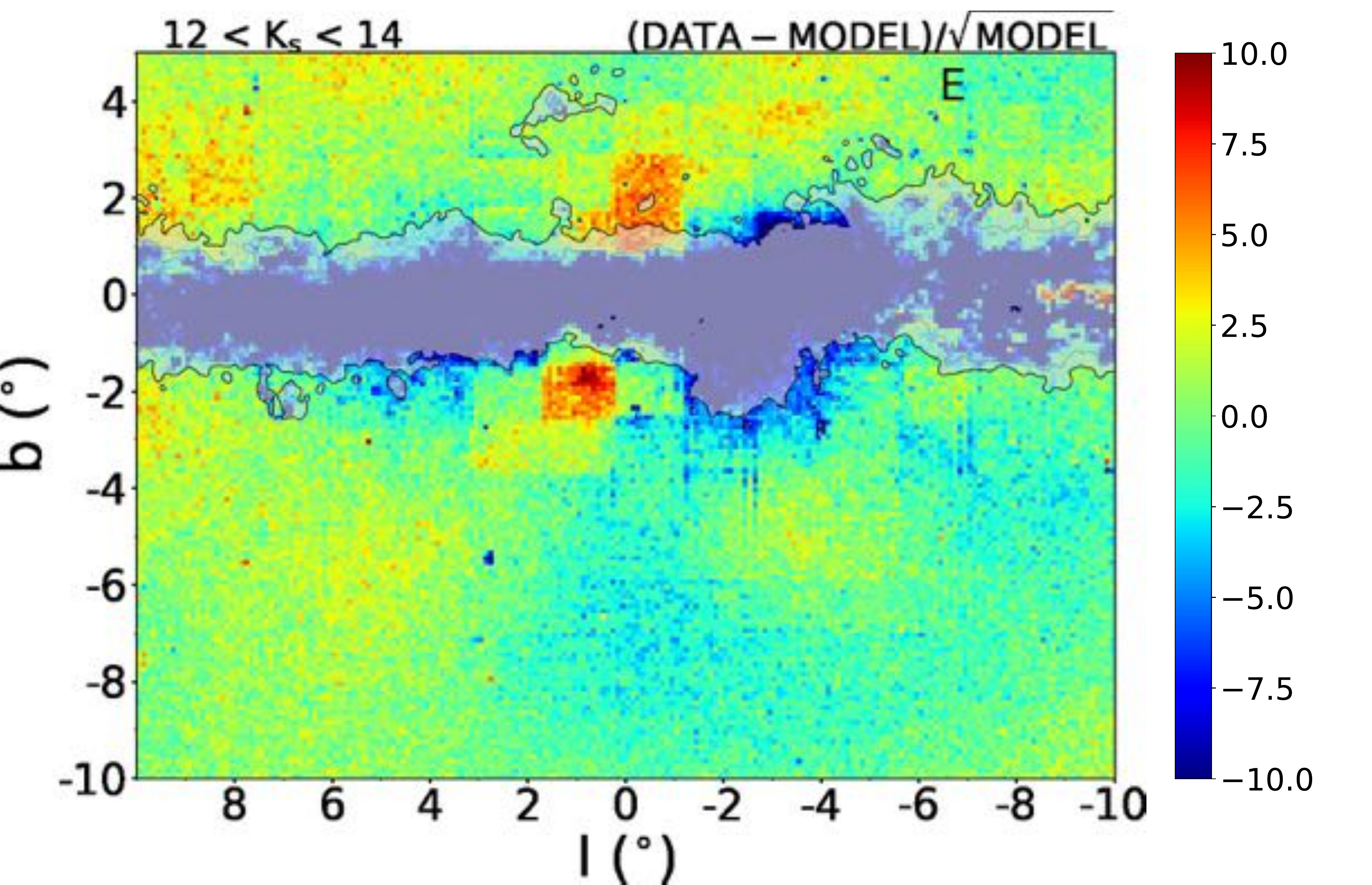}
\hspace*{-0.35in}
\includegraphics[scale=0.12]{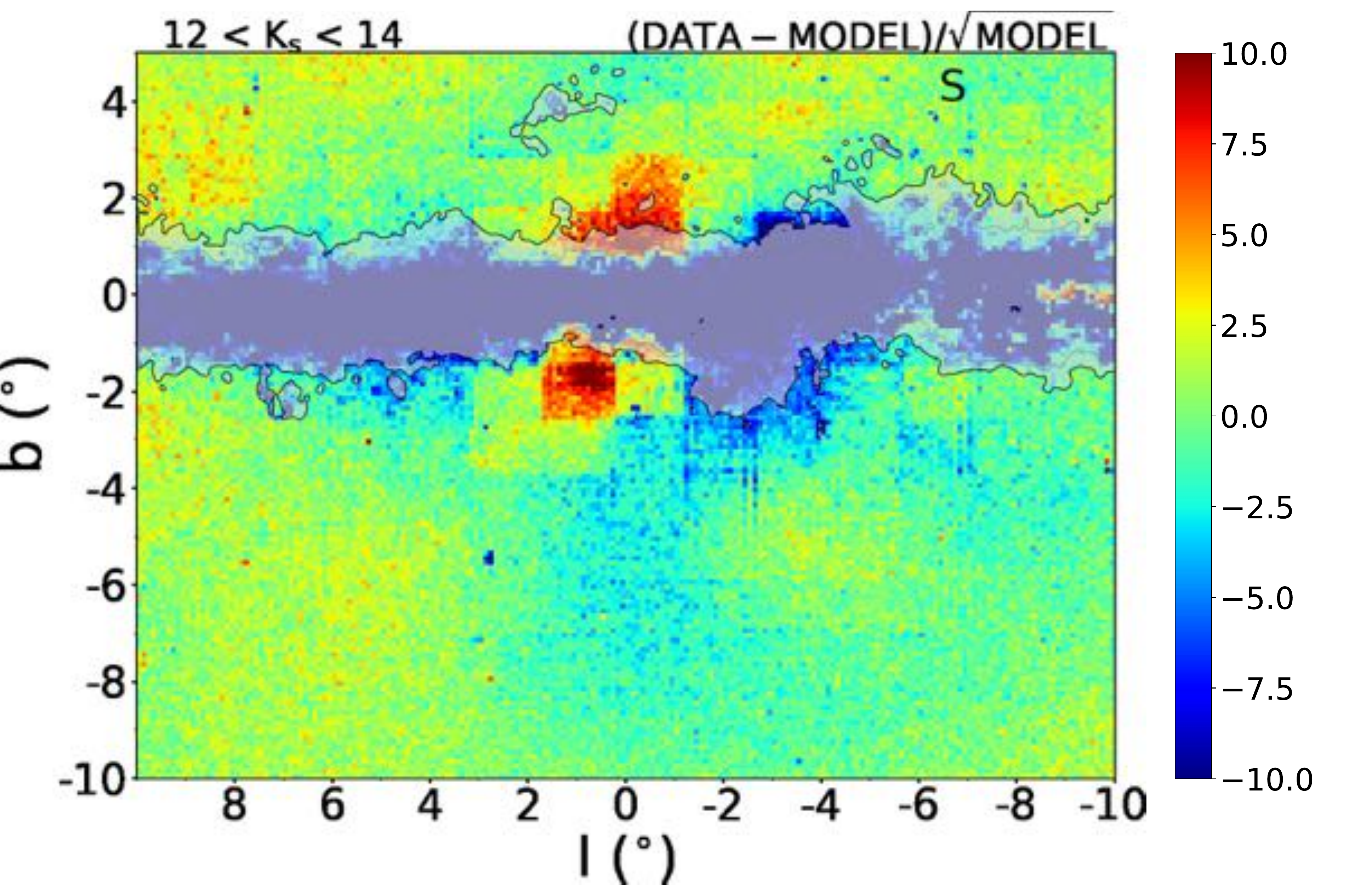}
\hspace*{-0.35in}
\includegraphics[scale=0.12]{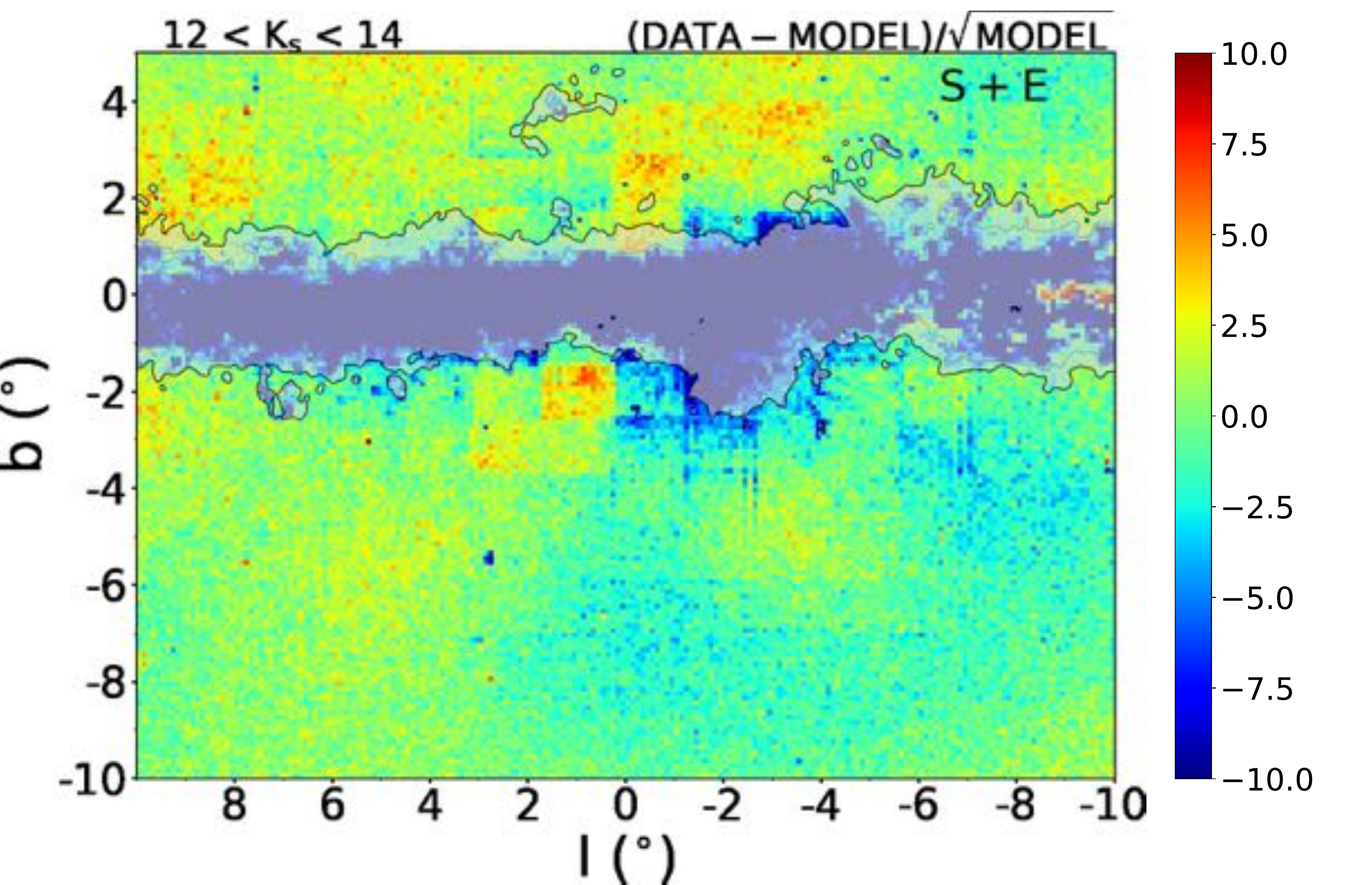} 
\caption[Residuals between data and best fit models in Galactic
  Coordinates]{\textit{Top row:} Square root density plots of the
  three best fit models in Table \ref{bestresults}, in
  Galactic coordinates for $12< K_{s}$/mag$<14$. \textit{Bottom row}: residuals between the data and each model
  in the top panels, divided by the Poisson noise in the model. Model
  $E$ performs slightly better than model $S$ in the central regions,
  while the two component fit $(S+E)$ provides the best description of
  the data. The arms of the X shape are clearly visible especially at
  negative latitudes where the VVV covers a large part of the sky (see also the
  residuals in figure 3 of \citealt{Ne16}). The
  fact that the arms on the far side of the Bulge are less visible than the
  ones on the near side is simply a projection effect.}
\label{res}
\end{figure*} 

\begin{figure*}
\hspace*{-0.19cm}
\includegraphics[scale =0.234]{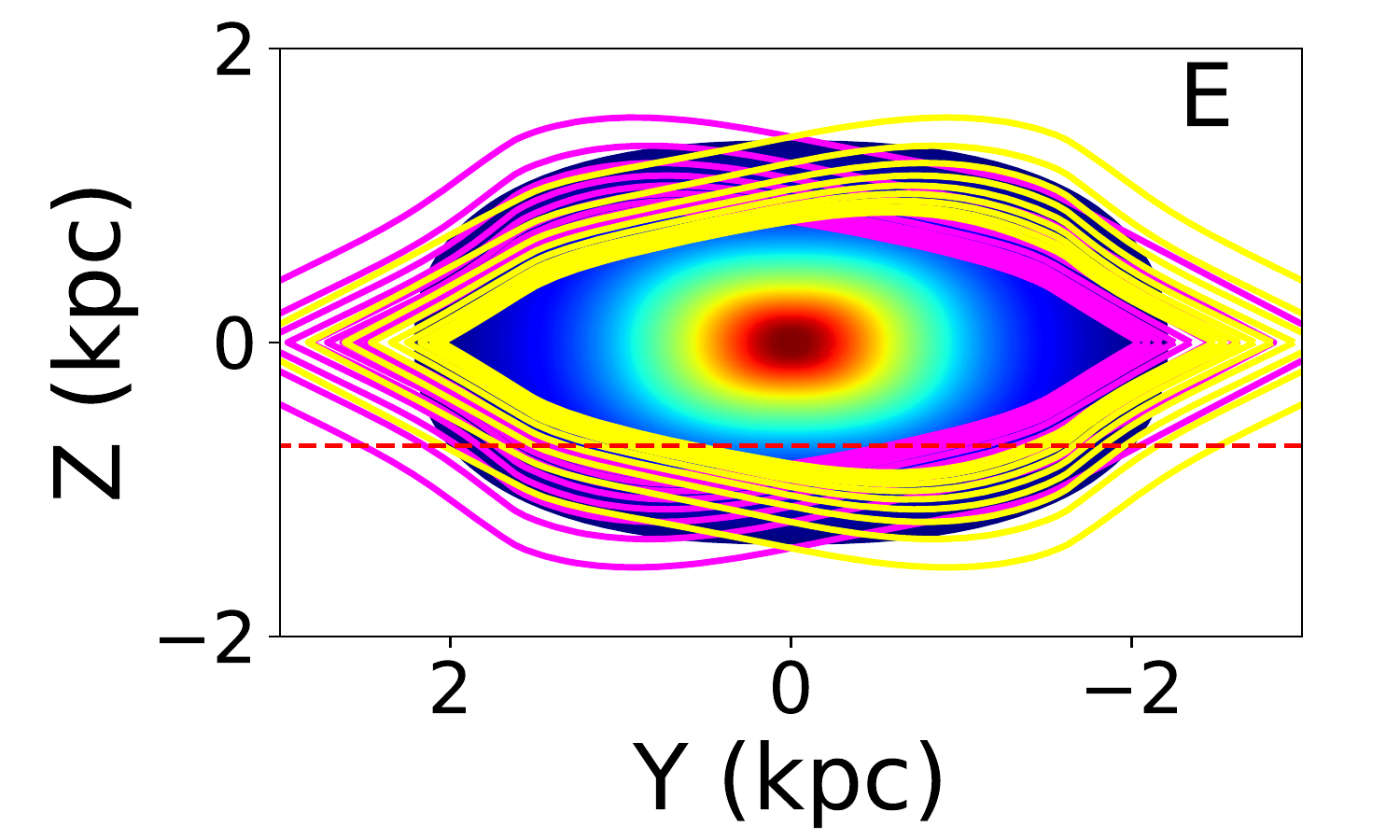}
\hspace*{-0.2cm}
\includegraphics[scale =0.234]{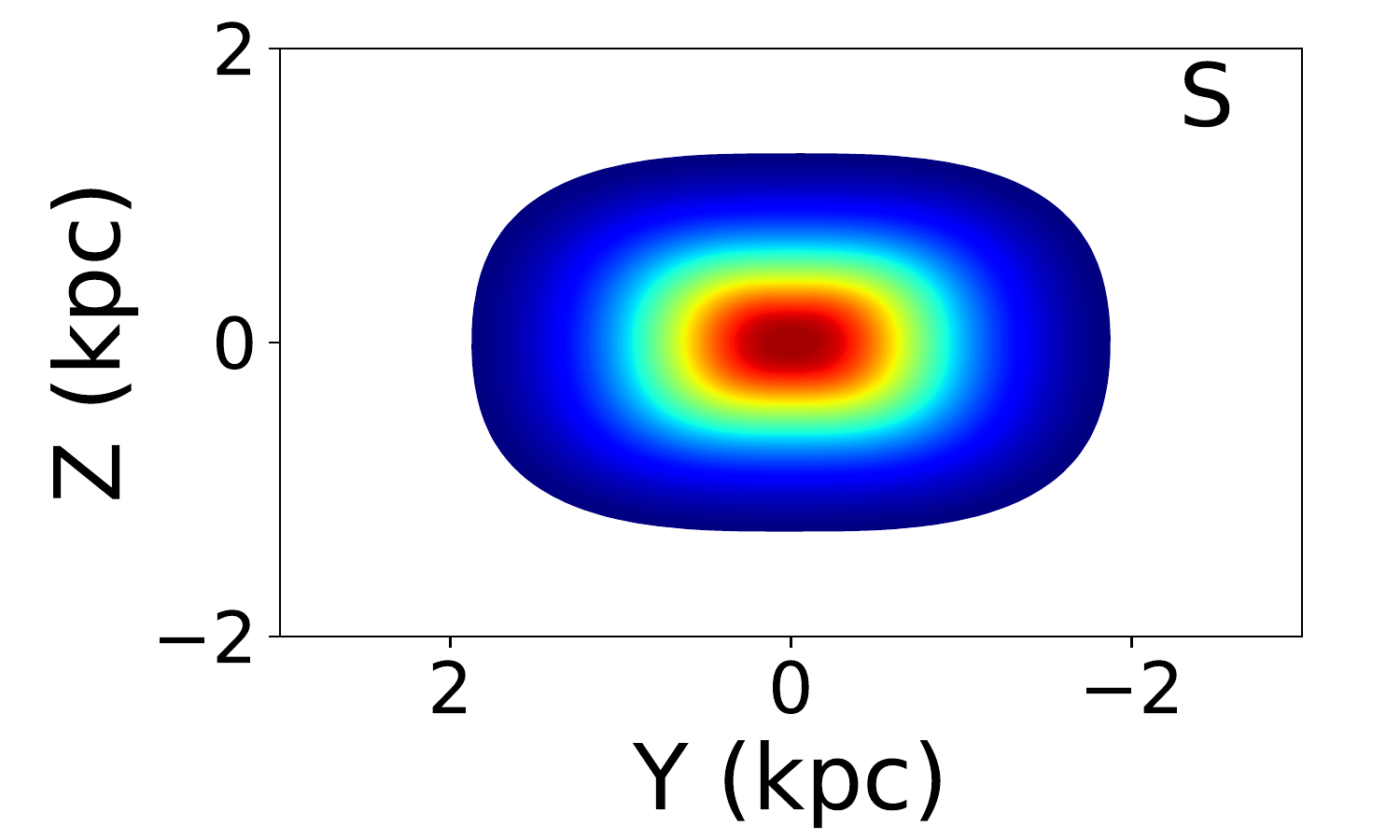}
\hspace*{-0.2cm}
\includegraphics[scale =0.234]{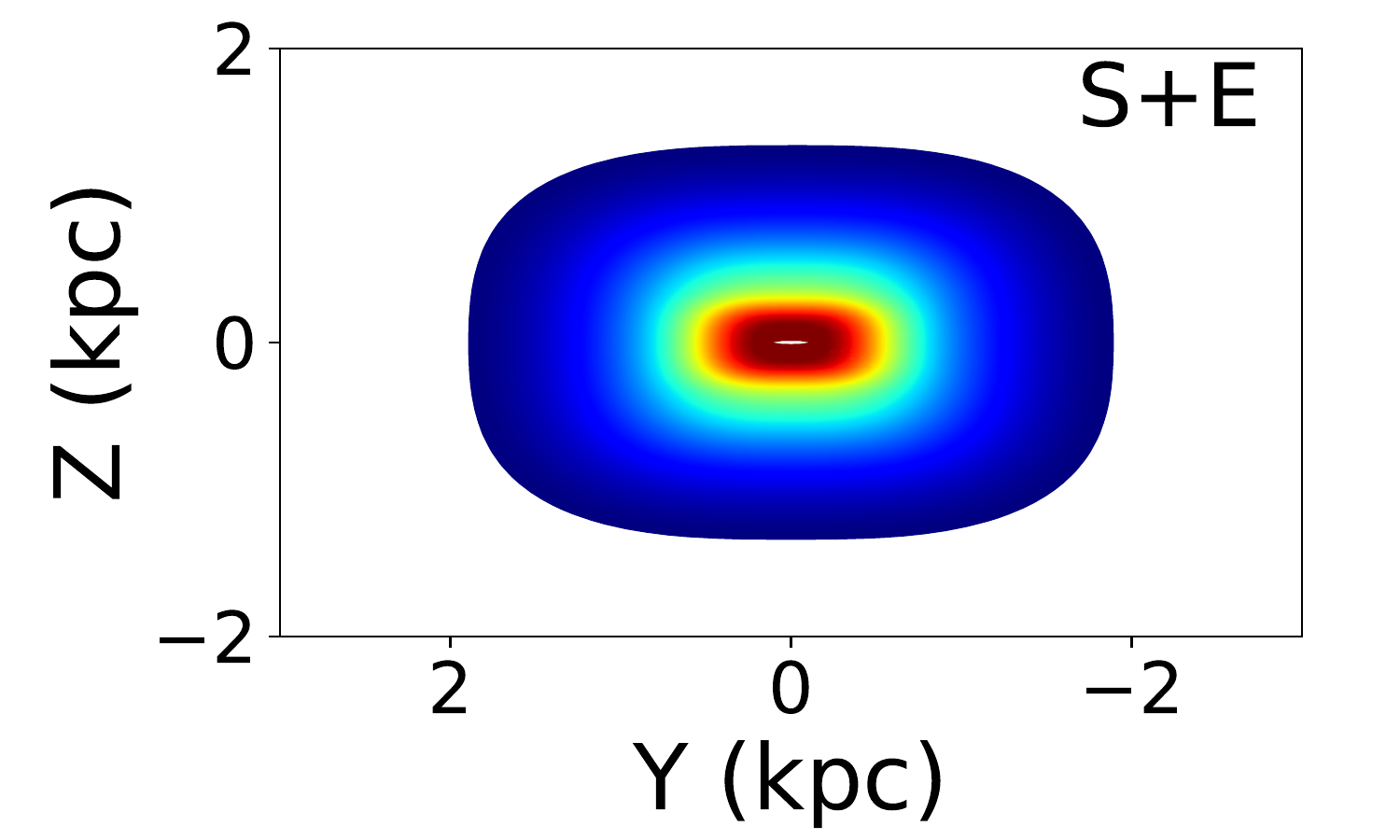}
\hspace*{-0.2cm}
\includegraphics[scale =0.234]{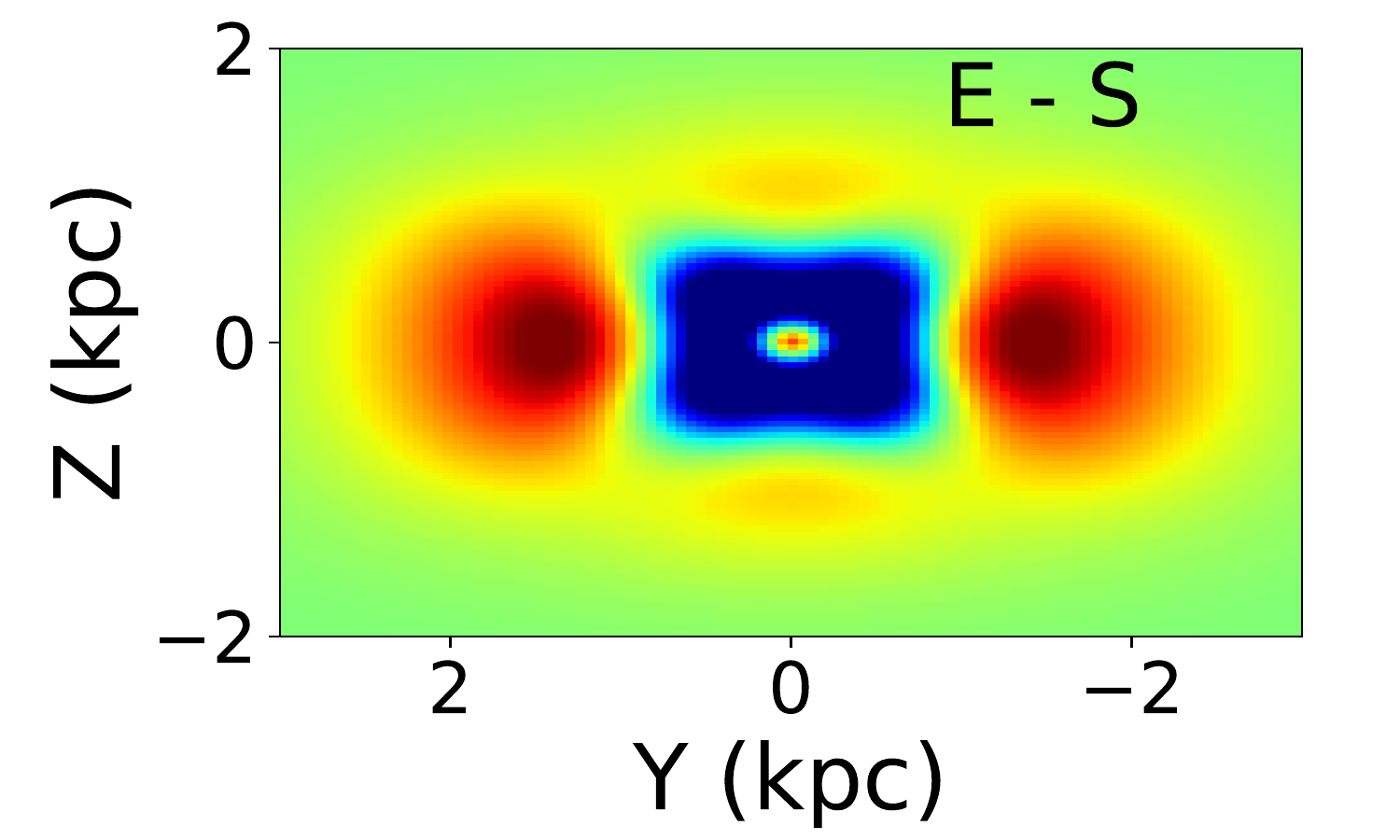} 
\hspace*{-0.2cm}
\includegraphics[scale =0.234]{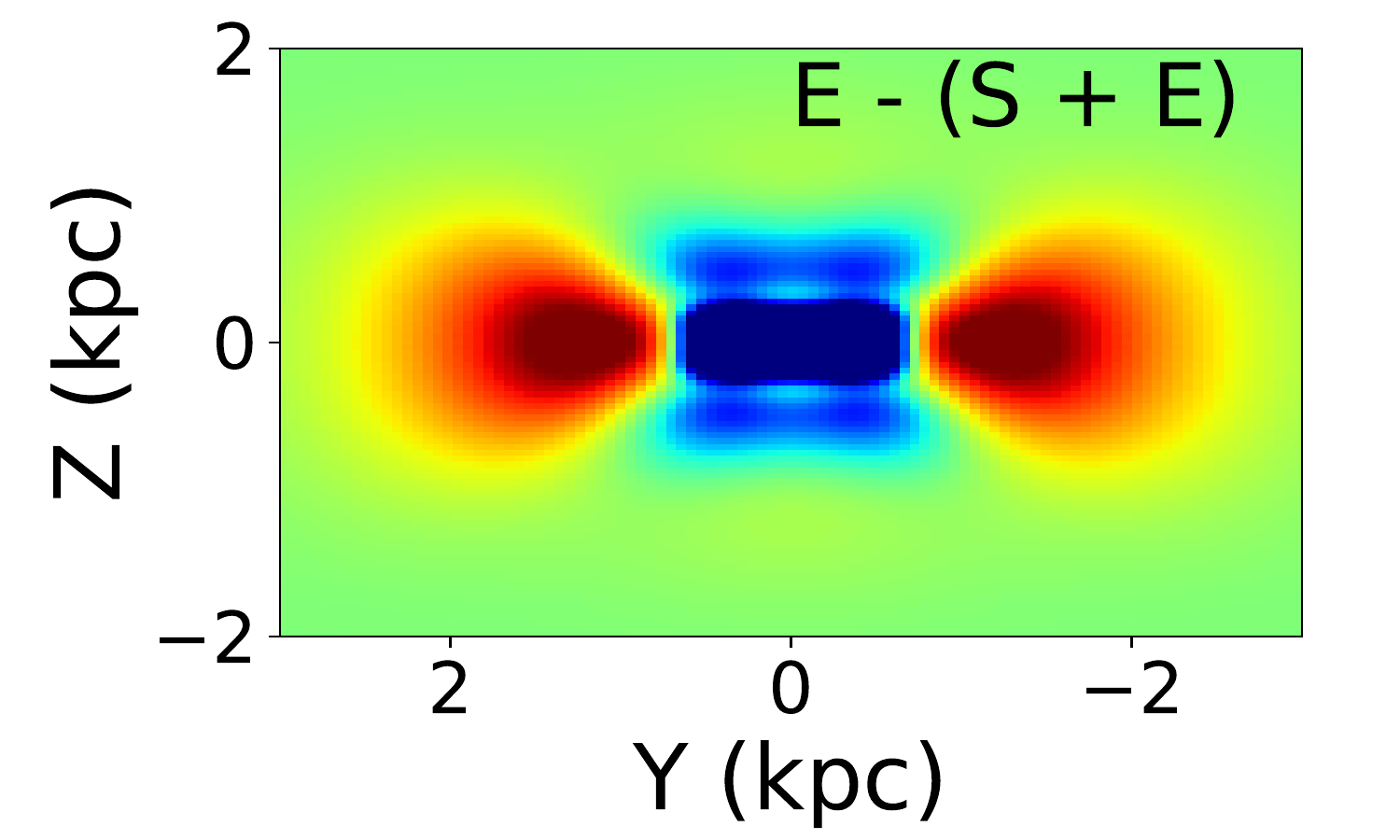} \\
\vspace{-0.3cm}
\includegraphics[scale =0.225]{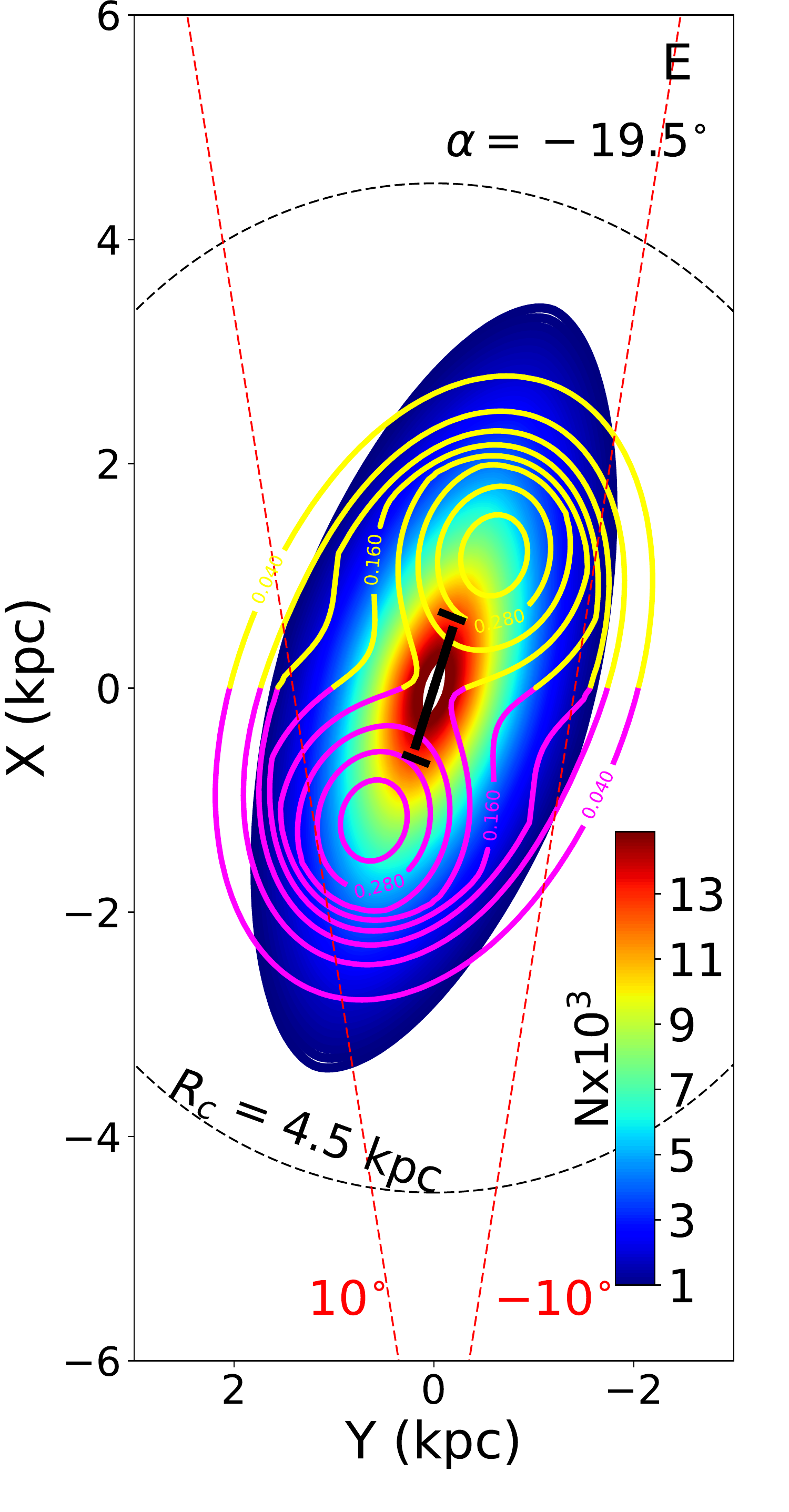}
\includegraphics[scale =0.225]{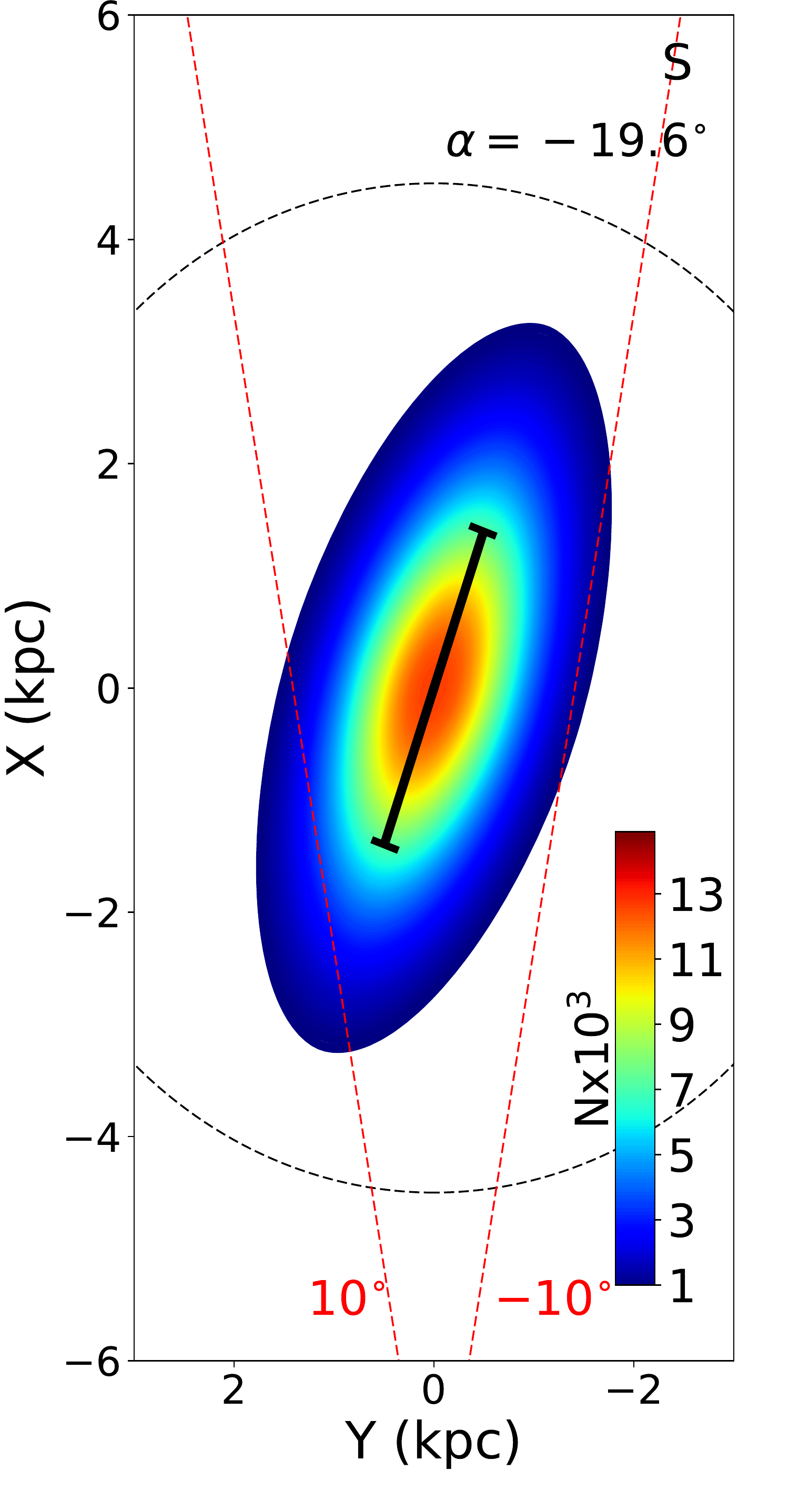}
\includegraphics[scale =0.225]{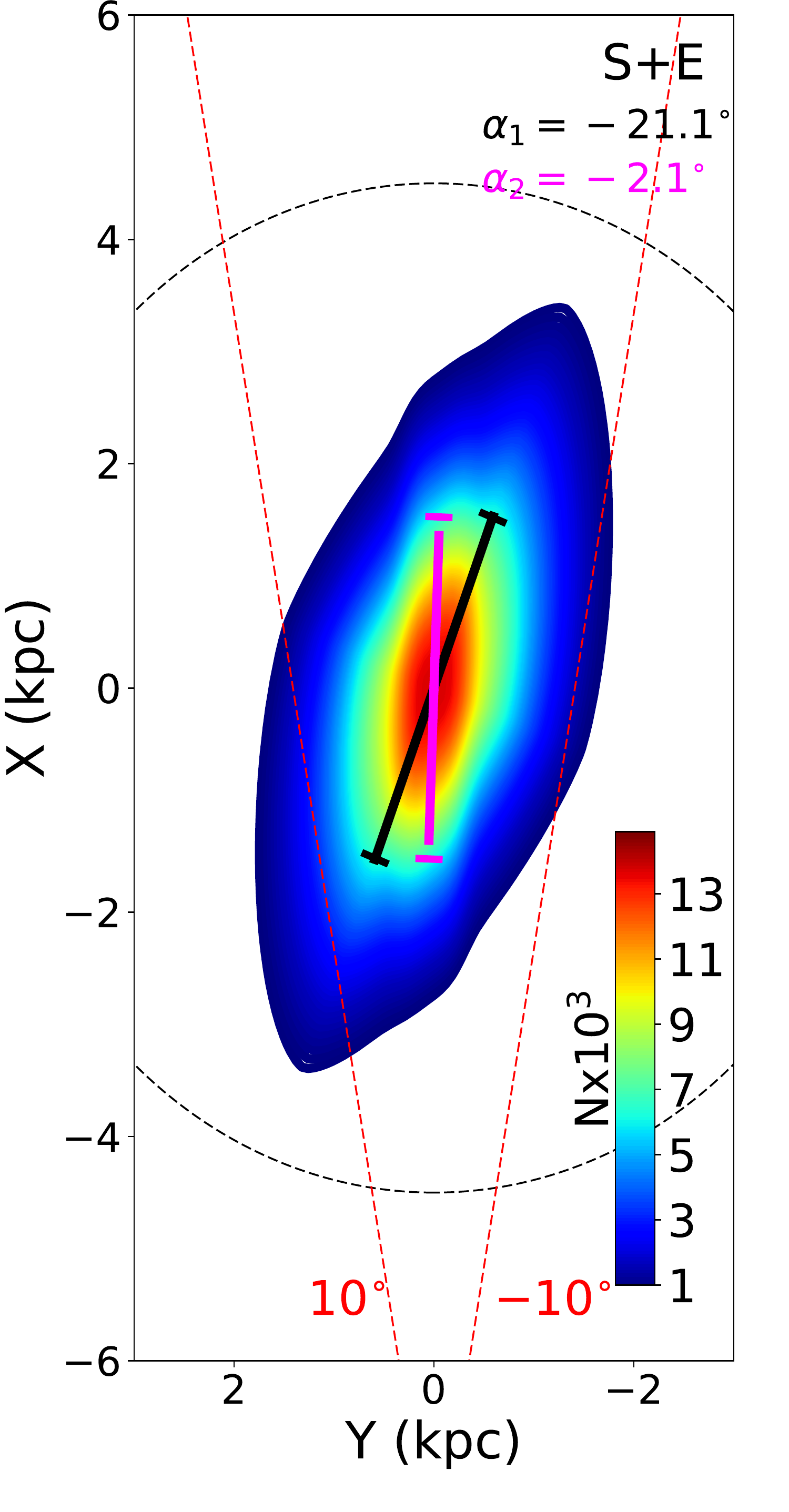}
\includegraphics[scale =0.225]{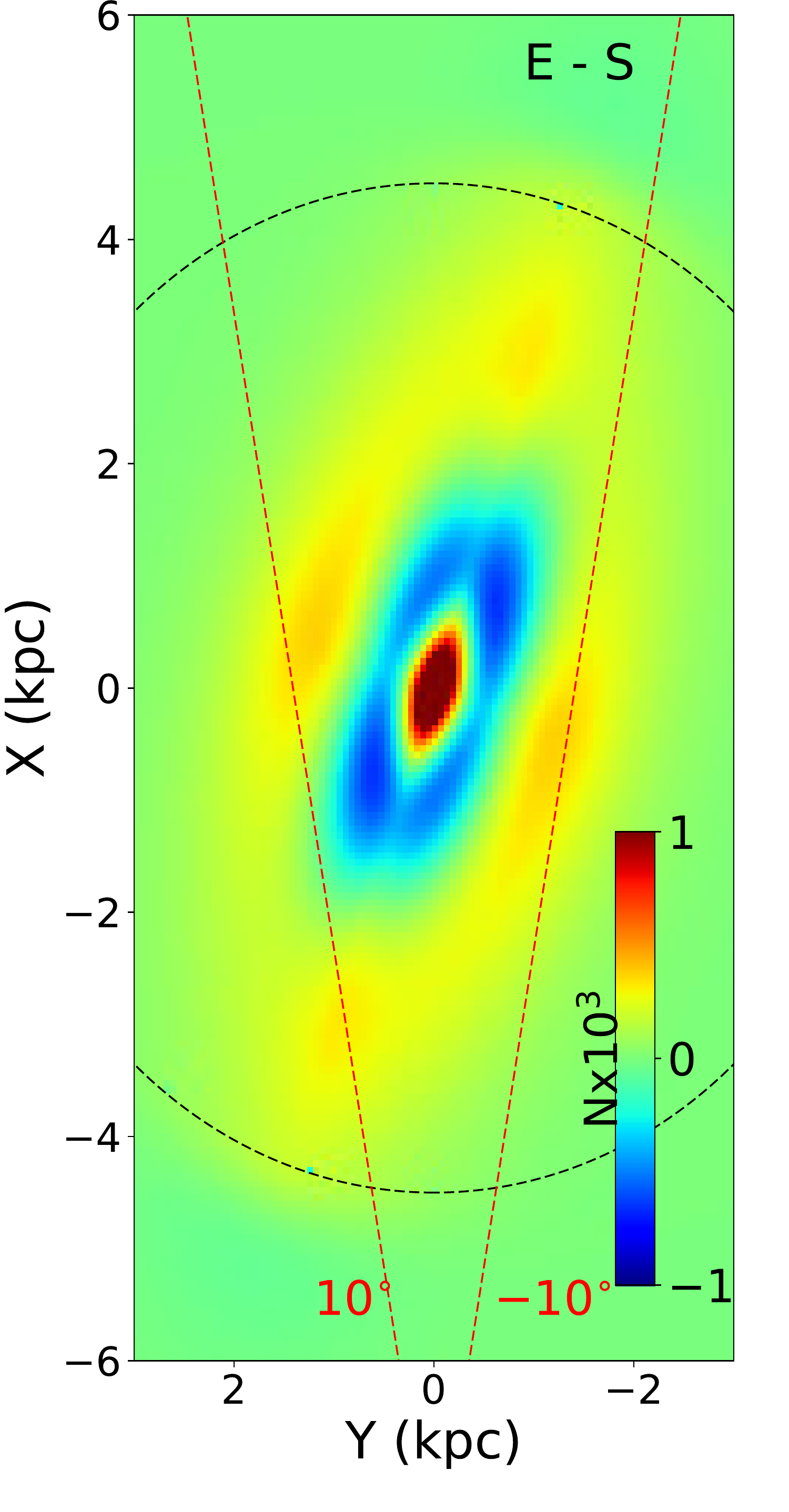}
\includegraphics[scale =0.225]{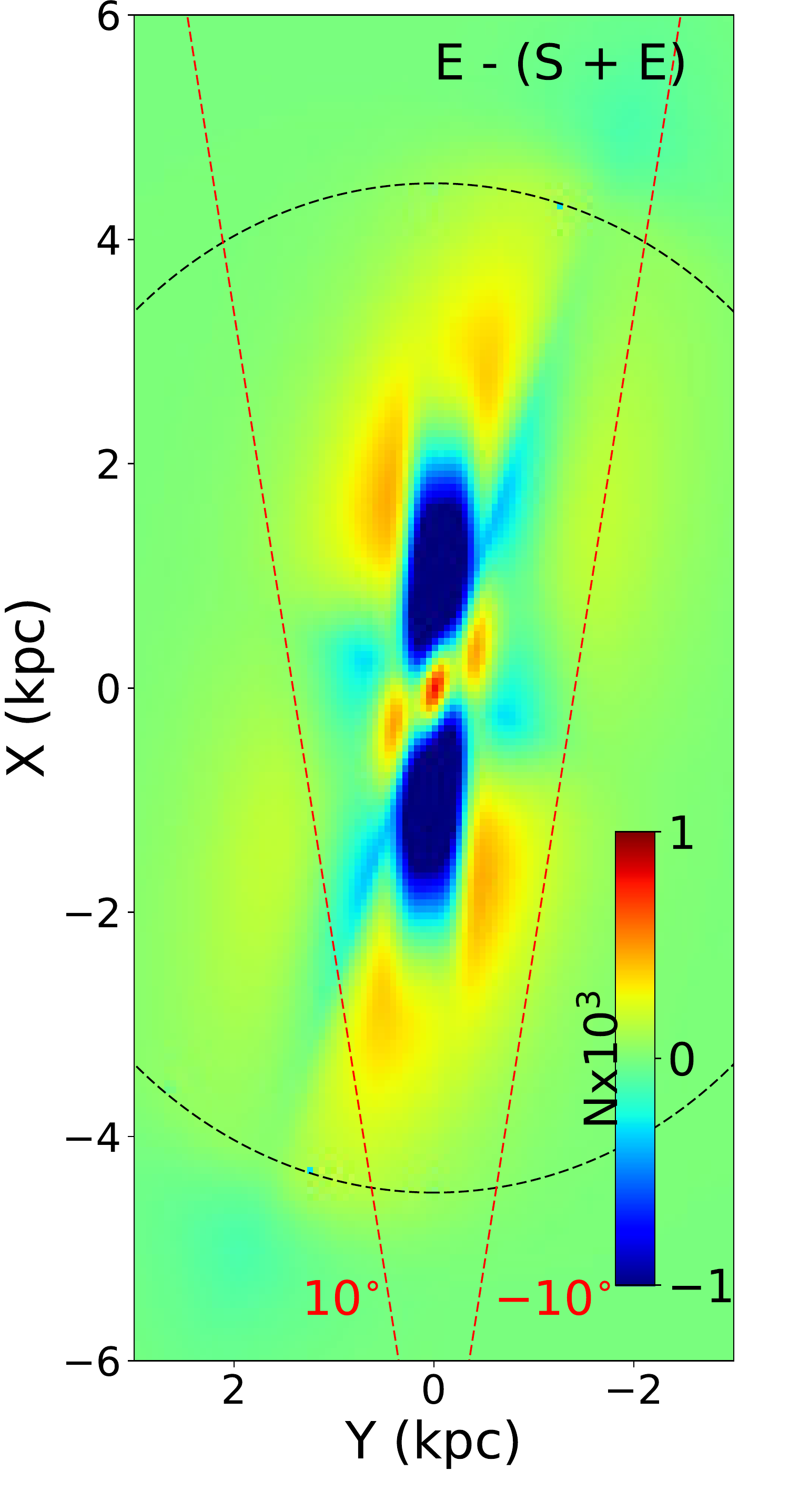}
\caption{The best fit 3D density law of the Milky Way
  Bulge projected along the Z-axis ($bottom$ $panels$) and X-axis ($top$
  $panels$) for three models, $E$, $S$ and $S+E$ (from left to right); the last 2 columns are the residuals $E$ -$S$ and $E$ - $(S+E)$. All the main components have an orientation of $|\alpha| \sim
  20^{\circ}$. As the difference between the $E$ and $S$ models
  illustrates, the $E$ density law is more centrally concentrated and
  this makes it a slightly better fit to the data. The $E$ - $(S+E)$
  residuals are irregular because $S$ and $E$ components have different viewing angles. The segments shown are the scale lengths along
  the major axis. Also marked, are the cut-off radius $R_{c}$ and the limiting longitudes of the VVV survey. Overlaid on model $E$, we add the X-shaped density law provided by  \citet{Lo2017}. We project on the X-Y plane only a slice of the density law within $-0.7<Z/$kpc$<-2.0$ (below the red line in the Y-Z plane) corresponding to $b<-5^{\circ}$. The density contours of the foreground RC (X < 0) are marked in violet and those of the background RC (X > 0) in yellow. The same colour scheme is adopted for the Y-Z plot projection where we do not make any Z cuts. } 
\label{xyz}
\end{figure*} 
%
%
\begin{figure*}
\centering
\includegraphics[scale =0.112]{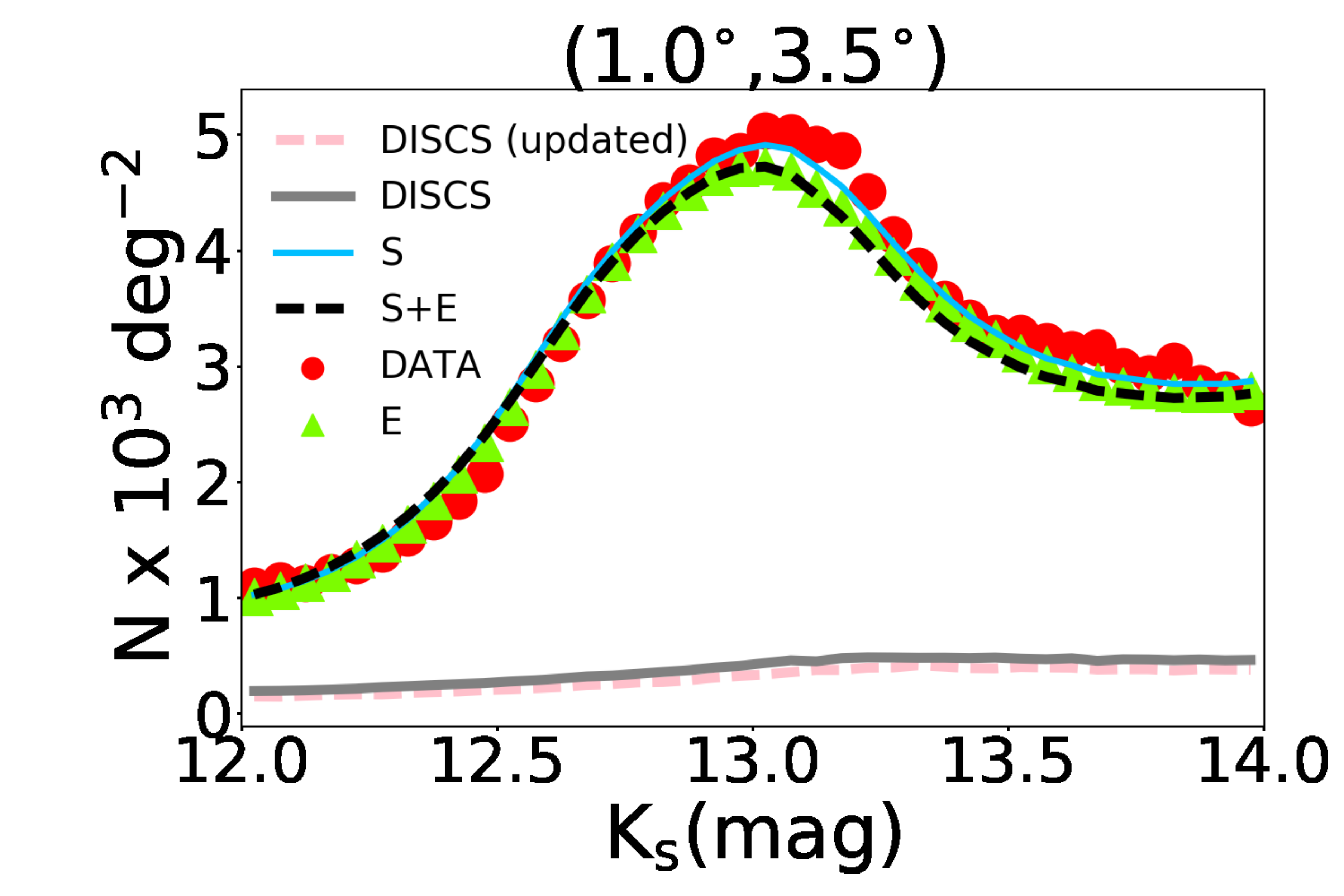}
\includegraphics[scale =0.112]{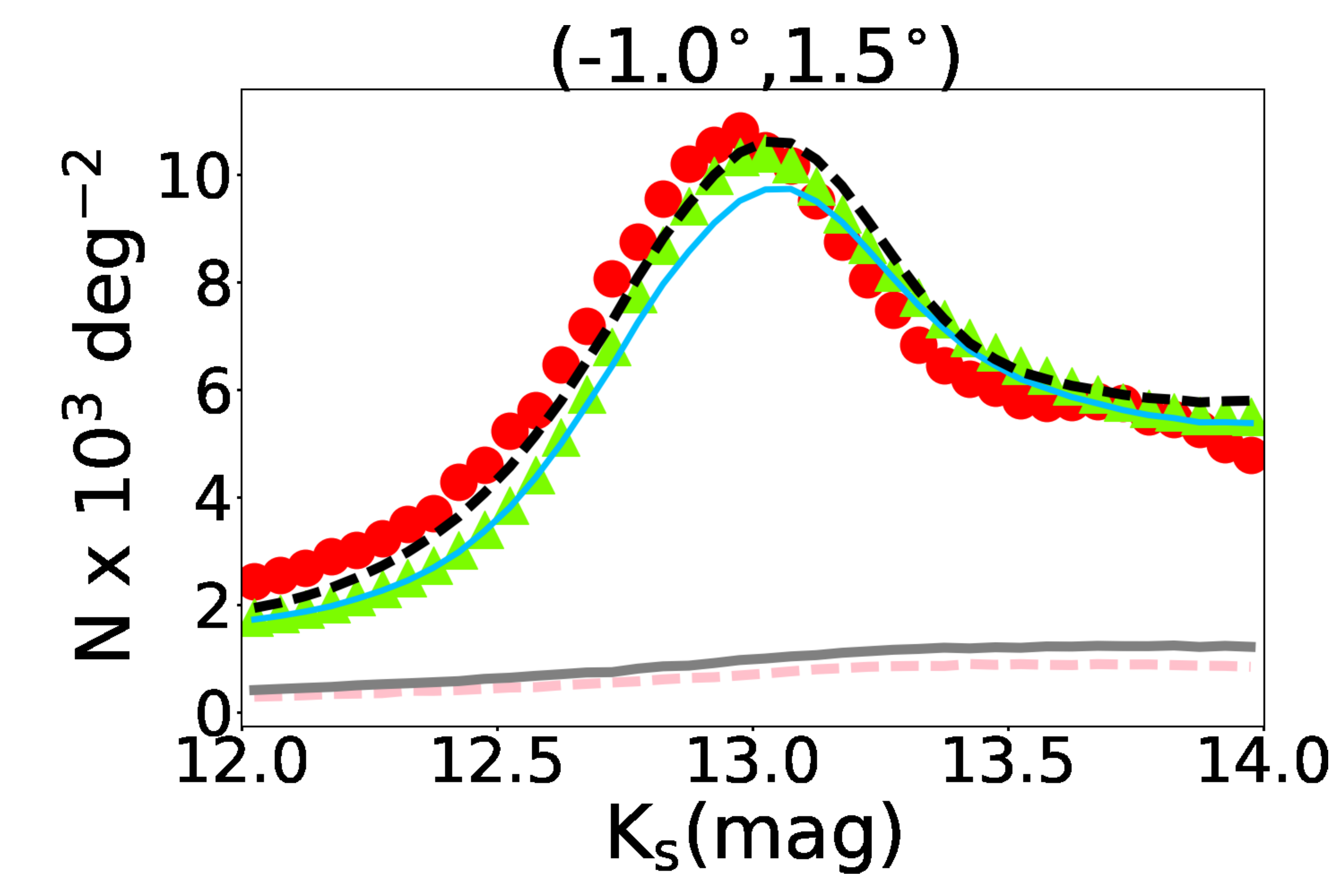} 
\includegraphics[scale =0.112]{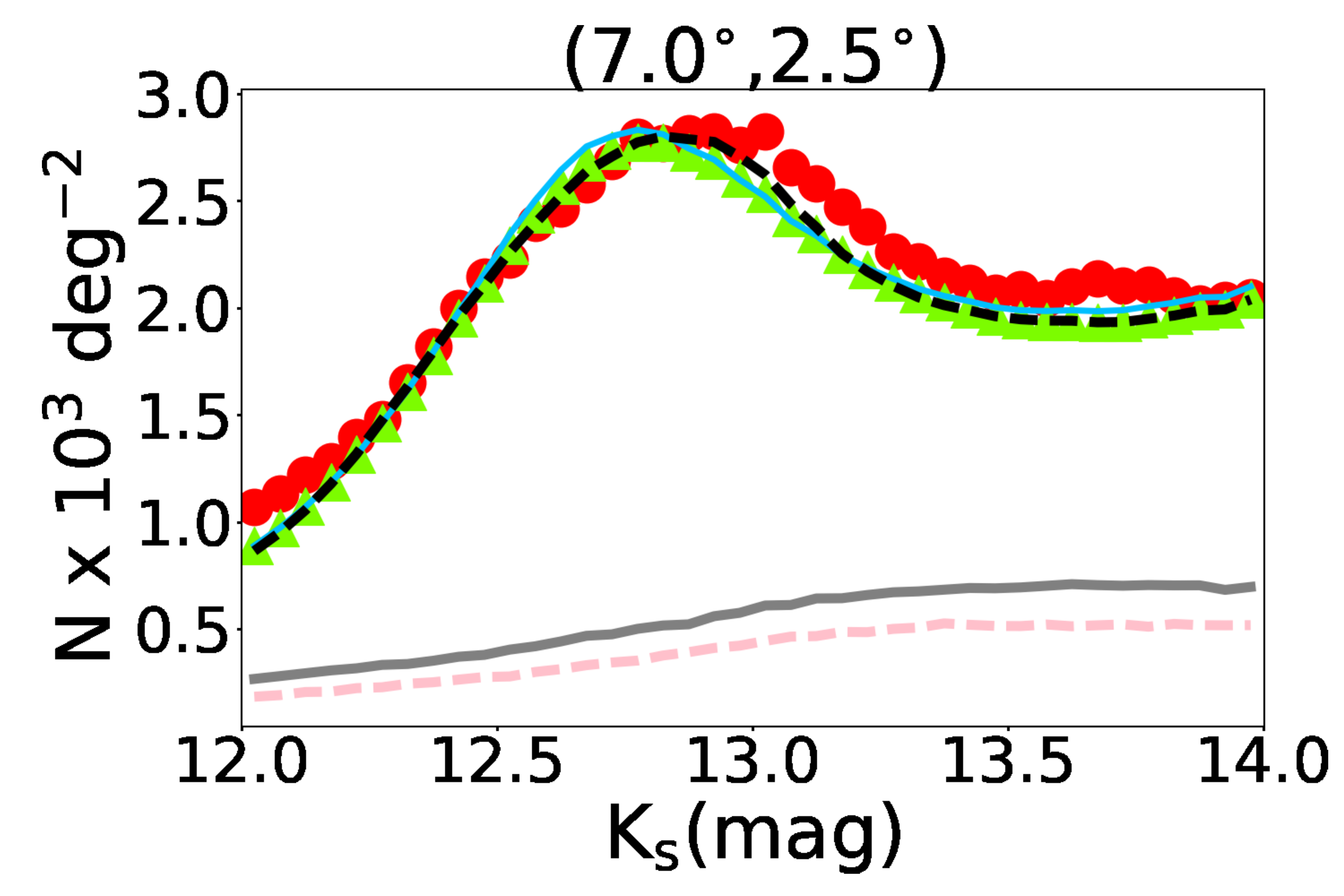}
\includegraphics[scale =0.112]{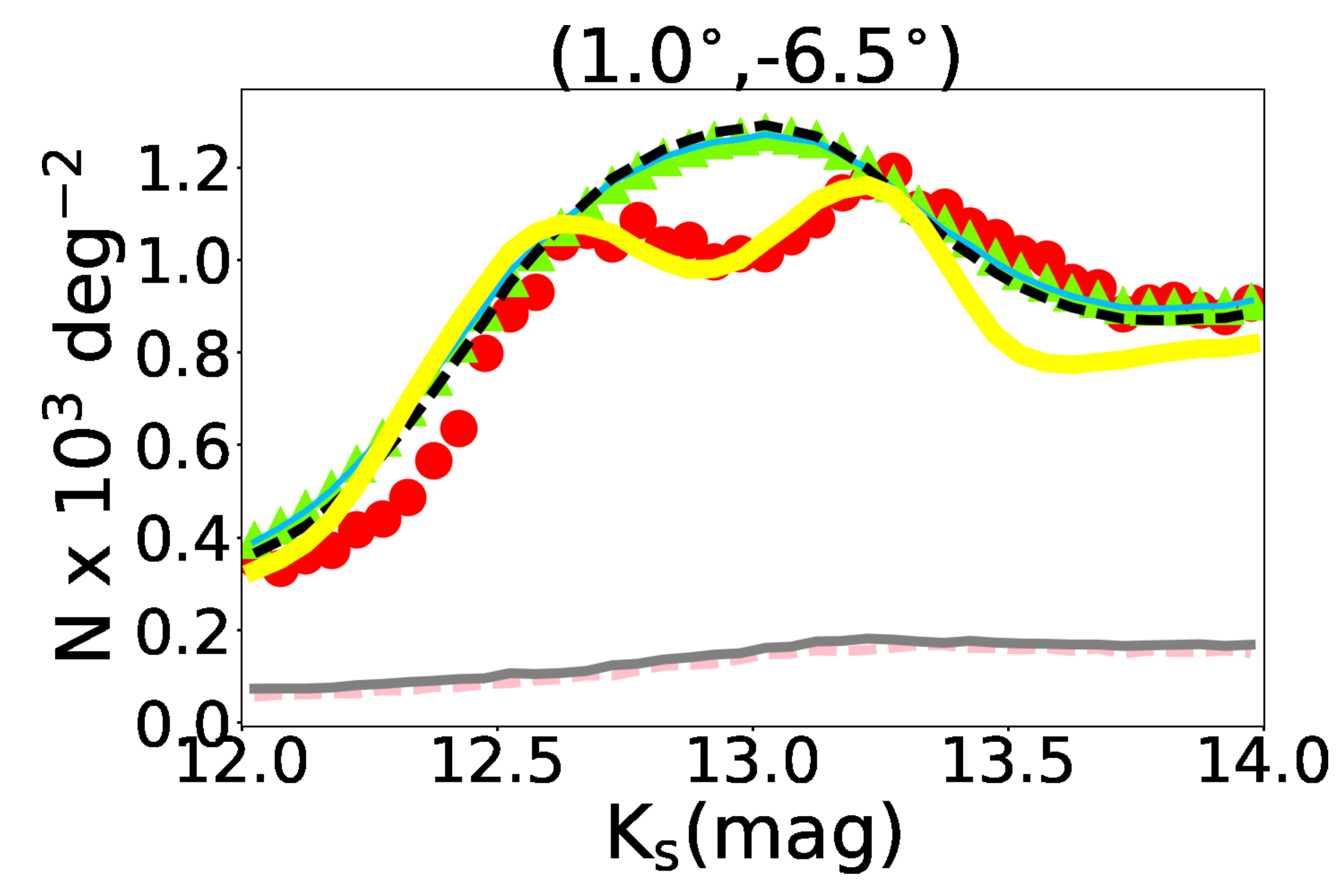} 
\subfloat[]{\includegraphics[scale =0.112]{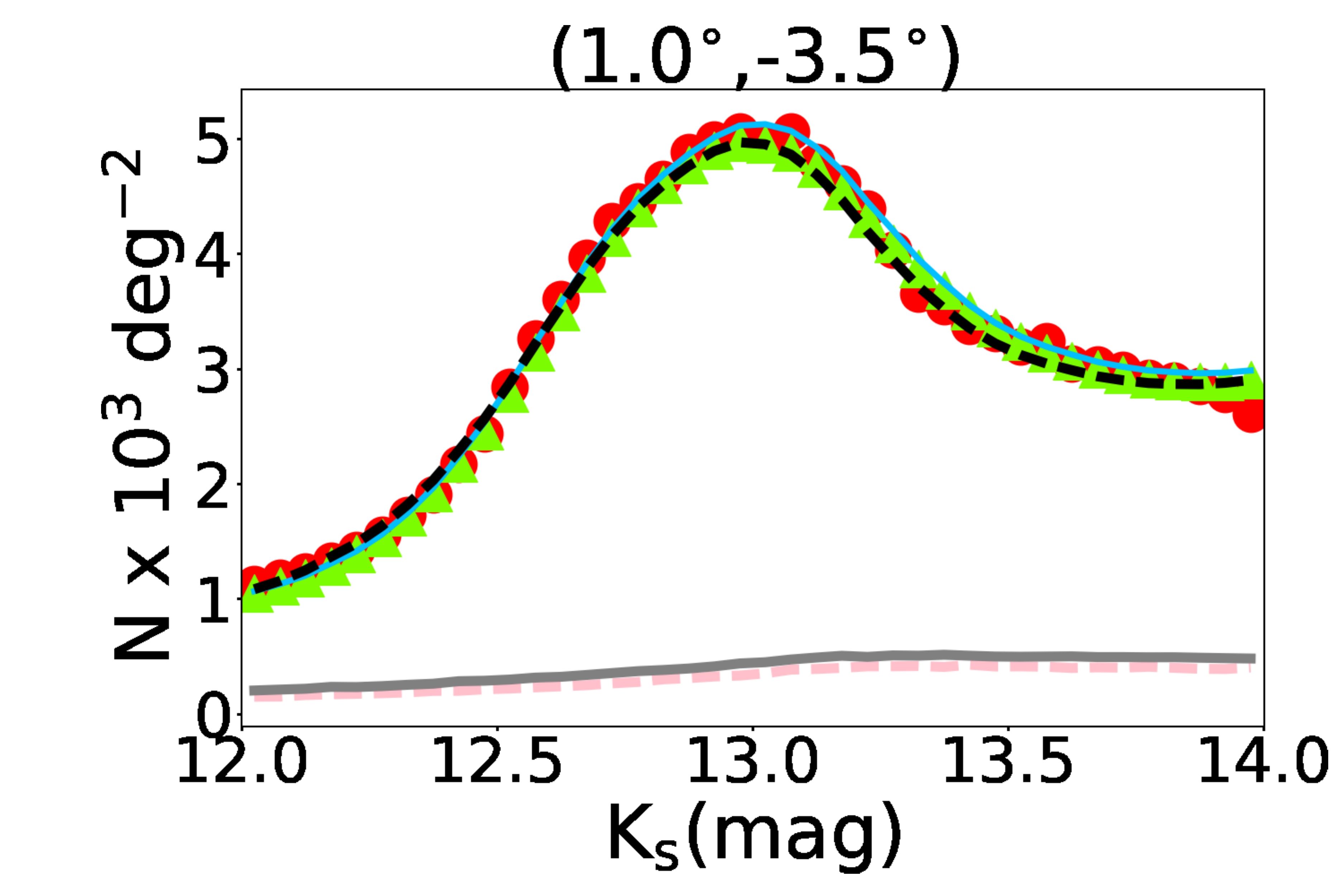}}
\subfloat[]{\includegraphics[scale =0.112]{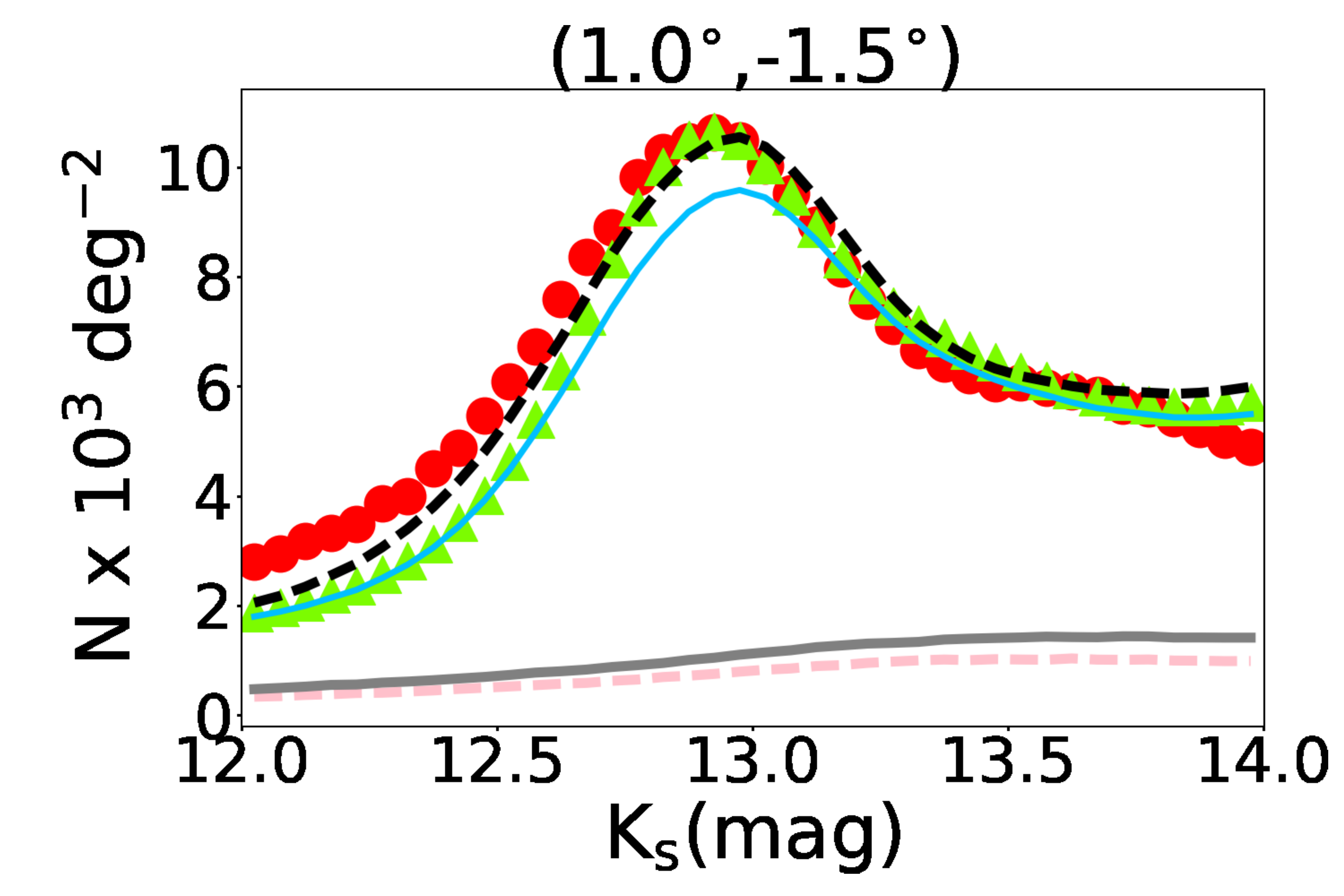}}
\subfloat[]{\includegraphics[scale =0.112]{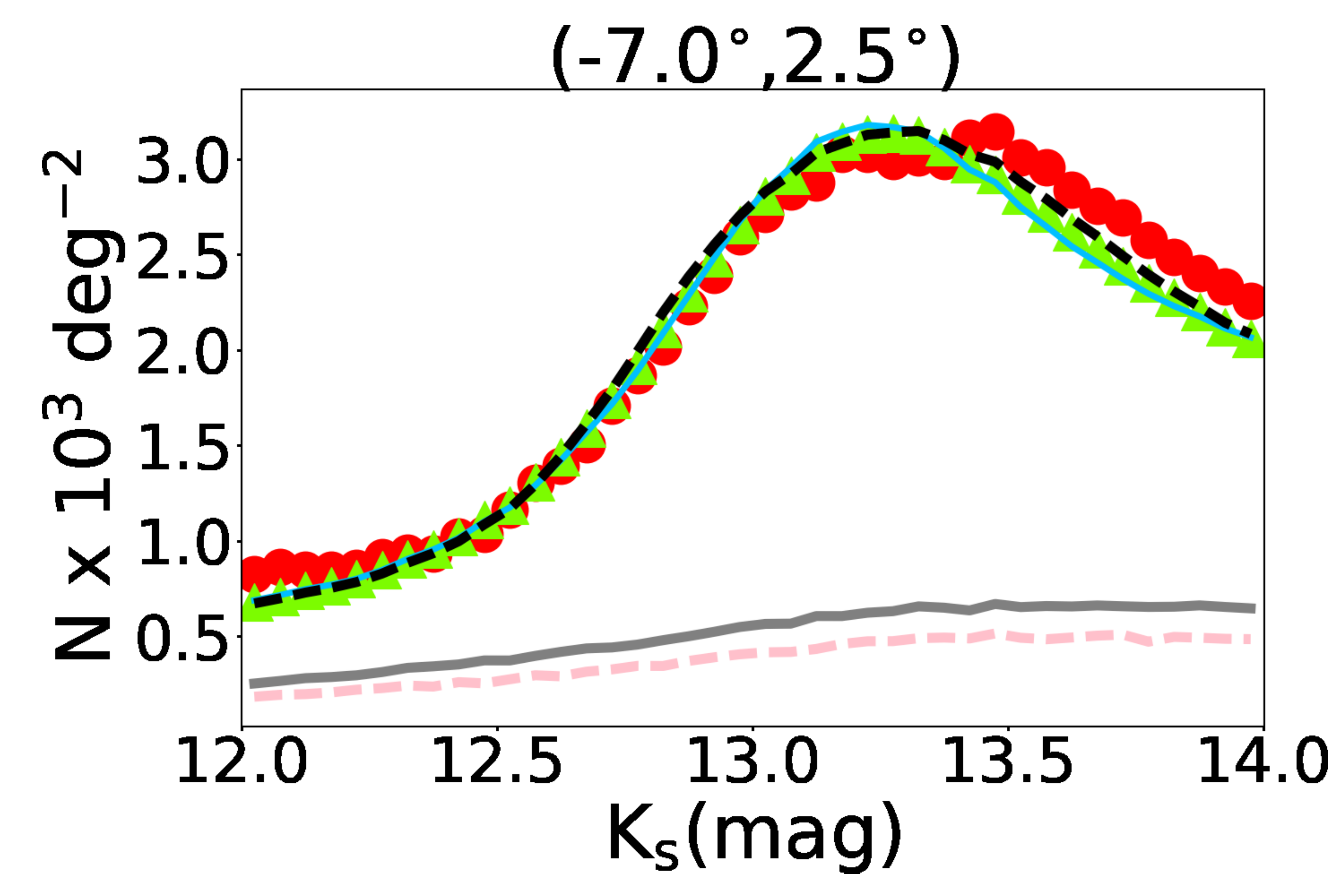}}
\subfloat[]{\includegraphics[scale =0.112]{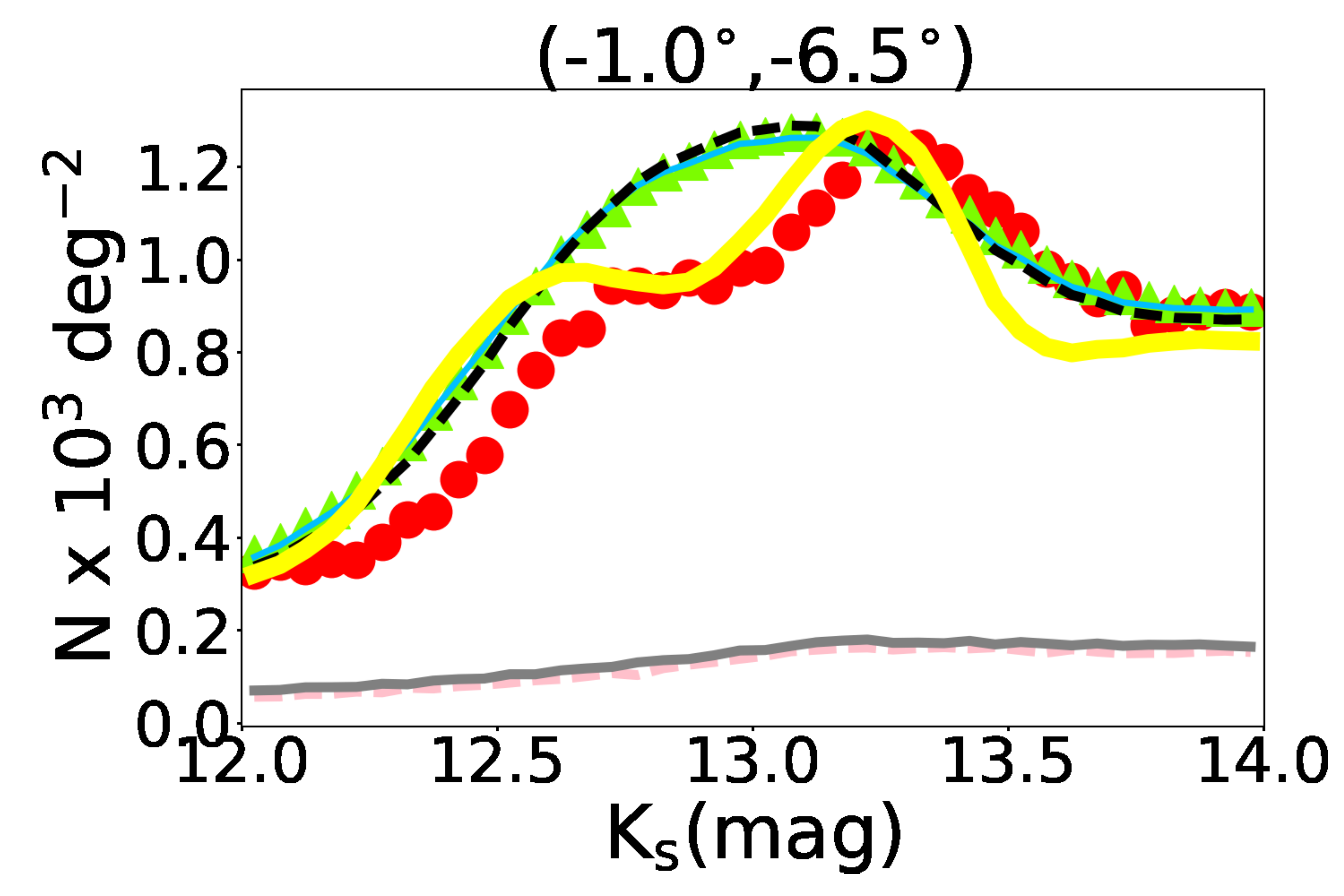}}
\caption[Apparent magnitude distributions for 8 different fields of
  $2^{\circ} \times 1^{\circ}$ for the data and the best fit
  models]{Apparent magnitude distributions for 8 different fields of
  $2^{\circ} \times 1^{\circ}$ (first plot also shown in
  Figure~\ref{LF_shift}), marked in the middle 
  panels of Figure~\ref{diff} with letters, where each letter corresponds to the two fields in a column.  Red circles are the VVV data counts in the 5 magnitude bins used
  in the fitting procedure, while the continuous red line is the
  magnitude distribution calculated on a finer resolution. The disc populations are
  shown in gray and the best fit models in green ($E$), blue ($S$) and
  black ($S+E$). $Column$ (a): the magnitude distributions are symmetric with
  respect to the Galactic Plane; also notice the good agreement
  between the data and the model. $Column$ (b): magnitude distributions in two
  tiles ($b306$ and $b347$) that pop up in the residuals in
  Figure~\ref{res}. The single component models do not fit the
  data very well in the first two bright magnitudes slices, which produce the
  overdensities observed in the top panels of
  Figure~\ref{diff}. Adding an extra component (black squares) partially solves the
  problem. $Column$ (c): magnitude distributions in symmetric
  fields with respect to $l = 0^{\circ}$. The peak of the RC is
  shifted between positive (bright RC) and negative (faint RC)
  latitudes, a consequence of the angle of the major axis with respect
  to the Sun-GC line. $Column$ (d): magnitude distributions at $b
  < -5^{\circ}$ where the double RC is clearly visible in the
  data. Our resolution here $\Delta K_{s} = 0.4$ mag does not allow us to
  fit for the double RC. We add the \citet{Lo2017} model in yellow, which predicts a double RC.}
\label{LF_all}
\end{figure*} 
\begin{figure*}
\centering
\includegraphics[scale =0.112]{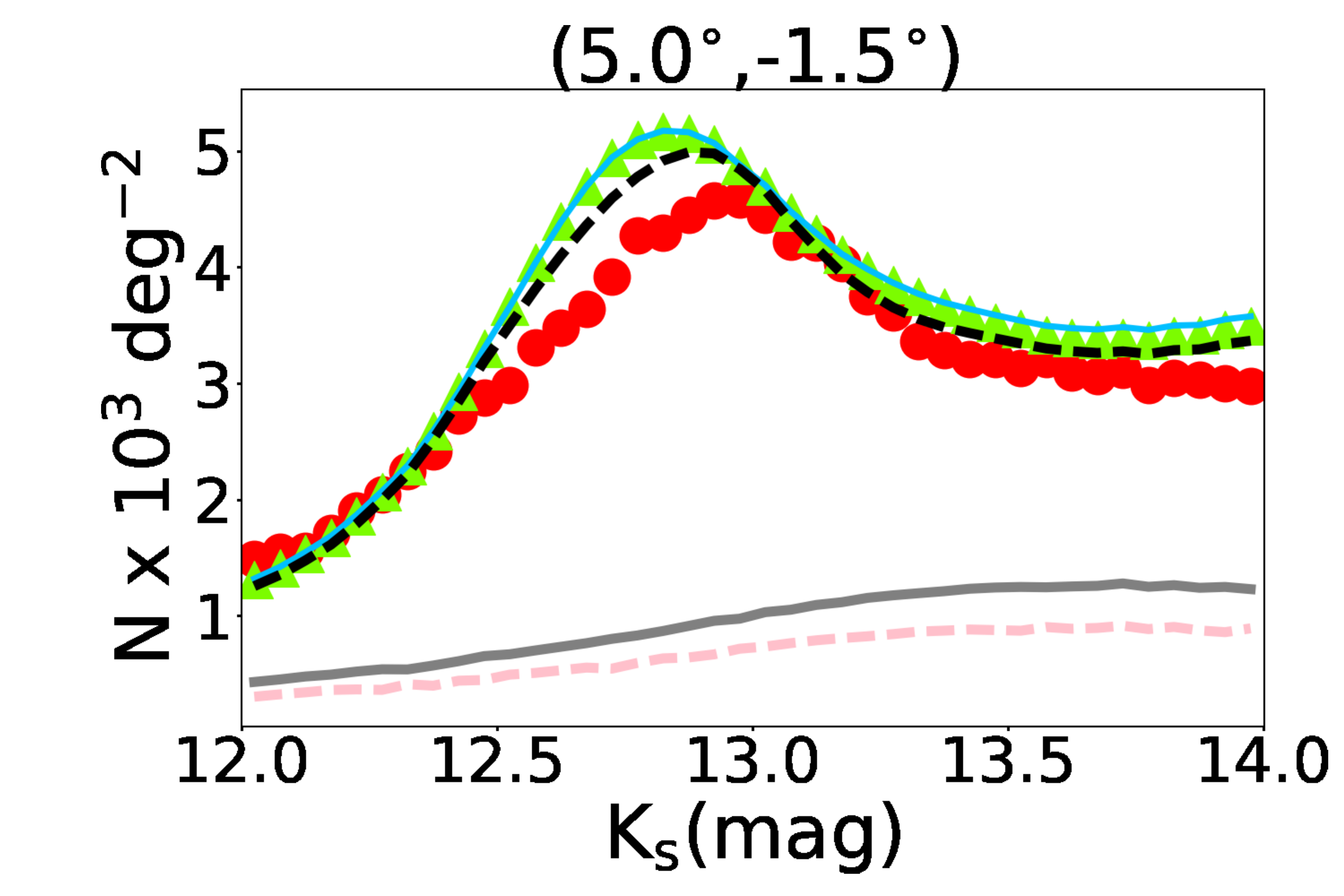} 
\includegraphics[scale =0.112]{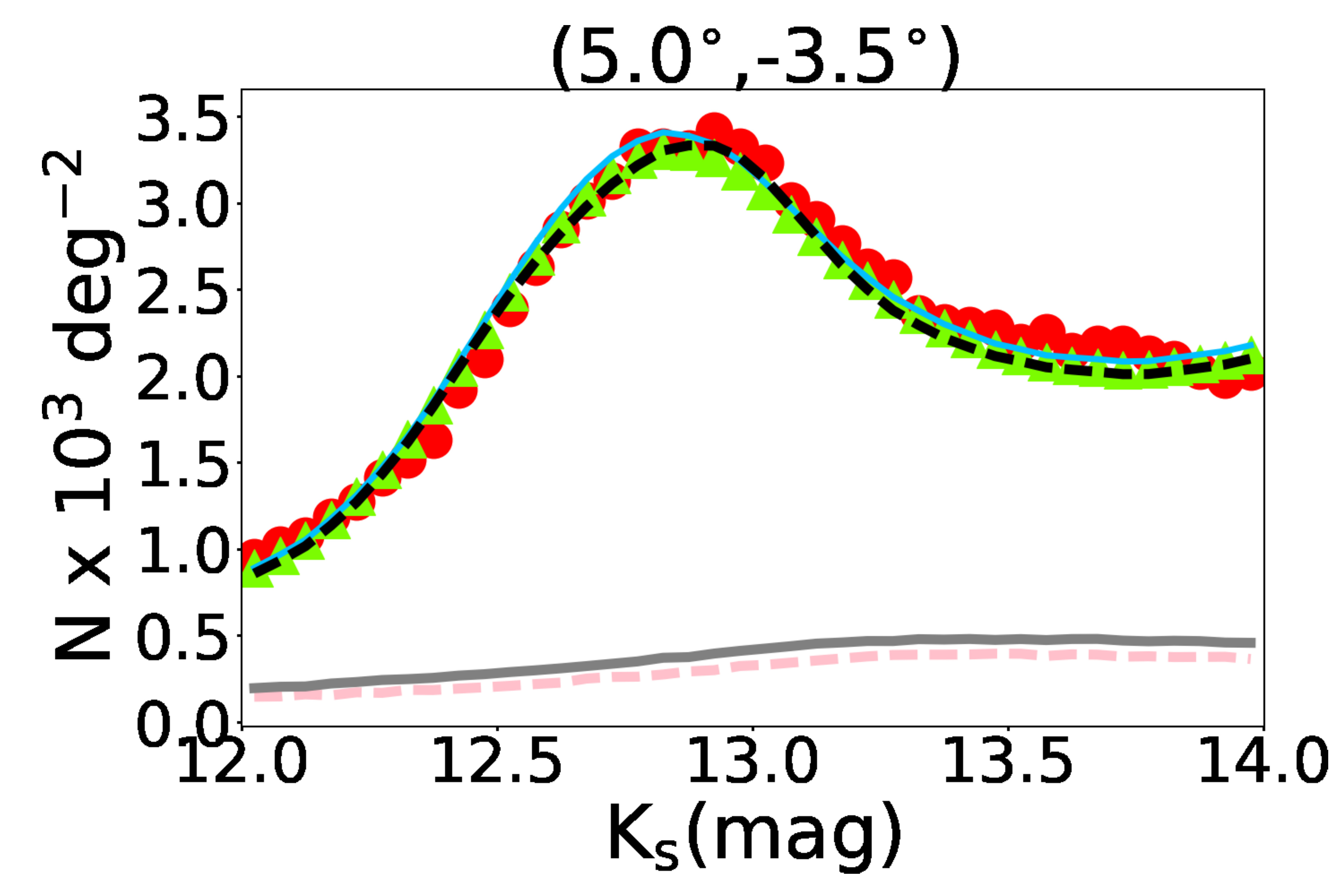} 
\includegraphics[scale =0.112]{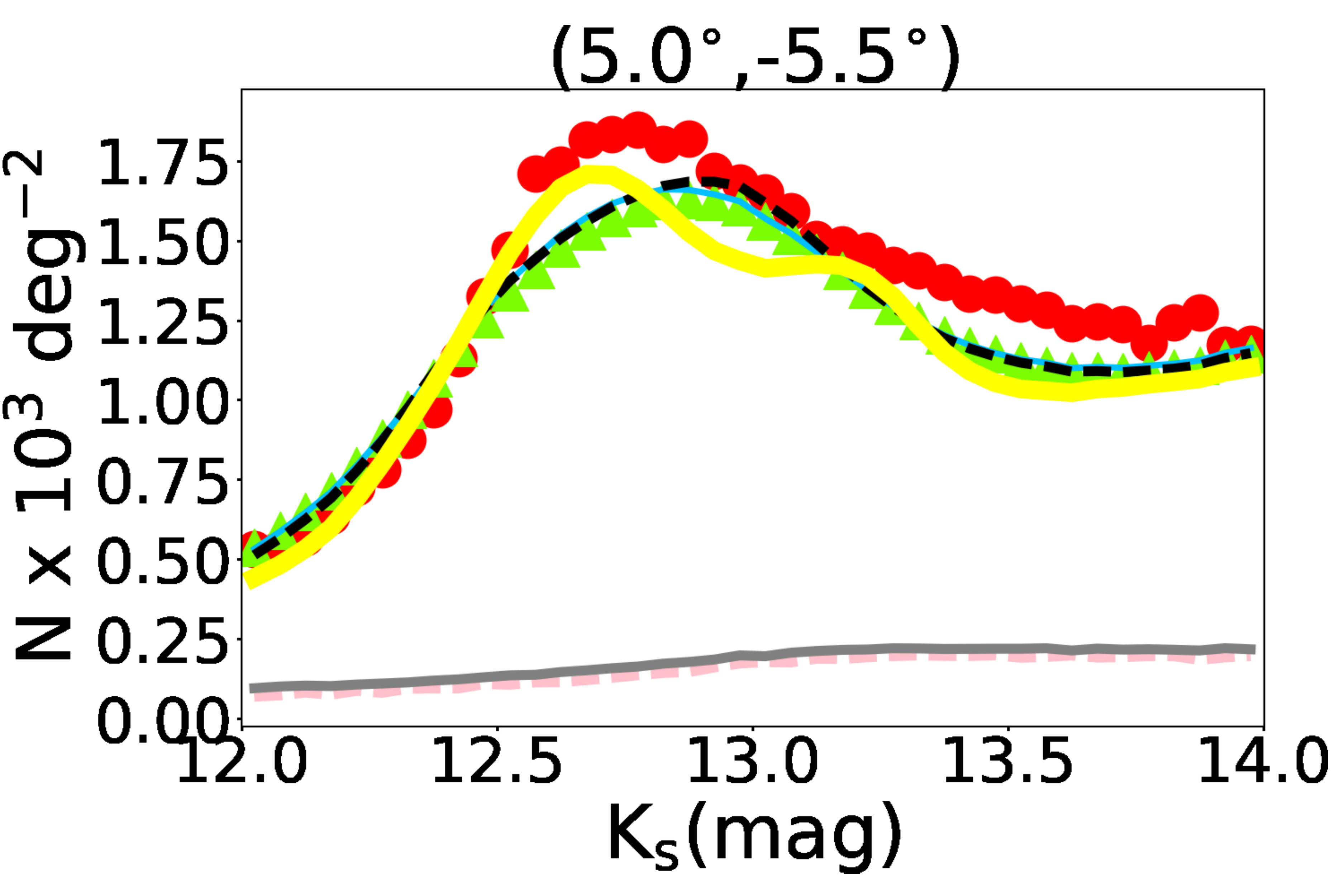}
\includegraphics[scale =0.112]{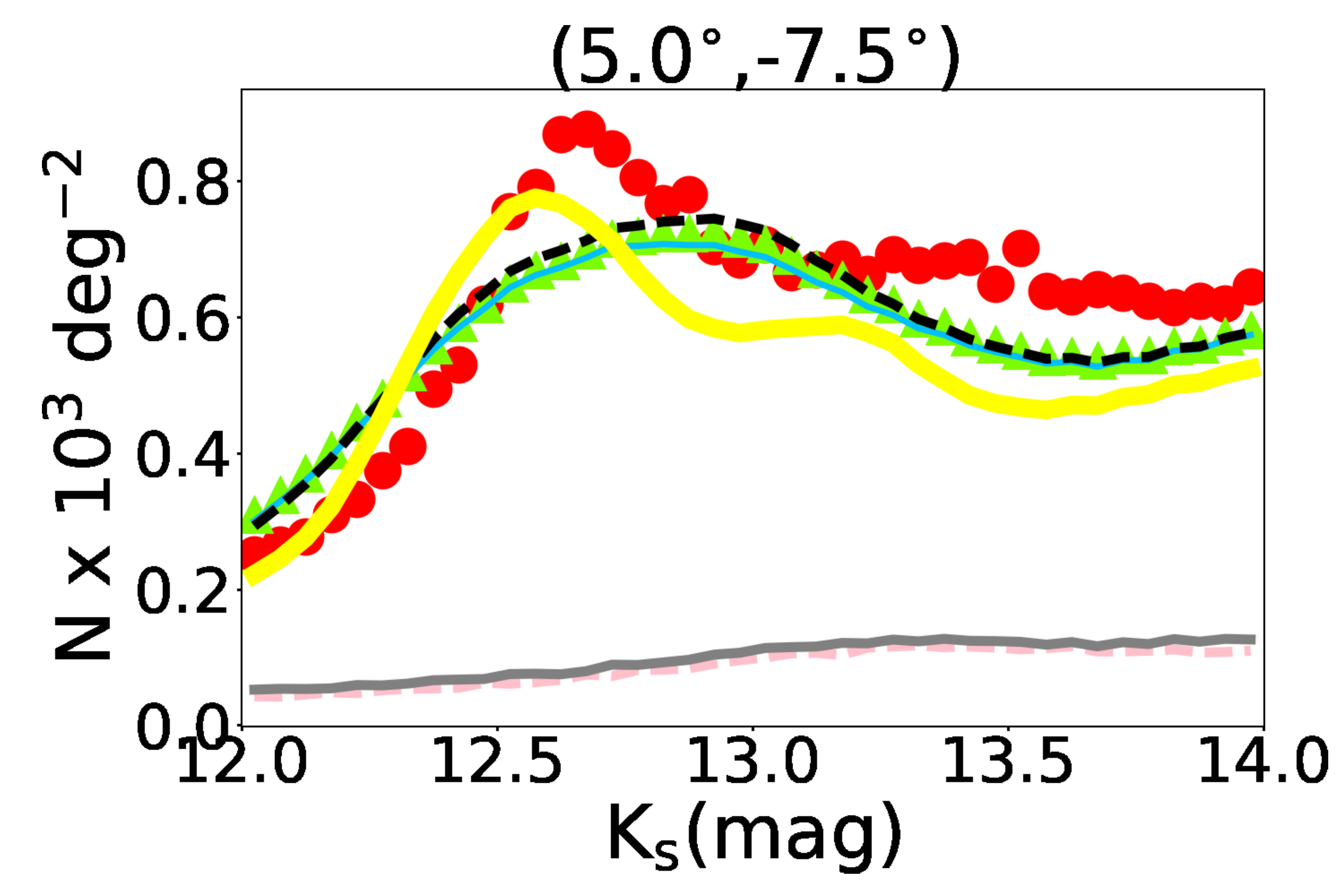} \\
\includegraphics[scale =0.112]{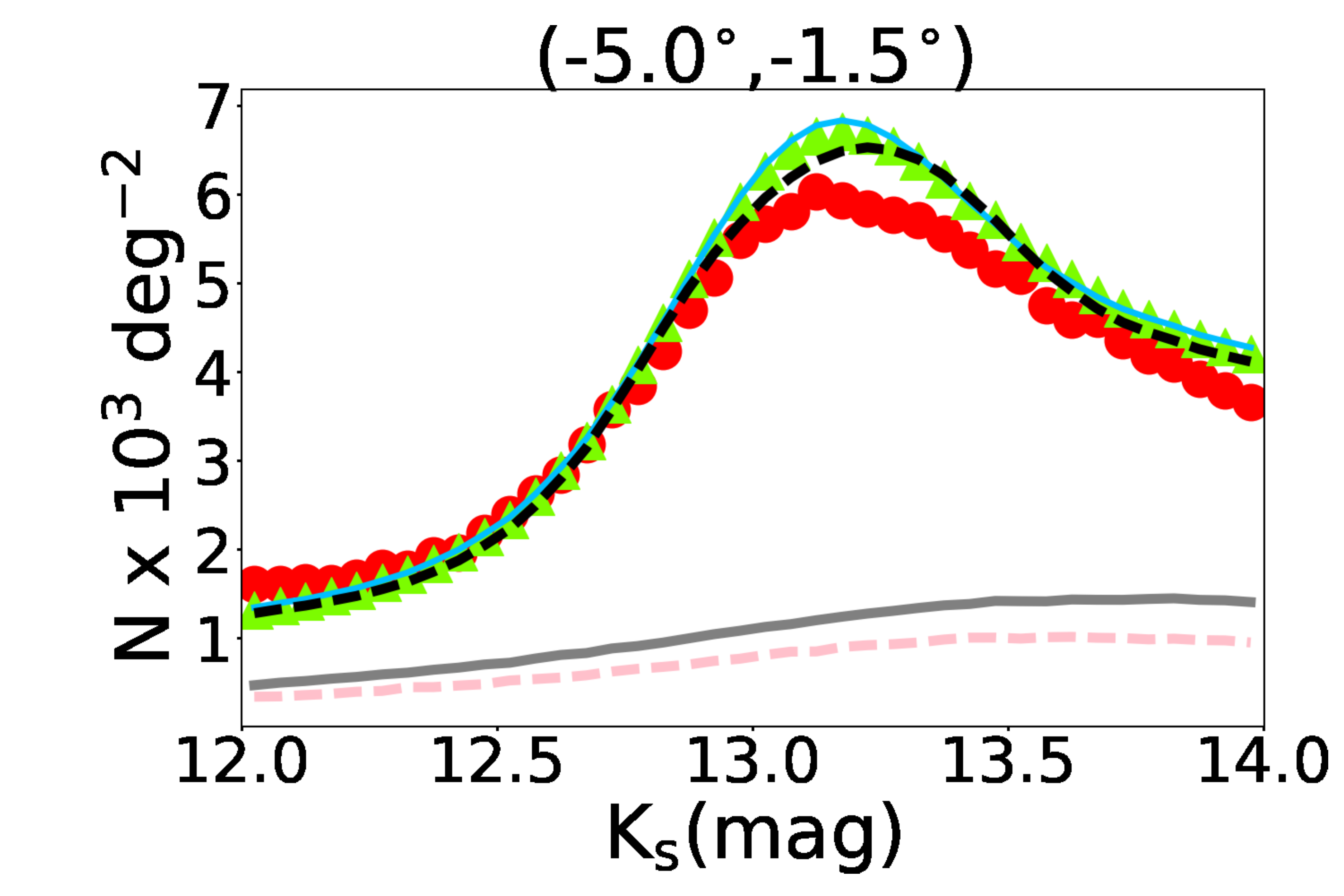} 
\includegraphics[scale =0.112]{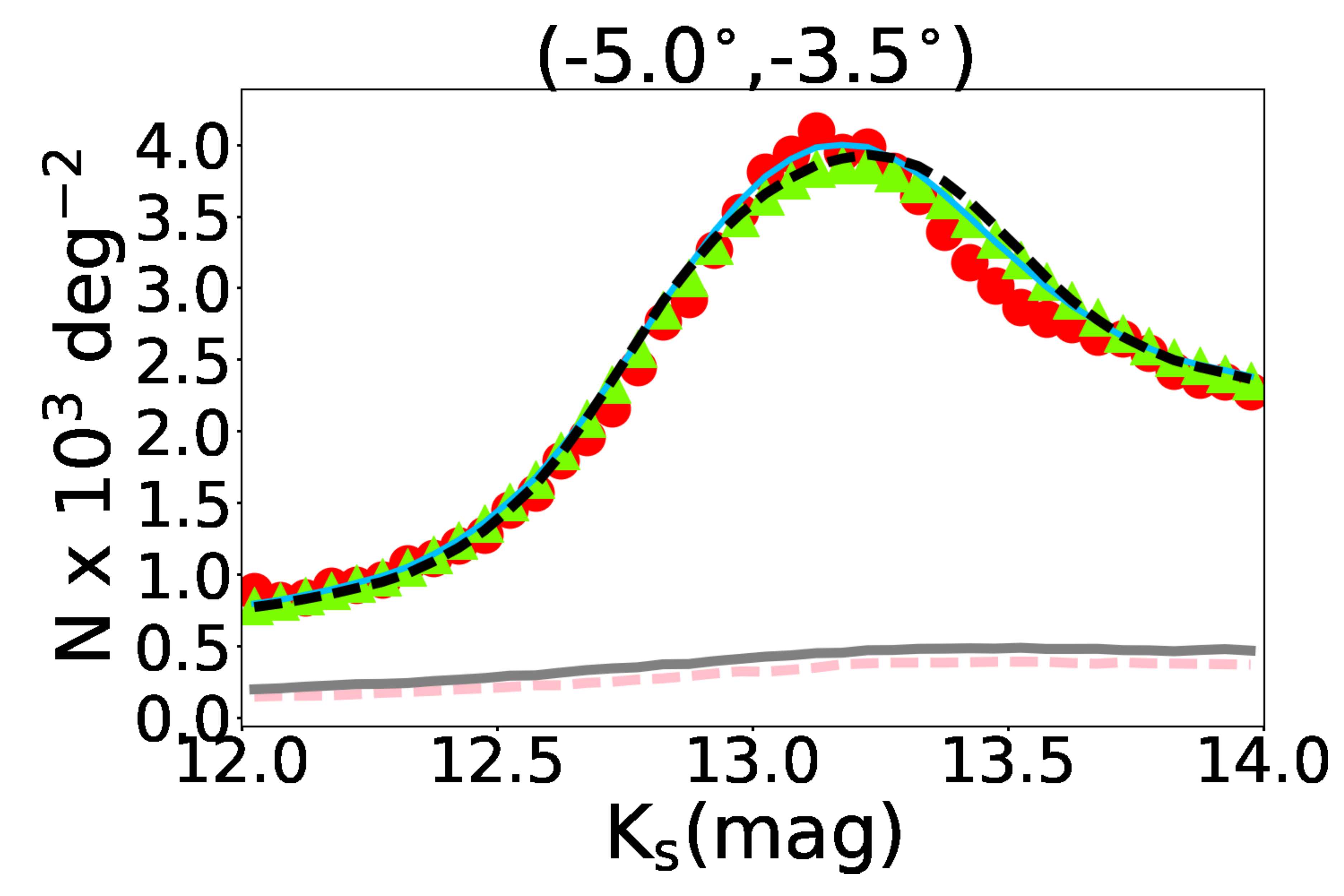} 
\includegraphics[scale =0.112]{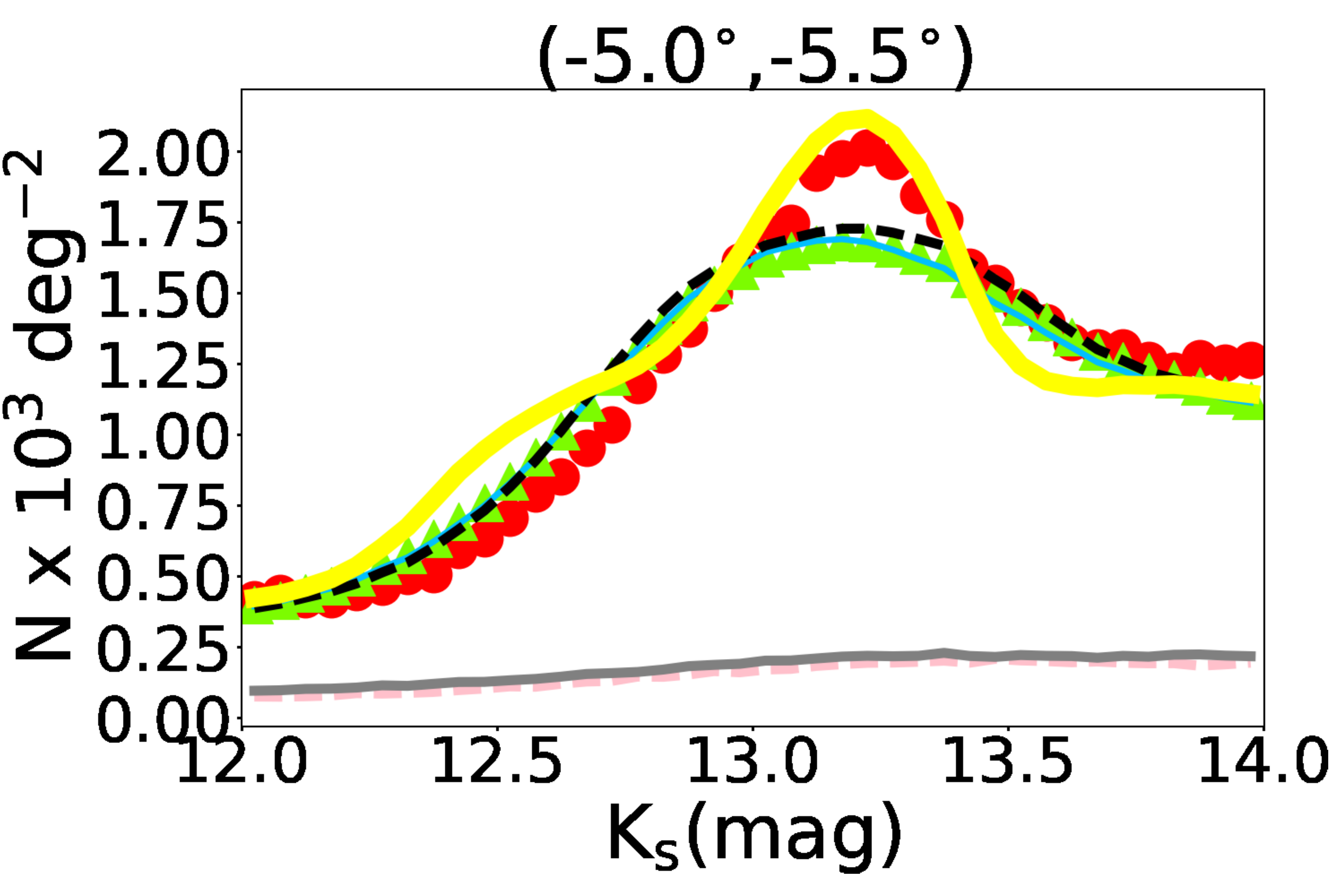}
\includegraphics[scale =0.112]{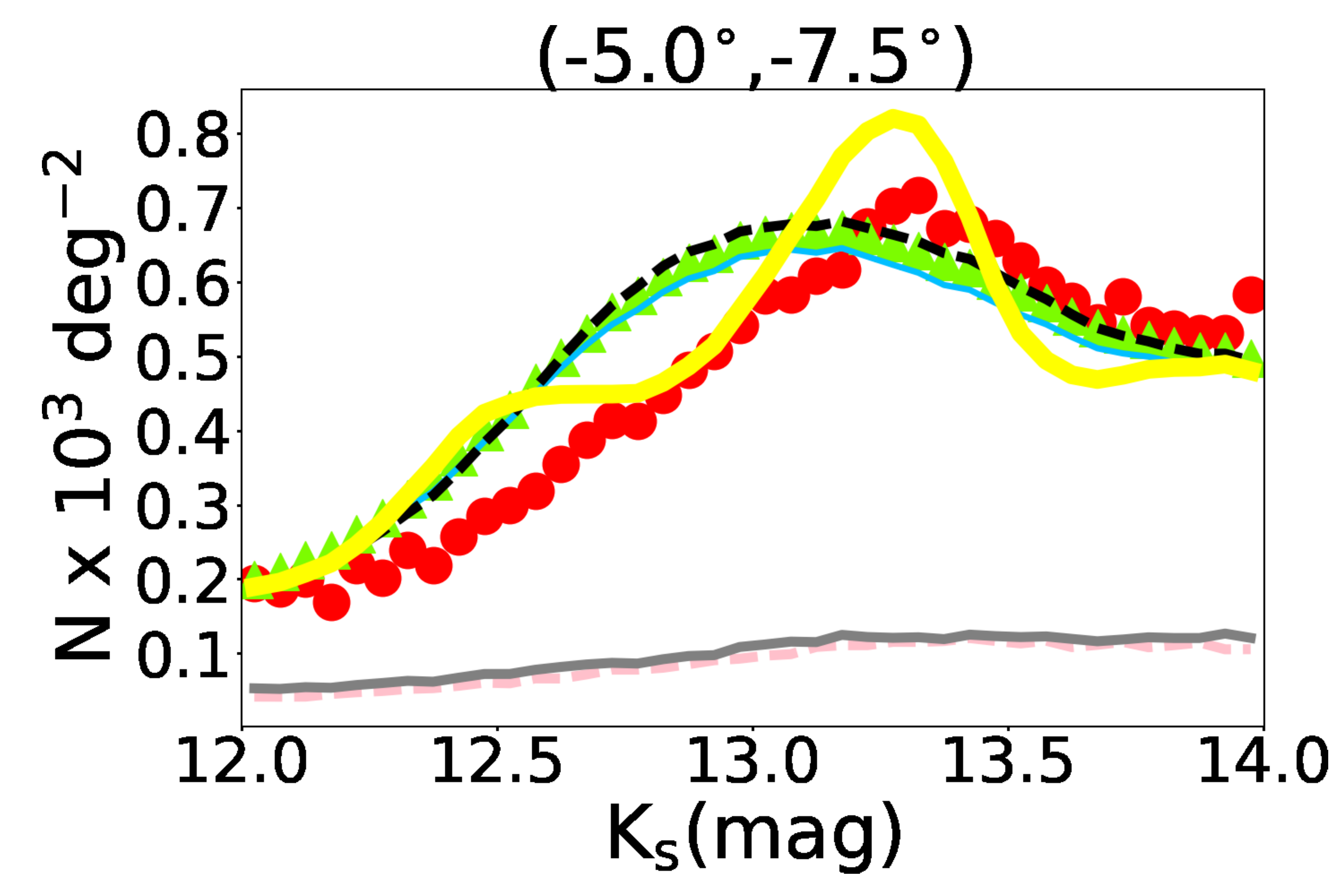} 
\caption{Apparent magnitude variation across the X-shaped arms at intermediate
  longitudes, the centres of the fields are marked with crosses in Figure~\ref{diff} (see also caption of Figure~\ref{LF_all}). Each row has constant longitude, the top row has $l = 5^{\circ}$ and the bottom row $l=-5^{\circ}$, with the latitude decreasing from left to right. A large portion of the fields close to the Galactic Plane ($first$ $column$, $-2^{\circ}< b <-1^{\circ}$) are masked out in the fitting
  procedure because of the high reddening and the amplitude of the RC
  is not well fit (we show the result of the global fit). The fit
  at intermediate latitude ($second$ $column$, $-4^{\circ}< b <-3^{\circ}$
  ) performs well, while at lower latitudes ($third$ $column$,
  $-5^{\circ}< b <-6^{\circ}$ ), where the X-shape arms become visible,
  we can clearly see that the amplitude of the RC peak in the data is
  not well matched by the model. At even lower latitudes ($fourth$ $column$,
  $-7^{\circ}< b <-7^{\circ}$) the discrepancy is more obvious at all
  magnitudes due to the lower weight the low number counts in these
  fields have in the overall fit. The \citet{Lo2017} model, in yellow, is shown only for fields with $b<-5^{\circ}$.}
\label{LF_all2}
\end{figure*} 
%

\subsection{Final Results and MCMC Checks}

The best fit parameters describing the Bulge density laws are listed in Table~\ref{bestresults} for each model
considered together with the maximum log-likelihood ln$(L_{\mathrm{P}})$ and the total number of stars in each component. The statistical errors
for all model parameters are negligible compared to the complex
systematic errors, thus they are not reported.

In Section~\ref{sec:fitting}, we tested our fitting procedure on a mock 
catalogue and proved that it successfully recovered the input values. We thus
expect, within the limitations of the model and of the luminosity
function assumptions, to have found the true values of the Bulge
density distribution. Nonetheless, we execute an additional check of
the results and of their statistical errors with a Markov chain Monte
Carlo (MCMC) in a Bayesian framework. In Figure \ref{cov} in the Appendix we compare the results (green dots) obtained using the L-BFGS minimisation method and an MCMC, for model $S$ (label `$\sigma^{\mathrm{RC}}$ free' in Table~\ref{bestresults}, see discussion in Section 7.1).

We have used the Affine Invariant MCMC
Ensemble sampler \textit{emcee} \citep{Fo13} to explore the full
posterior distribution with 30 walkers. The contours of the
two-dimensional projection of the posterior probability distribution
of the parameters of model $S$ are nearly Gaussian (Figure \ref{cov}), but there is a clear correlation between the
scale-lengths of the Bulge density law, and $\sigma^{\mathrm{RC}}$ and the viewing angle $\alpha$.
The marginalised distribution for each parameter is shown along the
diagonal and the 0.5 quantiles with the upper
and lower errors (16\% and 84\% quantiles) are given above each 1-D histogram. For the initial
state of the Markov chain we chose the best-fit results reported in
Table~\ref{bestresults} for model $S$, label `$\sigma^{\mathrm{RC}}$ free': with 30,000 iterations (20\% of the chain was
discarded as burn-in) and an acceptance fraction of 43\%, the final
results and errors on the parameters are in complete agreement with
the values found with the L-BFGS method. The statistical errors are
equally small and, as can be seen from Figure \ref{MCMC} showing a
sample of 10 chains for 8 parameters, they converge to a value close to the initial one
(the blue line in Figures \ref{cov} and \ref{MCMC}).
We note that \citet{We15} also report small
statistical errors: ``We have performed an MCMC to estimate errors on
both the model parameters (...) The resultant statistical errors are
extremely small, significantly smaller than the difference between
fits with different models. We therefore consider the statistical
errors to be negligible in comparison to systematic errors".

There may exist various causes for the systematic errors: the
extinction related dispersion and photometric errors (see
Figure~\ref{errK}); confusion of point sources in overcrowded regions;
clssification and calibration errors; and the fact that the model
is not an exact description of the data. In
addition, at the end of Section~\ref{sec:fitting}, we have shown that the 
choice of the Bulge density distribution alone can introduce systematic errors
larger than the standard errors.

The next subsections look into more detail at the models reported in Table
\ref{bestresults} and their suitability in describing the data.

\subsection{Data versus model comparison}

We here discuss Figures \ref{models}-\ref{LF_all2}, which
show the data and the best-fit models (labeled `main' in Table \ref{bestresults}) in different projections.

The log likelihood values in Table~\ref{bestresults} suggest 
model $E$ performs slightly better than model $S$; however, there are many similarities between the
two sets of results: for both, we obtain $\alpha \simeq -19.5^{\circ}$ and
axis ratios [1:0.44:0.31]. As we will see in the residual maps, a two component model 
($S+E$) provides a better description of the data 
especially in the inner regions, but at the expense of a
larger number of free parameters.

Using the first data release of the VVV, \citet{We13} fitted exponentials
to the density profiles of the major, intermediate and minor axes shown (see
their figure 15) and found scale lengths of 0.70, 0.44 and 0.18 kpc
respectively, corresponding to axis ratios [1:0.63:0.25]. In the
Discussions section we compare our results with other
literature estimates for the shape and viewing angle.

In Figure~\ref{models} we show the number density distribution of VVV
stars in five magnitude bins (left column) and the best-fit models $E$
and $(S+E)$ in Galactic coordinates (middle and right column
respectively). Model $S$ is very similar
to model $E$ and thus not shown. A magnitude slice is almost equivalent to selecting stars within a certain distance range, where the brightest magnitude slice $12<K_{s}<12.4$ selects the nearest stars, in front of the Bulge, and the faintest magnitude slice $13.6<K_{s}<14.0$, selects the furthest stars, behind the Bulge.

The middle and right panels of figure~\ref{diff} show the difference between 
the data and the $E$ and $S+E$ models, divided by the Poisson noise in the 
model, ($N_{i}$-$M_{i}$)/$\sqrt{M_{i}}$, for each magnitude bin. The addition of the second component $E$ (left column) improves the fit in the central region, especially in the bright magnitude slice (top right panels of Figure \ref{diff}). Model $S+E$ assumes 
a young population (5 Gyr), described by model $E$ and an old one (10 Gyr), 
described by model $S$, which is the main component. The younger component has 
a $0.1$ mag brighter peak than the 10 Gyrs old population (see Figure \ref{LF}
where $\mu_{5}^{\mathrm{RC}}=$ -1.64, hence $\mu_{5}^{\mathrm{RC}}=$ -1.74 after applying the 
$\Delta \mu^{\mathrm{RC}} = -0.1$ shift), which helps to provide a better fit in the 
central regions of the brighter magnitude bins. The LFs for 
both populations are shown in the right panel of Figure \ref{LF}, where the integrated area under each curve gives the total number of stars expected for each component 
for the best fit parameters. We provide several interpretations for the additional 
component in Section~\ref{sec:discussion}. 

To summarise our results, we show the residuals for all
three models in one magnitude slice in Figure~\ref{res}. As previously observed, models $E$ and $S$ do not provide a good match for the data in the area close
to the Galactic Centre, $|b|<3^{\circ}$ and $|l|<2^{\circ}$, though in
general they perform very well. The model $(S+E)$ matches the observed
distribution remarkably well (with a median percentage residual of 5 \%) apart from some discrepancies within the
central tiles $b306$, $b320$, $b347$.  It is not clear at this stage
whether this is entirely due to data processing or calibration issues or 
possibly some real feature present in the central data.

The best-fit density laws projected along the Z-axis (bottom
panels) and X-axis (top panels) for the models $E$, $S$ and $S+E$
(first three columns of the figure) are shown in Figure~\ref{xyz}. On the figure, we mark the cut-off radius $R_{c}$, the viewing angle, and the limiting longitudes of the VVV survey. Overlaid on model $E$, we add a density law that is a complex parametrisation of the Bulge X-shaped component, provided by  \citet{Lo2017} in their equation 4. This model predicts a double RC at intermediate latitudes. We only show a slice of the density law within $-0.7<Z/$kpc$<-2.0$ (below the red line in the Y-Z plane) corresponding to $b<-5^{\circ}$, projected onto the X-Y plane. The density contours of the foreground RC (X < 0) are marked in violet and those of the background RC (X > 0) in yellow. The same colour scheme is adopted for the Y-Z plot projection where we do not make any Z cuts, to show that the nearer RC will be dominant at positive longitudes while the further RC will be dominant at negative longitudes, symetrically above and below the plane. In addition, we can expect: 1) the contribution to the magnitude distribution of the background RC to diminish as the longitude to increase towards positive values, as the contribution of the foreground RC increases and 2) the separation between the two RCs in the magnitude distribution to increase as we look away from $l=0^{\circ}$. This separation will depend on the angles formed by the X-shape with the GP as well as on the Galactic coordinates or distance from the Galactic plane.

The Y-Z distributions display a 'boxy' Bulge with a flattened distribution along the Z direction. The difference between the $E$ and $S$ models (fourth column) illustrates that the $E$ density law is more centrally
concentrated. The $E$ - $(S+E)$ residuals in the X-Y plane (fifth column) are
irregular because the two components in the $S+E$ model have different viewing angles. 

In Figure~\ref{LF_all} we show 8 examples of magnitude distributions in
$2^{\circ} \times 1^{\circ}$ fields where the data points we fit are
the red circles and the best-fit models are represented with a black
dotted line (model $S+E$), green triangles (model $E$) and a thin blue line
(model $S$). The disc populations are shown in gray (the updated discs, discussed in Section 7.2, in pink).
The centres of the fields are marked with letters in the middle column of 
Figure \ref{diff}, where each letter represents a column of 
Figure~\ref{LF_all}. The panels are organised as follows:

\begin{itemize}
\item column (a) shows the magnitude distributions in two symmetric
  fields centred at $(l,b) = (1^{\circ},3.5^{\circ})$ and $(l,b) =
  (1^{\circ},-3.5^{\circ})$: the star counts are symmetric with
  respect to the Galactic Plane and there is good agreement between
  the data and the best fit model.
\item column (b) shows the magnitude distributions in two tiles
  ($b306$ and $b347$) that have pronounced residuals in
  Figure~\ref{res} with centres at $(l,b) = (-1^{\circ},1.5^{\circ})$
  and $(l,b) = (1^{\circ},-1.5^{\circ})$, though interestingly these are
  symmetric with respect to the Galactic Centre. The fit is poorest
  in the first two brighter magnitudes slices, which produce
  the overdensities observed in the top panels of
  Figure~\ref{diff}. As we have shown, adding an extra component
  partially solves the problem: model $S+E$ gives the
  best fit to the data in central regions, making the overall fit for
  the bright magnitude slices better, as can be seen in the top right
  panels of Figure \ref{res}.
\item column (c) shows the magnitude distributions in symmetric fields
  with respect to $l = 0^{\circ}$, centred at $(l,b) =
  (7^{\circ},2.5^{\circ})$ and $(l,b) = (-7^{\circ},2.5^{\circ})$. The
  peak of the RC is shifted between positive (bright RC) and negative
  (faint RC) longitudes, a consequence of the orientation of the Bulge with regard to the Sun-GC direction.
\item column (d) shows magnitude distributions in two fields at intermediate latitudes, centred at $(l,b) = (1^{\circ},-6.5^{\circ})$ and
  $(l,b) = (-1^{\circ},-6.5^{\circ})$. In these fields, the double RC is clearly
  visible in the VVV data as already shown by previous studies \citep[e.g.][]{Sa12}. Because the fields are symmetrical to $l=0^{\circ}$,  the two RCs contribute similarly to the magnitude distribution, even though the foreground RC will contributes slightly more to the distribution at $l=1^{\circ}$ compared to $l=-1^{\circ}$; for the background RC, the opposite is true. The X-bulge density law proposed by \citet{Lo2017} describes well this phenomenon: by looking at Figure~\ref{xyz}, we can clearly see how the $l=1^{\circ}$ line of sight will pass closer to foreground RC than to the background RC, but thanks to the viewing angle under which we observe the Bar, both RCs will be visible in the magnitude distribution, even though in different contributions.  The double peaked magnitude distribution shown in yellow is the \citet{Lo2017} model calculated using Equation \ref{model_total} where $\rho_{\mathrm{B}}$ is the X-shaped density distribution that we have shown in Figure \ref{xyz}. The model is successful at predicting the double RC, as expected.
\end{itemize}

We now explore the signal of the double RC at intermediate longitudes $l=\pm 5^{\circ}$ and varying distance below the GP by comparing the data with our models and the model based on the X-shaped density law (which should be accurate for $b<-5^{\circ}$) in Figure \ref{LF_all2}. The centres of the $2^{\circ} \times 1^{\circ}$ fields are marked with `x' in Figure \ref{diff}.
The top and bottom rows show the apparent magnitude variations at constant longitudes, $l=5^{\circ}$ and $l=-5^{\circ}$ respectively, with the latitude
decreasing from left to right between $-1.5^{\circ}$ and
$-7.5^{\circ}$.  

The fields close to the Galactic Plane (first column,
$-2^{\circ}< b <-1^{\circ}$) are largely incomplete because we mask the
high extinction regions, thus the amplitude of the RC is not well
fitted (please note that this is the result of the global fit). At $-4^{\circ}< b
<-3^{\circ}$  (second column) the models perform well but at lower latitudes 
$-5^{\circ}< b <-6^{\circ}$ (third column), where the double RC becomes visible,
we can clearly see the amplitude of the RC peak in the data is
not well matched by the model. The excess of stars supposedely belongs
to the peanut shape that is not well represented by a single simple
triaxial shape. Even further from the Galactic Plane, where the number
counts are much lower (fourth column, $-7^{\circ}< b <-7^{\circ}$),
the discrepancy is more obvious at all magnitudes. 

At $b<-5^{\circ}$, the model using the X-shaped density law \citep{Lo2017} is doing a better job at predicting the magnitude distribution. However, while the model predicts a double RC, the data only shows one. This could be explained if the second RC contributed less than predicted by the \citet{Lo2017} model at intermediate longitudes (this could easily be the case, e.g. imagine a $l=5^{\circ}$ line crossing the X-Y plane in Figure \ref{xyz}). From the magnitude variation of the peaks at constant longitude (left to right), it is clear that with increasing distance from the plane the foreground RC gets brighter (positive longitude, top row) and the background RC gets fainter (negative longitude, bottom row), behaviour specific to an X-shaped distribution.
\begin{table}
 \centering
 \hspace*{0cm}
 \begin{minipage}{76mm}
  \caption[]{Best fit parameters variation for model $S$ as a function of $\sigma^{\mathrm{RC}}$. We fitted model $S$ using LFs with increasing $\sigma^{\mathrm{RC}}$ in steps of 0.03 mag; in the last row, $\sigma^{\mathrm{RC}}$ is a free parameter. These values are shown in Figure \ref{sigmas}.}
   \label{bestresults2}
  \footnotesize
  \centering
  \begin{tabular}{@{}cccccc@{}}
ln($L_{P}$) & $\sigma^{\mathrm{RC}}$ &  $\alpha$ & $x_{0}$ , $y_{0}$, $z_{0}$, $c_{//}$, $c_{\perp}$  \\
    \hline
  36,731,464 & 0.06& -19.36 &1.48, 0.63, 0.47,  3.00, 1.88  \\
 36,731,560 &0.09 & -20.09 & 1.46, 0.62, 0.47,  3.06, 1.87  \\
 36,735,035 &0.12 &-21.14 &1.42, 0.61, 0.47,  3.13, 1.88  \\
36,737,262 & 0.15 & -22.64 &1.37, 0.59, 0.47,  3.31, 1.90 \\
 36,739,532 & 0.18& -24.69 & 1.30, 0.57, 0.46,   3.37, 1.95  \\
 36,741,544 & 0.21 &  -27.40 & 1.23, 0.54, 0.46,  3.52, 2.03  \\
 36,743,487 &  0.26 & -34.40 & 1.10, 0.45, 0.45, 3.58, 2.09  \\
 \hline
    \end{tabular}
 \end{minipage}
\end{table}
To roughly quantify what proportion of the VVV stars are found in the X-shaped residuals relative to the total number of Bulge stars, we consider the fields at
intermediate latitudes, outside $|b|< 4^{\circ}$. Here the feature is most
clearly visible in the VVV data, even though it may continue to lower 
latitudes. The model tends to overfit the region between the X-shape arms 
(see blue depression at $l=0^{\circ}$ in Figure \ref{res}) and underfit the 
X-shape arms themselves, hence we conjecture that the model has to be reduced 
by some fraction $<1$ such that only $\sim$5\% of the residuals are negative,
the positive excess that remains, would theoretically belong to the
arms. With this assumption and mirroring the region at $b <
-5^{\circ}$ to positive latitudes, $b > 5^{\circ}$, we find that $\sim
7$ \% of the stars belong to the X-shaped arms relative to the
entire modelled Bulge, within $-10^{\circ}< l< 10^{\circ},
-10^{\circ}< b< 10^{\circ}$. 
\begin{figure}
\hspace{-0.4cm}
\includegraphics[scale =0.131]{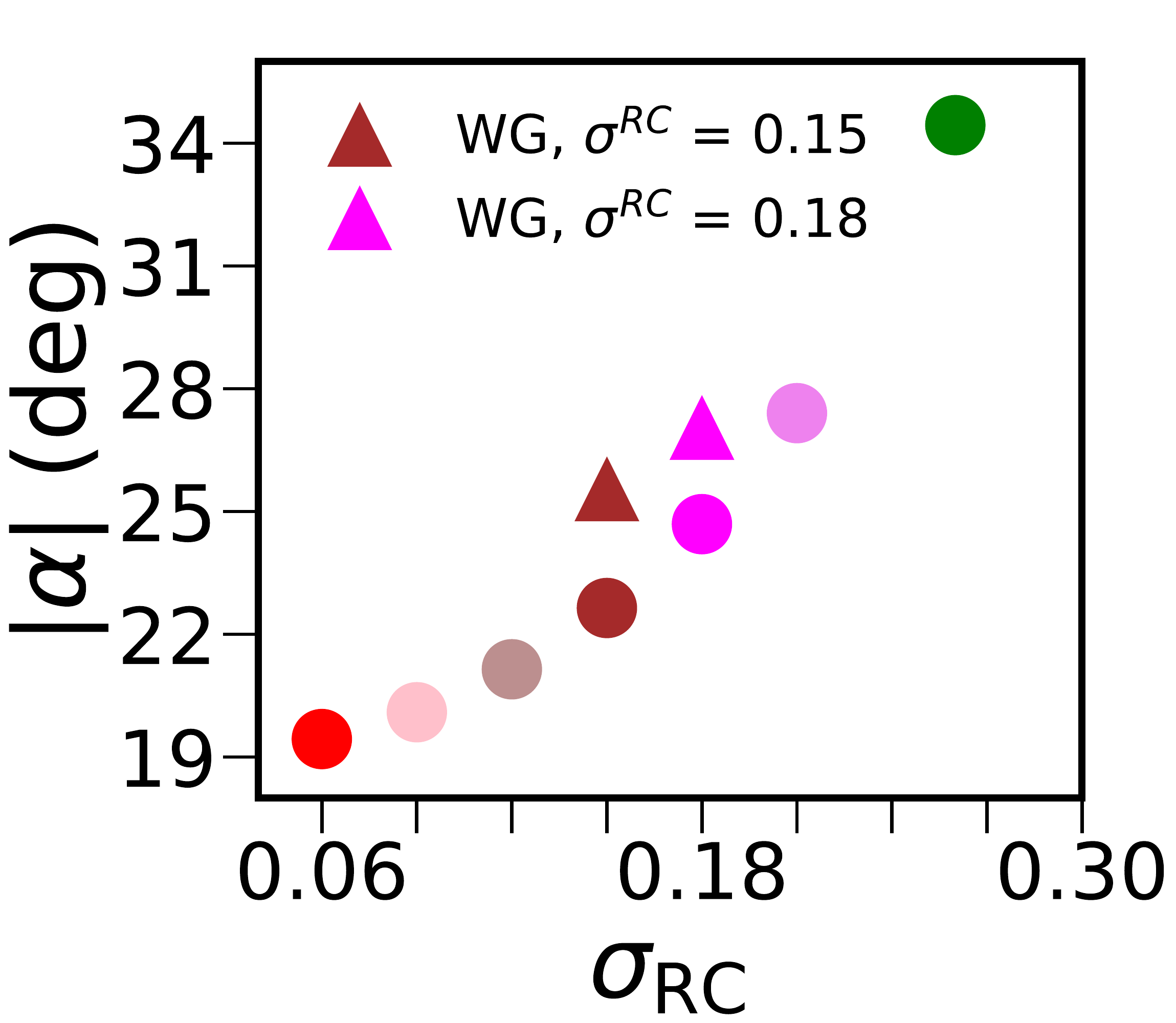}
\includegraphics[scale =0.131]{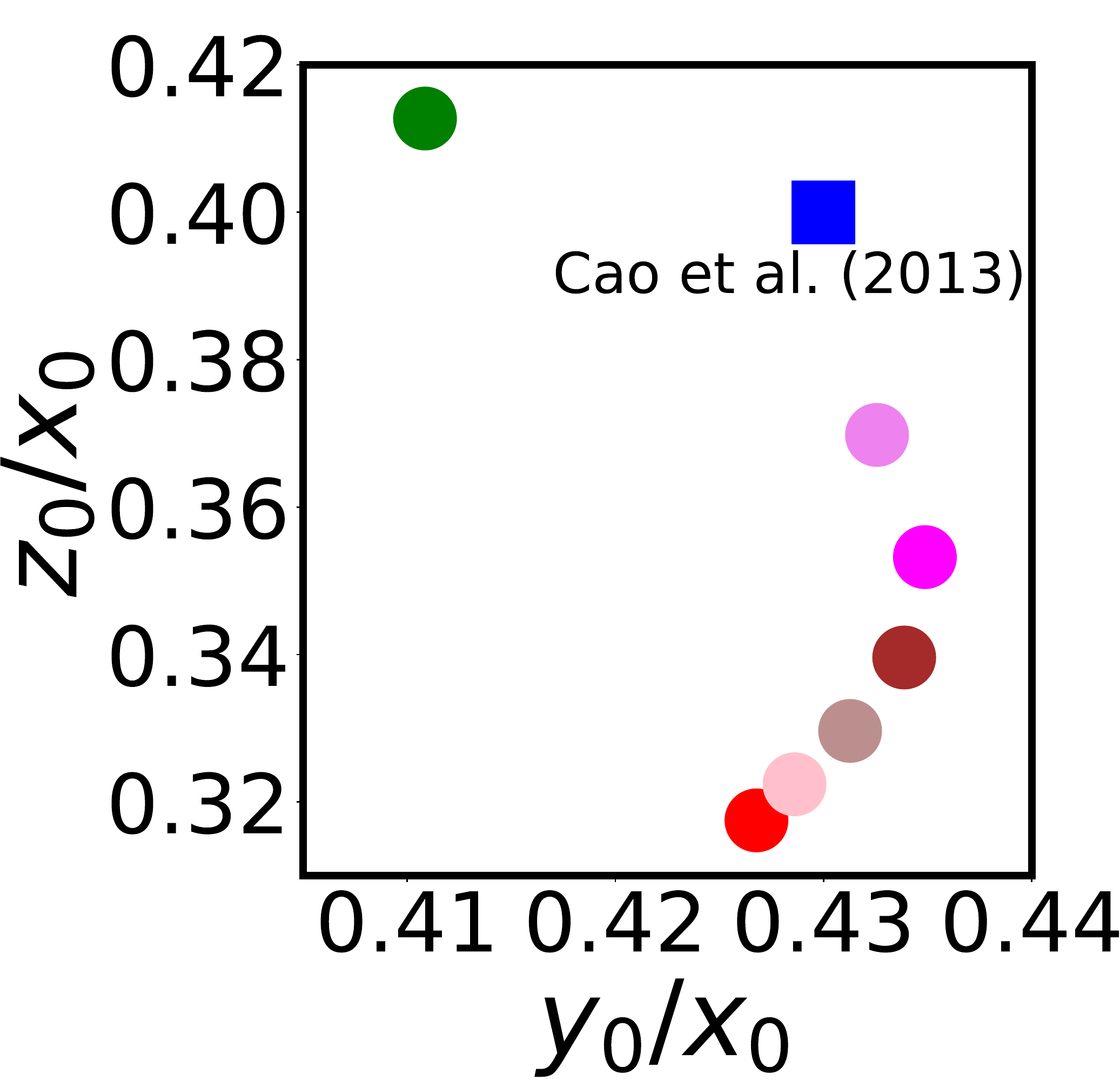} 
\includegraphics[scale =0.131]{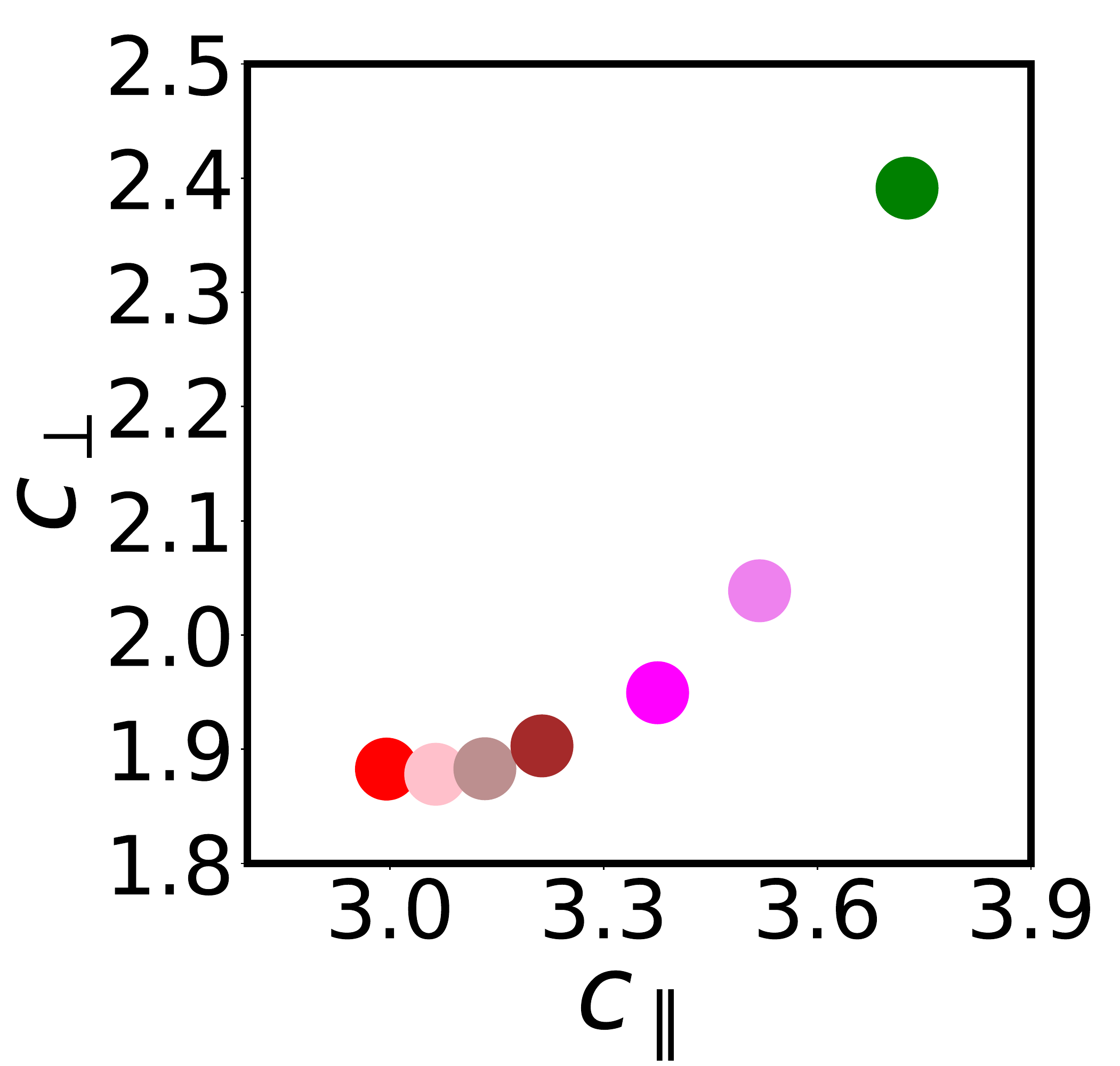} 
\caption{Best fit parameters variation as a function of increasing $\sigma^{\mathrm{RC}}$ with a step of 0.03 mag (each identified with a different colour), for model $S$. The result for $\sigma_{\mathrm{free}}^{\mathrm{RC}}=0.26$ mag (free parameter) is marked in green. $Left$ $panel$: as $\sigma^{\mathrm{RC}}$ increases, the fitted viewing angle $\alpha$ also increases. $\alpha$ is the most affected parameter by the $\sigma^{\mathrm{RC}}$ variation. $Middle$ and $right$ $panels$: the parameters controlling the shape of the density distribution are not
  affected by the choice of $\sigma^{\mathrm{RC}}$ even though a tiny
  trend can be observed. We also add the results of \citet{We13} (WG) and \citet{Ca13}. }
\label{sigmas}
\end{figure} 

\section{Discussion}
\label{sec:discussion}
\subsection{The Luminosity function}
\subsubsection{Variations of the LF}

The intrinsic LF of the Bulge is sensitive to both age and metallicity variations but in absence of accurate and consistent measurements of these quantities
 across the Bulge, we do not attempt to build a spatially varying LF. 
 
The main effect of not having a simple Bulge population as the one we assumed in Section 4.3, will be to blur the intrinsic LF. We therefore convolve the Galaxia generated LF with a Gaussian kernel of $\sigma$ as a free parameter (therefore $\sigma_{\mathrm{free}}^{\mathrm{RC}}$ $=\sqrt{\sigma^{2}+(\sigma^{\mathrm{RC}})^{2}}$ becomes a free parameter, where $\sigma^{\mathrm{RC}} = 0.06$ mag) and re-fit the VVV data with models $E$ and $S$. The best fit results are listed in Table \ref{bestresults} for both models and are labeled `$\sigma^{\mathrm{RC}}$ free'. 

Compared to the approach presented in this paper, \citet{We13} followed a different route and attempted to constrain the
Bulge properties non-parametrically.  Instead of estimating the
parameters of a Bulge density law, they deconvolve the observed RC
magnitude distributions to extract line-of-sight RC densities from
which they build a 3D map assuming eight-fold mirror symmetry. The result they obtained, $\alpha = (-26.5\pm2)^{\circ}$, falls in the middle of our extreme values for $\sigma^{\mathrm{RC}} = 0.06$ mag, $\alpha = -19.4^{\circ}$ (red dot) and $\sigma_{\mathrm{free}}^{\mathrm{RC}} = 0.26$ mag, $\alpha = -34.4^{\circ}$ (green dot).

In absence of a fully reliable calibration of the Bulge RC dispersion in literature (the latest measurement in the local neighbourhood with Gaia finds $\sigma^{\mathrm{RC}} = 0.17$ mag - see \citealt{Ha17}), we investigated the effect of the $\sigma^{\mathrm{RC}}$ variation on the density law parameters by convolving the Galaxia generated LF with a Gaussian of increasing $\sigma$ in order to obtain $\sigma^{\mathrm{RC}} = 0.09, 0.12,
0.15$, 0.18 and 0.21 mag for a 10 Gyrs population. The results shown in Figure \ref{sigmas} and listed in Table \ref{bestresults2} confirm the expectation that a higher $\sigma^{\mathrm{RC}}$ is responsible for
a higher viewing angle $\alpha$. This was also observed by \citet{We13} who found $|\alpha|$ reduces from 26.5$^{\circ}$ to 25.5$^{\circ}$ (triangles in Figure \ref{sigmas}) when $\sigma^{\mathrm{RC}}$ decreseas from 0.18 to 0.15 mag and explain `this is because increasing the line-of-sight geometric dispersion has given the illusion that the bar is closer to end on'. We find a stronger dependence of the viewing angle on $\sigma^{\mathrm{RC}}$, with $\alpha$ decreasing from 24.69$^{\circ}$  to 22.64$^{\circ}$  for the same $\sigma^{\mathrm{RC}}$ variation. Our work is probably more sensitive to the viewing angle as we do not need to assume eight-fold mirror symmetry thanks to continuous coverage in all quadrants of the newer VVV data release.

 While the viewing angle is strongly dependent on $\sigma^{\mathrm{RC}}$, the axis ratios do not vary significantly as seen in the middle and right panels of Figure~\ref{sigmas}. In fact, the parameters controlling the shape of the
density distribution are not significantly affected by the choice of
$\sigma^{\mathrm{RC}}$ even though a trend can be observed in the middle and
right panels of the figure: in the middle panel, as the $\sigma^{\mathrm{RC}}$
increases (the values are colour-coded as in the left panel) the bar
gets puffier in both the vertical and $Y$-axis directions but as it
reaches $\sigma^{\mathrm{RC}}=0.15$ mag it starts to get thinner again along
the $Y$-axis while it continues to get thicker in the vertical
directions. The parameters controlling the shape of the distribution
are also correlated, with $n$ decreasing as $\sigma^{\mathrm{RC}}$ increases
while the $c_{\parallel}/c_{\perp}$ ratio also increases. Notice that these
variations are tiny and do not change the shape of the
distribution significantly, the viewing angle $\alpha$ being the most
affected parameter.

\begin{figure*}
\hspace{-3.2cm}
\includegraphics[scale =0.25]{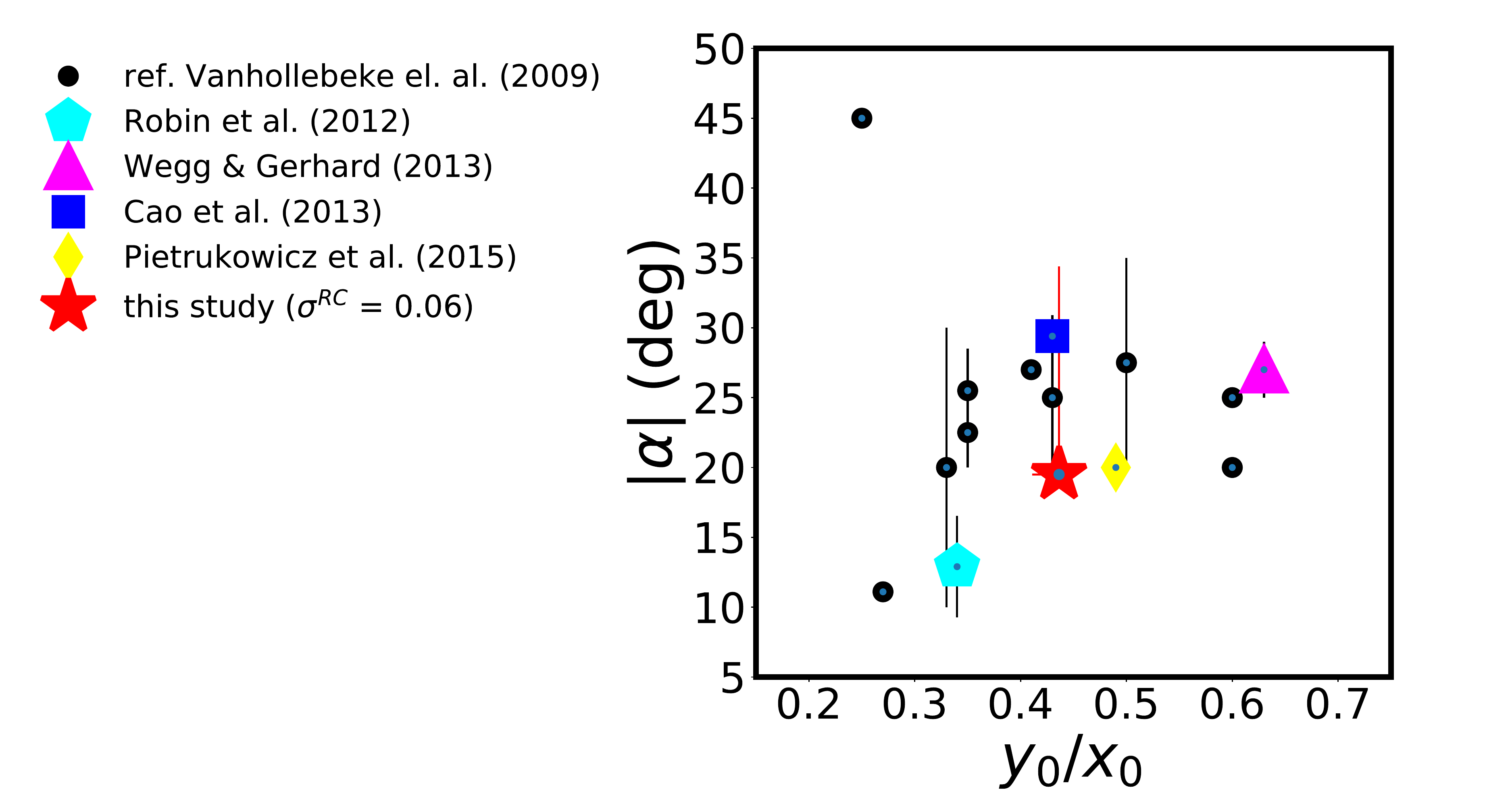}
\includegraphics[scale =0.25]{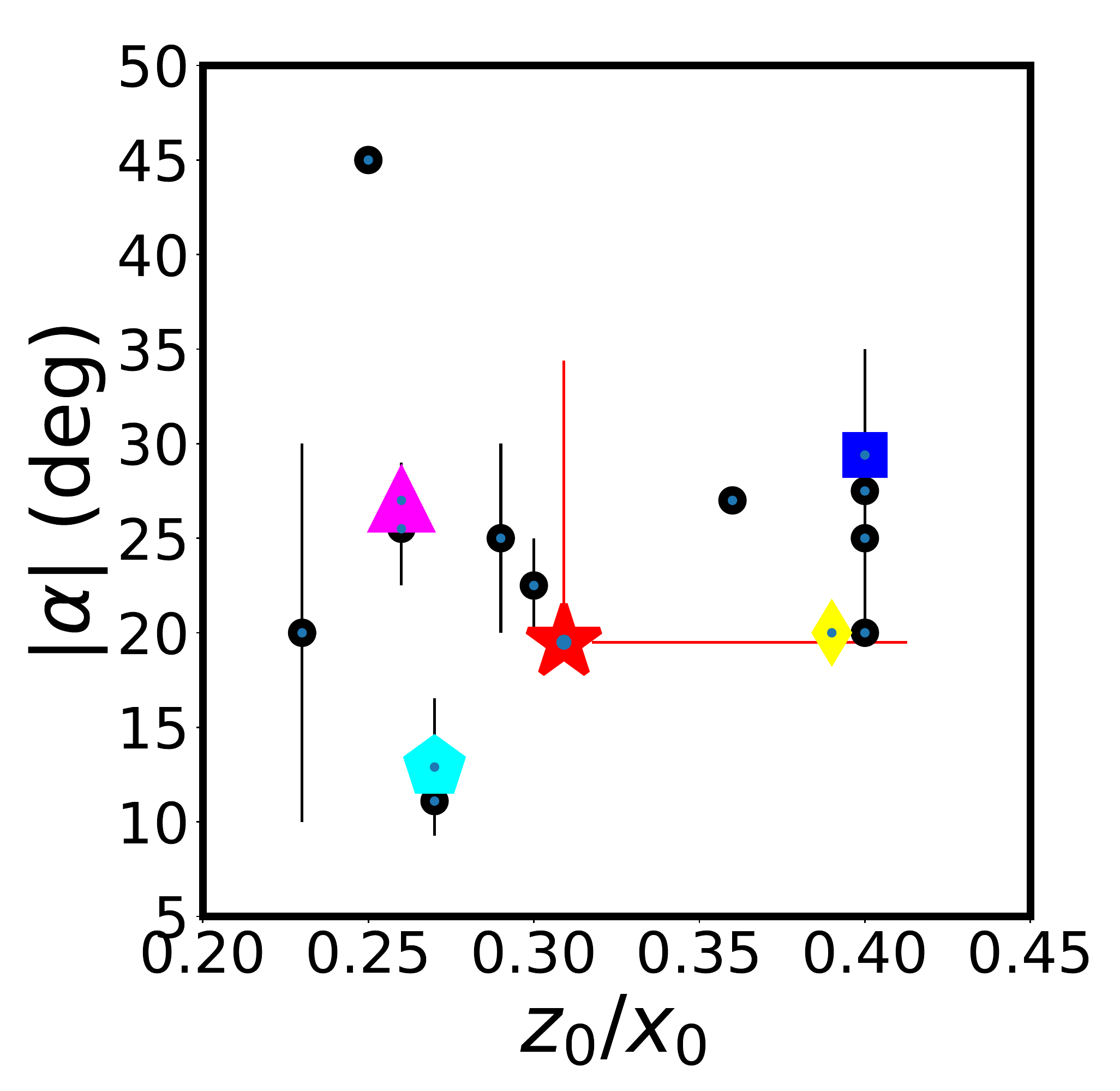} 
\caption{Comparison of the axis ratios ($y_{0}/x_{0}$ and $z_{0}/x_{0}$ in the $left$ and $right$ $panels$) and Bulge viewing angle $\alpha$ obtained in this work for a one component fit (model $E$, red star), with previous studies. The error bars (in red) mark the maximum variation produced in our results by varying $\sigma^{\mathrm{RC}}$ (see also Figure \ref{sigmas}). Literature results prior to 2009 \citep{Va09}, are
  marked with black dots and the error bars are shown when
  reported. More recent measurements are shown with different colours:
  the results from a non parametric Bulge study using VVV RC stars
  with magenta \citep[][]{We13}, those from a parametric study using
  2MASS giants with cyan \citep[][]{Ro12} and the results from two
  OGLE separate studies, with RR Lyrae \citep{Pi15} and RC stars
  \citep{Ca13}, with yellow and blue respectively.}
\label{comparisons}
\end{figure*} 

\subsection{The discs}
\subsubsection{The thin disc}

In \textit{Galaxia}, the thin disc density distribution has a scale
length of $R_{d} = 2.53$ kpc and a disc hole scale length of $R_{h} =
1.32$ kpc. More recently, \citet{Ro12} used 2MASS photometry to fit
simultaneously for the parameters of the Bulge population and the
shape of the inner disc. The updated parameters of the thin disc in
the best Bulge model (model $S+E$, see their table 2) are $R_{d}=
2.17$ kpc and $R_{h} =$ 1.33 kpc: the disc hole scale length is almost
identical and the disc scale length is shorter by 400 pc compared to
the previous values. 
\subsubsection{The thick disc}
\citet{Ro14} used photometric data at high and intermediate latitudes
from SDSS and 2MASS surveys to re-analyse the shape of the thick
disc. Their updates include a new IMF best fit, $dN/dm \propto
m^{-0.22}$, a slightly smaller scale height of $h_{Z} = ($535.2 $\pm
4.6$) pc and a shorter scale length of $h_{R} = ($ 2362 $\pm 25$) pc plus a new 
age best fit of 12 Gyrs.
\\
\\
We have performed all the fits described in the manuscript with the updated thin and thick discs versions and the results are reported in last rows of Table ~\ref{bestresults}. They are also shown in pink in Figures ~\ref{LF_all} and ~\ref{LF_all2}. Finally, we have decided to use the default \citep{Ro03} values throughout the paper, for the following reasons:
\begin{itemize}
\item In the fitting procedure we exclude the area close to the
  Galactic plane where the reddening $E(J'-K'_{s})$ is higher than 1
  mag and where the majority of the thin disc population lies. The
  $E(J'-K'_{s}) = 1.0$ contour extends, on average, to latitudes
  $-1.5^{\circ} < b < 1.5^{\circ}$ (see Figures~\ref{ext} and~\ref{discs}) 
  which is equivalent to a height perpendicular to the Galactic plane of $-0.2
  < Z$/kpc$ < 0.2$ at 8 kpc. This region is masked out in the fitting
  procedure described in Section~\ref{sec:datafit}. The scale height of the 
  oldest thin disc component is $\sim$ 300 pc, therefore a large part of the 
  thin disc population is excluded from the fit.

\item The fitting results confirmed that the Bulge density law
  parameters remain almost identical, regardless of the discs
  (original or updated version) assumed;
  
\item To avoid multiple modifications to the $Galaxia$ code.
\end{itemize}

\subsection{Stellar Mass of the Bulge}

\textit{Galaxia} relies on the isochrones table as input to the
code to assign a mass to each star produced as part of the mock
catalogue for a given IMF. \citet{Pi04} undertook a detailed analysis
of the outer Bulge stellar density and LF, assuming a 
\citet{Sa55} IMF, by fitting model parameters to a set of 94 windows in 
the outer Bulge situated at
$-8^{\circ} < l < 10^{\circ}$ and $b < |4|^{\circ}$ in the DENIS survey.
The same IMF was used in the Besancon model and implemented in $Galaxia$.  
The Bulge stars lie at
distances between 4 and 12 kpc, corresponding to absolute magnitudes
$M_{K}$ in the range -3.5 to 1 mag in the 12 to 14 apparent magnitude
range $K_{s}$. At these bright magnitudes, the intrinsic luminosity
function is not influenced by the choice of IMF therefore using a
different IMF would not change the results in Table \ref{bestresults}.

The stellar mass in the Bulge is however sensitive to the IMF choice. We calculate the mass of the Bulge assuming a bottom light Chabrier IMF, similar to a Kroupa IMF, which we have used throughout the paper. The slope of the latter becomes significantly shallower than the Salpeter IMF below 1 M$_{\odot}$ and describes more accurately observations in the Galactic Bulge (see also Section 
4.3). The majority of stars in the Bulge have masses $< 1M_{\odot}$ so we would expect to greatly overestimate the total mass of the Bulge using a Salpeter IMF. In
addition, \citet{Sh16} test the theoretical predictions of $Galaxia$
with asteroseismic information for 13,000 red giant stars observed by
Kepler and find that the distribution of the masses predicted by the
$Galaxia$ Galactic model overestimates the number of low mass stars. We therefore report the Bulge mass estimates using the Chabrier 
IMF which is similar to the Kroupa IMF, excluding the Salpeter IMF. \\

Using the best-fit model ($E$) in $Galaxia$ we can predict the fraction of
Bulge stellar mass in the RGB and the RGB luminosity that we should observe within the VVV coverage, with the colour-magnitude cuts $0.4 < J-K_{s}<1.0$ - $12< K_{s}< 14$, and hence use the observed population (and the measured fraction) to constrain the total stellar mass of the Bulge. For a given IMF the PARSEC isochrones 
enable us calculate the remnant population and therefore
the total mass of the original population. For the total MW Bulge mass, we 
obtain 
$M_{\mathrm{Bulge}}^{\mathrm{Chabrier}} = 2.36 \times 10^{10} M_{\odot}$,
\textit{including} the mass of the remnants (49\% of the
total mass of the Bulge) using a Chabrier IMF. 
For the two component $S+E$ best fit, we obtain 2.21 $\times 10^{10} $
M$_{\odot}$ ($S$ component) and 1.2 $\times 10^{9} $ M$_{\odot}$ ($E$
component), for Chabrier IMFs. If the low-latitude residuals are 
caused by the presence of an extra component (component $E$), then this 
component would contribute $\sim$5\% to the total Bulge mass budget. 

Other recent measurements of the Bulge stellar mass include:
\begin{itemize}
\item \citet{Ro12} who estimate the total mass of the Bulge by fitting a
  population synthesis model to 2MASS and for the two component best
  fit, find $M_{\mathrm{Bulge}}=6.36 \times 10^{9}$ M$_{\odot}$ where the mass
  of the $S$ component is 6.1 $10^{9} M_{\odot}$ and the mass of the
  smaller component $E$ is 2.6 $\times 10^{8}$ M$_{\odot}$.
\item \citet{Po15} construct dynamical models of the MW Bulge using
  RC giants and kinematic data from the BRAVA survey. They compute the total mass of
  the Galactic Box/Peanut Bulge and find a stellar mass $M_{Bulge} =
  1.25-1.6 \times 10^{10} M_{\odot}$, depending on the total amount of
  dark matter in the Bulge.
\citet{We15} use a Kroupa
IMF at 10 Gyrs and report a Bulge mass estimate of 1.81 $\times
10^{10} M_{\odot}$ that includes, in addition to a Bulge population, a thin and a
superthin bar. 
\item \citet{Va16} scale the total magnitude distribution obtained
  from all VVV fields (320 square degrees) to the one from
  \citealt{Zo03} (an 8$'\times$8$'$ field at $(l,b) =
  (0^{\circ},-6^{\circ})$), to obtain a stellar mass of $M_{Bulge} = 2
  \times 10^{10} M_{\odot}$. They find that a $\pm 0.3$ dex variation in their 
assumed IMF slope $\alpha = -1.33 \pm 0.07$ \citep{Zo2000}, would change the 
mass estimate by less than 12\%; moreover the authors assume constant disc 
contamination. In contrast we included in our model the disc contamination which is calculated as a function of field and magnitude.
\end{itemize}

\subsection{Structural comparison with previous results}

Figure~\ref{comparisons} compares the Bulge parameters measured in
this work to those obtained in other studies with a variety of
tracers, e.g. RR Lyrae, Mira variables and RC stars. Note that the
results of modelling the Galactic Bulge with an exponential and
sech$^{2}$ density laws as reported in Table~\ref{bestresults} are
indistinguishable and are marked in red in
Figure~\ref{comparisons}. The red error bars give the maximum
variation seen in the parameters as a result of varying $\sigma^{\mathrm{RC}}$:
only the viewing angle seems to be significantly affected 
(see Figure \ref{sigmas}).

We take advantage of a large catalogue of Bulge studies up to 2009 in
\citet{Va09} to show the range of the viewing angle values and the
axis ratios with black dots, adding the error bars when
present. Additionally, four more recent results are marked with
colours: one from a parametric model describing the 2MASS data
\citep[cyan,][]{Ro12}, one from a non-parametric study of VVV RC stars
\citep[magenta,][]{We13}, and two from the OGLE survey, with RRLyrae
\citep[yellow,][]{Pi15} and RC stars
\citep[blue,][]{Ca13}. \citet{Ro12} obtain an
orientation of about 13$^{\circ}$ for the main structure, in their
two-component fit. \citet{Ca13}, who use number counts of red clump giants from the OGLE III survey to fit triaxial parametric models for the Bulge, obtain axis ratios within the parameter space spanned by our results when testing different  $\sigma^{\mathrm{RC}}$; we mark their result in both Figures \ref{sigmas} and \ref{comparisons}. 

\subsection{Central region}

As in previous studies (e.g. \citealt{Ro12}), we find that
one-component Bulge models do not provide a good match to the data in
the central Galactic regions. We notice a clear overdensity at $|b|< 3^{\circ}$
and $|l|< 2^{\circ}$ (left and middle panels in Figure~\ref{res}). The match 
between the model and the data can be improved
by considering a combination of two populations (right panel in the
same figure), where the central one has a RC peak brighter than the RC
peak of the main population. 

The presence of the overdensity can be explained using several theories:
\begin{itemize}
\item \textbf{the presence of a `disky' pseudo-Bulge - } in addition
  to a `boxy' Bulge \citep{Dwek95}, our Galaxy might host a `disky'
  pseudo-Bulge, observed in many early and late type b/p Bulge
  galaxies (Bureau et al \nocite{Bu06} 2006). These types of Bulges
  grow `secularly' out of disc gas that was transported inwards along
  the bars \citep[e.g.][]{Ko04}. The centrally concentrated residuals
  with short vertical scale height we observe might be indicative
  of this type of disc-like, high density pseudo-Bulge, also seen in
  N-body models by \citet{Ge12}. In addition, a change-of-slope at
  $|l| \sim 4^{\circ}$ can be seen in the VVV RC star count maxima
  longitude profiles close to the Galactic plane (see \citealt{Ge12})
  while further from the Galactic Plane, the change in slope
  disappears. \citet{Ge12} argue that a pseudo-Bulge does not need the presence of 
  a secondary nuclear bar to explain the change of slope in the longitude 
  profile. 
\item \textbf{spiral arms / thin bar overdensity - } the bar and
  buckling instabilities in the stellar disc lead to a boxy Bulge
  which extends to a longer in-plane bar \citep{At05}. The bar couples
  with the spiral arms in the disk, giving rise alternatively to
  leading, straight, or trailing bar ends \citep{Ma06}. It is possible
  that a density enhancement produced by spiral arms along the Sun-GC
  line-of-sight might explain the residuals observed in Fig~\ref{diff}
  (the spiral arms are confined to the Galactic plane but a height of
  100 pc would subtend an angle of $\sim 1.5^{\circ}$ at 4 kpc, while
  the same vertical height would only subtend $\sim 0.7^{\circ}$ at 8
  kpc).
\item \textbf{the discs - } we have used a 2D reddening map to correct for 
  the effects of extinction, assuming all the sources are behind the dust. 
  Disc objects (e.g. nearby dwarfs) or Bulge stars that are not behind all 
  the dust (and are situated between the Sun and the Galactic Centre) will 
  be over-corrected by various amounts, which could possibly lead to an 
  apparent overdensity in the central regions of the Bulge in the bright 
  magnitude slices that contain the sources closest to the Sun. We have tested this hypothesis by adding 3D extinction to a sample of \textit{Galaxia} generated mock thin disc stars. We have then corrected for extinction using a 2D map, therefore considering only the highest value of the 3D map. If this method was accurate, the number density of the 2D extinction corrected sample should coincide with that of the initial mock sample. Instead, we found only a 4\% increase in the number of counts in the $12.0<K_{s}<12.8$ magnitude range after correction for the stars near the plane but for stars with $|b|>1.5^{\circ}$ no increase was found. We thus believe that using a 2D reddening map does not produce the central overdensity observed.

\item \textbf{younger central population -} altering the disc
  parameters does not improve the agreement between data and model in
  the magnitude range considered ($12< K_{s}<14$). We can only account
  for it by adding an extra component with a smaller size
  ([ $x_{0}, y_{0}, z_{0}$] = [1.52, 0.24, 0.27] kpc) and a brighter RC, with 
  $\Delta M =0.1$ between the main component and the secondary one. A younger 
  population, of 5 Gyrs and solar metallicity (as the main population), would 
  produce such a shift but it could also be achieved considering other 
  combinations of ages and metallicities between the two populations. The 
  presence of a young population inside the Bulge is not inconsistent with 
  simulations \citep[e.g.][]{Ne14} and Miras variable \citep[e.g.][]{Ca16} 
  observations. 
\end{itemize}
\begin{figure*}
\hspace*{-0.19in}
\includegraphics[scale=0.21]{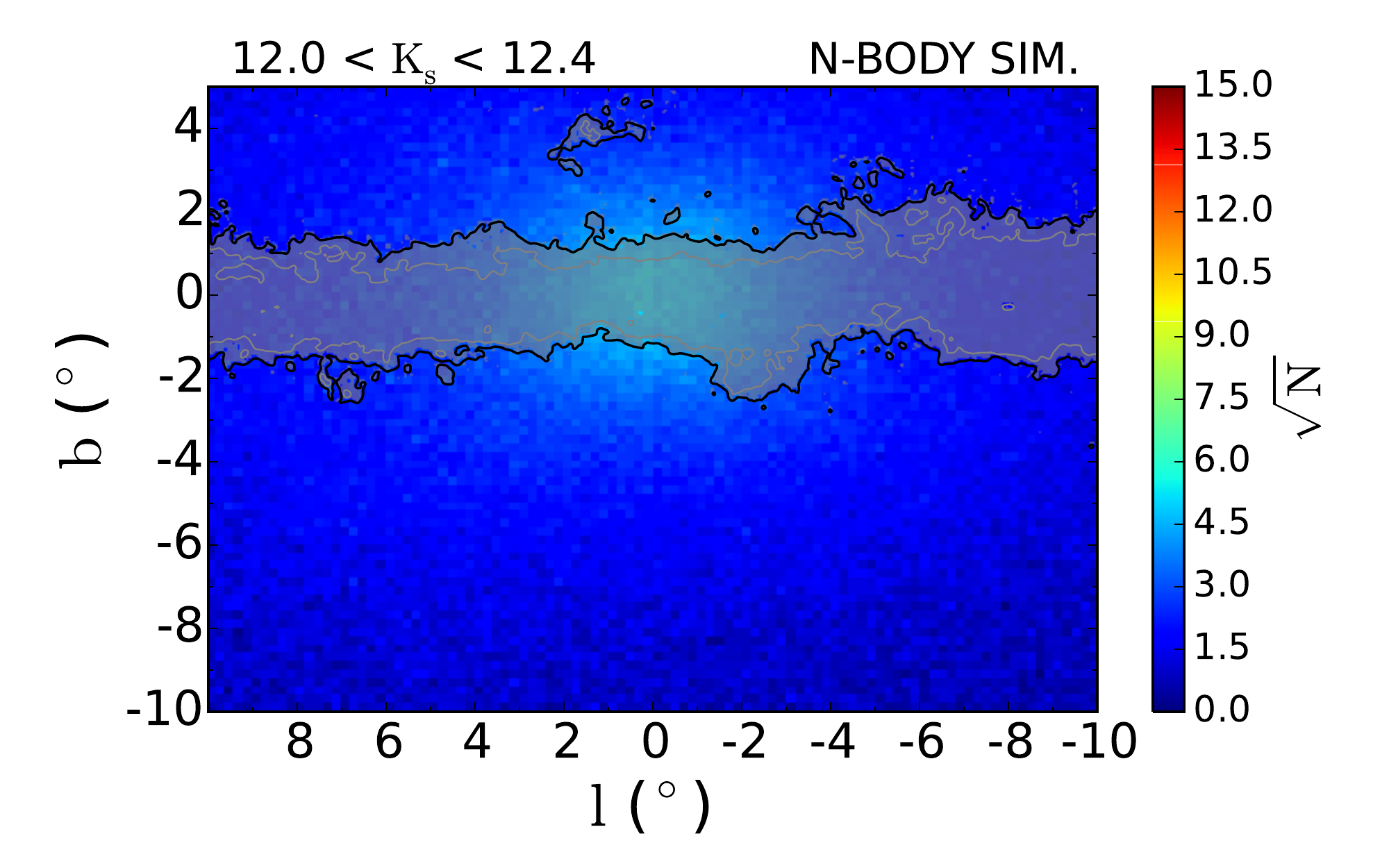}
\hspace*{-0.39in}
\includegraphics[scale=0.21]{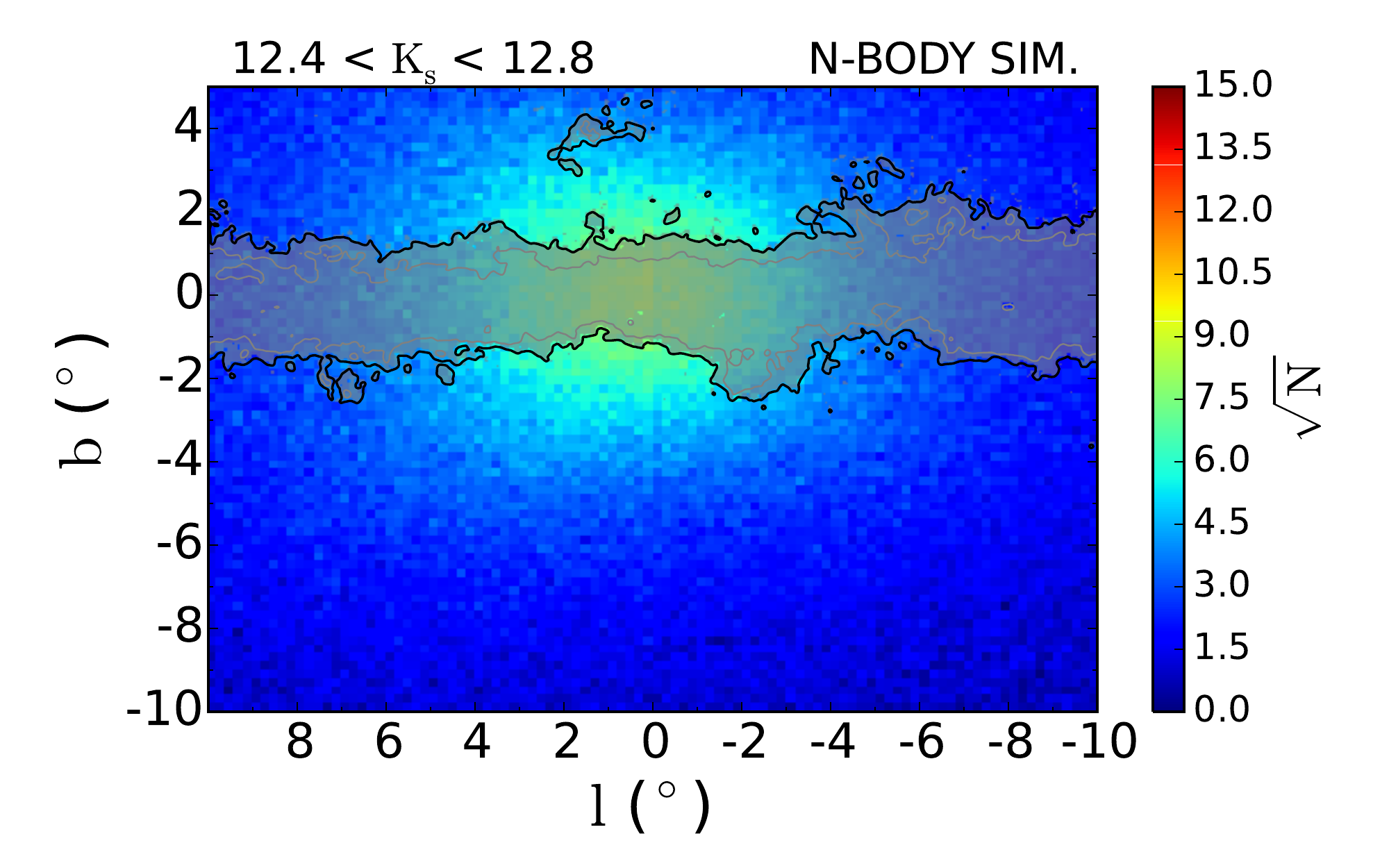}
\hspace*{-0.39in}
\includegraphics[scale=0.21]{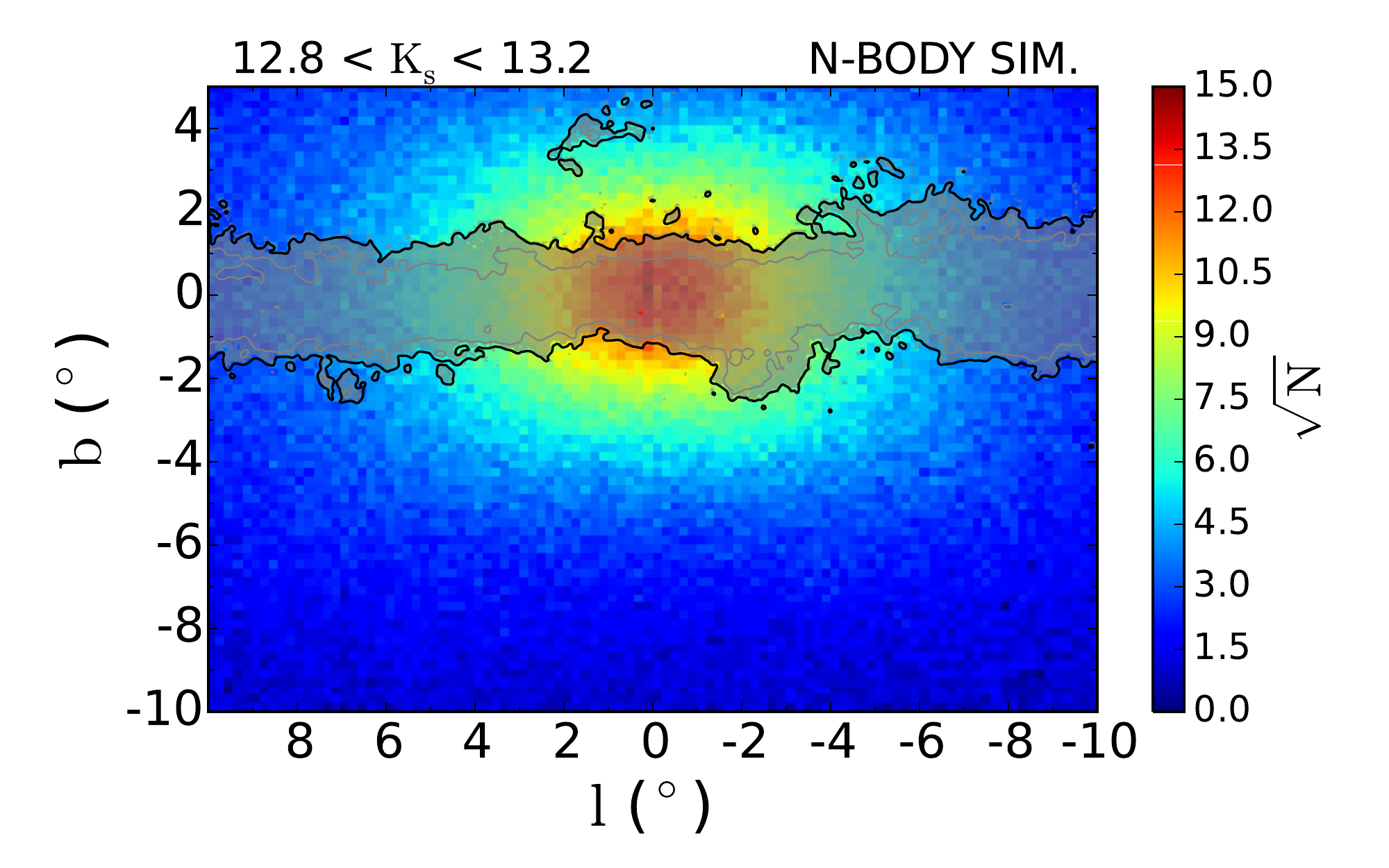}
\hspace*{-0.39in}
\includegraphics[scale=0.21]{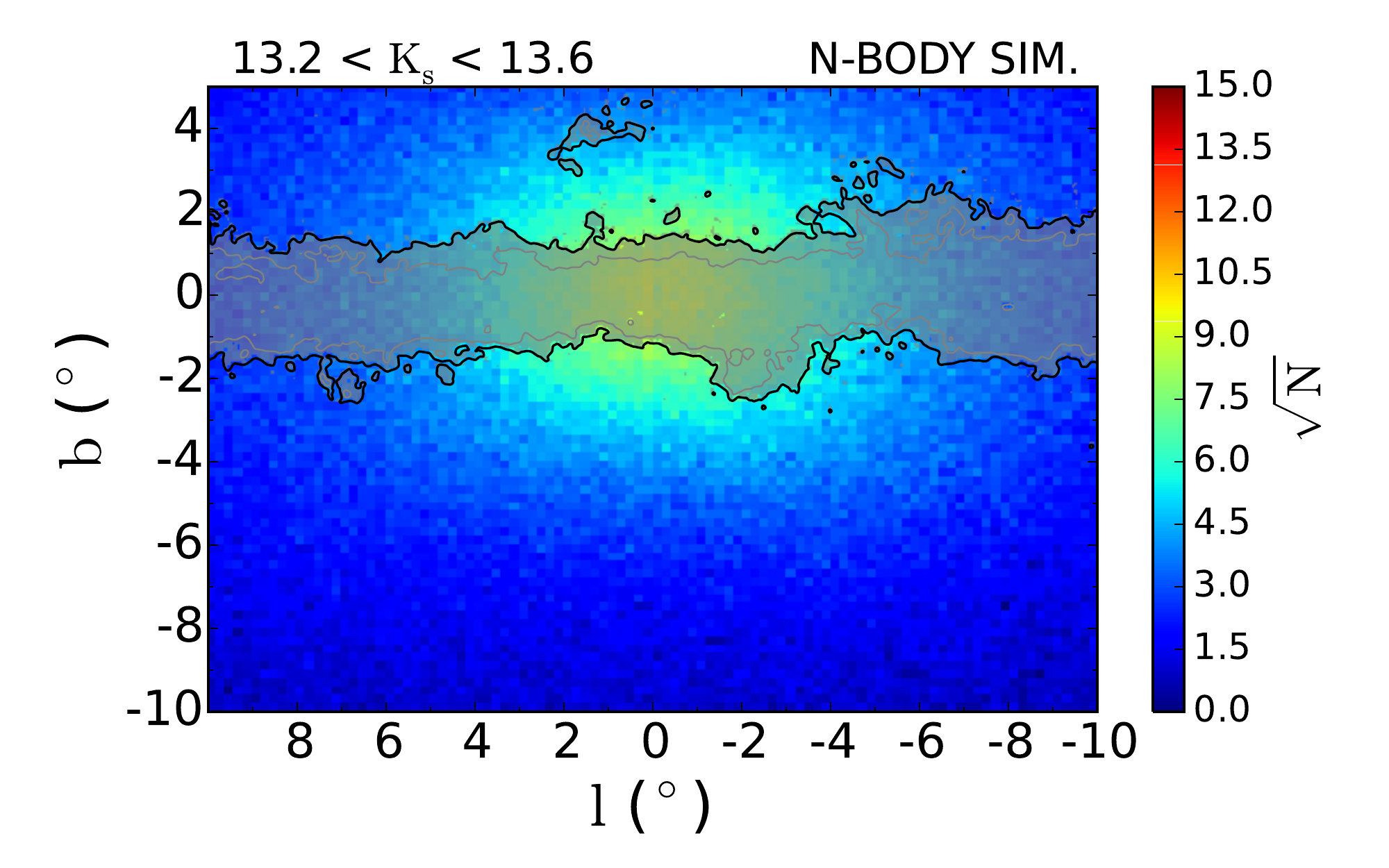}
\hspace*{-0.39in}
\includegraphics[scale=0.21]{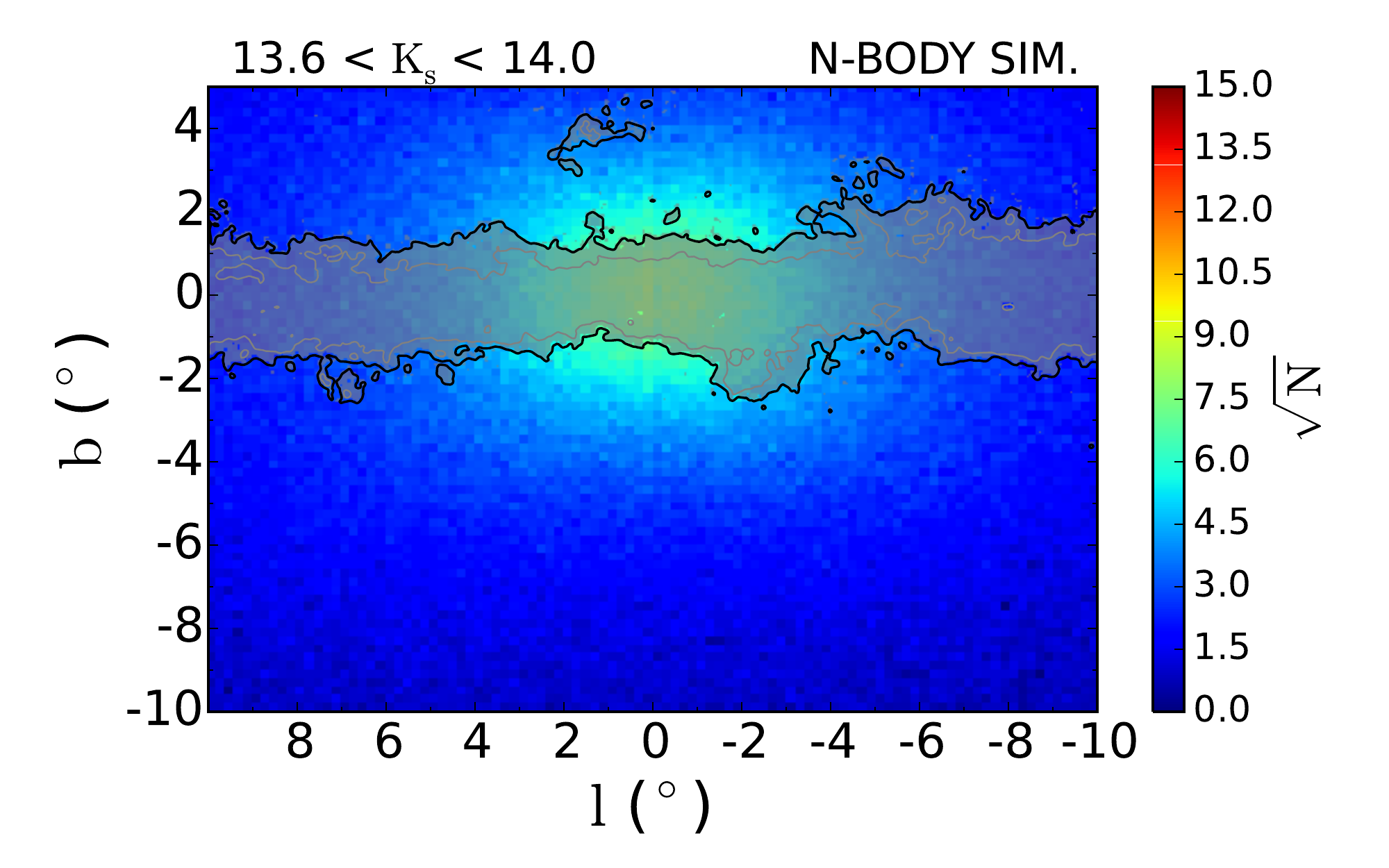}
\\
\hspace*{-0.19in}
\includegraphics[scale=0.21]{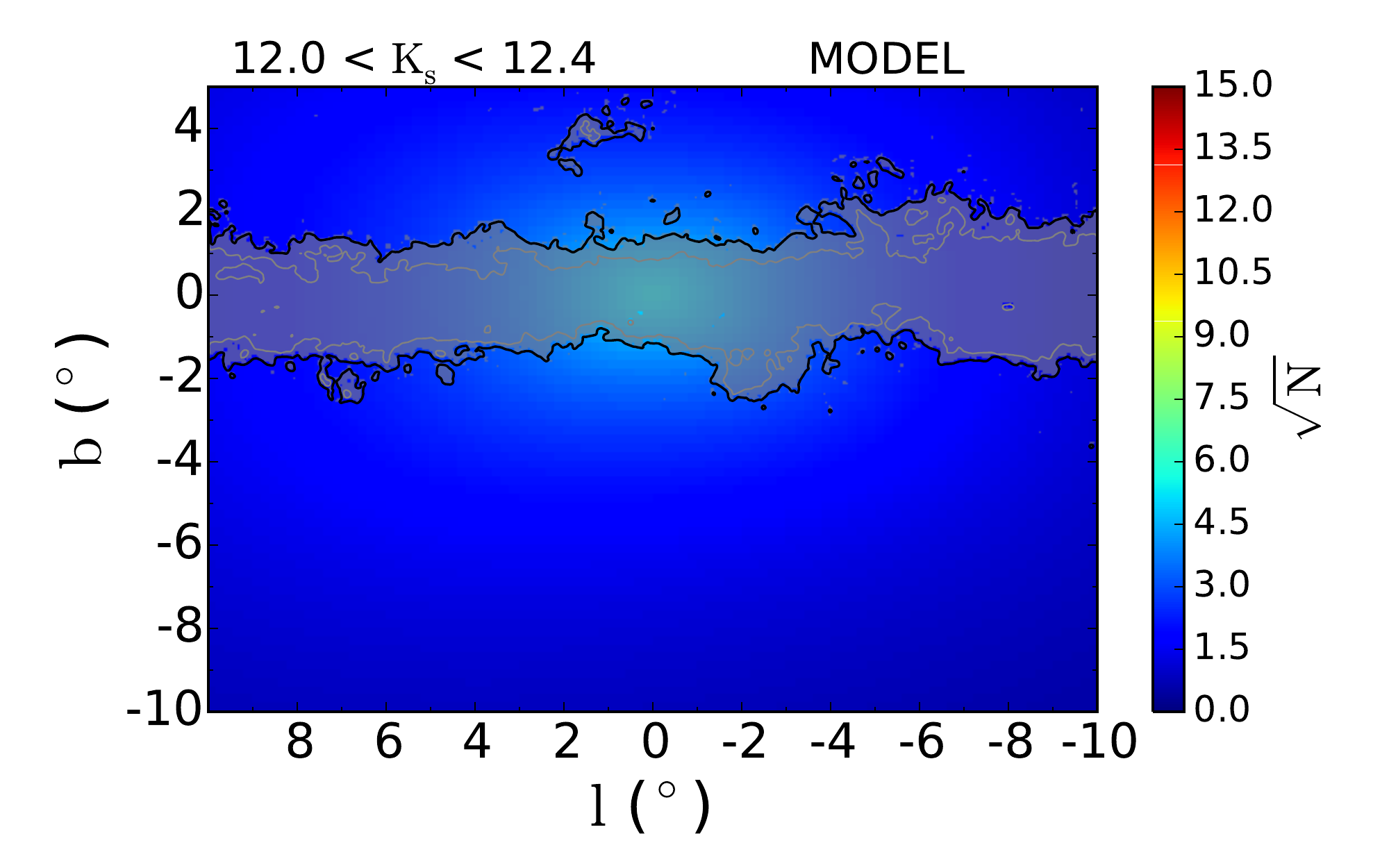}
\hspace*{-0.39in}
\includegraphics[scale=0.21]{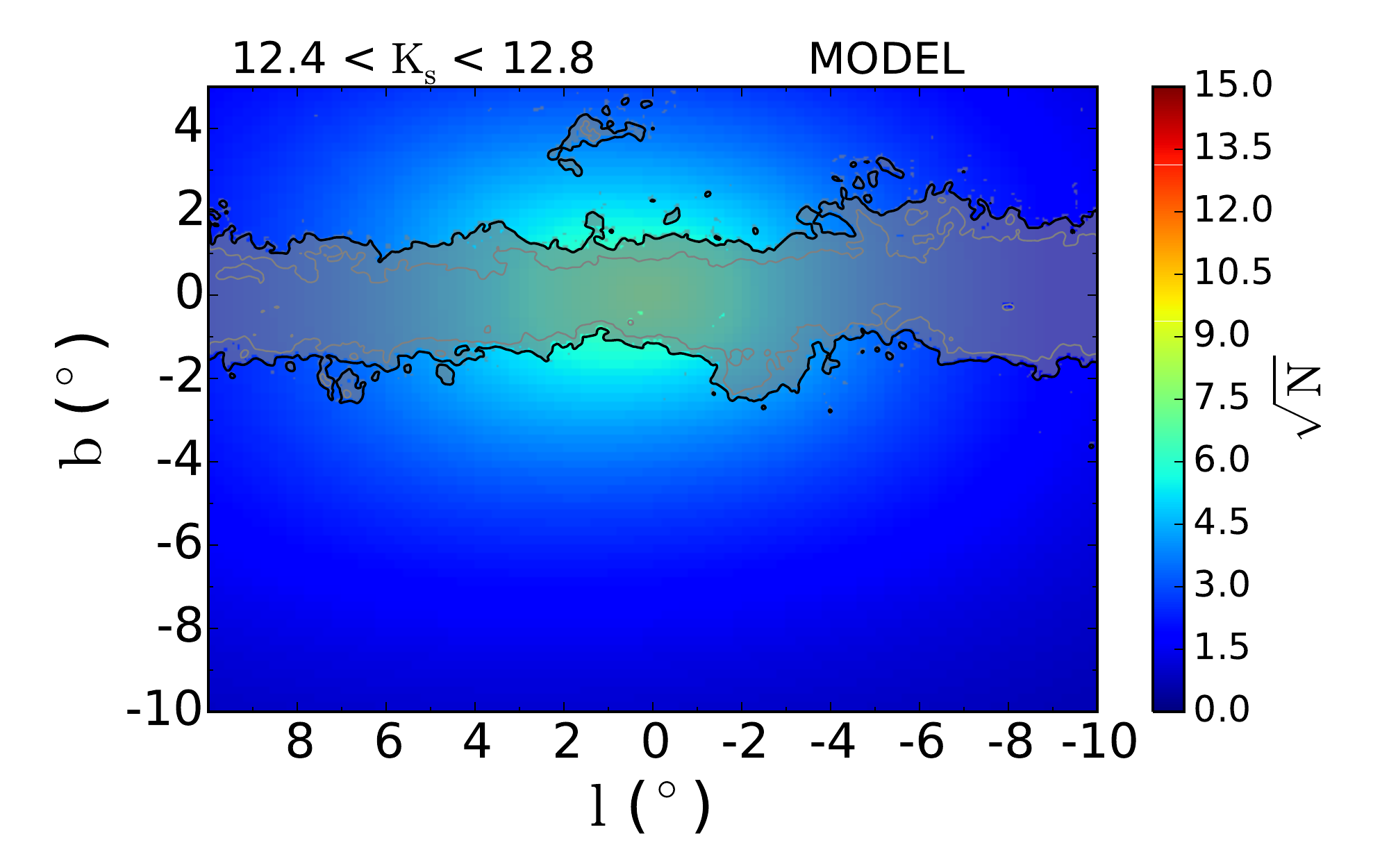}
\hspace*{-0.39in}
\includegraphics[scale=0.21]{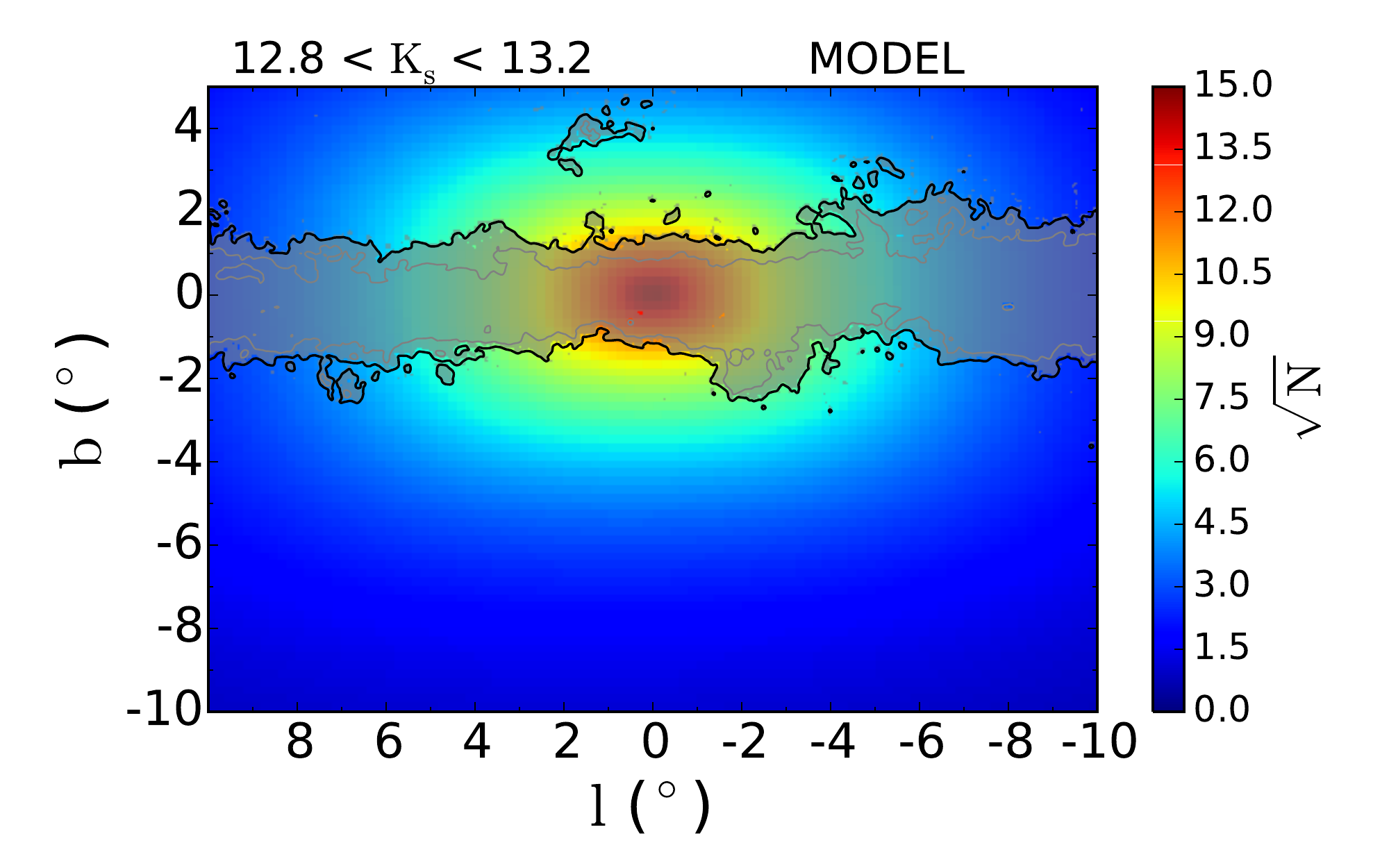}
\hspace*{-0.39in}
\includegraphics[scale=0.21]{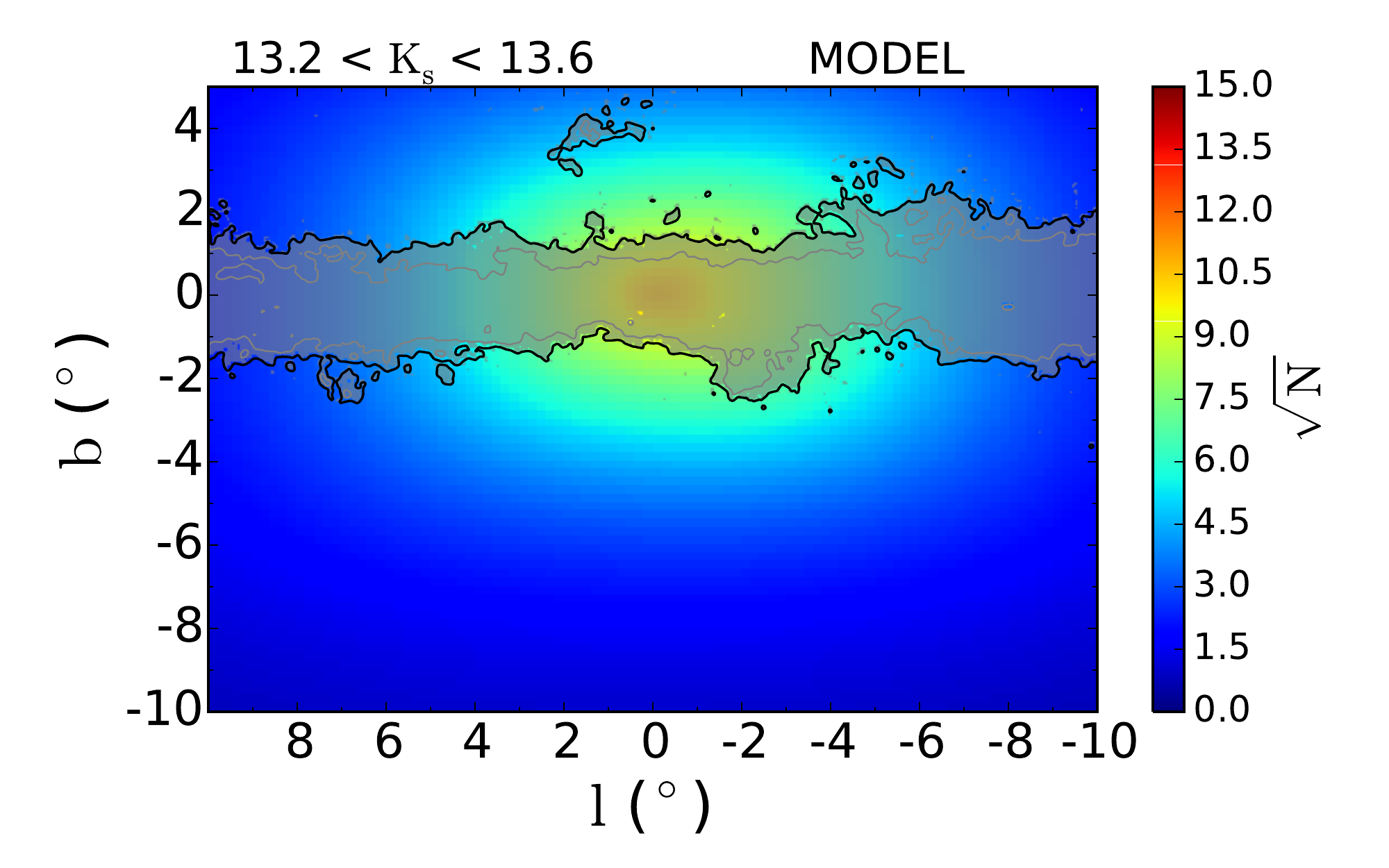}
\hspace*{-0.39in}
\includegraphics[scale=0.21]{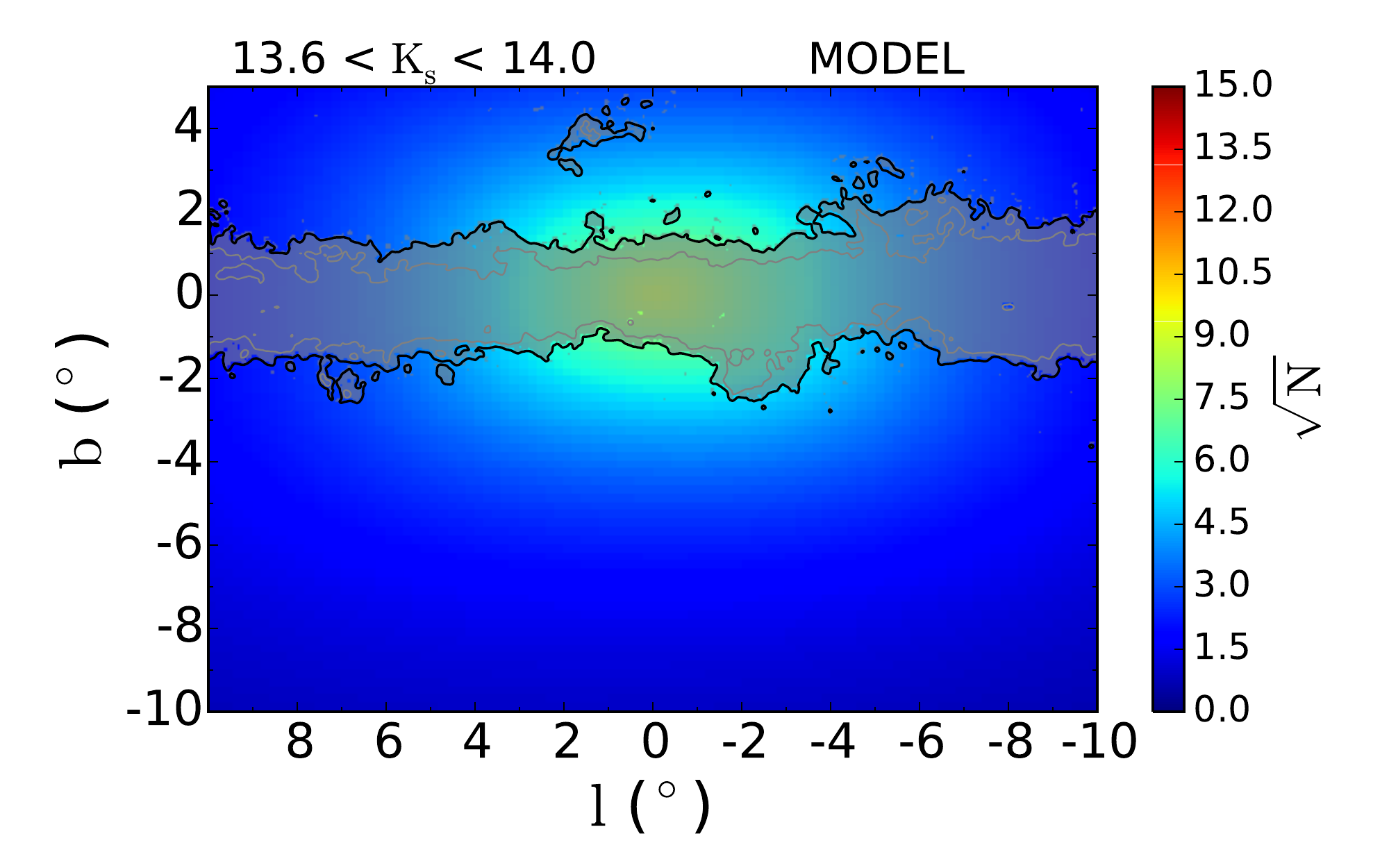}
\\
\hspace*{-0.19in}
\includegraphics[scale=0.21]{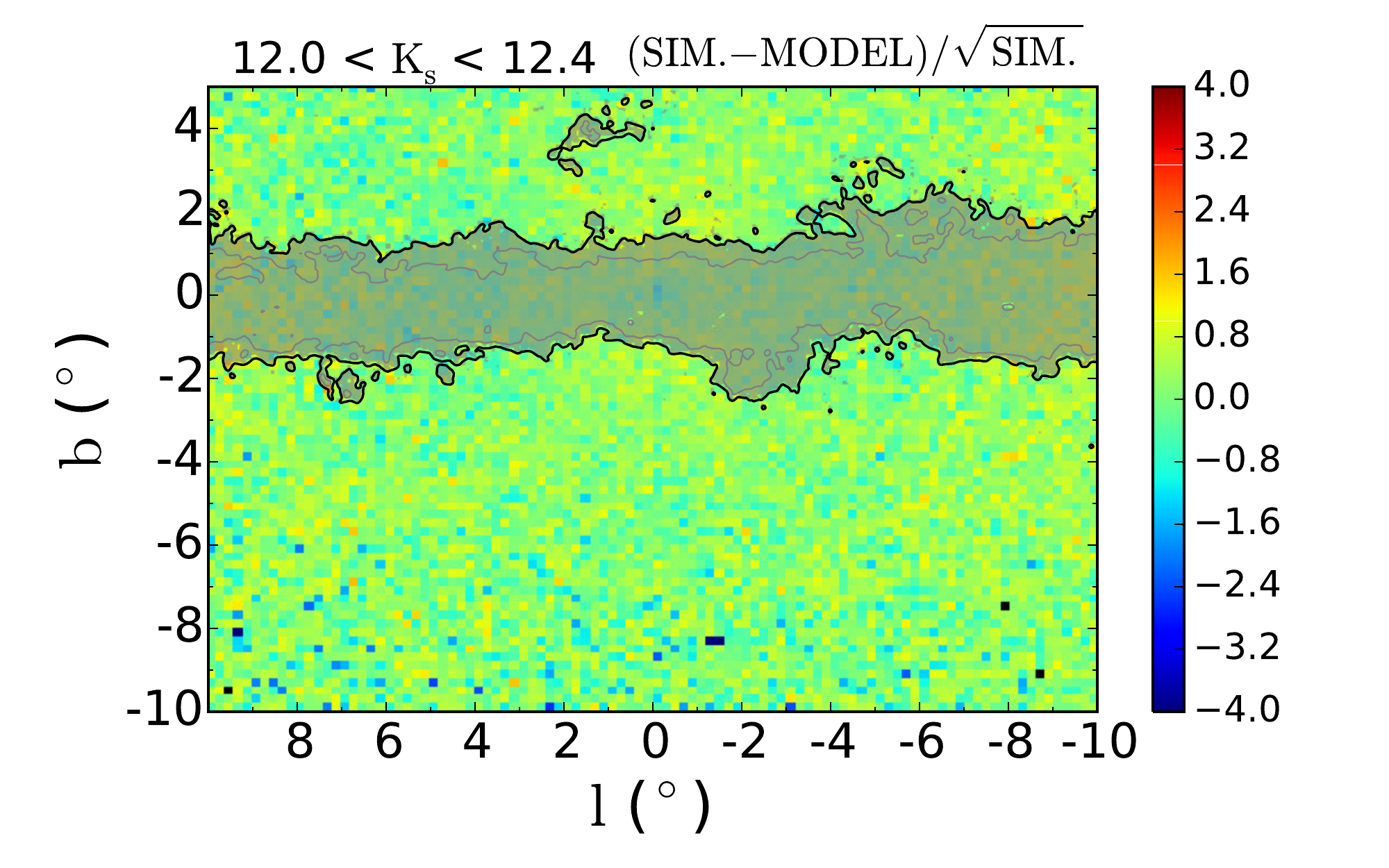} 
\hspace*{-0.39in}
\includegraphics[scale=0.21]{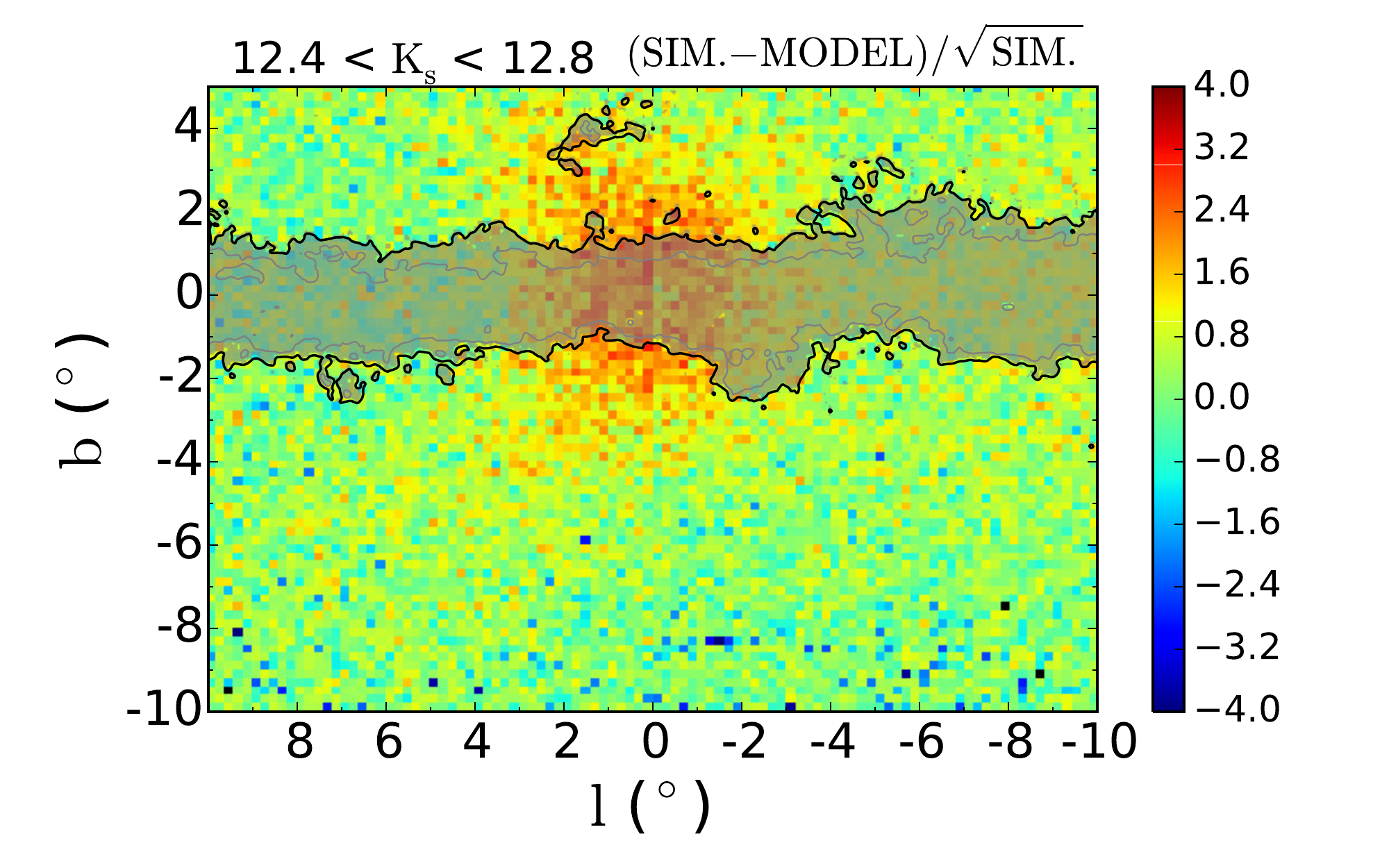} 
\hspace*{-0.39in}
\includegraphics[scale=0.21]{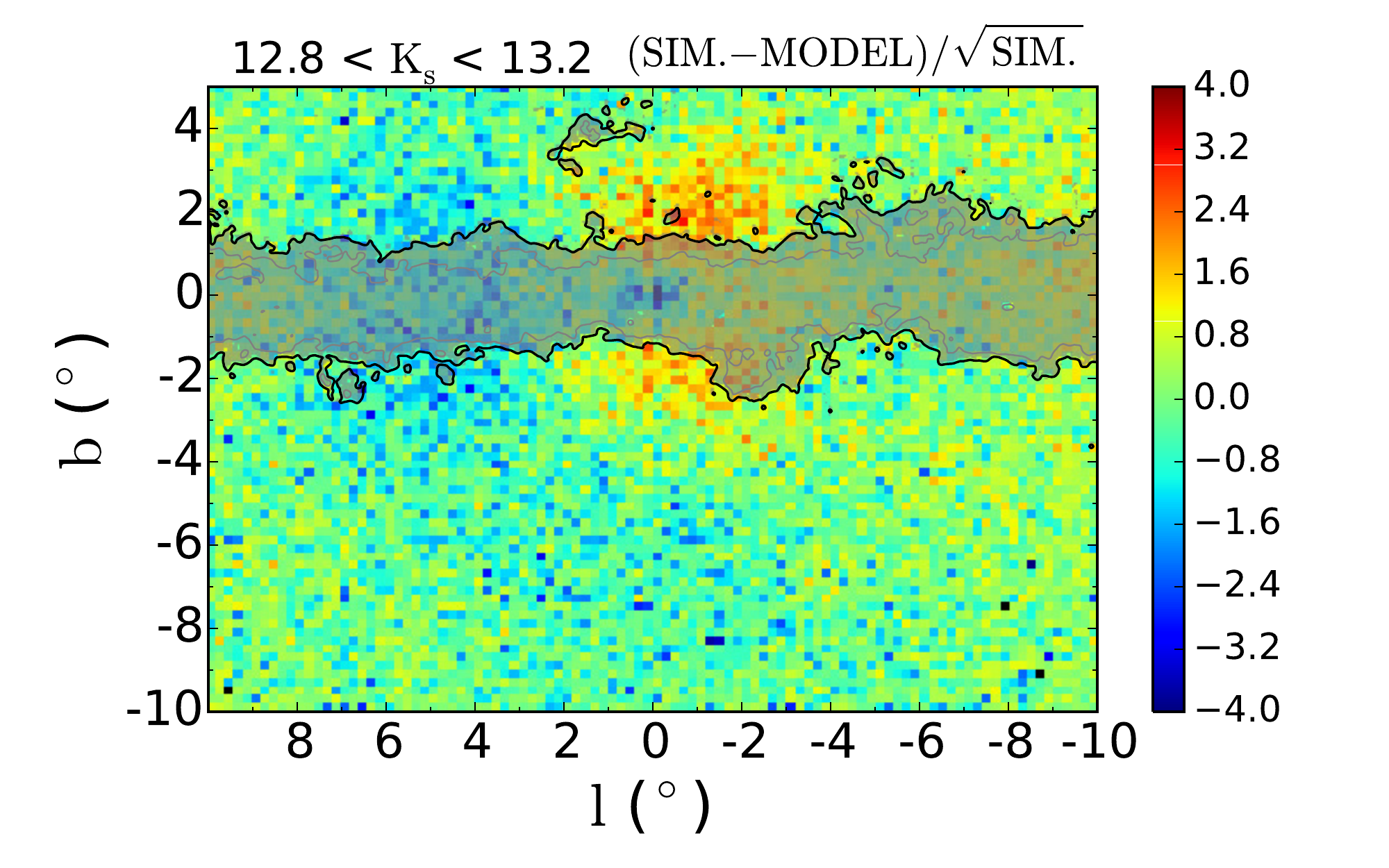} 
\hspace*{-0.39in}
\includegraphics[scale=0.21]{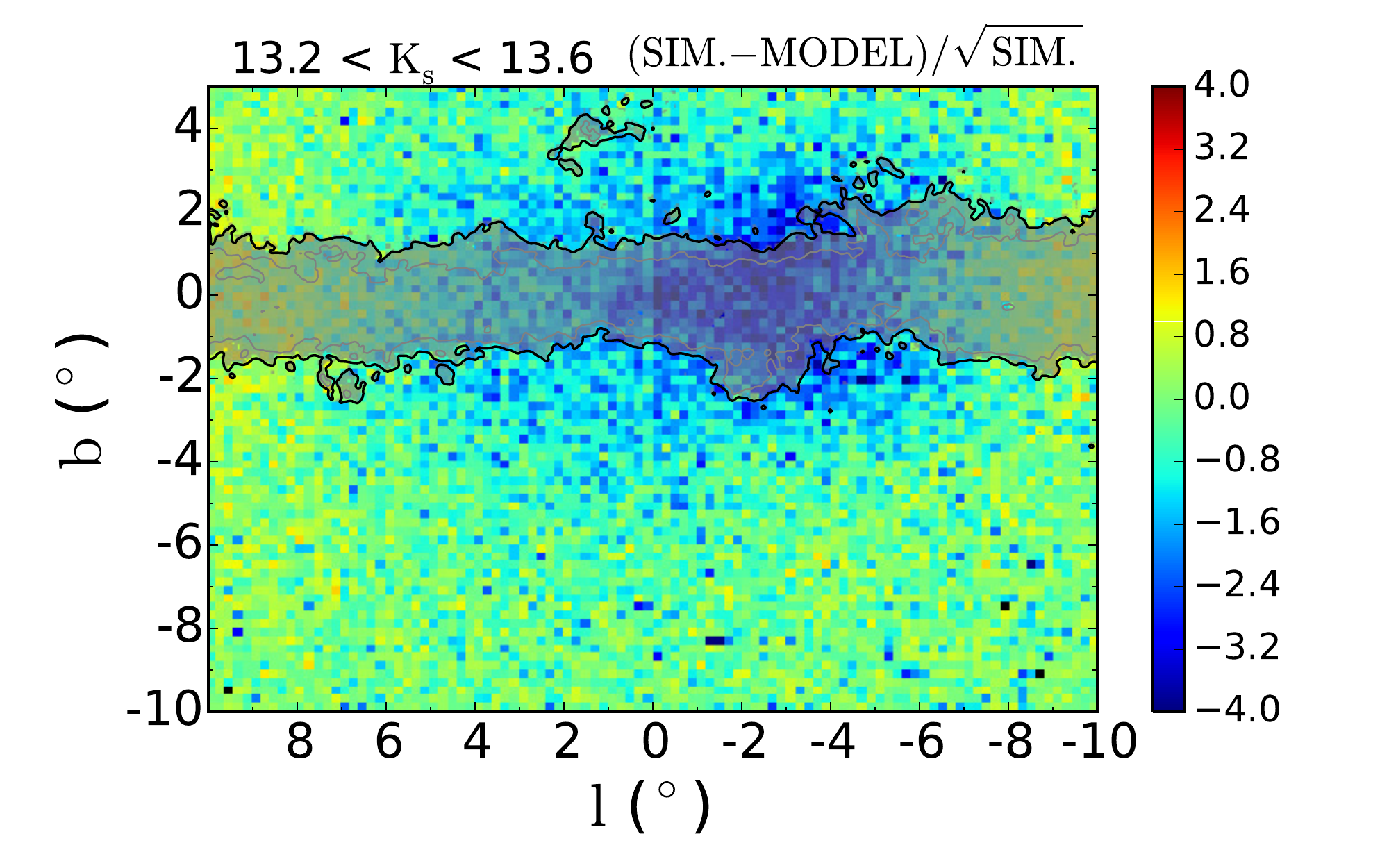} 
\hspace*{-0.39in}
\includegraphics[scale=0.21]{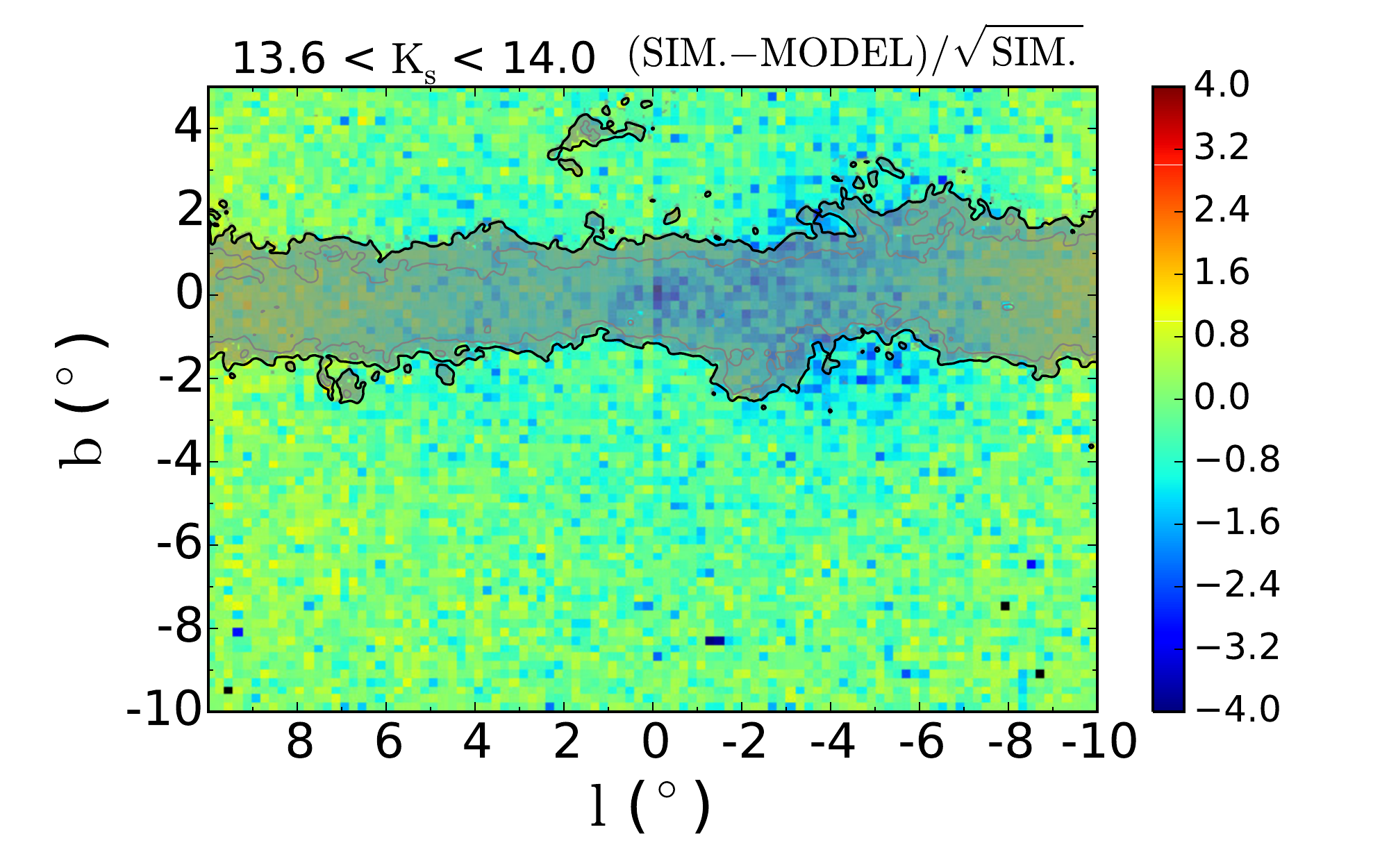} 
\caption{N-body simulation from \citet{Sh10} ($top$ $row$), model prediction ($middle$ $row$) and
  residuals between the simulation and the model ($bottom$ $row$). The size of the bin is 12' x 12', double the bin size used for the data, and the viewing angle of the model is 15 $^{\circ}$. The residuals in the brightest slice $12.0<$K$_{s}<12.4$ ($first$ $column$) do not show the central excess seen in the data (top row of Figure~\ref{diff}). The residuals in the second slice ($second$ $column$) $12.4<$K$_{s}<12.8$, show an overdensity in the simulation running diagonally from the centre towards the left, both at positive and negative latitudes; the same feature can be observed in the data residuals (second row of Figure \ref{diff}), but mainly at negative latitudes and positive longitudes. At fainter magnitudes, the residuals show the model overpredicts the number of counts on each side of the GC; however, in the last two slices, the overprediction is dominant at negative latitudes, near the GC, similarly to what is observed in the data (fourth row of Figure~\ref{diff}).
 } 
\label{simulation}
\end{figure*}

\subsection{The X-shape of the Bulge: testing the model on a Peanut 
Bulge N-body simulation}

In this work we have not implemented a parametric description of the X-shaped component which is
apparent in the RC population \citep[e.g.][]{Ne16}. To test if our model is capable of robustly recovering basic parameters such
as the orientation angle in realistic mock Bulges, we test its
performance on an N-body model \citep{Sh10} that has been shown to
predict the split Red Clump \citep{Li12}. \citet{Li12} found that 7\%
of the light in the whole boxy Bulge region resided in the X-shaped
structure. To use the same observables (the number
density distribution in different $K_{s}$ magnitude ranges) as in the VVV, we
convolve the N-body model with the Bulge LF,
$\phi_{B}$.

The N-body model has $\sim$1 million particles but within the VVV
footprint only $\sim$400,000 remain.  Because of the low number of
particles, we have performed two fits, one using 6$'$ $\times$ 6$'$ bins as
in the data, and one with double the bin size 12$'$ $\times$ 12$'$. The
simulation, the best-fit model and the residuals are shown in Figure
\ref{simulation}, for the larger bin size needed to satisfy our
assumption that the measurements are distributed according to a
Gaussian function. We also mask the high reddening regions to mimic
the data-fitting procedure, but exclude the discs as they do not
contribute significantly to the simulation. The model therefore,
contains only the analytic form of the Bulge. Using the larger bins
12$'$ $\times$ 12$'$ we find a viewing angle of 15 $^{\circ}$, on the
lower end of the expected value of $20_{-5}^{+5}$ degrees for the
simulation while with the smaller bins we find $\alpha =
20^{\circ}$. \citet{Na15} also test their model on this N-body
simulation, assuming four viewing angles  $\alpha = 0^{\circ},
15^{\circ}, 30^{\circ},45^{\circ}$ and find that $\alpha =
30^{\circ}, 45^{\circ}$ provide the best fit, with $\alpha =
15^{\circ}$ being nearly as consistent.

The fit in the central regions of the brightest slice
$12.0<$K$_{s}<12.4$ is good, suggesting that the overdensity 
observed in the data does not belong to a structure specific to the peanut
Bulge which we could not fit due to our simple assumption of a
triaxial Bulge. In the second slice, $12.4<$K$_{s}<12.8$, there is an
overdensity in the simulation running diagonally from the centre
towards the left, both at positive and negative latitudes; the same
feature can be observed in the data residuals (see Figure \ref{diff}),
but mainly at negative latitudes and positive longitudes (see alse the
residuals for the WISE survey, in figure 3, \citealt{Ne16}). Due to
the orientation of the Bulge, the arms of the X-shape appear asymmetric,
with the arms at positive latitudes more visible in the residuals
while the arms further from the Sun, behind the Galactic centre,
almost invisible. This is only a projection effect caused by the
orientation of the Bulge as discussed in the previous Section. At fainter
magnitudes the model overpredicts the number of counts at intermediate
longitudes on each side of the Galactic centre; however in the faintest
two slices the underdensity is observed mainly at negative latitudes,
$-2^{\circ}< l < -6^{\circ}$ and $-3^{\circ}< b < 3^{\circ}$, in the same 
region as in the data but slightly closer to the Galactic centre. The overdensity in the model is created by the redistribution of the total number of stars present in
the peanut Bulge into a boxy shape which the model is constrained to
have.

Other tracers such as F dwarfs ($<$ 5 Gyrs old,
\citealt{Lo16}), RR Lyrae \citep[$>$ 10 Gyrs old, e.g.][]{Ku16,
  Gr16} and short period Mira variables \citep[old stars, see][]{Ca16,Lo2017} do not 
trace the X-shaped component, thus emphasizing that the mechanism and the 
time of formation of the Galactic Bulge are not 100\% clear.

\section{Conclusions}
\label{sec:conclusions}
We have used a parametric approach to find the best analytic function
that describes the 3D Bulge density observed by the VVV survey in the
NIR. We test our modeling methodology on a range of mock datasets,
including those generated using self-consistent simulations of the
Milky Way Galaxy. We thus prove that the technique delivers robust
results and take care to quantify random and systematic uncertainties
associated with the method. Our best-fit model produces a faithful
representation of the stellar density field in the inner Galaxy, with
reassuringly low levels of residuals across the entire VVV field of
view.

With find the Bulge/Bar to be ``boxy'', with an axis ratio of
[1:0.44:0.31] and an angle between the major axis and the Sun-Galactic
Centre line of sight of $\sim$20 degrees (top row, Table
\ref{bestresults}). We show that there exists a strong degeneracy
between the viewing angle and the dispersion of the RC absolute
magnitude distribution. To reproduce the stellar line-of-sight
distribution at a range of locations on the sky, models with larger
intrinsic RC dispersions prefer larger viewing angles. We
demonstrate that as a result of the degeneracy, assuming
$\sigma^{\mathrm{RC}}=0.18$ leads to inferring viewing angles of
$\sim25^{\circ}$ \citep[closer to the value reported in][$\alpha = ($-26.5$\pm 2)^{\circ}$]{We13}. However, in the
analysis presented here, we also found evidence for a much larger intrinsic
dispersion of the RC absolute magnitudes in the Bulge area, namely $\sigma_{\mathrm{free}}^{\mathrm{RC}}
\sim 0.26$, larger than the latest measurment in the solar neighbourhood ($\sigma^{\mathrm{RC}} = 0.17$ mag, \citealt{Ha17}) where the younger thin disc population is dominant. We envisage that with new next-generation Bulge surveys it should be possible to measure the true
$\sigma^{\mathrm{RC}}$ across the Bulge and thus, at last, settle the viewing angle argument.

The most pronounced residuals of the one-component Bulge model appear
to cluster tightly in the central regions of the Galaxy. One
interpretation of this residual density distribution is that the core
portions of the Bulge are better described by a combination of two
triaxial components: the main old one and a younger smaller one (last
two rows, Table \ref{bestresults}). Apart from this additional
small-scale ellipsoidal component indicated by the low-latitude counts
of the VVV RC giants, the only other noticeable outlier is the pattern
induced by the split RC. We choose not to augment our Bulge model with
an ingredient responsible for the cross-shaped pattern, but provide an
estimate of the total stellar mass locked in this population. We show that $\sim 7\%$ of the total Bulge/Bar could be residing in 
this non-ellipsoidal constituent, in agreement with some of the
numerical simulations of the Galactic Bar \citep[
  e.g.][]{Sh10}. However, this is probably only a lower estimate of the total fraction of stars contributing to the X-shape as it does not include the stars on orbits that could be fit within the triaxial component.

Finally, the mass of the Bulge/Bar is a key ingredient in the recipe
for the total density distribution of the Milky Way. The Bulge mass
has to satisfy microlensing constraints \citep[see e.g][]{Sumi2013}
and ought to affect the Dark Matter density distribution as the halo
reacts to the presence of baryons in the inner Galaxy \citep[see
  e.g.][]{CB2016}. In light
of the most recent measurements of the Bulge MF
\citep[e.g.][]{Calamida2015}, we argue for a Milky Way Bulge with a total
mass of $\sim 2.3 \times 10^{10} M_{\odot}$, some $50\%$ of which is
contributed by dark stellar remnants.

\section*{Acknowledgements}
This work relies on DR2 of the VVV survey taken with the VISTA telescope 
under the ESO Public Survey program ID 179.B-2002, pipeline processed, 
calibrated and provided by the Cambridge Astronomical Survey Unit.
The authors are grateful for the enlightening feedback from all Cambridge Streams enthusiasts and the anonymous referee.
The research leading to these results has received funding from the European Research Council under
the European Union’s Seventh Framework Programme (FP/2007-2013) through the Gaia Research for European
Astronomy Training (GREAT-ITN) Marie Curie Network Grant Agreement n.
264895 and the ERC Grant Agreement n. 308024. SK thanks the United Kingdom Science and Technology Council (STFC) for the award of Ernest Rutherford fellowship (grant number ST/N004493/1). JS  acknowledges  support from the Newton  Advanced  Fellowship awarded by the Royal Society and the Newton Fund. ZYL is sponsored by Shanghai Yangfan Research Grant (no. 14YF1407700). 

\bibliographystyle{mn2e}
\bibliography{bibl} 
\appendix
\section{Completeness of the VVV fields as a function of latitude}
 \begin{figure}
\includegraphics[scale =0.11]{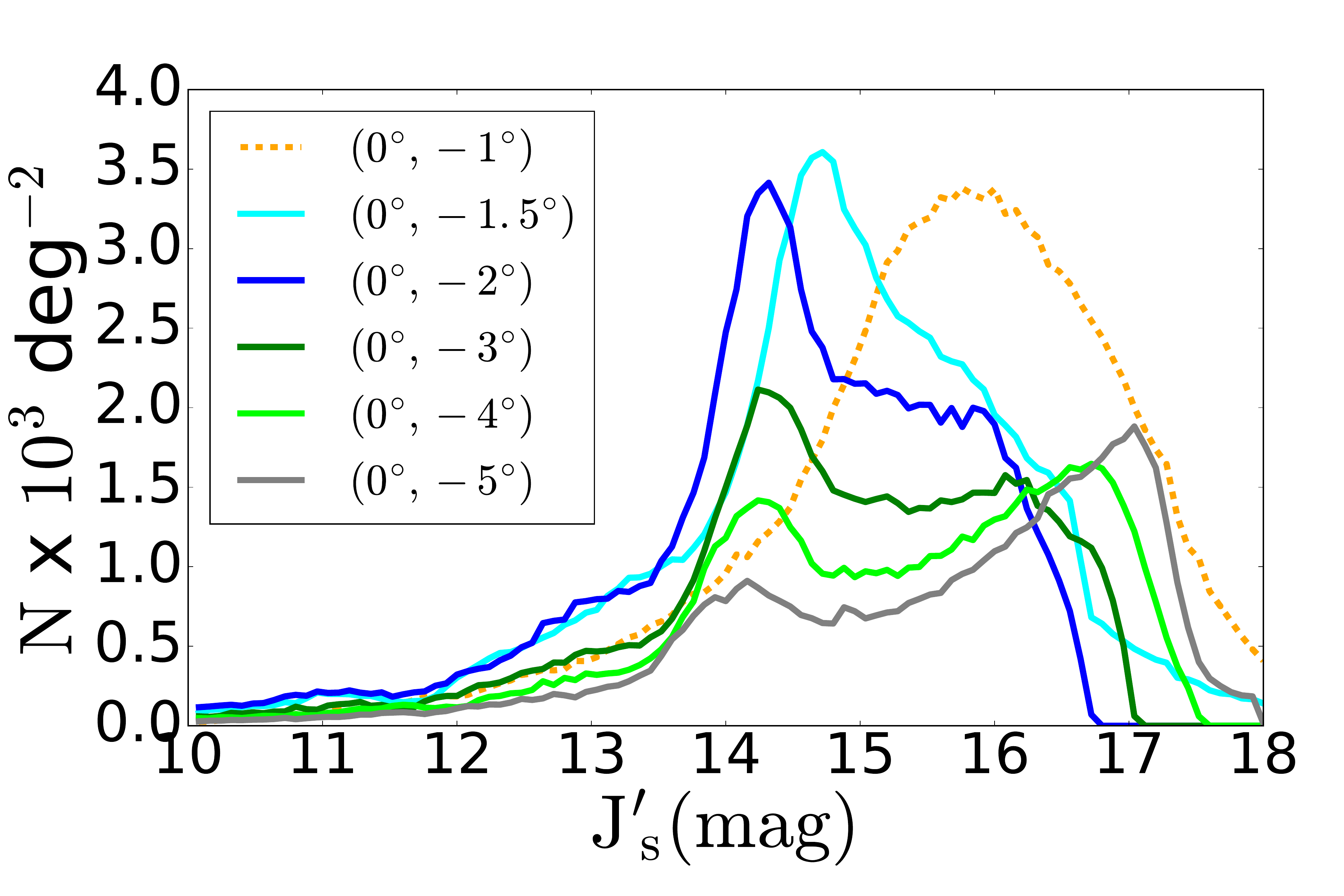}
\includegraphics[scale =0.11]{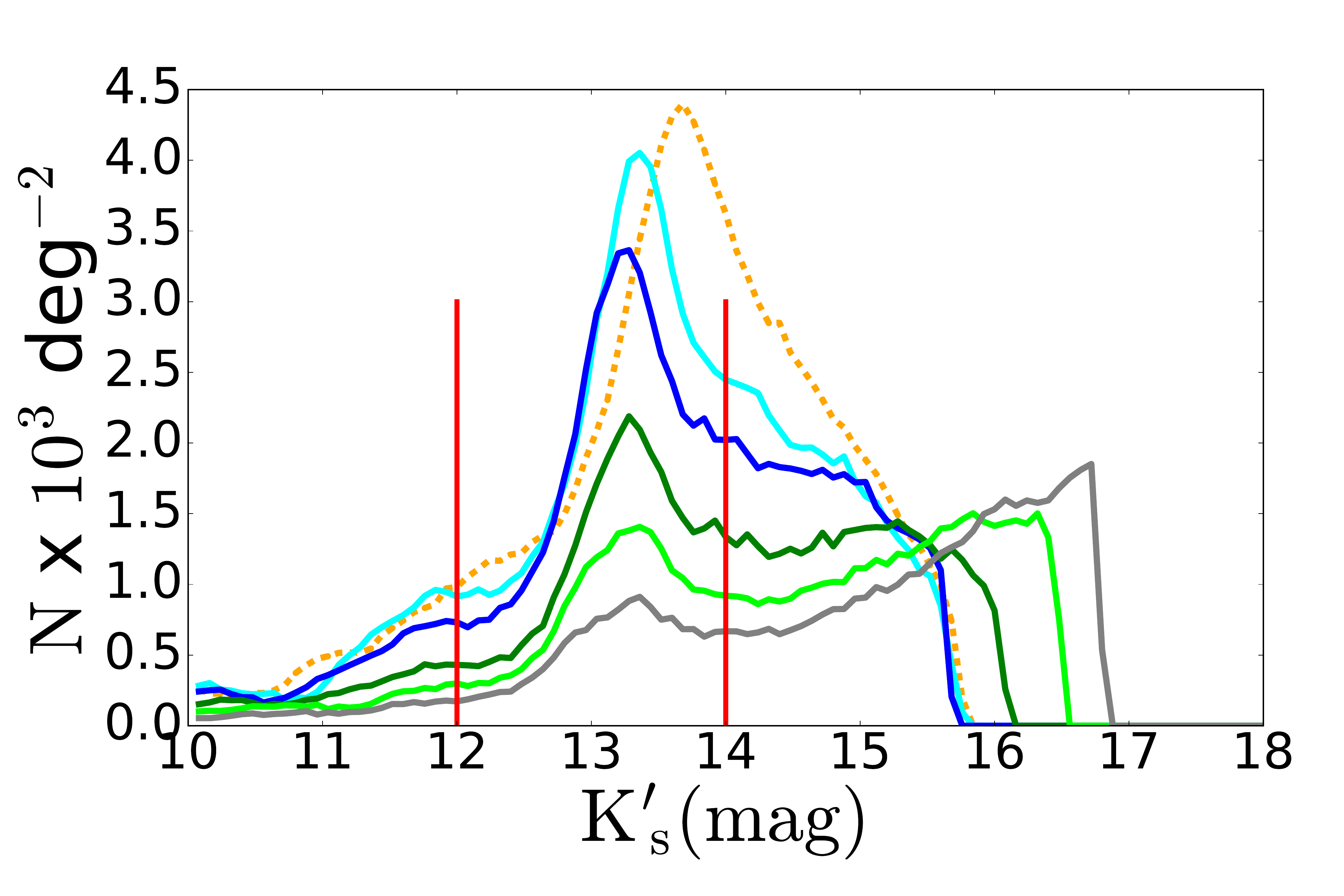} \\
\hspace{-0.5cm}
\includegraphics[scale =0.11]{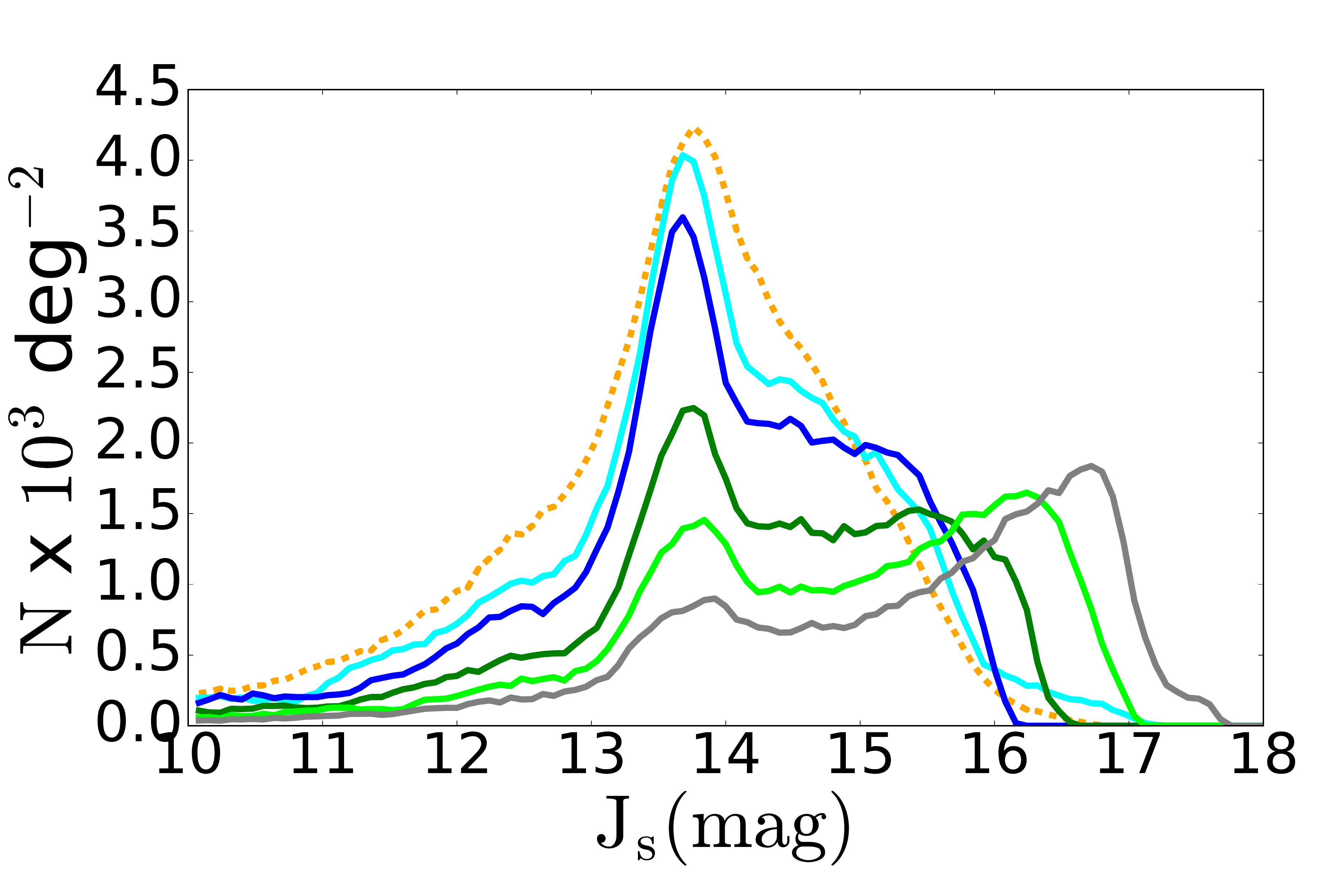}
\includegraphics[scale =0.11]{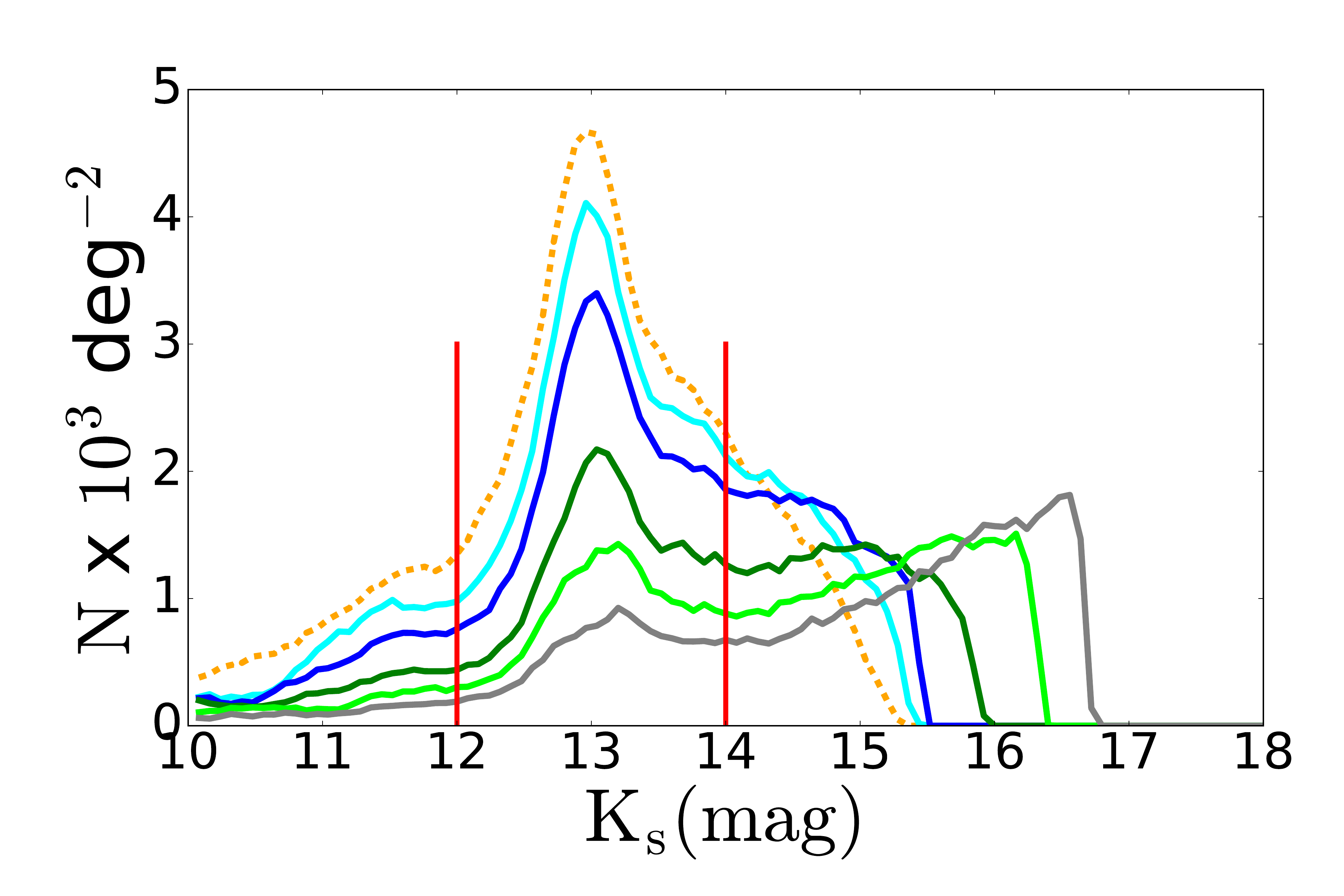} 
\caption[Completeness of the VVV survey]{$Top$ $panels$: observed magnitude
distributions for the $J'$ and $K'$ bands at several latitudes (marked with different colours) below the
Galactic Plane and $l=0^{\circ}$, in $1.0^{\circ} \times 0.4^{\circ}$ fields. $Bottom$ $panels$: same as above, but for the extinction corrected magnitudes
$J$ and $K$. The fields with $b < -3^{\circ}$ are complete up to
$K'_{s} \sim 15.2$ mag and $K_{s} \sim 15$ mag at least, making our working
magnitude range $12< K_{s}<14$ a conservative cut.}
\label{completeness}
\end{figure} 
\newpage
\section{MCMC results}
Figures on the next page.
\begin{figure*}
\centering
\vspace{-0.2cm}
\includegraphics[scale=0.16]{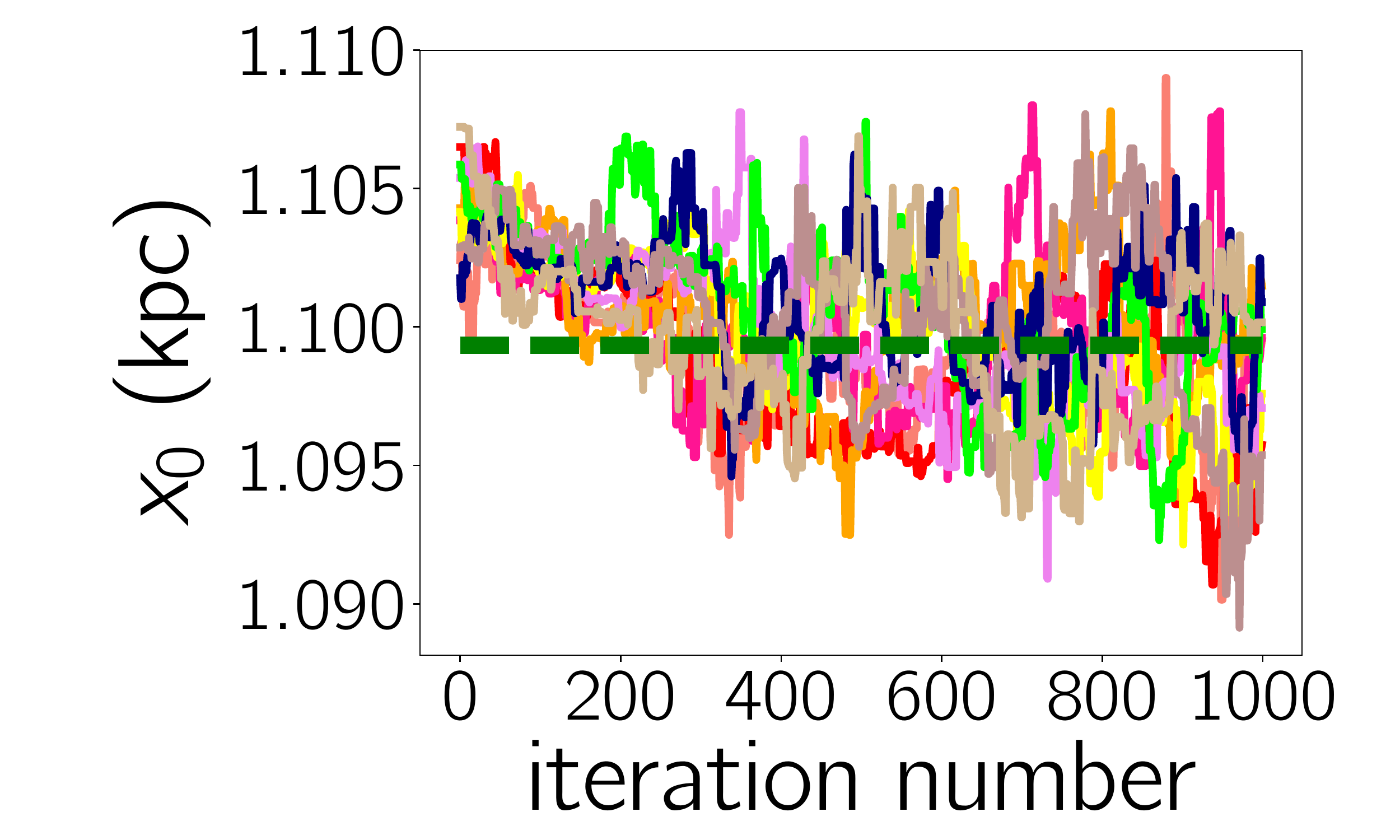}
\includegraphics[scale=0.16]{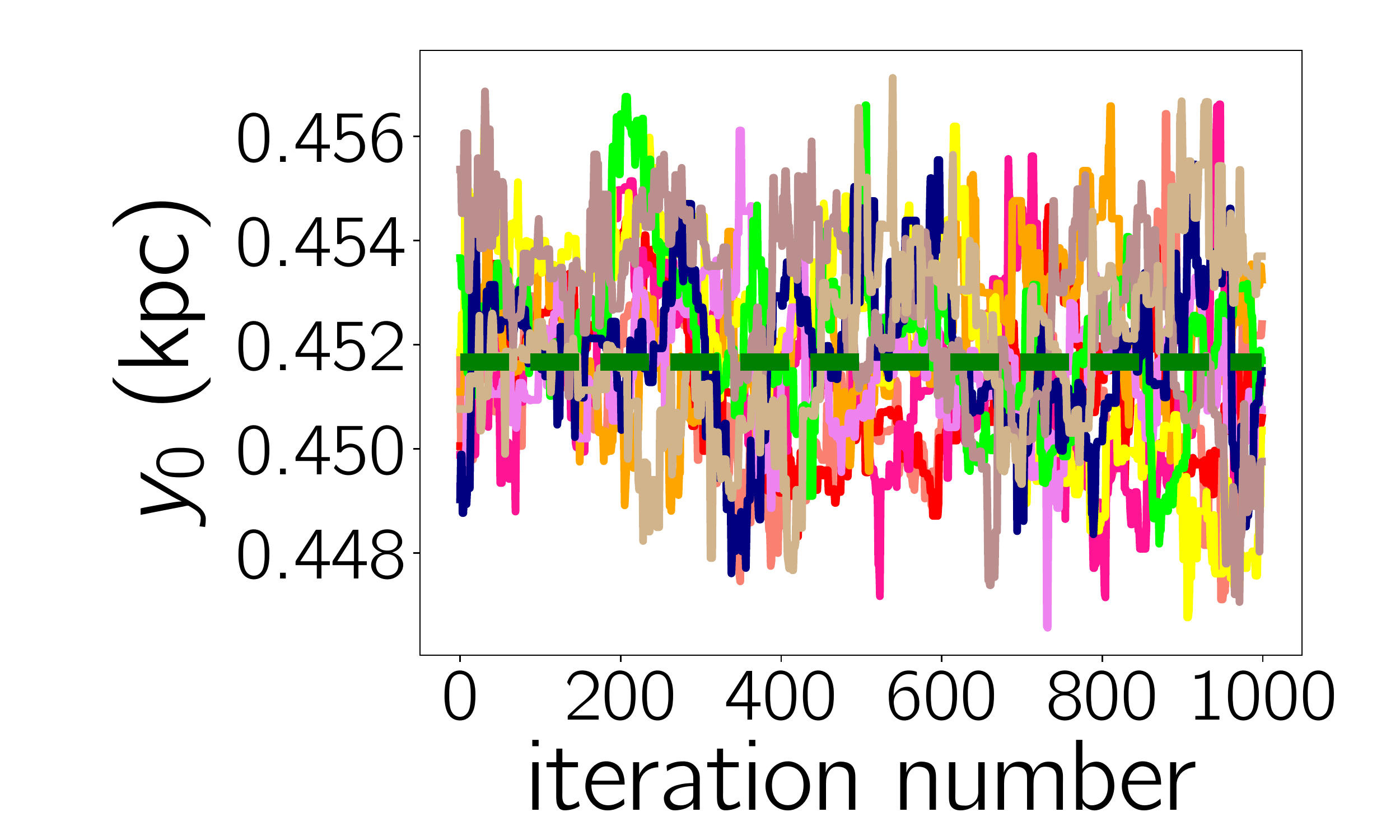}
\includegraphics[scale=0.16]{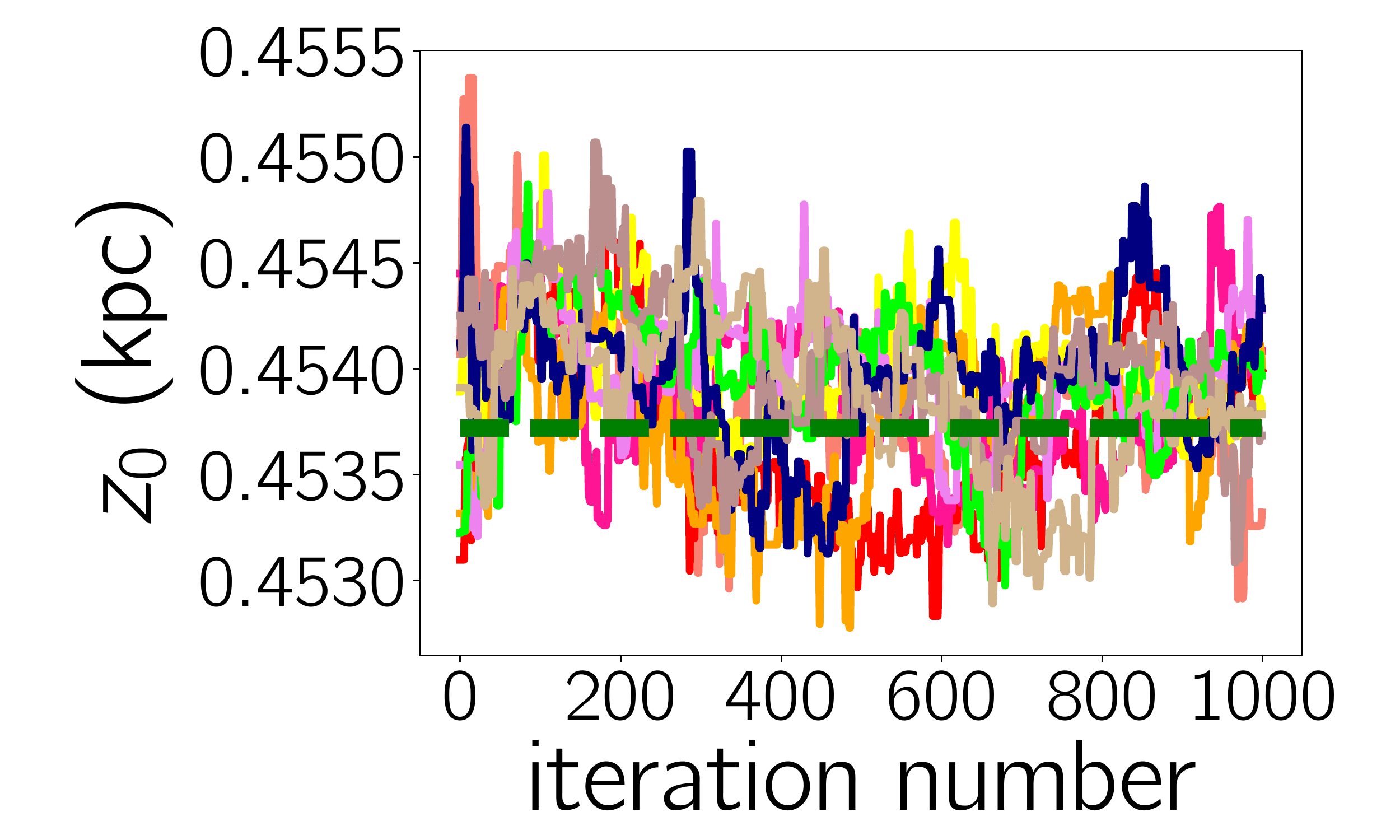}
\includegraphics[scale=0.16]{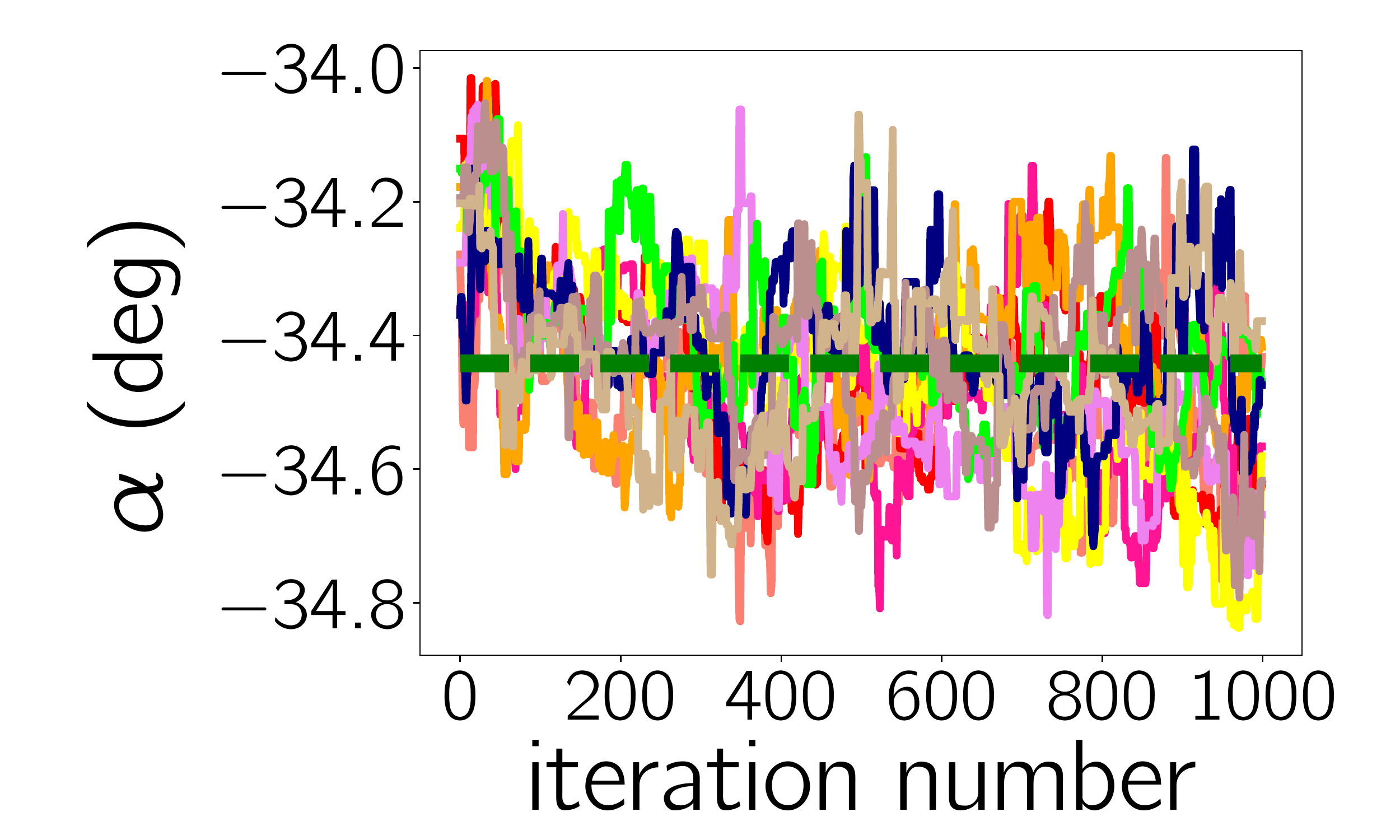}
\includegraphics[scale=0.16]{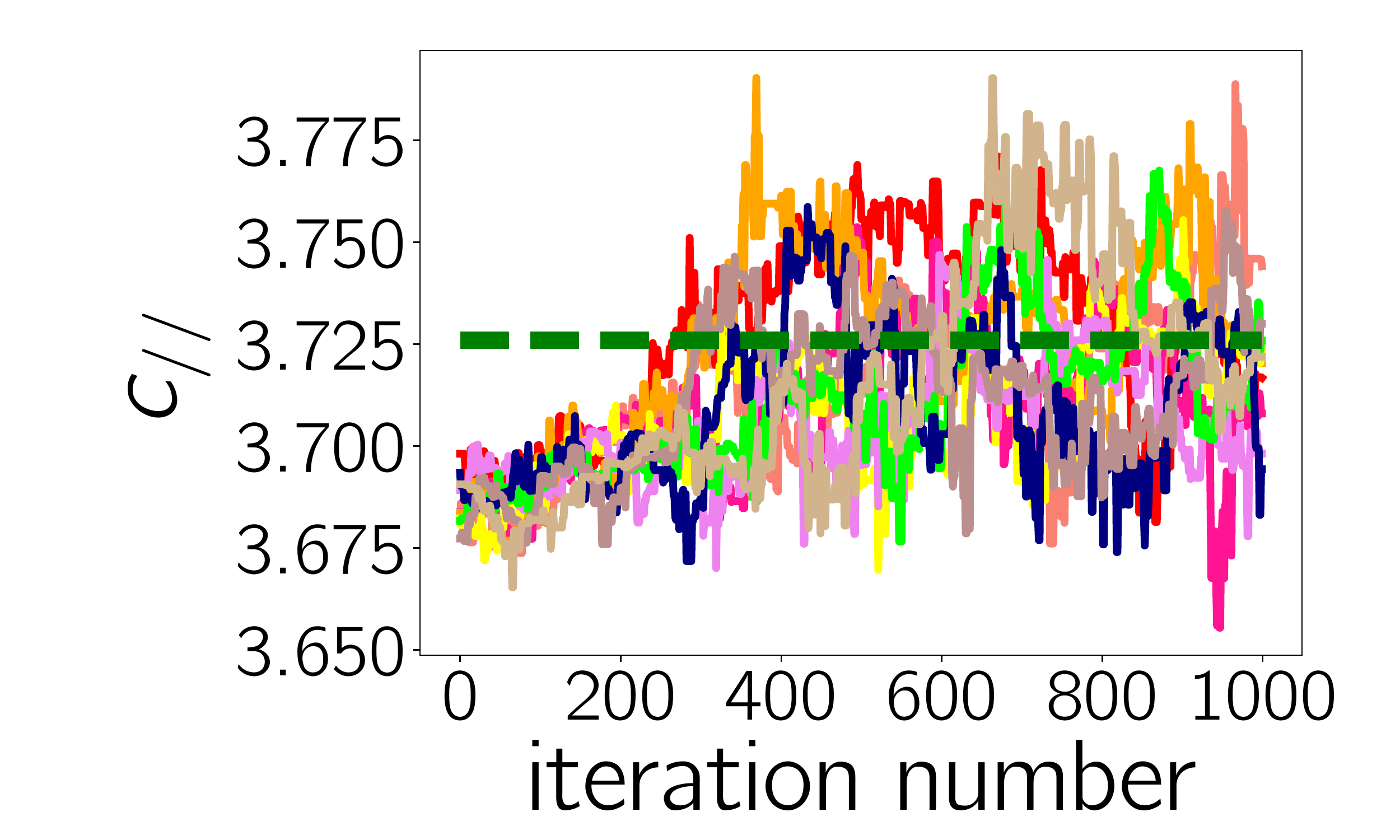}
\includegraphics[scale=0.16]{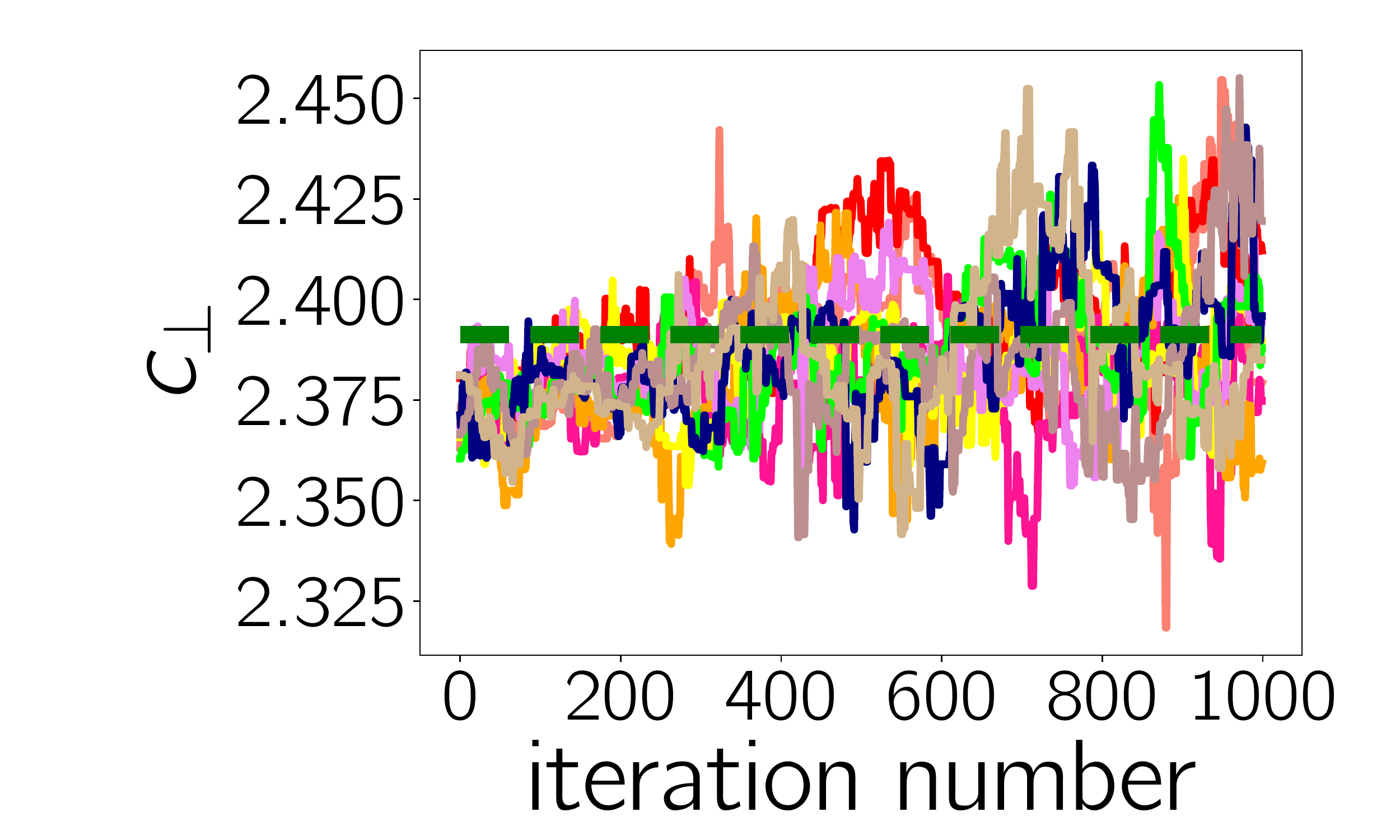}
\includegraphics[scale=0.16]{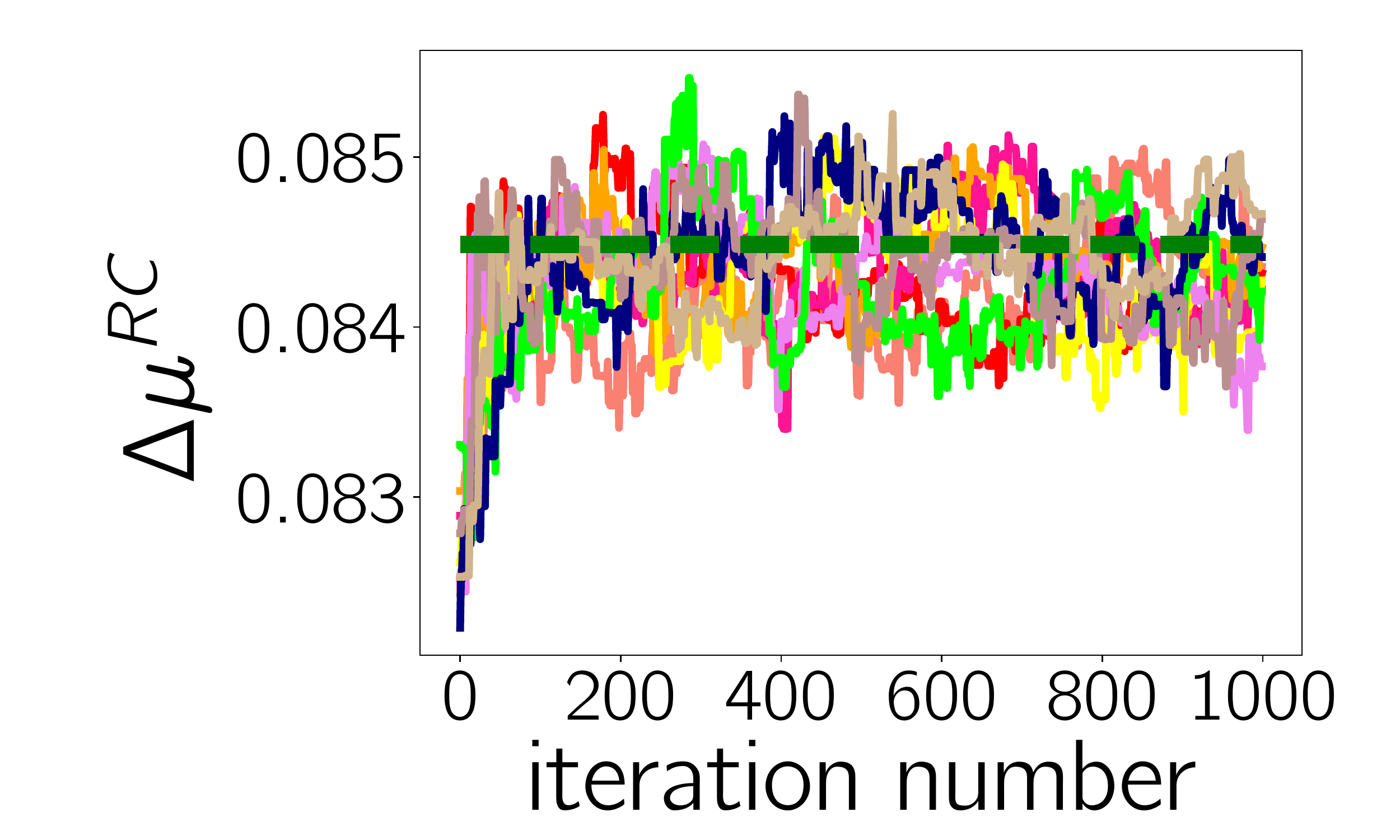}
\includegraphics[scale=0.16]{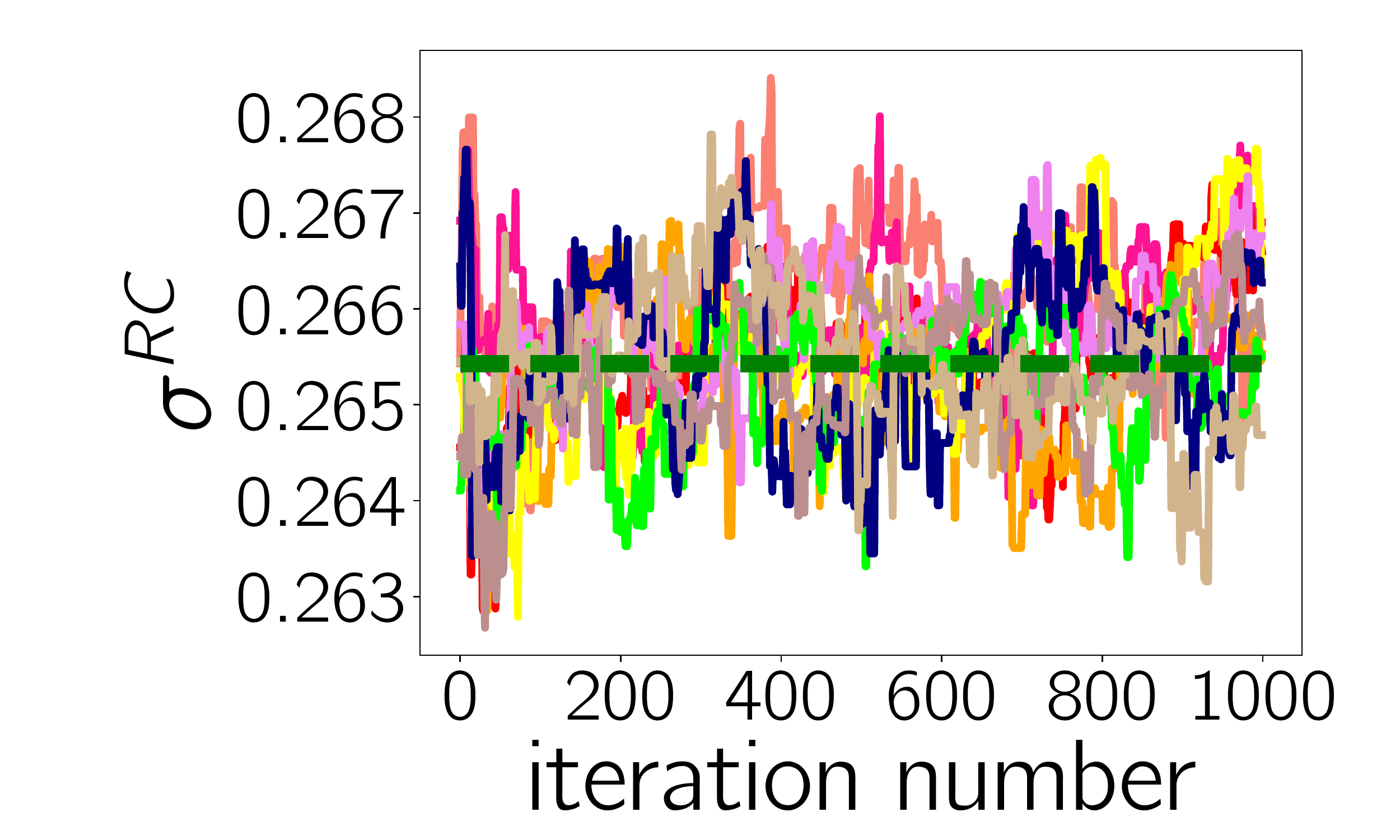}
\caption[Parameter values for model $S$ as a function of iteration number
for 10 walkers]{ Parameter values for model $S$ as a function of
iteration number for 10 walkers (each in a different colour) in the chain. All the walkers
converge within the first 100 accepted steps to a value close to the
starting point (green dotted line) chosen to be the result of the BFGS method
(see Table \ref{bestresults}, label `$\sigma^{\mathrm{RC}}$ free').}
\label{MCMC}
\end{figure*}

\begin{figure*}
\vspace{-0.0cm}

\includegraphics[scale=0.28]{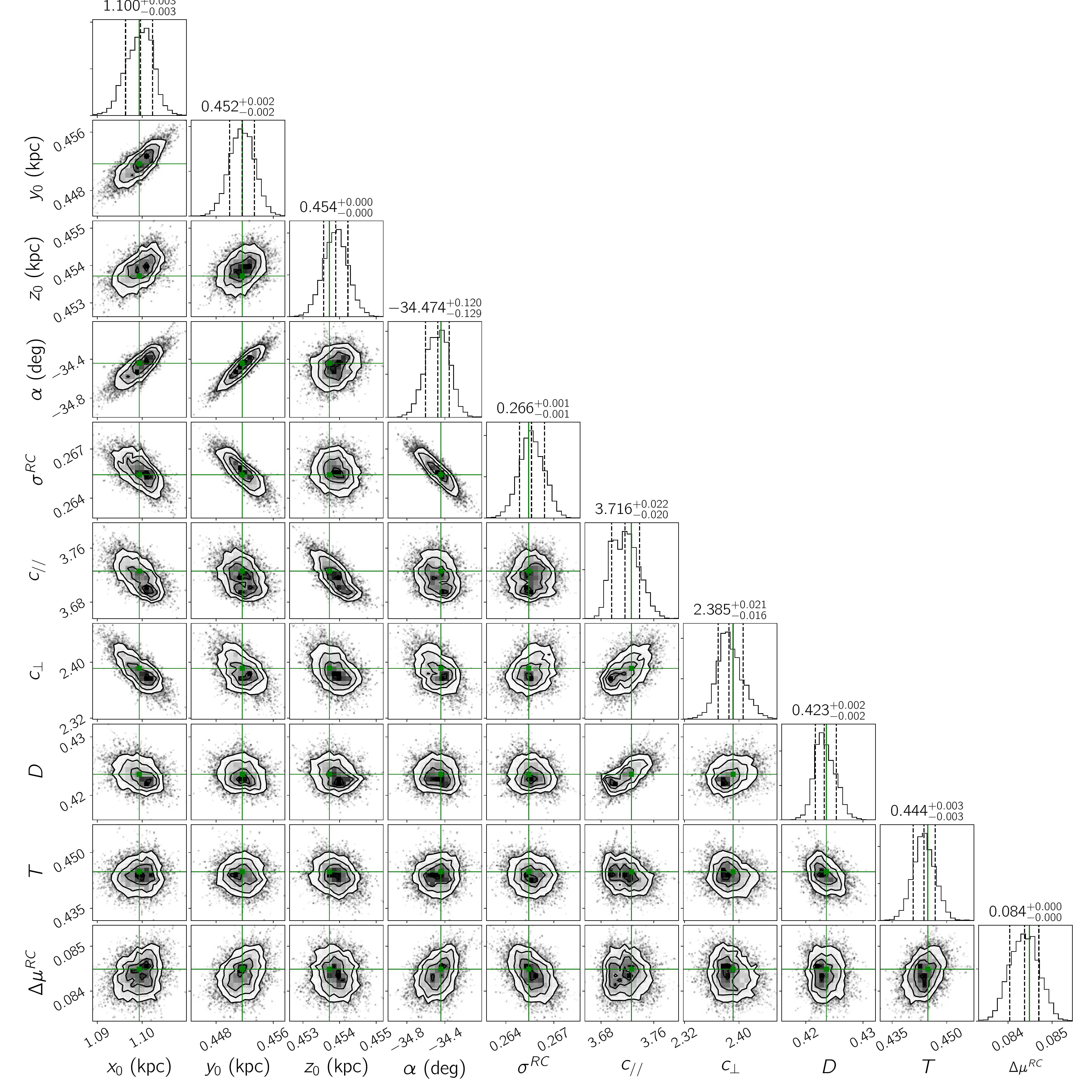} 

\caption[Two dimensional projections of the posterior probability distributions of the parameters of $model$ $S$]{One and two dimensional projections of the posterior probability distributions of the parameters of model $S$ (label `$\sigma^{\mathrm{RC}}$ free' in Table \ref{bestresults}), demonstrating the covariances between parameters. Along the diagonal, the 1-D marginalised distribution for each parameter is shown; on top of the distributions we give the 0.5 quantile values with the upper and lower errors (16\% and 84\% quantiles). For the initial state of the Markov chain we chose the best fit results reported in Table \ref{bestresults} (green dot). With 30,000 iterations and an acceptance fraction of 43\%, the final results and errors on the parameters are in complete agreement with the values found using the BFGS method.}
\label{cov}
\end{figure*} 
\end{document}